\documentclass[fleqn,usenatbib]{mnras}

\usepackage{newtxtext,newtxmath}
\usepackage[T1]{fontenc}

\usepackage{graphicx}
\usepackage{float}
\usepackage{longtable}
\usepackage{wrapfig}
\usepackage{amsmath}

\usepackage{bm}
\usepackage[font=small,labelfont=bf]{caption}
\usepackage{citesort}
\usepackage{threeparttable}
\usepackage{comment}
\usepackage[dvipsnames]{xcolor}
\usepackage{pdflscape}
\usepackage{adjustbox}
\usepackage{multicol}

\usepackage{xr}
\newcommand{\HI}{H\,{\sc i}}

\newcommand{\CI}{C\,{\sc i}}
\newcommand{\CII}{C\,{\sc ii}}
\newcommand{\ZnII}{Zn\,{\sc ii}}
\newcommand{\SII}{S\,{\sc ii}}
\newcommand{\PII}{P\,{\sc ii}}

\newcommand{\OI}{O\,{\sc i}}
\newcommand{\Meudon}{{\sc Meudon PDR}}

\newcommand{\SB}[1]{{\color{ForestGreen}SB: #1}}
\newcommand{\DK}[1]{{\color[rgb]{0.19,0.55,0.91}DK: #1}}

\title[\CI\ and H$_2$ in Magellanic Clouds]{Cold diffuse interstellar medium of Magellanic Clouds: II. Physical conditions from excitation of \CI\ and H$_2$}
\author[D. N.~Kosenko et al.]{D. N.~Kosenko$^1$\thanks{E-mail: kosenkodn@yandex.ru},
S. A.~Balashev$^1$\thanks{E-mail: s.balashev@gmail.com} ,
V.V. Klimenko$^{1,2}$
\\
$^1$Ioffe Institute, 26 Politeknicheskaya st., St.\ Petersburg, 194021, Russia\\
$^2$Department of Physics \& Astronomy, University of South Carolina, Columbia, SC 29208, USA
}

\date{Accepted XXX. Received YYY; in original form ZZZ}

\pubyear{2023}

\begin{document}

\label{firstpage}
\pagerange{\pageref{firstpage}--\pageref{lastpage}}
\maketitle

\begin{abstract}
%Recently we have revisited FUSE archival data obtained for the Large and Small Magellanic Clouds to get H$_2$ and HD column densities. For some of the systems we also found Hubble Space Telescope archival data.

We present a comprehensive study of the excitation of \CI\ fine-structure levels along 57 sight lines in the Large and Small Magellanic Clouds. The sight lines were selected by the detection of H$_2$ in FUSE spectra. Using archival HST/COS and HST/STIS spectra we detected absorption of \CI\ fine-structure levels and measured their populations for 29 and 22 sight lines in the LMC and SMC, respectively. The \CI\ column density ranges from $10^{13}$ to $10^{14}\,{\rm cm}^{-2}$ for the LMC and $10^{13}$ to $10^{15.4}\,{\rm cm}^{-2}$ for the SMC. We found excitation of \CI\ fine-structure levels in the LMC and SMC to be 2-3 times higher than typical values in local diffuse ISM. Comparing excitation of both \CI\ fine-structure levels and H$_2$ rotational levels with a grid of PDR Meudon models we find that neutral cold gas in the LMC and SMC is illuminated by stronger UV field than in local ISM ($\chi=5^{+7}_{-3}$ units of Mathis field for the LMC and $2^{+4}_{-1}$ for the SMC) and has on average higher thermal pressure ($\log p/k =4.2\pm0.4$ and $4.3\pm0.5$, respectively). Magellanic Clouds sight lines likely probe region near star-formation sites, which also affects the thermal state and \CI/H$_2$ relative abundances. At the same time such high measurements of UV field are consistent with some values obtained at high redshifts. Together with low metallicities this make Magellanic Clouds to be an interesting test case to study of the central parts of high redshift galaxies. 

%We present a comprehensive study of the excitation of \CI\ fine-structure levels in the Large and Small Magellanic Clouds. We used archival HST spectra to analyse line profiles of \CI\ fine-structure levels and derive their populations.  We also derived the metallicity from associated metal lines. Providing joint analysis of H$_2$ (obtained previously using FUSE spectra) and \CI\ we were able to estimate physical conditions, namely number densities and UV field intensities (or cosmic ray ionization rate assumed to be coscaled within used model) in the Magellanic Clouds. We found that average UV field intensities are 0.7 and 0.3 with standard deviation 0.4 and 0.5 in LMC and SMC, respectively. These values are higher than what is obtained near solar vicinity in our Galaxy, that is in line with previous studies. This may indicate that Magellanic Clouds sightlines probe region near star-formation sites, which is also affects the thermal state and \CI/H$_2$ relative abundances. At the same time such high measurements of UV field are consistent with some values obtained for high redshifts. Together with low metallicities this make Magellanic Clouds to be an interesting test case to study of the central part of high redshift galaxies. \SB{To be fit in 250 word limit}.
\end{abstract}

\begin{keywords}
galaxies: ISM; ISM: atoms; molecules; kinematics and dynamics
\end{keywords}

\section{Introduction}

%Galaxy evolution is in the tight connection with the interstellar medium (ISM), since it is in the tight connections with ISM physical state. Comparison of the abundances and level populations of the elements with the models allows to estimate physical parameters in the medium where they reside.
Study of the interstellar medium (ISM) is an essential part of unraveling the processes of galaxy evolution, as gas in the ISM supplies star formation process which defines galaxy evolution landscape. In turn, ongoing star-formation affects the gas in the ISM by ultraviolet (UV) radiation, cosmic rays, shock waves, stellar outflows and other processes. All of these processes may change the physical state of the gas leading to the ionization of atoms, dissociation and formation of molecules, excitation of energy levels of elements. Therefore, observations of interstellar clouds by measurements of elemental abundances and ionization states, and excitation of energy levels of species %\SK{excitation} \DK{tautology with the previous line} %population of their energy levels 
may constrain physical conditions in the ISM and shed light on galaxy evolution process. %{medium where they reside}. 

H$_2$ molecule, being the most abundant molecule in the Universe, is one of the main tracers of the cold phase of diffuse neutral ISM ($T\sim 100$ K). In the typical ISM, the cold phase likely have relatively high densities, $n\gtrsim 100$ cm$^{-3}$, which also favours the production of neutral carbon (\CI) due to recombination of \CII. The latter, however, is still the dominant ionization state of carbon in diffuse ISM.
%In general, carbon exists in the ISM in the form of \CII, but in the regions where abundance of H$_2$ is high enough to be self-shielded from UV radiation, one can detect a fraction of neutral carbon. %and H$_2$ is turn to be partly self-shielded, one can detect a fraction of neutral carbon.
%Observationally, 
A tight connection between presence of H$_2$ and \CI\ in the diffuse ISM is well confirmed by spectroscopic observations of UV resonant absorption lines of these species towards bright background sources both in local ISM \citep[e.g.][]{Jenkins2001, Burgh2010} and at high redshifts \citep[e.g.][]{Srianand2005, Noterdaeme2018}.  %In the case of sufficient amount of H$_2$ in the medium, one can detect neutral carbon, \CI\, (as its ionization potential, 11.26 eV, is close to the H$_2$ dissociation energy and due to H$_2$ self-shielding the radiation is attenuated). \SB{This is fully not correct. I also have the same comment following your talk on the GAISH conference.}
 %H$_2$ and \CI\, in the diffuse ISM can be studied by UV spectroscopic analysis of resonant absorption lines towards bright background sources (for H$_2$ dipole transitions are forbidden and hence emission is weak). 

A joint analysis of H$_2$ rotational and \CI\, fine-structure levels population allows to constrain conditions in the diffuse medium, namely, kinetic temperature, $T_{\rm kin}$, intensity of the ultraviolet (UV) field, $\chi$, and total hydrogen gas density, $n_{\rm H}$. This method have been applied to several systems at high redshifts \citep{Balashev2017, Klimenko2020, Balashev2019, Balashev2020b, Kosenko2021} and in our Galaxy and Magellanic Clouds \citep{Klimenko2020}. The latter is interesting since the physical conditions, %\SK{which one?} \DK{added} 
in particularly UV field intensity and metallicity in Magellanic Clouds is known to be different from ones in our Galaxy. Hence the analysis of \CI\ and H$_2$ provides an important view on how metallicity may affect the thermal state of the diffuse ISM.

Large and Small Magellanic Clouds (LMC and SMC, respectively) are one the closest dwarf galaxies to the Milky Way (MW). Their average metallicities are smaller then in our Galaxy ($Z\sim 0.5Z_{\odot}$ for the LMC and $Z\sim 0.2Z_{\odot}$ for the SMC, \citealt{Russell1992}). Metallicity of the SMC is also comparable to the mean metallicity of the high-redshift systems \citep[e.g.][and references therein]{Krogager2020}. Since the LMC and SMC are close to the Milky Way (50 kpc and 62 kpc, respectively, \citealt{Pietrzynski2019, Graczyk2020}), there is a unique possibility to study the ISM of low-metallicity galaxies with numerous sight lines.
%Therefore, due to their proximity to the Milky Way (about 50 kpc and 62 kpc for LMC and SMC, respectively, \cite{Pietrzynski2019, Graczyk2020}) we obtain a possibility to study low-metallicicty galaxied towards numerous sightlines. 
In our accompanying paper (\citealt{Kosenko2023}, here and after Paper I) we have revisited archival data, obtained by Far Ultraviolet Spectroscopic Explorer (FUSE, \citealt{Moos2000, Sahnow2000}) space telescope in Magellanic Clouds \citep{Blair2009, Welty2012}. We focused on identification of line transition of HD molecules and independently reanalysed H$_2$ absorption systems in 48 and 46 sightlines towards stars in the LMC and SMC, respectively. We found HD in 24 systems in Magellanic Clouds, 19 out of which have not been reported before. 
Some of sightlines have archival observations with a high spectral resolution using Hubble Space Telescope (HST) Cosmic Origins Spectrograph (COS, \citealt{Green2012}) and Space Telescope Imaging Spectrograph (STIS, \citealt{Woodgate1998}), which cover the strongest \CI\ transitions. This provide an opportunity to systematically study the physical conditions in diffuse medium of the LMC and SMC by joint analysis \CI\ fine-structure and H$_2$ rotational levels populations, that is presented in this paper.

The paper is organized as follows. In Section~\ref{sect:Data} we describe a sample of data which was used in the work. Section~\ref{sect:analysis} presents the method that we used to analyse observations. The results of data analysis are summarised in Section~\ref{sect:results}. We present constraints on the physical condition in the observed systems in Section~\ref{sect:phys_properties}. Section~\ref{sect:discussion} is devoted to discussion of our results, before we summarize in Section~\ref{sect:summary}.

\section{Data}
\label{sect:Data}

H$_2$ and \CI\, absorption lines fall into UV part of electromagnetic spectrum, which are limited for observation due to atmospheric absorption. The progress of H$_2$ and \CI\, observations in the Milky Way and in nearby galaxies was certainly attributed to availability of the space UV telescopes. The best resolution and quality observations of H$_2$ lines wavelength range were performed by FUSE (the wavelength range 907-1187 \AA\, and nominal spectral resolution $R \sim 20000$), HST/COS ( 900 - 2150 \AA\, and $R\sim 20000$) and STIS (1150-3100 \AA, and  $R\sim 30000-110000$). We described in details the analysis of FUSE data in Paper I. In this paper we focus on the HST data, which cover the strongest \CI\ and metal transitions. Among 94 sight lines discussed in Paper I we found 29 sight lines in the LMC and 28 sight lines in the SMC that have archival HST observations\footnote{\url{https://mast.stsci.edu/search/ui/\#/hst}} and are well suited for analysis of \CI\ and metal absorption lines. Several \CI\ transitions are covered also by FUSE observations, but these lines are weak and usually strongly blended. Additionally, spectral resolution of HST/STIS is much better than for FUSE. We have HST/STIS data for 21 systems out of 29 in the LMC and 23 out of 28 in the SMC. For the other we used HST/COS data, which have similar nominal resolution as FUSE, but better quality of the spectra, e.g. it is less affected by wavelength calibration issues.
%for some sightlines
%\SB{What is the fraction?} \DK{HST provides better resolution then FUSE (for FUSE $R\lesssim 20000$, for COS $R \sim 20000$), removed "some sightlines" from the text} \SB{In case of COS it is not. The question is how many fraction of COS data in comparison with STIS?} \DK{only COS data we have for 9 systems out of 30 in LMC and 6 out of 29 in SMC; also note that resolution of FUSE data is quite strongly decreased in almost all cases, so resolution of COS spectra is pretty higher}

%For 30 sightlines in LMC and 29 sight lines in SMC among the selected H$_2$-bearing systems detected in FUSE we found HST\footnote{\url{https://mast.stsci.edu/search/ui/#/hst}} archival data available for neutral carbon and metal absorption lines study, which is necessary for physical conditions estimate. 
Although the data were partially published \citep{Roman_Duval2021, Welty2016, Tchernyshyov2015, Jenkins2017, Jenkins2021}, we also used unpublished data (to our knowledge). All data used in this paper were collected from HST observations under the programs ID 16092,  16820, 11692, 16365, 16272, 16094, 16100, 12581, 7437, 16230, 16431, 16907, 13070, 13122, 13522, 13969, 13931, 14437, 14842, 14855, 14935, 15366, 15385, 15536, 15774, 16325, 16467, 16534, 12796, 13635, 14874, 15451,  7437, 5444, 14909, 15629, 16101, 9116, 12501, 11625, 9412, 16373, 16099, 4110, 8566. 
%SB{This numbers are only for unpublished, or also includes published?} \DK{unpublished and partly published in the works not related to this work}

%\SB{May be it is better to provide the references of the programs under such this observations were done, or the references where the data was partially published.} \DK{In most of the papers I've seen with HST data, they provided reference on the archive; also it seems that not all of the data have been published. But I can add  references on the programs and on the published articles} \SB{Yes, we can added a sentence like (that will provide a direct reference to the data): While the data were partially published in .... $<$list of papers and programs$>$, we used additionally apparently unpublished data as well collected in HST archives under the programs: .... $<$list of programs$>$.} \DK{added}

\section{Analysis}
\label{sect:analysis}
In spectra towards the stars in the LMC and SMC we usually detect two groups of absorption lines: the first is related to gas in Milky Way disk and halo and shifted by $\sim0\rm\,km\,s^{-1}$ with respect to Local Standard of Rest (LSR) and the second is related to gas in the LMC and SMC and shifted by 180-330 km\,s$^{-1}$ and 90-180 km\,s$^{-1}$, respectively.
%In spectra of stars in Magellanic Clouds there are typically two groups of the components: one from Milky Way disk and halo (around $0\rm\,km\,s^{-1}$ in Local Standard of Rest) and another from LMC or SMC (from about 180 km\,s$^{-1}$ to 330 km\,s$^{-1}$ for LMC components and from 90 km\,s$^{-1}$ to 180 km\,s$^{-1}$ for SMC) %\SB{Check and correct. Mb provide the ranges} \DK{added ranges}. 
In several cases we detect absorptions that have intermediate velocity shift and are likely related to high velocity cloud in Milky-Way halo or/and gas stream between the MW and the LMC and SMC. %\SB{How many?} \DK{I found 1 clearly seen. Maybe there are few more, but they are blended with LMC/SMC components and it is not possible to say whether it is HVC.} 
In most of the cases each group can be resolved into a few sub-components, therefore we used the line profile fitting technique to model absorption lines. 

\subsection{Method}

We model line profiles with standard multicomponent Voigt profile\footnote{We used the python package spectro (\url{https://github.com/balashev/spectro}).}. We used Bayesian framework to estimate posterior distribution functions for the absorption system parameters: column densities, $N$ (measured throughout the paper in cm$^{-2}$), Doppler parameters, $b$, and redshifts, $z$. To sample posteriors we used affine-invariant Monte Carlo Markov Chain (MCMC) sampler \citep{Goodman2010}. 
The continuum was independently constructed for each considered fit (\CI, \ZnII\, etc) by B-spline interpolation of the neighboring regions to absorption lines, clear from the evident absorptions. 
%We note that in the case of the saturated H$_2$ lines, which show Lorentzian wings this is a bit ambiguous task, therefore we iteratively adjusted continuum in the case of outlying lines. \SB{This is not for this paper.}

To report point and interval estimates on the parameters we use maximum aposteriori probability and highest posterior density $68.3\%$ credible intervals, respectively. Upper limits on column densities (where necessary) were found using one-sided 3$\sigma$ (99.7$\%$) credible interval. We note that reported uncertainties are statistical ones, derived in particular model assumptions. Systematic uncertainties, arisen from continuum placement, particular choice of fitting spectral pixels, and component decomposition in some cases may dominate the statistical ones, therefore the latter should be taken with caution, especially in some systems, where the reported uncertainties are found to be relatively small (in comparison to other systems). From the other side, since we used MCMC sampler to constrain the posterior distribution, we were able to explore a multimodal structure of the likelihood function and hence in some cases our constraints on column density are quite wide, since it includes multimodal distribution (e.g. both non-saturated and saturated solutions) for the line profiles. 

Using HST spectra we fit \CI\ absorption lines from three fine-structure levels (denoted below as \CI, \CI$^{*}$ and \CI$^{**}$) with an assumption of tied redshifts and Doppler parameters between them. This is reasonable approximation, since \CI\ levels are mostly populated by collisions and hence typically quite homogeneously populated within the cloud, unless density and temperatures do not drastically change. Since we study \CI\ in highly saturated H$_2$ clouds, where molecular hydrogen is already self-shielded, the temperature variations are likely small.
%[Moreover since we measured \CI\ column densities towards sightlines with relatively high H$_2$ abundances, the measured populations of \CI\ describe a bulk of the cloud, where the molecular hydrogen is self-shielded and temperature variations are likely small.]

An important ingredient for determination of the physical conditions in the gas phase is metallicity. We measure metallicity in the systems using \ZnII\ or (if possible) \PII\ absorption lines, which are believed to be weakly depleted \citep{DeCia2018}. In other cases 
%[If there were no \ZnII\ nor \PII,] 
we use \SII\ absorption lines. The column density of \HI\ was taken from  \citealt{Welty2012, Roman_Duval2019}. The metallicity was calculated as, $Z \equiv [{\rm X/H}] \equiv \log\left(N({\rm X})/N({\rm H})\right) - \log\left({\rm X/H}\right)_{\odot}$, where X - is an element used to derive metallicity, and solar abundances, $\left({\rm X/H}\right)_{\odot}$ are taken from \citealt{Asplund2009}.
%[Using constrained column densities of  metals and $N_{\rm HI}$ obtained by \cite{Welty2012, Roman_Duval2019}, and the Solar values from \cite{Asplund2009} we estimated metallicity as, $Z \equiv [{\rm X/H}] \equiv \log\left({\rm X/H}\right) - \log\left({\rm X/H}\right)_{\odot}$.] 

\subsection{Details on some individual sightlines}
\label{sec:individual}
Below we provide some details regarding several interesting sightlines.

\subsubsection{Sk-67 5}

The star is located in diffuse H\,{\sc ii} region on the Western edge of the LMC.
%and the system have previously been studied by \cite{Andre2004}, who used data, obtained by FUSE, HST/STIS and VLT/UVES, towards Sk-67 5 was also found high velocity component (high velocity cloud, HVC) from Na\,{\sc ii} and Ca\,{\sc ii} \citep{Andre2004} and Fe\,{\sc ii} \citep{Lehner2009}.\cite{Andre2004} found high value of Cl\,{\sc i}/H$_2$ and CO and concluded the presence of dense molecular clumps towards Sk-67 5. %They found $\log N({\rm H_2}) = 19.44\pm 0.05$ which is in agreement with our value $\log N({\rm H_2}) = 19.47\pm 0.01$. Their value of HD column density is $13.62^{+0.09}_{-0.12}$, while we obtained a wider range $15.5^{+0.3}_{-2.1}$, includes saturated solution for the fit (see Sect.~\ref{sect:analysis}). Indeed, to obtain HD column density \cite{Andre2004} used an average value between Apparent Optical Depth (AOD, \cite{Savage1991}) method and line profile fitting, however it is known that AOD is not appropriate for lines with large optical depth and/or low resolution (above it was noted than resolution of FUSE spectra may be degraded after data reduction), therefore HD column density given by \cite{Andre2004} may be underestimated.
\citealt{Tchernyshyov2015, Roman_Duval2021} exhaustively studied metals in this system, using HST data, and our measurement of \ZnII\ column density ($12.98^{+0.02}_{-0.01}$) is consistent with their results ($13.03\pm 0.02$). Additionally, \citealt{Roman_Duval2019, Roman_Duval2021} studied \HI\ and \CI\ (for \HI\ they got $21.02\pm 0.04$, which agrees well with value found by \citealt{Welty2012}, therefore we used a value given by \citealt{Welty2012}).
We got total $\log N(\mbox{\CI})$ to be $14.02^{+0.02}_{-0.01}$ (we fitted with 3 components in the LMC), slightly higher than obtained by \citealt{Roman_Duval2021}, $13.86^{+0.01}_{-0.01}$ (a similar value was also obtained by \citealt{Welty2016}). Meanwhile our relative abundance on \CI\ levels are well consistent with obtained by \citealt{Roman_Duval2021}, while their column densities $\log N(\mbox{\CI}) = 13.65\pm 0.02$, $\log N(\mbox{\CI}^{*}) = 13.32\pm 0.02$, $\log N(\mbox{\CI}^{**}) = 12.86^{+0.04}_{-0.05}$ resemble what we derived in the central component of \CI\ profile.  %\DK{looks like only our main LMC component}. 
%\cite{Welty2016} also studied \CI\ in this system and got results close to \cite{Roman_Duval2021}.

\subsubsection{Sk-67 105}

Sk-67 105 is one of the most massive eclipsing binary system \citep{Niemela1986} in the LMC, which probably reveals a contact configuration with 48.3$M_{\odot}$ and 31.4$M_{\odot}$ \citep{Ostrov2003}. From H$_2$ lines (mostly from saturated $\rm J = 0$ and 1) we found an evidence of partial coverage of absorption system towards Sk-67 105. To get more reliable results we corrected fit of H$_2$ lines for partial covering factor found to be $0.888^{+0.002}_{-0.002}$.

Also we used HST archival data to measure \CI\ fine-structure level population in this system and found that our results are consistent with the values obtained by \citealt{Roman_Duval2021}. Our result on \ZnII\ column density $\log N(\rm ZnII) = 12.87^{+0.05}_{-0.08}$ is also consistent with the result of \citealt{Roman_Duval2021} and therefore gives one of the lowest metallicity in our LMC sample. % which, however, may indicate a high depletion in this system. %\SB{The last prhase is a bit in air... Reference?} \DK{this system is included in the depletion study of Roman-Duval 2019, 2021, but they do not say anything specific about the depletion in this system} \SB{Zn is volatile, so is not significantly depleted}

\subsubsection{Sk-68 135}

The star is located in the north of 30 Dor complex, one of the most studied star-forming regions. %\citealt{Andre2004} found HVC towards Sk-68 135 with large amount of C\,{\sc ii} relative to Na\,{\sc i}. \citealt{Lehner2009} also found significant amount of Fe\,{\sc ii} in this HVC, which probably indicates a presence of a shock wave. \citealt{Andre2004} report a large radiation field in this system (several thousand times over a mean Galactic value). Therefore they suggested that the envelope of molecular cloud should be destroyed by radiation from the star or swept away by the shock wave. 
%Our HD column density estimate, $\log N({\rm HD}) = 14.0^{+0.6}_{-0.3}$ is consistent with the result obtained by \citealt{Andre2004}, $\log N({\rm HD}) = 14.15^{+0.11}_{-0.15}$, while we found H$_2$ column density to be a bit higher ($\log N(\rm H_2) = 19.99\pm 0.01$ versus $19.87\pm 0.05$; note that \cite{Andre2004} fitted H$_2$ assuming a single-component model, while we fitted H$_2$ with 2-components model). \cite{Roman_Duval2019} found \HI\ column density in this system to be significantly lower then \cite{Welty2012}.   \cite{Roman_Duval2019} used Ly$\alpha$ from HST, while \cite{Welty2012} used Ly$\beta$ line from FUSE spectra, where the difficulties in \HI\ column density estimation partly arises from large amount of H$_2$ absorption lines in the wings of Ly$\beta$. Additionally \cite{Welty2012} did not provide uncertainties of the $N(\rm HI)$ measurements. Analysis of Ly$\alpha$ provides more reliable result, therefore we took value from \cite{Roman_Duval2019}. %\SB{WHy? Need to explain here.} \DK{added}
Metals and \CI\ have been studied by \citealt{Tchernyshyov2015, Roman_Duval2021}. \citealt{Roman_Duval2021} found $\log N(\mbox{\CI}) = 14.17^{+0.04}_{-0.05}$, $\log N(\mbox{\CI}^{*}) = 13.97\pm 0.03$, $\log N(\mbox{\CI}^{**})=13.55\pm 0.04$ which agrees with our results, except \CI$^{*}$,  which we found to be slightly larger. 
Our result on \ZnII\ column density $13.35\pm0.01$ agrees with a value $13.36\pm 0.06$ found by \citealt{Tchernyshyov2015}.

\subsubsection{Sk-69 246}

This star belongs to 30 Dor complex and the system on the line-of-sight have previously been studied by \citealt{Bluhm2001} and \citealt{Andre2004}. %\citealt{Bluhm2001} and \citealt{Andre2004}
These authors reported CO molecules in this system, using FUSE spectra, however, we checked that HST data and do not confirm the detection of CO at reported column density.
%\SB{Need to be careful here! Is it only in FUSE. Than we should note, that HST spectrum do not confirm CO!}\DK{added}

%\cite{Bluhm2001} have already studied H$_2$ and HD, where they used curve-of-growth method to estimate column densities.
%, so to get self-consistent results we refitted H$_2$ and HD absorption lines.
%Nevertheless, we got $\log N({\rm H_2}) = 19.76\pm 0.01$ and $\log N({\rm HD}) = 13.89\pm 0.06$ which is in agreement with results obtained by \cite{Bluhm2001}. Comparing with \cite{Andre2004} we obtained that $\log N({\rm H_2})$ is close to their result ($19.66\pm 0.04$). They found HD to be lower then both values found by \cite{Bluhm2001} and us. \cite{Andre2004} used only L5, L8 and L14 HD lines, while we used L3, L5, L7 and W0 (we did not use neither L14, since spectra from 1alif channel do not cover this line, moreover this line is weak and noizy nor L8 since it is blended by H$_2$ and \SII\ absorptions). Also \cite{Andre2004} reported reported blend of L3 line, but we found that this line does not disagree with other HD lines so we included it in the analysis. 

Our results on \CI, \CI$^{*}$ and \CI$^{**}$ column densities are close to that of \citealt{Andre2004}. They used FUSE spectra, where almost all lines are blended, and HST/STIS (only lines near 1275 \AA), while we additionally used the strongest 1656, 1560 and 1328 \AA\ lines. Also we found an excellent agreement with results on \CI\ provided by \citealt{Roman_Duval2021}. \ZnII\ column density measured by us ($\log N(\mbox{\ZnII}) = 13.32^{+0.01}_{-0.02}$) is a bit lower then $13.38\pm 0.04$ found by \citealt{Roman_Duval2021}.

\subsubsection{AV 95}

The star AV 95 belongs to the bar of SMC and the system towards AV 95 have previously been studied by \citealt{Andre2004, Tchernyshyov2015, Jenkins2017}. 
%Both our HD and H$_2$ column densities are consistent with values $\log N({\rm H_2}) = 19.43\pm 0.04$ and $\log N({\rm HD}) = 13.82^{+0.96}_{-0.18}$) found by \cite{Andre2004}, respectively. 
For \CI, \CI$^{*}$ and \CI$^{**}$ we could place only upper limits on column densities due to low S/N of the spectrum and weakness of the lines. \citealt{Andre2004} reported measurements, but, as was noted above, they used only lines seen in the FUSE spectra and lines around 1275\,\AA\, from HST/STIS data, while we additionally used stronger carbon lines in HST high-resolution spectrum. Our estimate on \ZnII\ column density, $\log N (\mbox{ZnII}) = 13.08\pm0.02$, agrees with both values $13.09^{+0.05}_{-0.06}$ and $13.15\pm 0.03$ found by \citealt{Tchernyshyov2015} and \citealt{Jenkins2017}, respectively.

\subsubsection{AV 242}

In the system towards AV 242 we found an evidence of high velocity clouds (HVC), which contains H$_2$ molecules (see Paper I). To our knowledge, it is the first found HVC towards the SMC containing H$_2$ and it had not been discussed before. %\SB{Was it reported before? If no, we need to state it. We can also said that it will be analysed in separate paper.} \DK{I have not found this system individually discussed in the literature at all; we wrote about it in the HD paper} \SB{Then we should refer to our paper here!} \DK{done}
We also found \CI\, in this HVC $\log N_{\rm HVC}(\rm CI) = 12.91^{+0.05}_{-0.08}$ and $\log N(\rm ZnII) = 12.66^{+0.09}_{-0.10}$.
Metal lines have been studied by \citealt{Jenkins2017}, their reported \ZnII\, column density, $13.39^{+0.14}_{-0.21}$, is consistent with ours measured value $13.17^{+0.04}_{-0.05}$ summed over the components. However, as \citealt{Jenkins2017} used AOD method, they could not separate subcomponents seen in this sightline. 

\subsubsection{Sk 191}

We found one of the highest H$_2$ column density along the sightline towards Sk 191, while there was lack of \ZnII\ in this system. On the one hand, this can be explained by exceptionally low metallicity. The star is located close to Magellanic Bridge, which reveals abundances lower than in the LMC and SMC, therefore the medium in front of Sk 191 may have lower metallicity. But on the other hand, the environment may have an especially high depletion and a possible detection of \OI\ by \citealt{Jenkins2017} confirms this assumption. Also such a high H$_2$ column density may possibly be an evidence of high depletion, since H$_2$ forms mainly on the surface of dust grains and correlates with amount of dust \citep{Telikova2022}.

\section{Results}
\label{sect:results}
\CI\ absorption was detected towards the majority of the studied sightlines (22 out of 28 systems in the SMC and 29 out 29 systems in the LMC). The excitation of \CI\ fine-structure levels was measured in 23 systems in the LMC and 12 systems in the SMC. The fitting results are summarized in Tables~\ref{tab:CI_LMC} and \ref{tab:CI_SMC} for the LMC and SMC, respectively. Fit to the line profiles are shown in the Appendix~\ref{sect:CI_fit}.

\renewcommand{\arraystretch}{1.35}
\begin{table*}
    \caption{The fit results of \CI\ towards the LMC sight lines. %\SK{Do we need so many numbers for values in $z$ column. It looks like that uncertainty in some cases is less then 1 m/s.} \DK{changed redshifts to $v_{\rm LSR}$}\SK{I suggest use the same format of accuracy - like the fixed number of characters after point, no less than "0.1"} \SB{I disagree. we need to be consistent with statistical uncertainties}
    }
    \label{tab:CI_LMC}
    \begin{tabular}{lccccc}
    \hline 
    Star & $v_{\rm LSR}$ & $b$, km/s & $\log N(\mbox{\CI})$ & $\log N(\mbox{\CI}^{*})$ & $\log N(\mbox{\CI}^{**})$ \\
     \hline 
      Sk-67 2 & $266.9(^{+0.2}_{-0.1})$ & $3.7^{+0.4}_{-0.4}$ & $13.60^{+0.02}_{-0.02}$ & $13.43^{+0.03}_{-0.02}$ & $13.04^{+0.04}_{-0.04}$ \\
              & $276.3(^{+0.1}_{-0.2})$ & $1.5^{+0.1}_{-0.1}$ & $14.02^{+0.07}_{-0.08}$ & $13.88^{+0.05}_{-0.03}$ & $13.43^{+0.04}_{-0.03}$ \\
      Sk-67 5 & $279.8^{+0.4}_{-0.6}$ &  $5.7^{+1.5}_{-0.9}$ &  $12.96^{+0.05}_{-0.06}$ & $12.64^{+0.07}_{-0.13}$ & $12.17^{+0.22}_{-0.34}$ \\
              & $287.35^{+0.05}_{-0.05}$ & $1.2^{+0.1}_{-0.1}$ & $13.64^{+0.03}_{-0.04}$ & $13.29^{+0.02}_{-0.03}$ & $12.71^{+0.04}_{-0.06}$ \\
              & $293.795^{+0.002}_{-0.025}$ & $1.9^{+0.3}_{-0.1}$ & $13.09^{+0.02}_{-0.03}$ & $12.84^{+0.04}_{-0.04}$ & $12.24^{+0.11}_{-0.13}$ \\
     Sk-67 20 &$286.03^{+0.08}_{-0.11}$ & $4.2^{+0.2}_{-0.2}$ & $13.46^{+0.01}_{-0.02}$ & $13.44^{+0.01}_{-0.01}$ &  $12.97^{+0.03}_{-0.02}$ \\
     PGMW 3070 & $284.1^{+0.2}_{-0.2}$ & $4.4^{+0.5}_{-0.3}$ & $13.34^{+0.03}_{-0.02}$ &  $13.18^{+0.03}_{-0.03}$ & $12.73^{+0.10}_{-0.09}$ \\
     LH 10-3120 & $263.9^{+0.2}_{-0.4}$ & $1.0^{+0.5}_{-0.4}$ & $12.98^{+0.15}_{-0.08}$ & $12.93^{+0.06}_{-0.08}$ & $\lesssim 13.1$ \\
                & $281.3^{+0.4}_{-0.4}$ & $7.8^{+0.7}_{-0.6}$ & $13.32^{+0.02}_{-0.03}$ & $13.06^{+0.04}_{-0.06}$ & $12.77^{+0.09}_{-0.08}$ \\
     PGMW 3223 & $269.5^{+0.3}_{-0.6}$ & $6.4^{+1.0}_{-0.6}$ & $13.43^{+0.03}_{-0.05}$ & $13.35^{+0.04}_{-0.05}$ & $13.05^{+0.06}_{-0.07}$ \\
              & $287.2^{+0.9}_{-1.2}$ &  $4.0^{+3.5}_{-3.5}$ & $12.75^{+0.19}_{-0.13}$ & $12.75^{+0.18}_{-0.23}$ & $12.68^{+0.13}_{-0.13}$ \\
     Sk-66 35 & $271.3^{+0.4}_{-0.4}$ & $7.6^{+1.0}_{-0.6}$ & $13.35^{+0.03}_{-0.03}$ &  $13.15^{+0.04}_{-0.04}$ & $12.64^{+0.10}_{-0.16}$ \\
     Sk-66 51 & $302.6^{+18.2}_{-167.3}$ & $7.0^{+13.0}_{-6.5}$ 
     & $12.9^{+0.4}_{-1.0}$ & $\lesssim 13.5$ & $\lesssim14.4$  \\
              & $307.6^{+1.2}_{-0.7}$ & $0.7^{+4.8}_{-0.2}$ & $13.14^{+0.09}_{-0.11}$ & $13.07^{+0.08}_{-0.10}$ & $\lesssim 12.6$ \\
     Sk-70 79 & $232.2^{+0.1}_{-0.1}$ & $5.9^{+0.1}_{-0.1}$ & $13.77^{+0.01}_{-0.01}$ & $13.98^{+0.01}_{-0.01}$ & $13.71^{+0.01}_{-0.01}$  \\
     Sk-68 52 & $242.2^{+0.1}_{-0.1}$ & $1.8^{+0.3}_{-0.3}$ & $13.36^{+0.09}_{-0.08}$ & $13.66^{+0.05}_{-0.05}$ & $13.34^{+0.06}_{-0.04}$ \\
     Sk-71 8 & $215.5^{+0.6}_{-0.8}$ & $6.8^{+1.3}_{-1.0}$ & $13.06^{+0.07}_{-0.08}$ & $13.37^{+0.05}_{-0.05}$ & $13.23^{+0.05}_{-0.06}$ \\
             & $229.9^{+1.5}_{-1.5}$ & $4.4^{+2.8}_{-2.0}$ & $12.74^{+0.16}_{-0.15}$ & $12.60^{+0.27}_{-0.27}$ & $12.58^{+0.22}_{-0.29}$ \\
     Sk-69 106 & $249.7^{+0.1}_{-0.1}$ & $5.1^{+0.3}_{-0.2}$ & $13.30^{+0.02}_{-0.02}$ & $13.57^{+0.01}_{-0.01}$ & $13.44^{+0.01}_{-0.02}$ \\
     Sk-68 73 & $292.7^{+0.3}_{-0.4}$ & $8.7^{+0.5}_{-0.5}$ &  $13.71^{+0.03}_{-0.04}$ & $13.84^{+0.03}_{-0.06}$ & $13.59^{+0.05}_{-0.05}$ \\
              &  $295.4^{+0.1}_{-0.1}$ & $1.2^{+0.3}_{-0.2}$ & $13.76^{+0.13}_{-0.09}$ & $14.08^{+0.16}_{-0.09}$ & $13.70^{+0.11}_{-0.11}$ \\
     Sk-67 105 & $301.9^{+0.1}_{-0.1}$ & $1.5^{+0.3}_{-0.2}$ & $13.61^{+0.12}_{-0.13}$ & $13.73^{+0.06}_{-0.06}$ & $13.25^{+0.11}_{-0.05}$ \\
     BI 184 & $250.4^{+0.4}_{-0.3}$ & $1.8^{+0.4}_{-0.3}$ & $13.64^{+0.06}_{-0.18}$ & $13.44^{+0.04}_{-0.04}$ & $13.21^{+0.05}_{-0.07}$ \\ 
     Sk-71 45 & $243.7^{+0.6}_{-0.7}$ & $12.1^{+1.8}_{-1.2}$ & $13.34^{+0.03}_{-0.04}$ & $13.37^{+0.05}_{-0.04}$ & $13.12^{+0.03}_{-0.07}$ \\
     Sk-71 46 & $249.7^{+1.2}_{-1.5}$ & $10.1^{+0.3}_{-0.3}$  & $13.38^{+0.04}_{-0.04}$ &  $13.42^{+0.04}_{-0.09}$ & $13.14^{+0.08}_{-0.03}$ \\
              & $230.6^{+0.8}_{-2.0}$ & $10.0^{+0.1}_{-0.2}$ &  $13.26^{+0.06}_{-0.04}$ & $13.54^{+0.03}_{-0.04}$ &  $12.75^{+0.13}_{-0.19}$ \\
    Sk-69 191 & $237.4^{+0.2}_{-0.2}$ & $1.0^{+2.0}_{-0.1}$ &  $12.95^{+0.05}_{-0.04}$ &  $12.88^{+0.03}_{-0.03}$ & $12.32^{+0.11}_{-0.19}$ \\
    BI 237 & $290.5^{+0.7}_{-0.9}$ & $11.6^{+2.5}_{-1.1}$ &  $13.51^{+0.04}_{-0.03}$ & $13.36^{+0.06}_{-0.03}$ & $\lesssim 13.2$ \\
    Sk-68 129 & $278.3^{+0.1}_{-0.3}$ & $13.4^{+0.5}_{-0.4}$ & $13.72^{+0.01}_{-0.01}$ & $13.69^{+0.01}_{-0.01}$ & $13.08^{+0.06}_{-0.06}$ \\
    Sk-66 172 & $282.3^{+0.4}_{-0.3}$ & $4.8^{+0.6}_{-0.7}$ & $13.13^{+0.04}_{-0.04}$ & $13.28^{+0.04}_{-0.03}$ & $12.87^{+0.07}_{-0.08}$ \\
              & $299.6^{+0.3}_{-0.4}$ & $5.7^{+0.6}_{-0.7}$ & $13.17^{+0.03}_{-0.04}$ & $13.39^{+0.04}_{-0.03}$ & $12.96^{+0.06}_{-0.05}$ \\
    BI 253 & $275.0^{+1.0}_{-0.8}$ & $14.8^{+0.2}_{-1.3}$ &  $13.68^{+0.21}_{-0.69}$ &  $13.4^{+0.3}_{-0.5}$ & $13.1^{+0.4}_{-1.0}$\\
    Sk-68 135 & $268.0^{+0.2}_{-0.1}$ & $0.4^{+0.3}_{-0.2}$ & $13.38^{+0.25}_{-0.40}$ & $13.44^{+0.17}_{-0.12}$ & $12.88^{+0.15}_{-0.14}$  \\ 
        & $276.5^{+0.1}_{-0.1}$ & $5.8^{+0.3}_{-0.2}$ &  $14.00^{+0.02}_{-0.02}$ & $13.93^{+0.02}_{-0.02}$ & $13.45^{+0.02}_{-0.01}$ \\
    Sk-69 246 &  $277.7^{+0.3}_{-0.2}$ & $5.0^{+0.4}_{-0.5}$ & $13.72^{+0.05}_{-0.07}$ & 
    $13.69^{+0.04}_{-0.06}$ & $13.33^{+0.02}_{-0.03}$ \\
              &  $286.9^{+0.9}_{-1.8}$ & $4.1^{+2.2}_{-1.1}$ & $13.25^{+0.16}_{-0.11}$ & $13.07^{+0.20}_{-0.15}$ & $\lesssim 13.0$ \\
    Sk-68 140 & $277.7^{+0.1}_{-0.2}$ & $8.5^{+0.4}_{-0.4}$ & $13.66^{+0.02}_{-0.02}$ & $13.78^{+0.01}_{-0.01}$ & $13.41^{+0.03}_{-0.02}$ \\
    Sk-71 50 & $231.2^{+0.6}_{-0.7}$ & $5.9^{+1.6}_{-1.0}$ & $12.83^{+0.08}_{-0.10}$ & $13.19^{+0.05}_{-0.05}$ & $12.80^{+0.12}_{-0.11}$ \\
             & $273.9^{+0.4}_{-0.2}$ & $5.6^{+0.6}_{-0.4}$ & $13.52^{+0.02}_{-0.04}$ & $13.20^{+0.04}_{-0.06}$ & $\lesssim 12.7$ \\
    Sk-69 279 &  $276.3^{+0.4}_{-0.3}$ & $2.0^{+0.3}_{-0.5}$ & $13.83^{+0.52}_{-0.29}$ & $13.30^{+0.07}_{-0.05}$ & $12.52^{+0.10}_{-0.17}$ \\
    Sk-68 155 &  $289.9^{+0.3}_{-0.3}$ & $12.9^{+0.6}_{-0.6}$ &  $13.67^{+0.02}_{-0.01}$ & $13.62^{+0.01}_{-0.02}$ & $\lesssim 13.2$ \\
    Sk-70 115 & $220.72^{+0.03}_{-0.04}$ & $2.1^{+0.1}_{-0.1}$ & $13.56^{+0.02}_{-0.02}$ & $13.52^{+0.01}_{-0.01}$ & $13.04^{+0.02}_{-0.02}$ \\
    \hline 
    \end{tabular}
    
    \begin{tablenotes}
    \item Upper limits were constrained from $3\sigma$ confidence interval  
    \end{tablenotes}
\end{table*}

\renewcommand{\arraystretch}{1.35}
\begin{table*}
    \caption{The fit results of \CI\ towards the SMC sightlines.}
    \label{tab:CI_SMC}
    \begin{tabular}{lccccc}
    \hline
    Star & $v_{\rm LSR}$ & $b$, km/s & $\log N(\mbox{\CI})$ & $\log N(\mbox{\CI}^{*})$ & $\log N(\mbox{\CI}^{**})$  \\
     \hline 
      AV\,6 &  $128.6^{+1.2}_{-0.9}$ & $14.9^{+0.1}_{-1.3}$ & $\lesssim 13.02$ & $\lesssim 12.89$ & $\lesssim 12.84$ \\
      AV\,15 & $132.4^{+0.7}_{-0.6}$ &$7.9^{+1.2}_{-1.0}$ & $13.20^{+0.05}_{-0.05}$ & $13.07^{+0.06}_{-0.08}$ & $12.69^{+0.14}_{-0.16}$  \\
      AV\,26 & $122.8^{+0.1}_{-0.1}$ & $0.35^{+0.1}_{-0.1}$ & $13.78^{+0.06}_{-0.07}$ & $13.89^{+0.05}_{-0.06}$ & $13.53^{+0.04}_{-0.04}$ \\
             & $125.02^{+0.05}_{-0.07}$ & $0.22^{+0.1}_{-0.1}$ &  $15.16^{+0.06}_{-0.09}$ & $14.59^{+0.11}_{-0.09}$ &$13.82^{+0.10}_{-0.07}$ \\ 
      AV\,47 & $121.7^{+1.5}_{-1.5}$ & $13.4^{+1.7}_{-2.2}$ & $12.83^{+0.09}_{-0.11}$ & $13.00^{+0.06}_{-0.09}$ & $\lesssim 12.8$ \\
      AV\,69 & $127.7^{+0.9}_{-1.4}$ & $7.4^{+2.6}_{-6.9}$ &  $12.81^{+0.08}_{-0.09}$ & $12.77^{+0.11}_{-0.10}$ & $\lesssim 12.5$ \\
      AV\,75 & $117.8^{+0.9}_{-0.9}$ & $6.2^{+1.2}_{-1.1}$ &  $12.95^{+0.09}_{-0.07}$ & $13.02^{+0.17}_{-0.18}$ & $13.06^{+0.09}_{-0.07}$ \\
      AV\,80 & $116.3^{+0.2}_{-0.5}$ & $8.9^{+0.9}_{-0.8}$ & $13.35^{+0.04}_{-0.03}$ &  $13.16^{+0.04}_{-0.06}$ & $12.63^{+0.11}_{-0.24}$ \\
      AV\,81 & $149.6^{+0.9}_{-1.2}$ & $6.3^{+2.3}_{-1.7}$ & $13.06^{+0.07}_{-0.10}$ & $12.63^{+0.27}_{-0.52}$ & $\lesssim 12.8$  \\
      AV\,95 & $122.3^{+0.9}_{-1.5}$ & $5.2^{+2.7}_{-1.1}$ & $\lesssim 13.0$ & $\lesssim 12.8$ & $\lesssim 12.9$ \\
      AV\,104 & $116.6^{+2.4}_{-0.9}$ & $<8$ & $\lesssim  14.56$ & $\lesssim 13.72$ & $\lesssim 13.82$  \\
      AV\,170 & $128.6^{+0.3}_{-0.5}$ & $0.5^{+0.1}_{-0.1}$ & $13.56^{+0.32}_{-0.26}$ & $\lesssim 13.4$ & $\lesssim 13.2$ \\
      AV\,175  & $136.4^{+1.8}_{-1.5}$ & $22.3^{+2.9}_{-2.5}$ & $13.20^{+0.05}_{-0.09}$ &$\lesssim 13.4$ & $\lesssim 13.4$ \\
      AV\,207 & $164.0^{+1.0}_{-0.7}$ & $1.2^{+4.4}_{-0.8}$ & $12.99^{+0.25}_{-0.17}$ &  $13.25^{+0.12}_{-0.13}$ & $\lesssim 14.5$ \\
      AV\,210 & $162.8^{+0.9}_{-0.6}$ &  $5.4^{+1.5}_{-1.6}$ & $13.07^{+0.07}_{-0.07}$ & $13.07^{+0.08}_{-0.09}$ & $\lesssim 13.0$ \\
      AV\,215 & $127.0^{+0.9}_{-0.8}$ & $6.4^{+1.4}_{-1.5}$ & $13.35^{+0.06}_{-0.08}$ & $12.74^{+0.26}_{-0.45}$ & $\lesssim 12.8$ \\
              & $142.7^{+1.5}_{-1.8}$ &  $7.8^{+4.7}_{-1.0}$ & $13.16^{+0.12}_{-0.12}$ & $13.03^{+0.17}_{-0.06}$ & $12.66^{+0.27}_{-0.17}$  \\
      AV\,216 & $142.9^{+0.4}_{-0.5}$ & $2.3^{+1.0}_{-0.8}$ & $12.88^{+0.06}_{-0.07}$ & $12.76^{+0.08}_{-0.12}$ & $12.49^{+0.17}_{-0.19}$ \\
      AV\,243 & $128.9^{+1.5}_{-1.2}$ & $1.0^{+1.1}_{-0.9}$  & $\lesssim 15.7$ & $\lesssim 13.7$ & $\lesssim 14.7$ \\
      AV\,242 & $92.9^{+0.6}_{-0.7}$ & $3.6^{+1.0}_{-1.5}$ & $12.70^{+0.06}_{-0.07}$ & $\lesssim 12.7$ & $\lesssim 12.4$ \\
              & $158.4^{+0.4}_{-0.8}$ & $4.1^{+1.4}_{-1.9}$  & $12.86^{+0.13}_{-0.12}$ & $\lesssim 12.8$ & $\lesssim 12.4$ \\
      AV\,266 & $127.6^{+0.7}_{-0.6}$ & $6.8^{+1.4}_{-0.9}$ & $13.11^{+0.04}_{-0.04}$ & $12.73^{+0.11}_{-0.09}$ & $\lesssim 12.6$ \\ 
      AV\,372 & $143.2^{+0.3}_{-0.2}$ & $5.7^{+0.5}_{-0.5}$ & $13.40^{+0.02}_{-0.03}$ & $12.95^{+0.04}_{-0.08}$ & $\lesssim 12.5$ \\
      AV\,423 & $141.2^{+1.8}_{-3.3}$ & $9.7^{+0.3}_{-3.5}$ & $12.87^{+0.10}_{-0.17}$ &  $\lesssim 13.1$ & $\lesssim 12.6$ \\
              &  $159.2^{+1.5}_{-3.5}$ & $9.7^{+0.3}_{-9.2}$  & $12.81^{+0.10}_{-0.30}$ &  $\lesssim 13.1$ &  $\lesssim 12.7$ \\
      AV\,440 & $167.3^{+6.0}_{-7.2}$ & $0.5^{+1.6}_{-0.1}$ & $\lesssim 15.84$ & $\lesssim 13.41$ & $\lesssim 16.66$ \\
              & $116.6^{+2.4}_{-0.9}$ & $0.5^{+7.3}_{-0.1}$ & $\lesssim 16.87$ & $\lesssim 15.40$ & $\lesssim 17.29$ \\
      AV\,472 &  $126.8^{+1.8}_{-1.8}$ & $0.7^{+0.2}_{-0.2}$ & $\lesssim 15.7$ & $\lesssim 13.8$ & $\lesssim 16.6$  \\
      AV\,476 & $159.2^{+1.5}_{-1.8}$ & $12.9^{+1.3}_{-1.8}$ & $13.15^{+0.10}_{-0.07}$ & $13.07^{+0.05}_{-0.09}$ & $13.03^{+0.07}_{-0.10}$ \\
              & $168.7^{+0.3}_{-0.1}$ & $1.0^{+0.2}_{-0.2}$ & $15.40^{+0.50}_{-0.60}$ & $14.39^{+0.29}_{-0.30}$ &  $13.97^{+0.25}_{-0.12}$ \\
      AV\,479 & $158.5^{+0.1}_{-0.2}$ & $1.9^{+0.3}_{-0.3}$ & $12.71^{+0.04}_{-0.03}$ & $12.61^{+0.05}_{-0.06}$ & $12.07^{+0.10}_{-0.18}$ \\
              & $164.1^{+0.2}_{-0.4}$ & $1.4^{+0.8}_{-0.5}$ & $12.24^{+0.10}_{-0.11}$ & $12.36^{+0.08}_{-0.09}$ & $\lesssim 12.2$ \\
      AV\,488 & $143.2^{+0.4}_{-0.5}$ & $5.9^{+0.8}_{-0.9}$ & $13.08^{+0.04}_{-0.04}$ & $12.59^{+0.11}_{-0.14}$ & $\lesssim 12.5$ \\
      AV\,490 & $133.5^{+0.1}_{-0.1}$ & $4.9^{+0.2}_{-0.1}$ & $13.50^{+0.01}_{-0.01}$ & $12.94^{+0.02}_{-0.02}$ & $12.15^{+0.09}_{-0.07}$ \\ 
      Sk\,191 & $134.6^{+1.5}_{-1.8}$ & $9.2^{+3.8}_{-3.4}$  & $12.88^{+0.09}_{-0.12}$ & $\lesssim 13.10$ & $\lesssim 12.74$ \\
              & $153.7^{+0.4}_{-0.3}$ & $4.5^{+1.0}_{-0.7}$ & $13.34^{+0.03}_{-0.05}$ & $13.42^{+0.03}_{-0.06}$ & $13.05^{+0.07}_{-0.05}$ \\
    \hline 
      
    \end{tabular}
    \begin{tablenotes}
     \item Upper limits were constrained from $3\sigma$ confidence interval
        
    \end{tablenotes}
\end{table*}

\subsection{Metallicity}
%\SK{Is it gas phase metallicity measurement, right?} \DK{yes} \SB{added in Sect. 3}
The measurements of metallicity are given in the second column of Tables~\ref{tab:LMC_phys_cond} and \ref{tab:SMC_phys_cond} for the LMC and SMC, respectively and lines are shown in the Appendix~\ref{sect:metal_fit}.
The average metallicity in our sample is $\log Z_{\rm LMC} \sim -0.66$ with  dispersion of 0.22 and $\log Z_{\rm SMC} \sim -0.99$ with standard deviation of 0.19 for the LMC and SMC, respectively. 
It is less by $-0.3$\,dex and $-0.7$\,dex than the average gas phase metallicity  measured in local MW ISM \citep[e.g.,][]{Ritchey2023}. These values are consistent with previous values reported in literature ($\log Z^* \sim -0.2$ for the LMC and $\log Z^* \sim -0.6$ for the SMC relative to local ISM, \citealt{Russell1992}). The difference in absolute values are partially due to the difference in solar abundances between \cite{Anders1989} and \cite{Asplund2009}. Also it can be explained by depletion of metals by dust and usage of the different species, or selection effects, since
our targets probe a cold phase of the ISM. Meanwhile, even in local ISM the value of metallicity is debatable and possibly has large variations \citealt{DeCia2021}, but see also \cite{Ritchey2023}.

%[These estimates are slightly lower than previous values reported in literature ($\log Z \sim -0.3$ for the LMC and $\log Z \sim -0.7$ for the SMC  \citealt{Russell1992}) \SK{I converted both values to log scale to make them consistent.}
%\SK{Please, check. I see that Russel1992 measured metallicity relative to  both the local ISM and Sun. They found "The interstellar medium (ISM) of the LMC has a mean metallicity 0.2 dex lower than the local Galactic ISM, and the metallicity of the SMC is 0.6 dex lower". Local ISM has metallicity about -0.3 dex, therefore they have actually -0.5 for the LMC and -0.9 for the SMC, right?} \DK{no, they assumed metallicity of local ISM is about -0.1 (see their figures 1 and 2)} \SK{Check the difference in Solar metallicity between Asplund2009 and ones used by Russel, and you will see that it is about +0.1-0.3 units. So I suggest to coampare metallicity with local ISM values.} that can be explained by depletion of metals by dust and usage of the different species, or selection effects, since our targets probe a cold phase of the ISM. Meanwhile, even in local ISM the value of metallicity is debatable and possibly has large variations \citep{DeCia2021}, but see also \cite{Ritchey2023}.] 
%[some systematic effect in our sample]. 

\subsection{\CI/H$_2$ abundances}\
\label{sec:CI_H2_abund}

\begin{figure*}
    \begin{minipage}{0.49\textwidth}
         \center{\includegraphics[width=1\linewidth]{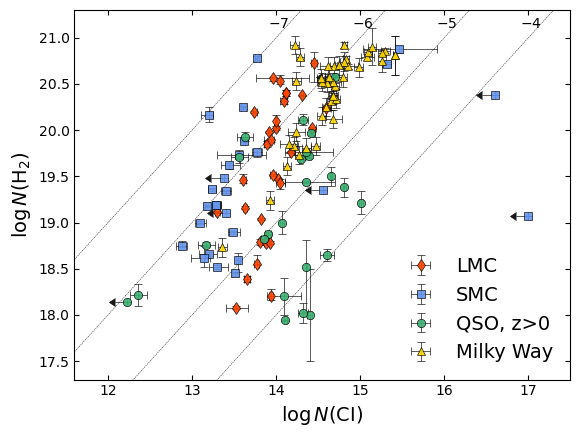}}
    \end{minipage}
    \hfill
    \begin{minipage}{0.49\textwidth}
        \center{\includegraphics[width=1\linewidth]{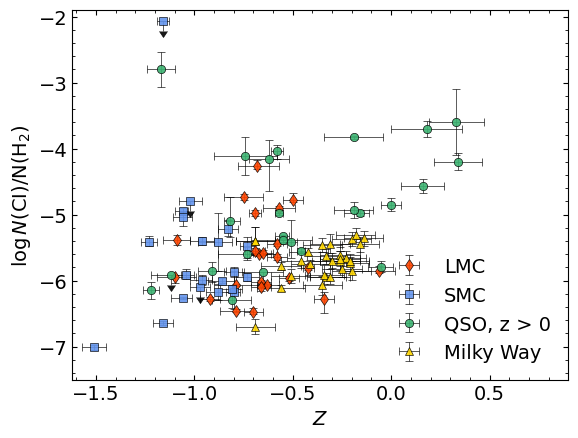}}
    \end{minipage}
    \caption{{\sl Left panel:} H$_2$ and \CI\ column densities. The blue squares, red diamonds, yellow triangles and green circles show the measurements in SMC, LMC, Milky Way \citep{Rachford2002, Rachford2009, Burgh2010, Jenkins2011, Shull2021, Ritchey2023} and high redshift DLAs \citep{Noterdaeme2018, Balashev2019, Srianand2005, Ledoux2003, Jorgenson2009, Jorgenson2010}, respectively. Dashed lines show \CI/H$_2$ ratios equal to -4, -5, -6, -7. {\sl Right panel:} $\log N({\rm CI})/N({\rm H_2})$ as a function of gas-phase metallicity. The symbols are the same as in left panel.
    %\SK{Please, add lines of constant CI/H2 ratio on the left panel. Why do I see the difference in CI/H2 between MW, LMC, SMC in the left panel, and there is no difference in the right? Points follow the same value of CI/H2 ratio? Then I can say that N(CI) correlates with Z, while N(H2) not.} \DK{Yes, on the right figure the can be clearly seen a difference between LMC and SMC for \CI, but note that most of SMC systems have lower $N_{\rm H_2}$ than LMC (which also arise from difference in the metallicities), so CI/H$_2$ ratio for LMC and SMC are close} \SB{Well, the trends on the left panels are certainly not 1-to-1 ratio. Which is evidently shown on the right (CI/H2). Actually, CI very slowly increased, while H2 sharply increased, so there is a large dispersion. Aslo the difference in CI/H2 are complex and depends not only on Z, but here since we are discussion basic measured properties, we can only show Z.}
    %\DK{left panel - something like this?} \SB{Yes, it is fine!} 
    %\SK{Maybe plot CI vs H2 in the left panel? It will be easy to understand carbon transition in this case.}
    }
    \label{fig:CI_H2}
\end{figure*}

In the left panel of Fig.~\ref{fig:CI_H2} we show the column densities of \CI\ and H$_2$ in the LMC and SMC and their comparison with measurements in the Milky Way \citep{Jenkins2011} and in high redshift DLAs \citep[see, e.g.][and references therein]{Klimenko2020}. 
We found that on average \CI\ column densities
%abundance \SK{[what is abundance - is it column density?]}\DK{yes} 
in the SMC is less than in the LMC and both are less than in the MW and DLAs. The difference in \CI\ abundance (and \CI/H$_2$ relative abundance) between the LMC, SMC and MW can be due to the difference in gas phase metallicity. A higher \CI\ abundances in high-z DLAs can be due to the selection effect (many \CI\ bearing high-z DLAs were preselected by the presence of strong \CI\ in SDSS spectra, see e.g. \citealt{Noterdaeme2018}) and/or distinct physical conditions. Both these factors likely explain large dispersion of \CI/H$_2$ ratios measured at high-z.  %[in LMC is systematically higher than SMC, which can be explained by the difference in the metallicity and hence \CI\, gas phase abundance. %However, typically \CII\ is the dominant ionization state in this diffuse clouds, and hence the \CI\ abundance depends on the local physical conditions \citep[e.g.][]{Wolfire2008}. \SB{Do we need some comparison of the Metallicity versus CI/H2? As we will add more data points for high-z.} \DK{yes; put it in Figure 1. Also added other high redshift systems}
%Also Milky Way show higher \CI\ abundances than both LMC and SMC, which also can be explained by higher metallicity, while QSO measurements show quite large dispersion, arisen from the distinctions in the physical conditions in each system.]

The right panel of Fig~\ref{fig:CI_H2} shows \CI/H$_2$ relative abundance as a function of metallicity in different samples.
%[LMC and SMC (red diamonds and blue squares, respectively). We also compare values, found in this paper with the values measured in the high redshift systems (green circles) and in our Galaxy (yellow triangles)]. 
%\SK{There is a trend of the increasing \CI\ to H$_2$ ratio with metallicity. [Can we estimated the probability?]} [One can see that \CI\, to H$_2$ ratio increases with the increase of metallicity.] \SB{To tell the truth, now, I do not see any signiicant trend for LMC, SMC and MW} \DK{me too, only for QSO}.
While the \CI\ abundance is evidently scaled with the total carbon abundance in the medium, and hence metallicity\footnote{The system towards Q\,0347-3819 \citep{Srianand2005} has an exceptionally high \CI/H$_2$ ratio among low metallicity systems due to low H$_2$ abundance.}, it also strongly depends on the chemistry 
%\SK{[ionization balance?]} \SB{This includes ionizationbalance}
of the clouds, since \CI\ is not the dominant form of the carbon in diffuse clouds. Indeed the formation rate of \CI\ depends on the metallicity, 
%\SK{[it looks like we already said that it depends on metallicity]} \SB{This was for the total carbon abundance}
since it is produced from \CII\ either by recombination with electron or small dust grains \citep{Wolfire2008}. Both electron and dust abundances, interactively depend on the metallicity, especially at the low values of the latter. Moreover H$_2$ abundance scales with the metallicity as well, since the main formation channel of H$_2$ molecules is the formation on the surface of dust grains, which abundance is proportional with the metallicity. Therefore dependence of \CI/H$_2$ ratio on the metallicity is relatively complicated and requires comprehensive modelling. We will exhaustively consider \CI/H$_2$ abundance in a separate study (Balashev\&Kosenko in prep.).  

%\SK{Maybe to say: While the \CI\ abundance is evidently scaled with the total carbon abundance in the medium, and hence metallicity, the formationrate of \CI\ also strongly depends on the metallicity. Indeed \CI\ is produced from \CII\ either by recombination with electron or on the small dust grains \citep{Wolfire2008}. Both electron and dust abundance, likely intricatively depends on the metallicity, especially at the lower end. We will discuss the \CI\ abundance in the separate study (Balashev\&Kosenko  in prep.)} \SB{I do not see difference with what is written}

\subsection{Kinetic temperature}
\label{sec:kin}

\begin{figure}
    \centering
    \includegraphics[width=\linewidth]{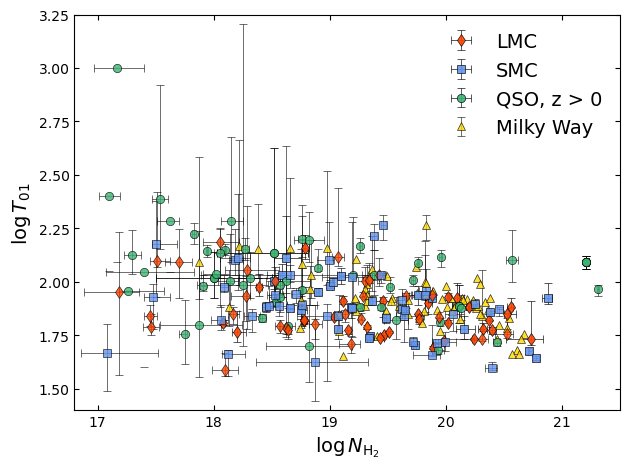}
    \caption{H$_2$ excitation temperature as a function of H$_2$ column density. The symbols are the same as in Fig.~\ref{fig:CI_H2}, except yellow triangles show Milky Way values from \citealt{Savage1977, Gillmon2006} (see discussion in Sect.~\ref{sec:kin}). %\SK{Lets calculate correlation coefficient and power law slope for different samples?} \DK{correlation is weak, about -0.3 for both LMC and SMC} \SB{To provide values in the Section 4.3} \DK{done}
    }
    \label{fig:H2_T}
\end{figure}

\begin{figure*}
    \centering
    \includegraphics[width=1.0\linewidth]{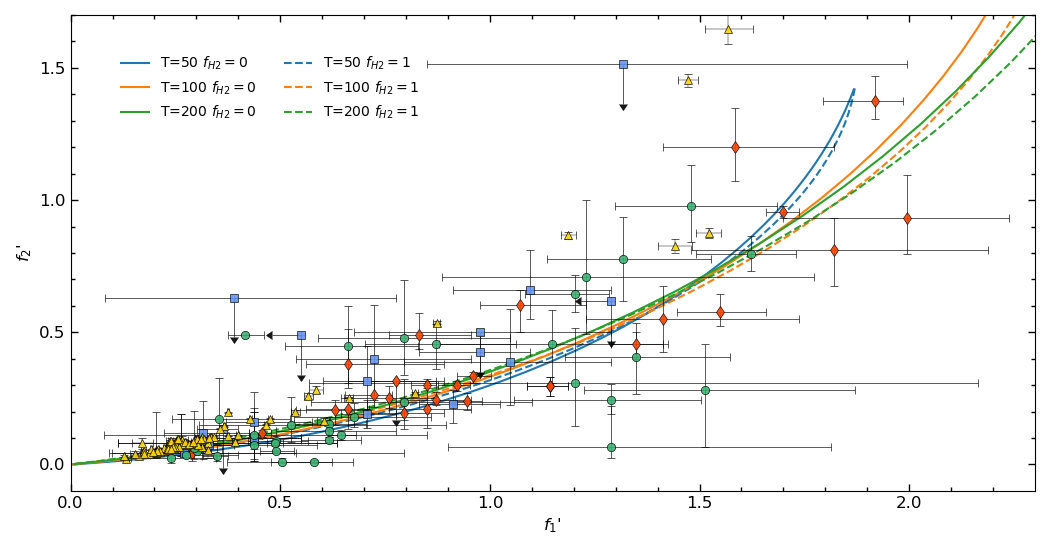}
    \caption{$f_1' = N(\mbox{\CI}^*) / N(\mbox{\CI})$ vs $f_2' = N(\mbox{\CI}^{**}) / N(\mbox{\CI})$. Here red diamonds are measurements in the LMC, blue squares are measurements in the SMC, green circles are QSO data (from the references provided in caption of Fig.~\ref{fig:CI_H2}) and yellow triangles are data in the MW \citep{Jenkins2011}. Model curves are calculated for the temperatures $T = 50$, 100 and 200
    \,K, and number densities in the range $\log n = 0 - 5$ assuming only atomic hydrogen ($f_{\rm H_2} = 0$, solid curves) and full molecularized case ($f_{\rm H_2} = 1$, dashed curves). %\SB{I think we can safely trim the x and y axis at the lower end, e.g. x> -0.8 and y >-1.6} \DK{errorbars for several points will be truncated, looks badly} \SB{Then may be we should try decimal scale. Since there is too many attention to points with large uncertainties, while the most of the well measured points are grouped at one quarter of the plot.} \DK{for me it looks worse then logs, moreover Sk-66 51 is very high (here it is truncated)} \SB{Sk-66 is an upper limit only, so it is fine. Even more, you can cut upper limit at the top} \SB{Also check, since the ratio in dex can not be negative! two errorbars are in $<0$ region!} \DK{corrected} \SB{Good!}
    %\SK{May add MW data here too. If they have lower UV excitation, we will see the difference in f1,2 values, right?} \SB{In principle, this is the case, MW have lower f1 and f2. So it can be reasonable to try. The problem is that the points can be too crowded along the curves.} \DK{I agree, it will be too complicated to distinguish points from Magellanic Clouds}\SK{Can you try to add it?} \SB{and what about high-z? For completeness...} \SB{You may make figure wider (2column), since the data is elongated along horizontal...}
    %\DK{added MW and DLA; hope, I haven't made a mistake while recalculating Jenkins2011 data} 
    %\SB{THanks! looks good. Can you try make figure  (2column) and trim y axis by $<1.7$?} \DK{something like that?} 
    %\SK{looks good. Please, reduce the font size to make it consistent with font size in others figures.} \SB{yes, and that will be all fine} \DK{reduced}
    }
    \label{fig:f1_f2}
\end{figure*}

It is well known that bulk of H$_2$ J=0 and J=1 reside in the inner, self-shielded part of the medium \citep[see e.g.][]{Balashev2009, Klimenko2020} reflecting its thermal state. As it was done previously in many studies, we used H$_2$ ortho-para (determined by J=0, 1) excitation temperature, $T_{01}$ to constrain the kinetic temperature of cold phase of ISM, probed by H$_2$/\CI\ absorptions.
Fig.~\ref{fig:H2_T} shows H$_2$ excitation temperature as a function of H$_2$ column density, measured in Magellanic Clouds (taken from paper I), %(red diamonds and blue squares for LMC and SMC, respectively) 
in our Galaxy \citep{Savage1977, Gillmon2006}, and at high-redshifts \citep[see e.g.][and references therein]{Balashev2019}. One can see that Magellanic Cloud sample, consistent with both high-redshift and MW sample over most of the range of $N_{\rm H_2}$ column density. However, we note that recently an extensive survey of H$_2$ in our Galaxy based on FUSE data by \citealt{Shull2021} reveals higher $T_{01}$ temperatures, that were previously derived by \citealt{Savage1977, Gillmon2006, Rachford2002, Rachford2009}. This discrepancy if it is real, devotes an additional study. %\SB{Is Shull2021 say something about it?} \DK{They just say that their average T is lower than Savage1977 found ($88\pm 20$ vs $77\pm 17$) and in the figure 7 they also show results of Rachford 2002, 2009, and results of Shull2021 are systematically higher than that of Rachord 2002, 2009, but it is not discussed} 
Apart from this, we definitely see that our measurements confirm the trend of decreasing $T_{01}$ with increase of H$_2$ column density for the Magellanic Cloud sample \citep[previously reported for the MW and high-z samples by][]{Muzahid2015, Balashev2017, Klimenko2020}.  However, due to large scatter, the correlation for Magellanic Clouds is weak, but significant %(Pearson correlation coefficient and $p$-values are for both LMC and SMC are about -0.3 with p-values 0.018 for LMC and 0.017 for SMC).
(Pearson correlation coefficients are  about $-0.16$ and -0.14 with p-values $5.3\times10^{-5}$ and $3.7\times 10^{-5}$ for LMC and SMC, respectively). We find a similar correlation for the sample of high-redshift DLAs ($r = -0.14$ with $p$-value of 0.0006), which cover the same range of H$_2$ column densities. The correlation in the MW sample (using temperatures reported by \citealt{Savage1977, Gillmon2006} data) is stronger: $r=-0.49$ with $p$-value $1.4\times 10^{-7}$. However the MW sample probe more saturated H$_2$ gas clouds, than other samples. %\DK{do we really need to put it here?} \SK{I rephrased it a little.} \SB{I think it is fine!}

\subsection{Excitation of fine structure levels}

%\SB{A Jenkins plot, n$_2$/n$_0$ vs n$_1$/n$_0$? Plus temperature? Or n$_2$/n$_0$ vs T?} \DK{added}
%\SK{Please, check the jenkins plot, they defined $f_1 = N(CI∗)/N(CI_{total})$ and $f_2 = N(CI∗∗)/N(CI_{total})$, but not as a ratio J=1,2 to J=0 level. Why do you plot other parameters?} 
%\DK{it was more convenient when we plotted this figure, and we did not want to redraw it, since it will take some time. In the text the difference with Jenkins notation is stated}

In the Fig.~\ref{fig:f1_f2} we plot \CI\ column density ratios $f_1' = N(\mbox{\CI}^*) / N(\mbox{\CI})$ vs $f_2' = N(\mbox{\CI}^{**}) / N(\mbox{\CI})$\footnote{Note that \citealt{Jenkins2001, Jenkins2011} used $f_1 = N(\mbox{\CI}^{*}) / N(\mbox{\CI}_{\rm tot})$ and $f_2 = N(\mbox{\CI}^{**}) / N(\mbox{\CI}_{\rm tot})$ in their works. However, the plots $f_1$-$f_2$ and $f_1'$-$f_2'$ look quite similar.}. 
measured in Magellanic Clouds, Milky Way and high-redshift galaxies. One can see that measurements at high redshifts show large dispersion, as Magellanic Clouds values, while measurements in our Galaxy reveal systematically lower excitation of the upper fine-structure levels. The possible explanation is lower UV field intensity in Milky Way (as the main sources of \CI\, excitation are collisions and excitation by UV radiation and CMB) and it will be discussed below.
We compare these measurements with model curves that were calculated %assuming the \CI\ fine-structure excitation by cosmic microwave background (CMB) radiation and collisions
for $T = 50, 100, 200$\,K in the range of number densities $n = 1 - 10^{5}\,{\rm cm^{-3}}$. Regarding collisional partner we considered two limiting cases: atomic and molecular hydrogen, denoted it by the molecular fraction parameter, $f_{\rm H_2}=2 n({\rm H_2}) / (2 n({\rm H_2}) + n({\rm HI}))$. One can see that measured values follow model tracks, indicating that we likely have no drastic misfits of the line profiles. Importantly, the relatively large uncertainties do not allow to constrain temperature, number density and molecular fraction, using only the measurements on $f_1'$ and $f_2'$ ratios, since the parameter space is degenerate. %\cite[see e.g.][]{Jenkins2001, Jenkins2011}.  Nevertheless \CI\, excitation allows constraining physical conditions in the system. 
\citealt{Jenkins2001, Jenkins2011} described a method how to constrain number density and fraction of low-pressure gas from \CI\ fine-structure level population, using the additional estimates on the temperature and UV field intensity. To get constraints on $\chi$ their model requires \CII\, column densities (using ionization balance for carbon and hydrogen together, see Equations 1 and 2 in \citealt{Jenkins2011}), since carbon is mostly ionized by UV radiation.
%\SB{We should describe that they used \CII, that different from  ours} \DK{added}. 
In our work, we followed a different approach, where the supplemental data was obtained from fit to H$_2$ rotational levels population \citep[see][]{Klimenko2020}. This allows us to get constraints on $n_{\rm H}$ and $\chi$, omitting additional calculations of \CII\ column density (where we have lack of the direct measurement) and of low and high pressure gas fractions. This is discussed in the next Section. %\SB{Check if this correct} \DK{everything is correct}
 
\section{Physical conditions}
\label{sect:phys_properties}

Number density and UV field strength
%\footnote{in the following we provide the UV field strenght, $\chi$, in the Mathis field units, \citealt{Mathis1983}} 
can be constrained from comparison of the measured populations of H$_2$ rotational and \CI\ fine-structure levels with ones calculated on the grids of models using \Meudon\ code \citep{Klimenko2020}. It was applied systematically to systems at high redshifts \citep{Balashev2019, Balashev2020b, Klimenko2020, Kosenko2021} and also to few systems in the Milky Way, LMC and SMC \citep{Klimenko2020}.   

In this work we used only two lower H$_2$ rotational levels ($\rm J \leq 1$), which contains the most of H$_2$ and basically reproduce the thermal balance in the medium since these levels are predominantly populated by collisions and \footnote{Ortho-para ratio of H$_2$ is set by collisions with H$^{+}$, H, H$_2$, He and H$^{+}_3$ \citep{LePetit2006}, %\SB{give ref to Le Petit.} \DK{added; also \HI, H$_2$, \HeI\, and H$^{+}_3$ influence on the OPR included in \Meudon},
$J=2$ level in most cases populated by collisions with H and H$_2$ only in the self-shielded part of the H$_2$-bearing medium \cite[see e.g.][]{Balashev2009}}. The higher H$_2$ rotational levels ($\rm J \ge 2$) can be significantly populated by UV pumping and therefore may provide a direct measurement of the UV field. However, UV pumping is very sensitive to saturation of H$_2$ resonant lines (as well as the shielding does), and therefore is very sensitive to the exact geometrical model, which is impossible to constrain in observations towards one sightline. In turn, excitation of two lower rotational levels of H$_2$ is sensitive to the thermal balance in the bulk of the medium, which roughly  $\propto \chi / n_{\rm H}$ and/or $\zeta / n_{\rm H}$, where $n_{\rm H}$ is a hydrogen gas density, and $\zeta$ is the Cosmic Ray Ionization Rate (CRIR). For high H$_2$ column densities, considered in this study, bulk of the medium represents region of the cloud, where H$_2$ is self-shielded, and hence significantly less biased by the exact geometry of the cloud and anisotropy of the UV field. Additionally, the column density of ground ortho and para levels usually significantly higher than $\rm J \ge 2$ levels, and typically show Lorentzian wings. This makes constraint of the column density less affected by choice of the exact velocity structure of the absorber (which is complicated in FUSE data) and the degeneration with the Doppler parameter. Finally, description of high H$_2$ rotational levels might require dynamical models in contrast to static \Meudon\ models used here. Indeed, $\rm J \ge 2$ can be significantly excited in the outer layers of the cloud \citep[see e.g.][]{Abgrall1992, Balashev2009}, where H$_2$ is not yet self-shielded, and due to hydrodynamical motions this region can be larger then static models predict, see detailed discussion in \citealt{Klimenko2020}. In general, we found that taking into account H$_2$ ($\rm J \ge 2$) induces systematic increase in derived $\chi$, that requires additional detailed studies. 

\begin{figure}
    \centering
    \includegraphics[width=\linewidth]{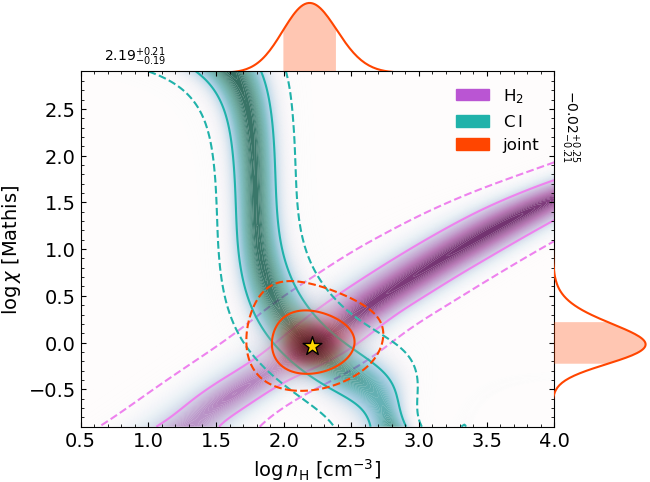}
    \caption{Estimate on the hydrogen gas density and UV field intensity for the system towards Sk-67 5 in the LMC. The green, violet and red color show the constraints obtained using from \CI, H$_2$ and joint analysis. Solid and dashed curves show 0.683 (1$\sigma$) and 0.954 (2$\sigma$) credible intervals for the posterior distribution function, respectively. %\SB{Check axis labels. In the text we used $n$ and $\chi$. May be we should used $n_{\rm H}$, throught the text.} \DK{made new pictures, I think it is simplier then correcting the text} 
    On the top and left axis we show the marginalized 1d distribution of $n_{\rm H}$ and $\chi$, respectively, where filled regions correspond to 1$\sigma$ estimates, and provide the estimated value in the attached text labels. %\SB{Mb, shrink x-axis to 4.2? To hide the edge of H2 region, that is due to smoothing} \DK{corrected}
    }
    \label{fig:n_chi_Sk675}
\end{figure}

Following \citealt{Klimenko2020, Kosenko2021} we obtained a region of $n_{\rm H}$ and $\chi$ values corresponding to fit to the measured column densities of $\rm J=0,1$ levels of H$_2$ levels and \CI\ fine-structure populations. We calculated two grids of constant hydrogen density models with metallicities corresponding to the average metallicities on the LMC and SMC. For each grid we varied $\log n_{\rm H}$ (in units cm$^{-3}$) and $\log \chi$ (in units of Mathis field \citep{Mathis1983}) in the ranges [0, 4.5] and [-1, 3], respectively, with the steps 0.5 for both $\log n_{\rm H}$ and $\log\chi$. Note that \Meudon\ used the total hydrogen gas density $n_{\rm H}=n({\rm HI})+2n({\rm H_2})$, which close to the gas number density $n_{\rm gas}\approx n({\rm HI})+n({\rm H_2}) + n(\rm He)$ at low molecular fraction, but can be 2 about times higher for the fully molecular region.
%The strength of UV field is measured in units of Mathis field \citep{Mathis1983}. 
We also assumed the CRIR to be linearly scaled with UV field as $\log\zeta_{-16} = \log\chi$, where $\zeta_{-16}$ is a primary ionization rate of the hydrogen atom in the units of $10^{-16}\rm\,s^{-1}$. We made this assumption since \citealt{Bialy2019, Balashev2022} showed that cosmic rays may impact much on the thermal state of the diffuse medium, especially at low metallicity. The linear dependence was taken for simplicity assuming that UV field and cosmic rays are both produced in the star-formation region and likely on average scaled with star-formation rate. 
For \CI\ we used a relative population of levels, since the ionization state of carbon depends on the chemistry (and hence several physical parameters) and dust properties. In turn, excited \CI\ fine-structure levels, i.e. \CI$^*$ and \CI$^{**}$, are mostly populated by the collisions (which determined by $n_{\rm H}$ and $T$), excitation by the CMB and pumping by UV radiation. However, the latter is typically dominated only at quite high values of UV field \citep[see e.g.][]{Balashev2019}. An example of the constraints on $n_{\rm H}-\chi$ derived from population of \CI, low rotational levels of $\rm H_2$ and joint analysis is shown in the Fig.~\ref{fig:n_chi_Sk675}. %\SB{Plot the Figure?} \DK{added; I think we can add other figures in the appendix} \SB{Or supplemental material :)}. 
One can see that while individual constrains from \CI\ and $\rm H_2$ are significantly degenerated, they have different dependence in $n_{\rm H}-\chi (\zeta)$ parameter space, and hence the joint constraint is much tighter than individual ones. 

We applied the described method to the systems in the LMC and SMC to get constraints on $n_{\rm H}$ and $\chi(\propto\zeta)$. We used results of H$_2$ fit in spectra obtained with FUSE telescope with resolution $R = \lambda/\Delta\lambda \lesssim 20000$, which is much less then resolution for the most HST data used in this work. Therefore FUSE data does not allow to accurately resolve velocity structure as HST data does and we cannot unambiguously associate H$_2$ and \CI\, components. Hence we used column densities summarized over all of the Magellanic Clouds components in our analysis.
The results are summarize in Tables~\ref{tab:LMC_phys_cond} and \ref{tab:SMC_phys_cond}, and shown in Fig.~\ref{fig:n_chi}, while the constraints on individual systems are presented in Appendix~\ref{sect:phys_constr}.
%\SK{Is it correct that we apply this method to the CI column densities summarized over components? E.g. AV26 consists of 2 components and has only one value of physical conditions in Table 4.} \SB{We need to say it here explicitly, that we used total column densities, since FUSE data is not enough to accurately resolve velocity structure...} \DK{agree, added}
We found that the hydrogen gas densities derived in the LMC and SMC are consistent with the values corresponding to the diffuse cold ISM, $\sim 30-10^3\,\rm cm ^{-3}$, and with values measured in other H$_2$-bearing absorption systems in the MW and high-$z$ DLAs. The average values of hydrogen gas densities in both the SMC and LMC are close, $\log n_{\rm H}\simeq2.5$ with dispersion 0.4 for the LMC and $\log n_{\rm H}\simeq2.6$ and with dispersion 0.6 for the SMC, respectively. %\SB{with dispersion ...} \DK{added; used logarithm of values, since distribution of decimal values is highly asymmetric} \SB{fine!}
Interestingly, that both the SMC and LMC sightlines indicate systematically higher UV field, than ones measured near Solar vicinity, which is in line with previous studies of Magellanic clouds \citep[e.g.][]{Bernard2008, Sandstrom2010, Welty2016, Roman_Duval2021}. The mean value of UV field intensity is $\log\chi \sim 0.3$ with a dispersion 0.5 for the SMC sightlines. This is lower than the mean value in the LMC sightlines $\sim 0.7$ with dispersion 0.4.

Using obtained H$_2$ excitation temperatures and hydrogen densities we found that the thermal pressures\footnote{One should bear in mind the above remark about density notation in \Meudon\, since to get pressures one should use the gas number density, $n_{\rm gas}$. We derived the gas number density as an average over the Meudon model (which assumes constant hydrogen gas density) at particular H$_2$ column density and physical parameters. 
%in the systems and \citealt{Sternberg2014, Bialy2016} model
} are on average $\log p/k \simeq 4.2$ and 4.3 with dispersion of about 0.4 and 0.5 for the LMC and SMC, respectively.% \DK{added recalculated values of pressure}.
This is consistent with the values found by \citealt{Klimenko2020} and by \citealt{Jenkins2021} for both Magellanic Clouds (their $\log p/k\simeq 4.1\pm 0.5$ and $\simeq 4.0\pm 0.2$, respectively) and \citealt{Roman_Duval2021} for the LMC sample (their $\simeq 3.9\pm 0.5$). 

%\SK{The next part is moved to Sect 6.2}
%[These values are much higher than the average value of thermal pressure in diffuse gas in local ISM, $\log p/k=3.6\pm0.2$ \citep{Jenkins2011}. Such a high value was observed in local ISM only in strong CO absorption systems (with $N({\rm CO})>10^{16}\,{\rm cm^{-2}}$, see, e.g., \citealt{Welty2020, Federman2021}), which are related to translucent and dense phase, where carbon is mainly in molecular state (CO) and the cooling is dominated by \CI\ and CO emission. Such high values of density and pressure in the diffuse phase in the LMC and SMC can be explained by high strength of UV field and CRIR, and lower metallicity. Indeed the phase diagram of the neutral medium strongly depends on aforementioned quantites \citep{Bialy2019, Balashev2022}. \SB{I disagree with below. It is likely explained by dependence of the phase diagram on UV, CRIR a nd metallicity} While gas density is already high \SB{But H2 content is not so}, strong UV field does not allow  \CII\ to convert in \CI\ and CO. Indeed, we see in Fig.\,\ref{fig:CI_H2} that on average \CI\ column density in Magellanic Clouds is less than in the Milky-Way meaning that \CI\ absorptions in the LMC and SMC still likely probe diffuse phase (where the cooling is dominated by \CII\ emission).

%\SK{Can we estimate carbon fraction in \CI\ in SMC and LMC? Maybe it is already classified as "dense" phase? if it higher than 0.01}

%\SB{Should we put parts of this Section with  comparison into comparison sections, that are below????}

\renewcommand{\arraystretch}{1.35}
\setlength{\tabcolsep}{3pt}
\begin{table}
    \centering
     \caption{Physical conditions derived in the absorption systems associated with the LMC}
     \begin{tabular}{cccccc}
      \hline 
      Star & $\rm [X/H]$ & $\rm X$ & $\log n_{\rm H} [\rm cm^{-3}]$ & $\log \chi$ & $\log\alpha G$ \\
      \hline 
      Sk-67 2 & $-0.34^{+0.05}_{-0.05}$ & Zn & $2.30^{+0.21}_{-0.18}$ & $1.15^{+0.25}_{-0.37}$ & $0.88^{+0.30}_{-0.48}$ \\
      Sk-67 5 & $-0.58^{+0.05}_{-0.05}$ & Zn & $2.19^{+0.21}_{-0.19}$ & $-0.02^{+0.25}_{-0.21}$ & $-0.11^{+0.30}_{-0.32}$\\
      Sk-67 20 & -- & -- & $2.59^{+0.22}_{-0.20}$ & $0.33^{+0.23}_{-0.21}$ & $-0.24^{+0.29}_{-0.33}$ \\
      PGMW 3070 & -- & -- & $2.49^{+0.25}_{-0.22}$ & $0.18^{+0.23}_{-0.24}$ & $-0.29^{+0.31}_{-0.38}$ \\
      LH10 3120 & $-0.75^{+0.10}_{-0.10}$ & Zn & $2.46^{+0.20}_{-0.19}$ & $-0.06^{+0.19}_{-0.0.21}$ & $-0.38^{+0.26}_{-0.32}$ \\
      PGMW 3223 & $-0.57^{+0.08}_{-0.08}$ & Zn & $2.72^{+0.29}_{-0.24}$ & $0.29^{+0.28}_{-0.24}$ & $-0.33^{+0.36}_{-0.49}$ \\
      Sk-66 35 & $-0.06^{+0.08}_{-0.20}$ & S & $2.16^{+0.22}_{-0.21}$ & $0.38^{+0.37}_{-0.32}$ & $0.17^{+0.40}_{-0.45}$ \\
      Sk-66 51 & -- & -- & $2.26^{+0.74}_{-0.32} $ & $\lesssim 1.2$ & $\lesssim 0.96$ \\
      Sk-70 79 & $-0.63^{+0.06}_{-0.05}$ & Zn & $2.76^{+0.26}_{-0.23}$ & $1.50^{+0.27}_{-0.34}$ & $0.85^{+0.34}_{-0.50}$ \\
      Sk-68 52 & $-0.69^{+0.05}_{-0.05}$ & Zn & $3.09^{+0.42}_{-0.34}$ & $0.97^{+0.33}_{-0.29}$ & $0.01^{+0.48}_{-0.57}$ \\
      Sk-71 8 & $-0.69^{+0.03}_{-0.03}$ & S & $3.02^{+0.44}_{-0.31}$ & $1.11^{+0.29}_{-0.27}$ & $0.22^{+0.48}_{-0.51}$ \\
      Sk-69 106 & -- & -- & $3.55^{+0.45}_{-0.28}$ & $1.03^{+0.27}_{-0.24}$ & $-0.50^{+0.48}_{-0.44}$ \\
      Sk-68 73 & $-0.58^{+0.05}_{-0.05}$ & Zn & $2.83^{+0.34}_{-0.27}$ & $1.15^{+0.36}_{-0.29}$ & $0.42^{+0.44}_{-0.48}$ \\
      Sk-67 105 & $-1.09^{+0.07}_{-0.07}$ & Zn & $2.82^{+0.33}_{-0.29}$ & $0.54^{+0.28}_{-0.25}$ & $-0.08^{+0.39}_{-0.46}$ \\
       BI 184 & $-1.10^{+0.11}_{-0.09}$ & P & $2.57^{+0.29}_{-0.24}$ & $0.46^{+0.30}_{-0.26}$ & $0.09^{+0.38}_{-0.42}$ \\
      Sk-71 45 & $-0.50^{+0.05}_{-0.05}$ & Zn & $2.89^{+0.29}_{-0.25}$ & $0.51^{+0.25}_{-0.25}$ & $-0.30^{+0.35}_{-0.42}$ \\
      Sk-71 46 & -- & -- & $2.54^{+0.24}_{-0.21}$ & $1.13^{+0.36}_{-0.28}$ & $0.61^{+0.40}_{-0.41}$ \\
      Sk-69 191 & $-0.42^{+0.03}_{-0.03}$ & S & $2.37^{+0.24}_{-0.21}$ & $0.36^{+0.27}_{-0.27}$ & $0.05^{+0.33}_{-0.40}$ \\
      BI 237 & $-0.79^{+0.09}_{-0.08}$ & Zn & $2.08^{+0.36}_{-0.29}$ & $1.04^{+0.44}_{-0.45}$ & $1.11^{+0.50}_{-0.72}$ \\
      Sk-68 129 & $-0.70^{+0.05}_{-0.05}$ & Zn & $2.24^{+0.17}_{-0.17}$ & $1.22^{+0.20}_{-0.33}$ & $1.11^{+0.24}_{-0.42}$ \\
      Sk-66 172 & $-0.68^{+0.11}_{-0.10}$ & Zn & $3.21^{+0.32}_{-0.27}$ & $0.37^{+0.27}_{-0.30}$ & $-0.72^{+0.38}_{-0.50}$  \\
      BI 253 & $-0.66^{+0.05}_{-0.05}$ & Zn & $2.40^{+0.55}_{-0.36}$ & $0.98^{+0.57}_{-0.45}$ & $0.70^{+0.67}_{-0.84}$ \\
      Sk-68 135 & $-0.67^{+0.06}_{-0.05}$ & Zn & $2.24^{+0.19}_{-0.17}$ & $0.98^{+0.50}_{-0.25}$ & $0.86^{+0.51}_{-0.34}$ \\
      Sk-69 246 & $-0.65^{+0.05}_{-0.05}$ & Zn & $2.31^{+0.22}_{-0.19}$ & $0.65^{+0.39}_{-0.30}$ & $0.46^{+0.41}_{-0.41}$ \\
      Sk-68 140 & $-0.92^{+0.05}_{-0.06}$ & Zn & $2.59^{+0.23}_{-0.20}$ & $1.21^{+0.33}_{-0.28}$ & $0.79^{+0.37}_{-0.40}$  \\
      Sk-71 50 & $-0.52^{+0.05}_{-0.05}$ & Zn & $2.11^{+0.20}_{-0.19}$ & $0.66^{+0.48}_{-0.33}$  & $0.63^{+0.49}_{-0.44}$\\
      Sk-69 279 & $-0.85^{+0.08}_{-0.07}$ & Zn & $1.62^{+0.27}_{-0.42}$ & $0.54^{+0.38}_{-0.42}$ & $0.94^{+0.42}_{-0.91}$ \\
      Sk-68 155 & $-0.66^{+0.05}_{-0.05}$ & Zn & $2.20^{+0.38}_{-0.30}$ & $1.10^{+0.43}_{-0.49}$ & $1.02^{+0.50}_{-0.80}$ \\
      Sk-70 115 & $-0.79^{+0.15}_{-0.11}$ & Zn & $2.47^{+0.22}_{-0.22}$ & $0.50^{+0.29}_{-0.26}$ & $0.18^{+0.33}_{-0.40}$ \\
      \hline
    \end{tabular}
    \begin{tablenotes}
     \item The columns are: (i) name of the star; (ii) estimated metallicity; (iii) species that is used to derive metallicity; (iv) the hydrogen gas density; (v) the UV field strength in the units of Mathis field; (iv) $\alpha G$ parameter (see Sect.~\ref{sec:thermal}).
     \item Upper limits were constrained from $3\sigma$ credible interval
    \end{tablenotes}
    \label{tab:LMC_phys_cond}
\end{table}

\begin{table}
    \centering
    \caption{Physical conditions derived in the absorption systems associated with the SMC.}
    \begin{tabular}{cccccc}
         \hline 
      Star & $\rm [X/H]$ & $\rm X$ & $\log n_{\rm H} [\rm cm^{-3}]$ & $\log \chi$ & $\log\alpha G$ \\
      \hline 
      AV 15 & $-1.06^{+0.03}_{-0.04}$ & S & $2.70^{+0.32}_{-0.28}$ & $-0.47^{+0.31}_{-0.29}$ & $-0.97^{+0.40}_{-0.50}$ \\
      AV 26 & $-0.96^{+0.06}_{-0.06}$ & Zn & $2.13^{+0.21}_{-0.20}$ & $-0.25^{+0.36}_{-0.34}$ & $-0.20^{+0.39}_{-0.36}$ \\
      AV 47 & $-0.83^{+0.05}_{-0.05}$ & Zn & $\gtrsim 1.8$ & $0.08^{+0.91}_{-0.38}$ & $\lesssim 0.44$ \\
      AV 69 & $-1.04^{+0.04}_{-0.03}$ & S & $2.60^{+0.72}_{-0.43}$ & $0.23^{+0.61}_{-0.59}$ & $-0.38^{+0.80}_{-1.47}$ \\
      AV 75 & $-1.23^{+0.04}_{-0.04}$ & S & $3.72^{+0.35}_{-0.44}$ & $0.92^{+0.32}_{-0.33}$ & $-0.58^{+0.42}_{-0.77}$ \\
      AV 80 & $-1.16^{+0.05}_{-0.05}$ & Zn & $2.33^{+0.22}_{-0.20}$ & $0.59^{+0.43}_{-0.42}$ & $0.47^{+0.45}_{-0.56}$ \\
      AV 81 & -- & -- & $2.86^{+0.74}_{-0.48}$ & $\lesssim 1.36$ & $\lesssim 0.63$ \\
      AV 207 & $-0.88^{+0.07}_{-0.06}$ & Zn & $\gtrsim 1.82$ & $1.34^{+0.41}_{-0.84}$ & $\lesssim 1.69$ \\
      AV 210 & $-0.73^{+0.06}_{-0.06}$ & Zn & $2.98^{+0.55}_{-0.47}$ & $0.05^{+0.57}_{-0.42}$ & $-0.80^{+0.67}_{-1.03}$ \\
      AV 215 & $-0.96^{+0.05}_{-0.06}$ & Zn & $2.29^{+0.27}_{-0.23}$ & $0.23^{+0.60}_{-0.39}$ & $0.12^{+0.61}_{-0.55}$ \\
      AV 216 & $-0.80^{+0.05}_{-0.05}$ & Zn & $2.95^{+0.46}_{-0.31}$ & $0.24^{+0.21}_{-0.63}$ & $-0.56^{+0.47}_{-1.10}$ \\
      AV 266 & -- & -- & $2.11^{+0.41}_{-0.30}$ & $\lesssim 1.4$ & $\lesssim 1.42$ \\
      AV 372 & $-1.06^{+0.05}_{-0.05}$ & Zn & $2.60^{+0.22}_{-0.78}$ & $\lesssim 1.3$ & $\lesssim 0.90$ \\
      AV 476 & $-0.88^{+0.07}_{-0.10}$ & Zn & $1.60^{+0.46}_{-0.37}$ & $0.26^{+0.40}_{-0.57}$ & $0.83^{+0.53}_{-1.15}$ \\
      AV 479 & $-0.86^{+0.06}_{-0.07}$ & Zn & $2.75^{+0.29}_{-0.29}$ & $0.05^{+0.37}_{-0.32}$ & $-0.54^{+0.42}_{-0.54}$ \\
      AV 488 & $-0.81^{+0.05}_{-0.06}$ & Zn & $1.93^{+0.45}_{-0.29}$ & $\lesssim 1.4$ & $\lesssim 1.62$ \\
      AV 490 & $-1.06^{+0.05}_{-0.06}$ & P & $1.83^{+0.18}_{-0.16}$ & $-0.19^{+0.99}_{-0.33}$ & $0.18^{+0.99}_{-0.41}$ \\
      Sk 191 & $-1.51^{+0.06}_{-0.06}$ & Zn & $3.08^{+0.34}_{-0.27}$ & $0.63^{+0.33}_{-0.28}$ & $-0.20^{+0.42}_{-0.47}$ \\
      \hline 
    \end{tabular}
    \begin{tablenotes}
     \item The columns are: (i) name of the star; (ii) estimated metallicity; (iii) species that is used to derive metallicity; (iv) the hydrogen gas density; (v) the UV field strength in the units of Mathis field; (vi) the cosmic ray ionization rate;  (iv) $\alpha G$ parameter (see Sect.~\ref{sec:thermal}).
     \item Upper limits were constrained from $3\sigma$ credible interval
    \end{tablenotes}
    \label{tab:SMC_phys_cond}
\end{table}

\begin{figure}
    \includegraphics[width=1\linewidth]{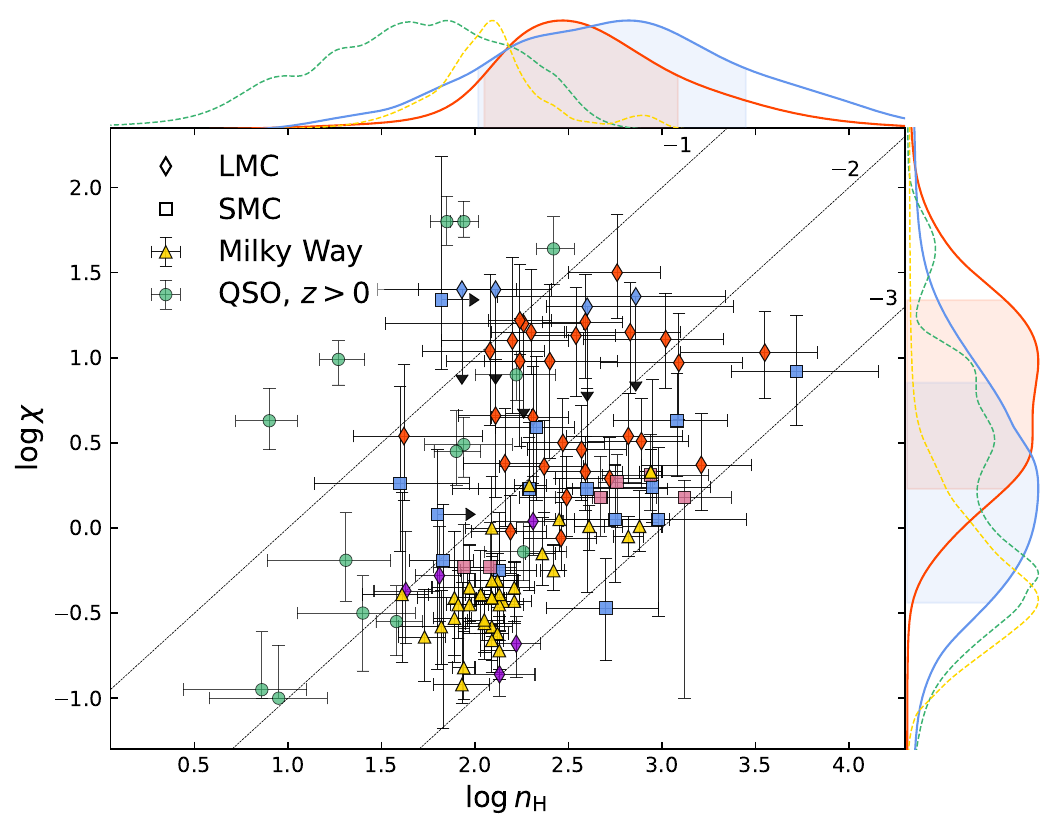}
    \caption{Estimated hydrogen gas density, $n_{\rm H}$, and UV field intensity, $\chi$ (in Mathis units), in the LMC and SMC. The blue and pink squares show measurements in SMC found in this work (see Table~\ref{tab:SMC_phys_cond}) and previously by \citealt{Klimenko2020}, respectively. The red and violet diamonds correspond to measurements in the LMC found in this work (see Table~\ref{tab:LMC_phys_cond}) and previously by \citealt{Klimenko2020}, respectively. The yellow triangles represent the measurements in the Milky Way (from \citealt{Klimenko2020} and re-analysed data of H$_2$/C\,{\sc i} excitation from \citealt{Jenkins2011}, see Sect.~\ref{sec:compar_MW}), while the green circles show constraints for high-redshift systems \citep[][]{Klimenko2020, Kosenko2021}. The lines represent constant $\log\chi/n_{\rm H}$ ratios from $-1$ to $-3$ from top to bottom, respectively. The curves at the top and left sides show kernel density estimations of the samples of values measured in the LMC and SMC (red and blue curves, respectively, with shaded regions correspond to 0.683 ($1\sigma$) confidence intervals), Milky Way (yellow curves) and at high redshifts (green curves). 
    %\SB{Increase the text of the axis labels a bit. To make it consistent with Fig.6}
    %\SK{Good figure. Please make the font size bigger a little.}    \SK{Can you remove the MW point at nh=2.7 and uv=-1 (HD210121\_0 - we have only CO data for this system)? What are the points near nh=2.3-2.5 and uv =-0.9? I don't see them in my file, as well as in Fig. 6}     \SK{Let's add levels of the constant chi/nh ratio (-3,-2,-1)? }\DK{Done; the points were from the previous version of the file you send me, corrected then}
    %\SK{Can you add measurements in MW to this plot?} \DK{they are shown by yellow triangles}
    %\SB{It will be good for discussion to plot 1d KDE for the distributions of n and chi above top axis and left from left axis. see e.g. Figs. from Balashev+2019.}\SK{Add graph with kinetic temperature. Will be gas in LMC and SMC hotter than in MW due to higher UV intensity?} \SB{Yes, it will be good to provide it separately, in Sect. 4!} \DK{added UV vs T01 in figure \ref{fig:T_UV}, but I cannot see any trend there} \SB{May be you should try UV/n as a function of T? Also you can try to plot T as a function of NH2 in Section 4.}     \DK{uv/n vs T is shown in figure ~\ref{fig:uv_n_T}, but it is hard to see any trend there too. H2 vs T is shown in figure ~\ref{fig:H2_T}, but I doubt that it is necessary here}
    }
    \label{fig:n_chi}
\end{figure}

\begin{comment}
    \begin{figure}
    \centering
    \includegraphics[width=\linewidth]{figures/T_UV.pdf}
    \caption{UV field intensity vs H$_2$ excitation temperature \DK{maybe remove this figure?} \SB{I agree, since we have UV/n figure, which is more reasonable}
    }
    \label{fig:T_UV}
\end{figure}
\end{comment}

\begin{figure}
    \centering
    \includegraphics[width=\linewidth]{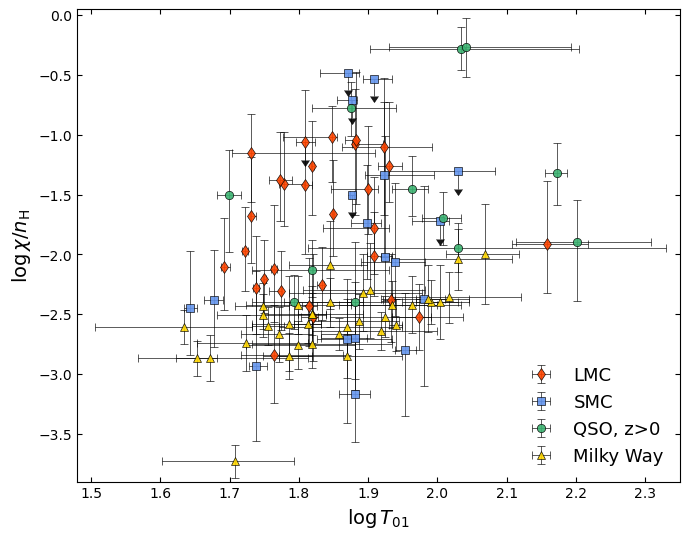}
    \caption{$\chi/n_{\rm H}$ vs H$_2$ excitation temperature estimated in various absorption systems. The symbols are the same as in Fig.~\ref{fig:n_chi}, with the data of $T_{01}$ for MW is taken as in Fig.~\ref{fig:H2_T}.  
    %for LMC (red diamonds), SMC (blue squares) measured in this work, and MW (yellow triangles) and high-redshift systems (green circles). 
    %\SK{Can you add MW data here? May be plot also dependence of chi/nH on metallicity? Since we already do not see any difference in T01 in Fig. 2, so what we expect to see here?}     \DK{But we see differences in $\chi$, therefore we can study impact on the thermal state of Magellanic Clouds, see section ``Thermal state''.  I think MW data will contaminate the figure, so I'd like not to put it here} \SB{The MW data indeed will complicate the figure, since yellow is already present. Moreover we can not trust $T_{01}$ in our galaxy.}     \DK{added MW and QSO}
    %\SB{Can shrink x-axis to 2.35. It will save space.} \DK{for B1444 there are large uncertainties on T, so it will be truncated} \SB{Daria, this large uncertainty (here and in some other Figures) has little meaning, but it takes one-third of the Figure width, making other important points crowded together. In principle, you can replace this point by lower limit or remove it from the figure.} \DK{removed the point}
    \label{fig:uv_n_T}
    }
\end{figure}

\begin{figure}
     \center{\includegraphics[width=1\linewidth]{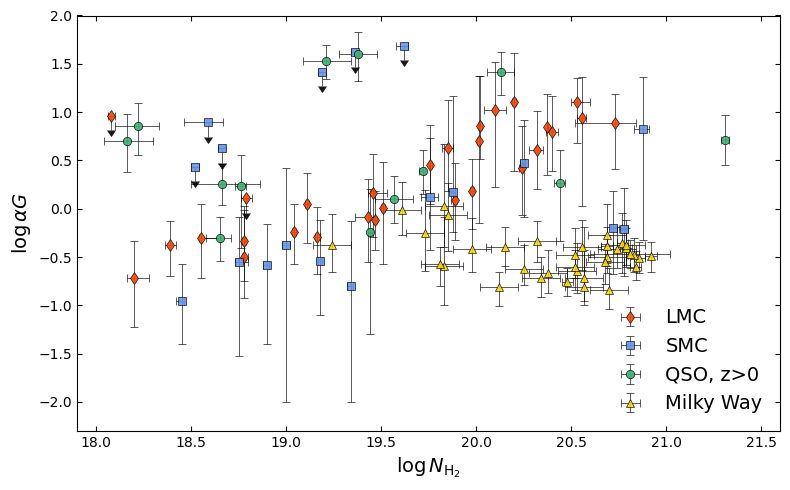}}
    \caption{$\alpha$G parameter (see Sect.~\ref{sec:thermal}) vs H$_2$ column densities estimated in various absorption systems. The symbols are the same as in Fig.~\ref{fig:n_chi}. %\SB{Increase a font a bit to make it constsistent with Fig.6}
    %for LMC (red diamonds), SMC (blue squares), MW (yellow triangles) and high-redshift systems (green circles).
    %\DK{Here I used approximation of $\alpha$G from \citealt{Bialy2016}, their formula (22)} %\SB{Provide the chosen value of R}. \DK{added about $\alpha G$ in the section 6.4} \SB{Right panel, should be N(CI) in axis label.} \DK{corrected} 
    %\SB{What we can say about increase of alphaG with H2??? Do we probe sites close to star-forming regions??} \DK{in general, $\alpha G$ parameter defines \HI/H$_2$ transition point, i.e. at  low $\alpha G$ self-shielding dominates and transition occurs at lower optical depth, than for high $\alpha G$ (which arises from high UV), where dust absorption is significant. We also can use $\alpha G$ to calculate SF threshold (as Bialy2016 done), but I am not sure we can judge about closeness to SF regions. Nevertheless most of our sightlines really seem to be close to the SF regions}
    }
    \label{fig:alphaG}
\end{figure}

\section{Discussion}
\label{sect:discussion}

\subsection{Comparison with previous the LMC and SMC measurements}
\label{sec:compar_MC}

Most of the systems in the LMC have been studied by \cite{Roman_Duval2019, Roman_Duval2021}, using HST data (in their course of METAL program). For almost all systems our measurements of \CI\ and metals column densities
%[results on column densities of \CI\ and metals[] 
agree with results of previous studies (some of disagreements are discussed above in Sect.~\ref{sec:individual}). \cite{Roman_Duval2021} also estimated number density and $\chi$ in these systems. They used approach described by \cite{Jenkins2001, Jenkins2011}, which  is based on the analysis of the location of the measured ratios $N(\mbox{\CI}^{*})/N(\mbox{\CI})_{\rm total}$ and $N(\mbox{\CI}^{**})/N(\mbox{\CI})_{\rm total}$ on the model tracks that themselves depend on $n_{\rm gas}$, $\chi$, $T$ and a fraction of low-pressure gas. In addition, the initial assumption about $\chi$ requires knowledge the \CII\ column density, which is difficult to constrain from observations, as HST spectra cover only two \CII\ lines, at 1334\AA\, and 2334\AA, where the first is strongly saturated and the second is weak. Therefore \citealt{Roman_Duval2021} estimated $N(\mbox{\CII})$ using $N(\mbox{\HI})$, which may lead to an additional uncertainty in UV field estimation because the fraction of \HI\ associated with cold \CI-bearing phase is not well known. %\SB{So what is the conslusion?} \DK{it seems that it makes this method more uncertain} \SB{Can be compare these values? Additional plot?}

Following the approach used in this work \cite[suggested by][]{Balashev2019, Klimenko2020}, it requires to know only H$_2$ rotational and \CI\, fine-structure level populations, which were measured in HST and FUSE spectra and directly related to cold \CI-bearing phase. In almost all cases our results on UV field intensities are consistent with estimates obtained by \citealt{Roman_Duval2021} (except few systems), but number densities in about half of the sample have been found much higher. Some discrepancies in the UV field intensities may arise from different estimates of \CI\, population of fine-structure levels, e.g. in the systems towards Sk-68 129, BI 253 and Sk-68 140 column densities of the \CI\, ground state seem to be overestimated in \citealt{Roman_Duval2021} therefore leading to the lower estimate of $\chi$ than ours. However, the discrepancy in the number density estimates is probably arisen from the difference of the methods. The molecular fraction for most of the systems in our sample is not large ($f_{\rm H_2} \lesssim 0.3$) so in the most cases difference between $n_{\rm H}$ (which is obtained by us) and $n_{\rm gas}$ (which is obtained by \citealt{Roman_Duval2021}) does not significantly bias the results\footnote{see note on the difference between $n_{\rm H}$ and $n_{\rm gas}$ in Section~\ref{sect:phys_properties} and previous footnote in the following Section.} and can not explain observed discrepancy.
Also one should note that we obtained H$_2$ column densities systematically higher than obtained by \citealt{Welty2012} (see discussion in the Paper I), which may lead to the different excitation temperatures, since \citealt{Roman_Duval2021} also used H$_2$ rotational temperature in their model and it may influence on their final results.

\subsection{Comparison with Milky Way}
\label{sec:compar_MW}
%\SB{Slava suggested to add this section.}

In Fig.~\ref{fig:n_chi} we compare our results on 
$\chi$ and $n_{\rm H}$ with values found in the Milky Way, that were obtained by reanalysis the observed excitation of \CI\ and H$_2$ following our method (Klimenko et al. in prep.)\footnote{We find that on average values of $\chi$ derived by our method are 0.5\,dex less than ones derived by \citealt{Jenkins2011}, while estimates on $n_{\rm H}$ are well consistent. 
%We also note, that at the same time \citealt{Jenkins2011} gave higher values of UV field intensity in MW (about 0.57 with the dispersion 0.34), which is consistent with  Magellanic Clouds values. 
However above we already discussed some limitations of the model, presented by \citealt{Jenkins2001, Jenkins2011}, and to be consistent in the comparison between the samples, we reanalysed sightlines from \citealt{Jenkins2011} by method used in this paper. %In Fig.\,\ref{fig:n_chi} we show values of $\chi$ and $n$ derived from fit to H$_2$ and \CI\ excitation in systems studied by \citealt{Jenkins2011}. 
}. 
One can see that values of UV field intensity in both the LMC and SMC are higher than in our Galaxy, 
% It can be naturally explained by the difference in the mean UV field between MW and Magellanic clouds, 
which is in line with previous studies \citep[e.g.][]{Bernard2008, Sandstrom2010, Welty2016, Roman_Duval2021}. Moreover the dispersion in Magellanic Clouds as well are higher, than in the MW. This can be explained since the MW sightlines mostly probe a solar vicinity away from the active star-formation regions. In turn, for Magellanic Clouds there may be a selection effect -- most of the stars are from star-forming regions, and hence if absorption system arisen from the nearby medium then it will be enhanced by local UV field. Also one can note that Magellanic Cloud systems probe the wider range of $\rm H_2$ column densities, $\log N({\rm H_2}) > 18$, while in case of the MW we are limited mostly by systems with $\log N({\rm H_2}) > 20$. Finally, sightlines in Magellanic Clouds probe different metallicity than the MW ones. These difference may affect the heating/cooling balance\footnote{Indeed, the cooling of the CNM is mostly by fine-structure line emission which is linearly scaled with gas phase elemental abundance, that depends on metallicity. The heating mostly determined \citep[see e.g.][]{Bialy2019} by cosmic ray heating, whose rate doesn't depend on the metallicity, and photoelectric heating, whose rate is scaled with dust to gas ratio, that can be non-linearly scaled with metallicity \citep{RemyRuyer2014, Balashev2022}. Finally, the decrease of dust to gas ratio increase the gas phase abundance of elements.}.  

%\SB{Actually the difference can be real, since the galaxies are different!}
%It can be partly explained that Milky Way values probe systems with high H$_2$ column densities (most of the systems have $\log N_{\rm H_2} > 20$), while Magellanic Cloud systems have wider range of $\log N_{\rm H_2} > 18$. One note that these values have been obtained with \citealt{Klimenko2020} model.  Also as metallicity in Milky Way is systematically larger than metallicity in both LMC and SMC, therefore leading to the higher heating rate (as the main source of heating in the diffuse ISM is grain photoelectric heating, which depends on the dust distribution and hence on the metallicity) and larger UV field intensities in the low-metallicity medium are needed to obtain the same temperatures as for system with higher metallicity. 
%\SB{It is a bit more complex.}

Our measurements indicate that thermal pressures in the LMC and SMC close to $4.1\pm 0.4$ that was obtained by \citealt{Klimenko2020} in local ISM. 
%and lower than values obtained by \citealt{Jenkins2011} (their $3.58\pm 0.18$).
However, 
%due to the difference in the strength of UV field between the MW and LMC and SMC, 
note that sightlines from the sample of \citealt{Klimenko2020} likely probe translucent phase of the cold ISM, while systems in our sample are likely related to diffuse phase, in terms of molecular carbon fraction and the cooling function.
The thermal pressures in the reanalysed MW sample from \citep{Jenkins2011} is found to be $\sim 3.7\pm 0.2$, that is slightly lower than what we found in the SMC and LMC samples. 

\begin{figure}
    \centering
    \includegraphics[width=\linewidth]{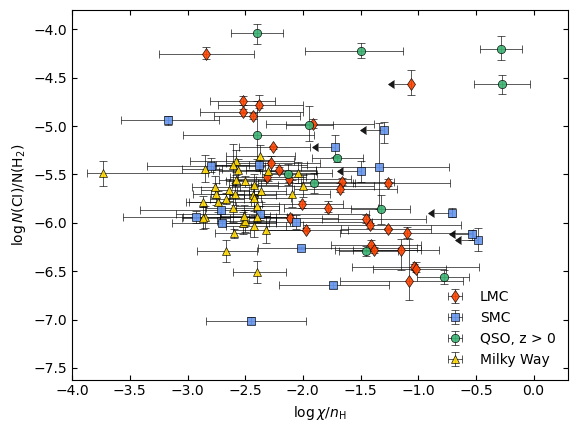}
    \caption{\CI/H$_2$ relative abundance vs ratio of UV field intensity to number density. The symbols are the same as in Fig.~\ref{fig:uv_n_T}. %estimated in LMC (red diamonds), SMC (blue squares), MW (yellow triangles) and high-redshift systems (green circles).
    %\SB{Need to add description of the figure here.} \DK{do we need to use this figure here? I do not understand where to put its description in the text} \SB{I agree that here, it is a bit out of scope to describe the chemical abundances. moreover ci/h2 ratio depends on other parameters. } \SB{Can you try ci/h2 vs Chi/n? The dependence should be even stronger} \DK{added; also added correlation coefficients in the text}
    }
    \label{fig:CI_H2_uv_n}
\end{figure}

\subsection{Comparison with $\rm H_2$-bearing DLAs at high z.}
\label{sec:compar_highz}

While the enhanced values of UV field that we found in Magellanic Clouds are consistent with some values measured at high redshifts DLAs, on average the hydrogen gas density and UV field strength determined at high redshifts are systematically lower than what we obtained in the SMC and LMC. This is likely due to the selection effects. First, high redshift DLAs probe the population of the field galaxies, which may have lower star-formation rate, than SMC and LMC. Second, in case of the SMC and LMC we probe the central star-forming region of the galaxies, while in case of quasar absorption lines we mostly probed the periphery of the distant galaxies due to large cross-section of these regions. Indeed, it was found that in a case of H$_2$-bearing DLAs thermal pressures and number densities increase with the increase in the total hydrogen column density \citep{Balashev2017, Balashev2019}, which likely anticorrelates with the impact parameter \citep{Krogager2017, Krogager2020} and therefore the distance to star-forming regions. 
%\SK{anticorrelates?} %\SB{Right!}
In that sense the LMC and SMC data should be compared with the hydrogen gas densities measured in the extremely saturated DLAs and towards GRB sightlines \citep{Ranjan2020}. The latter are represented by only in a few cases \citep[see e.g.][]{Balashev2017, Ranjan2018},  where the measurements of $n_{\rm H}$ are consistent with average values in the LMC and SMC. With addition of relatively low metallicities, this indicates that the LMC and SMC may be used as an interesting test case for studies of the central parts of the high-redshift galaxies.

\subsection{Thermal state}
\label{sec:thermal}
In static equilibrium the temperature of the ISM is determined from the thermal balance of the heating and cooling rates. Consequently in most simplest situation the temperature should depend on the ratio of $\log\chi/n_{\rm H}$ (or $\zeta/n_{\rm H}$, note, that we tight together $\chi$ and $\zeta$). 
In Fig.~\ref{fig:uv_n_T} we compare the obtained $\chi/n_{\rm H}$ and $T_{01}$ in different samples. Previously, \cite{Klimenko2020} reported that observationally $\chi/n_{\rm H}\propto T^{\alpha}$ and actually there may be a correlation between $T_{01}$ and $\chi/n_{\rm H}$, for which we get a power law index $\alpha = 1.6^{+1.3}_{-1.1}$ and $2.2^{+1.4}_{-1.3}$ for the LMC and SMC, respectively. %\SK{SI it a new results?} \DK{yes, in our sample}
However, the correlation is very weak (if it is real) -- Pearson correlation coefficients are 0.08 with p-value 0.69 and 0.14 with p-value 0.54 for the LMC and SMC, respectively. %\SK{Please, add p-value, since there is a high dispersion and just correlation coefficient can be a statistical fluctuation.} \DK{added} %\SB{Any correlation to report?} \DK{added}

Interestingly, that $\alpha G$ parameter\footnote{Following \citep{Sternberg2014, Bialy2016}, $\alpha$ parameter is a ratio of H$_2$ destruction rate (neglecting self-shielding) to the H$_2$ formation rate and $G$ defines the shielding of H$_2$ from UV radiation} that was introduced for description of the \HI/H$_2$ transition also depends on $\chi/n_{\rm H}$ ratio, apart from the metallicity dependence. Since the metallicities are also constrained in our sample, we provide $\alpha G$ parameter\footnote{To estimate $\alpha G$ parameter we used approximation of \citealt{Bialy2016}, where we used H$_2$ formation rate to be $R = 3\times 10^{-17}\varphi_{\rm g}Z$ cm$^3$s$^{-1}$, where $\varphi_{\rm g}$ depends on dust properties and close to unity \citep{Sternberg2014}} for the LMC and SMC samples in Tables~\ref{tab:LMC_phys_cond} and \ref{tab:SMC_phys_cond}, respectively. In Fig.~\ref{fig:alphaG} we compare the dependence of $\alpha G$ on H$_2$ column densities, measured in different systems. One can see that in case of Magellanic Cloud sample we obtained a large range of $\alpha G$: $\sim0.2-13$ and $\sim0.2 - 50$, for the LMC and SMC, respectively, while in the MW sample it is constrained to lie within $\sim0.1-1$. Higher values of $\alpha G$ in the LMC and SMC samples are consistent with $\alpha G$ that we measured in the high-redshift systems, $\sim0.5 - 40$. This can be connected with a higher value of UV field intensity and lower metallicity in both Magellanic Clouds and distant galaxies, leading to a lower H$_2$ abundance. 
Interestingly, that we also found a strong correlation of $\alpha G$ parameter with H$_2$ column density in our sample: Pearson 
%\SK{Spearman?} \SB{No, it is Pearson}
correlation coefficients are 0.67 with p-value $2\times 10^{-4}$ and 0.39 with p-value $0.05$  for the LMC and SMC, respectively. This likely indicates that higher H$_2$ column density systems in Magellanic Clouds probe the gas closer to the star-formation regions, where UV field is enhanced. This is directly confirmed by considering estimates of UV field obtained in our sample. On the other hand, Fig.~\ref{fig:CI_H2_uv_n} shows that in the LMC and SMC samples, the  relative abundance \CI/H$_2$ anti-correlates with the strength of UV field, and even more strongly with the ratio of $\chi/n_{\rm H}\propto \alpha G$. The correlation coefficient is $-0.66$ and p-value $2\times 10^{-4}$ for the LMC and $-0.19$ and p-value $0.35$ for the SMC. 
%\SB{Check both values seems to be too high from Fig.8!} \DK{I used here points estimations so it is so high. I can calculate it assuming uncertainties}. \SB{Ok}
This is in general in line with chemical models of diffuse ISM \citep[e.g.][]{Wolfire2008,Liszt2015}, and will be comprehensively explored in forthcoming paper (Balashev et al. in prep). 

%\SK{From figure it is not seen.} 

%\SK{We measure $\alpha G$ is in the range $xx-xx$ in the LMC and $xx-xx$ in the SMC, while it is found to be $0.1-1$ in the local ISM systems. High values of $\alpha G$ are consistent with what we measure in DLA systems, $\alpha G = xx-xx$. This can explained by ...}
%\SK{Fig. XX shows the correlation of $\alpha G$ and column density of H$_2$. We find that systems in different galaxies likely follow one relation: xxx.}
%One can see the increase of this parameter with the increasing of $\chi/n_{\rm H}$ and it is not surprising, since $\alpha G\propto \chi/n_{\rm H}$.

\begin{figure*}
    \centering
    \includegraphics[width=\linewidth]{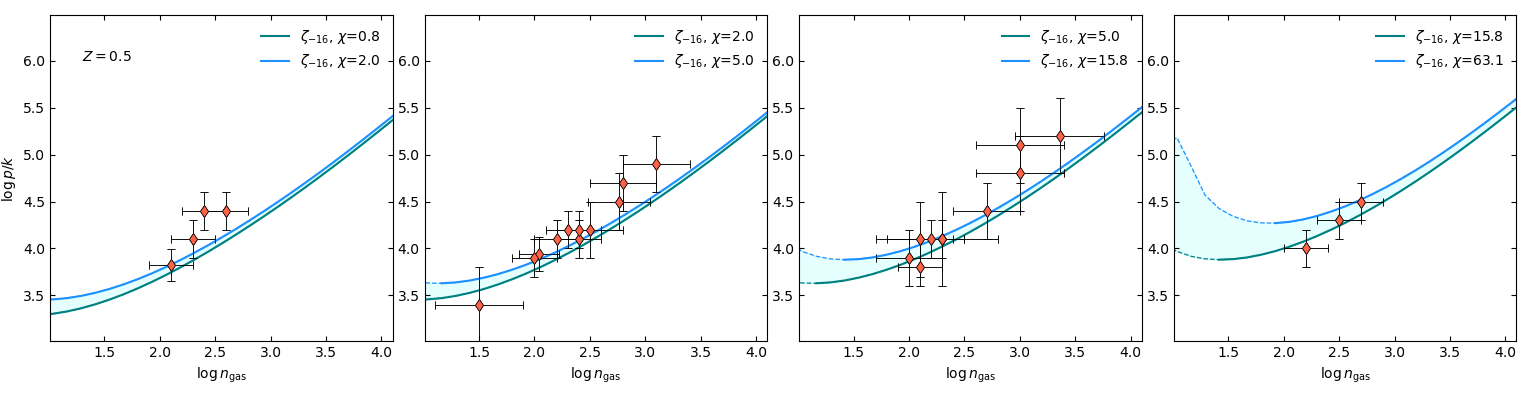}
    \caption{Pressure vs density (phase diagram) at the metallicity $Z = 0.5$, corresponding to the mean LMC metallicity. Data is divided into four bins relative to UV field intensity. The solid lines and the light-blue shaded area show model curves calculated respectively to the bins and indicated in the legend.  Red diamonds are values obtained for the LMC.  %\DK{added corrected values of p in the figures}
    %\SB{the colors of lines are a bit vivid, please change} \DK{changed and made more pale}
    %\SK{I don't see dashed lines, reported in the legend, in the Figure. Please, check it} \DK{corrected}
    }
    \label{fig:phase_LMC}
\end{figure*}

\begin{figure*}
    \centering
    \includegraphics[width=\linewidth]{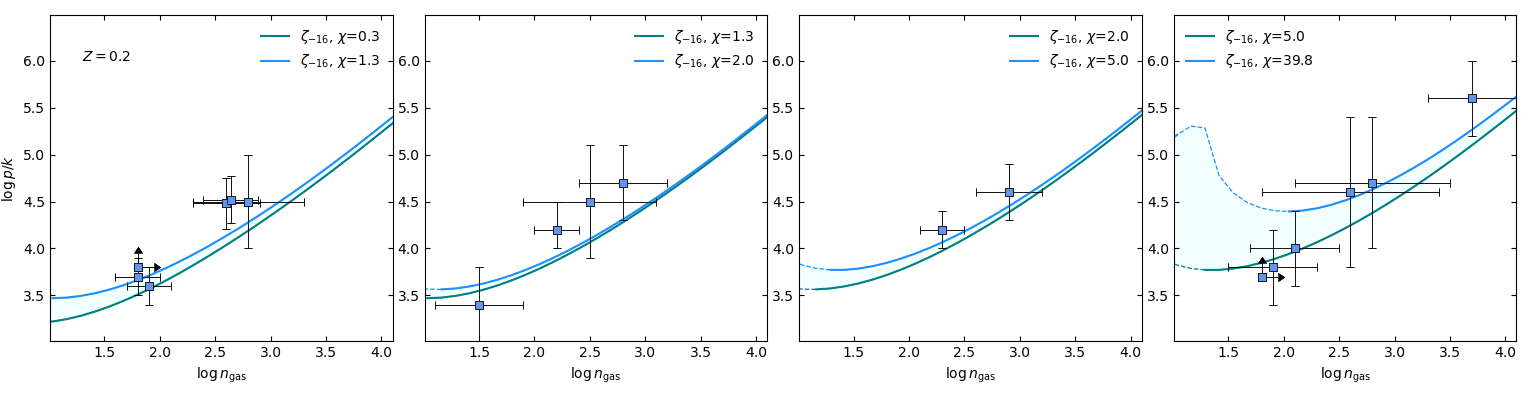}
    \caption{Pressure vs density (phase diagram) for the metallicity $Z = 0.2$, corresponding to the mean SMC metallicity. The graphic description is the same as in Fig.~\ref{fig:phase_LMC}. Blue squares are values obtained for the SMC.  
    %Data is divided into four bins relative to UV field intensity, model curves are calculated respectively to the bins.
    }
    \label{fig:phase_SMC}
\end{figure*}

We also show how our estimated results on physical condition are located on the phase diagram of the thermal state of the cold ISM. The phase diagrams for the four bins of $\chi, \zeta$ and associated measured values are shown in Figs~\ref{fig:phase_LMC} and \ref{fig:phase_SMC}, for average values of metallicity in the LMC and SMC, respectively. In fact, to derive $n_{\rm H}$ and $\rm UV$ flux we already used calculation of thermal state using \Meudon\ code. Therefore these phase diagrams are mostly shown for an illustrative purpose only, to highlight that the thermal state of the medium depends on the physical parameters, like $\chi$ and $\zeta$, as well as on the properties and abundance of dust. To calculate the phase diagram we used the same code and similar assumptions on the heating and cooling sources of the medium as it was recently done in \citealt{Balashev2022}. One can see that our calculations in general agree with \Meudon\ ones that are indirectly reflected by the location of the measured points. On average, \Meudon\ gives slightly higher temperature of the ISM, at lower values of $\chi$ and higher metallicity. This may be due to different scaling of the dust to gas ratio parameter depending on metallicity, the actual behaviour of which is very hard to constrain from observations \citep[see e.g.][]{RemyRuyer2014}.

\section{Summary}
\label{sect:summary}

%We have provided a systematic search of HD molecules in the LMC and SMC, the closest low-metallicity galaxies, using FUSE archival data. While the quality of the FUSE data often allow to obtain in the most cases only upper limits on HD column densities, we detected HD towards 24 sightlines (including 19 new detection). We refitted H$_2$ to constrain populations of rotational levels in the systems. 

%We used HST archival data to analyse \CI\ absorption lines to get population of fine-structure levels and to find metallicity from metal lines.

%Populations of \CI\ fine-structure and H$_2$ rotational levels then were used to find constraints on $n$ and $\chi$ in the systems.

Using HST archival data we analysed \CI\, absorption lines and obtained populations of fine-structure levels in 21 and 23 systems in the LMC and SMC, respectively. In the most of these systems we also analysed metal lines and measured metallicities to be $\log Z = -0.66\pm 0.22$ and $-0.99\pm 0.19$ (providing an average value and standard deviation) for the LMC and SMC, respectively. 

Using the obtained \CI\ fine-structure level populations and populations of H$_2$ rotational levels previously reported in accompanying paper \citep{Kosenko2023} we constrained physical conditions in the systems, namely, hydrogen gas density and UV field intensity (or cosmic ray ionization rate assumed to be coscaled with UV field). The average values of hydrogen gas densities are $\log n_{\rm H}\sim 2.5$ and $\log n_{\rm H}\sim 2.6$, with standard deviation 0.4 and 0.6, for the LMC and SMC, respectively. %\SK{Is it correct?}
The mean intensities of UV field are $\log\chi_{\rm LMC} \sim 0.7$ and $\log\chi_{\rm SMC}\sim 0.3$ (in units of Mathis field) with standard deviation of 0.4 and 0.5, for the LMC and SMC, respectively. We also estimated the average thermal pressure in our sample to be $\log p_{\rm LMC} \sim 4.2$ and $\log p_{\rm SMC}\sim 4.3$ with standard deviation 0.4 and 0.5 for the LMC and SMC, respectively. 

We compared the obtained hydrogen gas densities and thermal pressures in Magellanic Clouds with measurements in the Milky-Way and high-$z$ DLAs. The average thermal pressure in Magellanic Clouds is half an order of magnitude higher than thermal pressure measured in \CI\ absorption systems in the Milky-Way $\log p_{\rm MW}\sim 3.7\pm 0.2$. %\SB{Is this Jenkins+2011 or our reanalysed value?} \DK{it is Jenkins2011}. %\SB{But, now, in our paper we do not used Jenkins values, replacing them by reanalysed. So need to provide this ones.} 
Such high values of thermal pressure were observed in strong CO-bearing MW clouds \citep{Welty2020, Federman2021}, which represent translucent/dense phase of the ISM. This difference can be explained by lower metallicities and higher UV field/CRIR intensity in the Magellanic Clouds than in the MW samples, which is directly confirmed by our data and in line with previous studies \citep[e.g.][]{Bernard2008, Welty2016}. In that sense, Magellanic Clouds absorption systems probe ISM at sufficiently different regime than MW do and provide interesting test-case for the studies of high redshift galaxies. This also agree with some similarities of obtain physical parameters in the Magellanic Clouds and high redshift absorption systems, although the latter likely predominantly probe an outskirts of the galaxies due to their higher cross section. %by much higher intensity of UV field, which we found in our samples to be $\log \chi_{LMC} = xx$ and $\log \chi_{LMC} = xx$.  

%[to be typical for diffuse ISM in the Milky Way and in the distant galaxies, but estimated UV field intensities in both LMC and SMC are found to be higher than in our Galaxy,]

%\SB{Disagree with the commented below. metallicities do not depend on the UV field. Even more in most low metallicities galaxies, the star formation rate is expected to be lower. MC is likely encounter an star formation burst, due to interaction with our galaxy, which is confirmed by tidal Magellanic stream. The high redshifts systems are a bit different, since the environment of these galaxies are different - more gas, more interactions,...}
%A high intensity of UV field is probably connected with difference in the metallicities. At the same time such high $\chi$ are consistent with some measurements at high redshifts, which shows that Magellanic Clouds may be used as a proxy to study distant galaxies. \SK{As well, it can be result of selection effects, since studied sight lines in Magellanic Clouds probe mainly gas near star-forming regions.}

\section*{Data availability} 
The results of this paper are based on open data retrieved from the FUSE and HST telescope archives. These data can be shared on reasonable requests to the authors.
 
\section*{Acknowledgements} 
This work was supported by RSF grant 23-12-00166.

\bibliographystyle{mnras}
\bibliography{references.bib}

\clearpage
\appendix
\section{Details on \CI\, fit }
\label{sect:CI_fit}
In this section we show fit of the \CI\, absorption lines in each system. Here we do not show fully blended (e.g. \CI$^*$1560.70, \CI$^{**}$ 1561.36) and the most weak lines.

\subsection{Large Magellanic Cloud}

\begin{figure*}
    \centering
    \includegraphics[width=\linewidth]{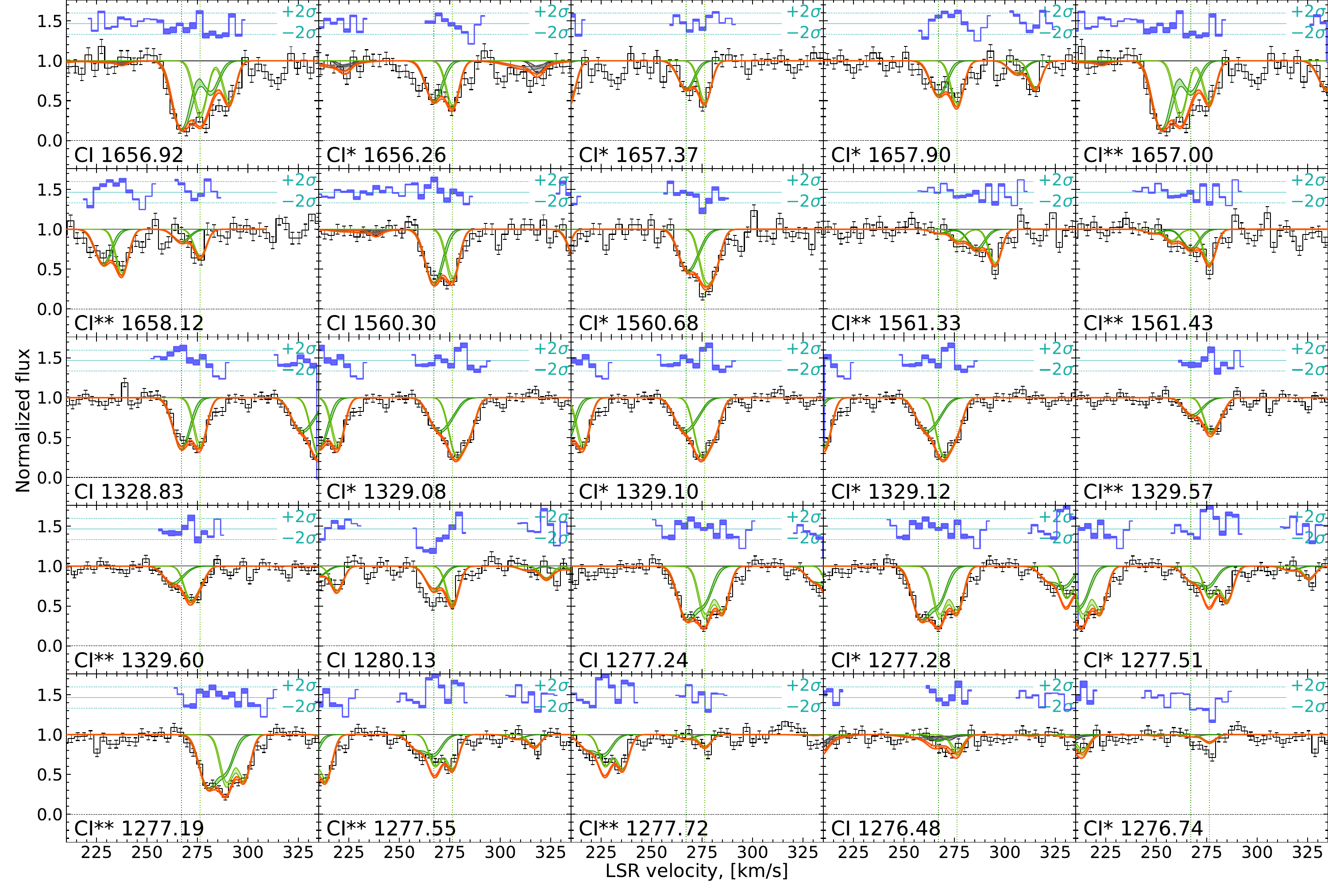}
    \caption{\CI\, absorption lines fit in the system towards Sk-67 2 in the LMC. Here black line show spectrum, coloured bands show synthetic spectrum, sampled from 0.683 credible interval of the posterior probability distribution of the fitting parameters. Red line represent total line profile, while grey and green lines show Milky Way and Magellanic Cloud components, respectively. Blue points at the top of each panel show residuals. Here we show only components found in the Magellanic Clouds.}
    \label{fig:Sk67_2_CI}
\end{figure*}

\begin{figure*}
    \centering
    \includegraphics[width=\linewidth]{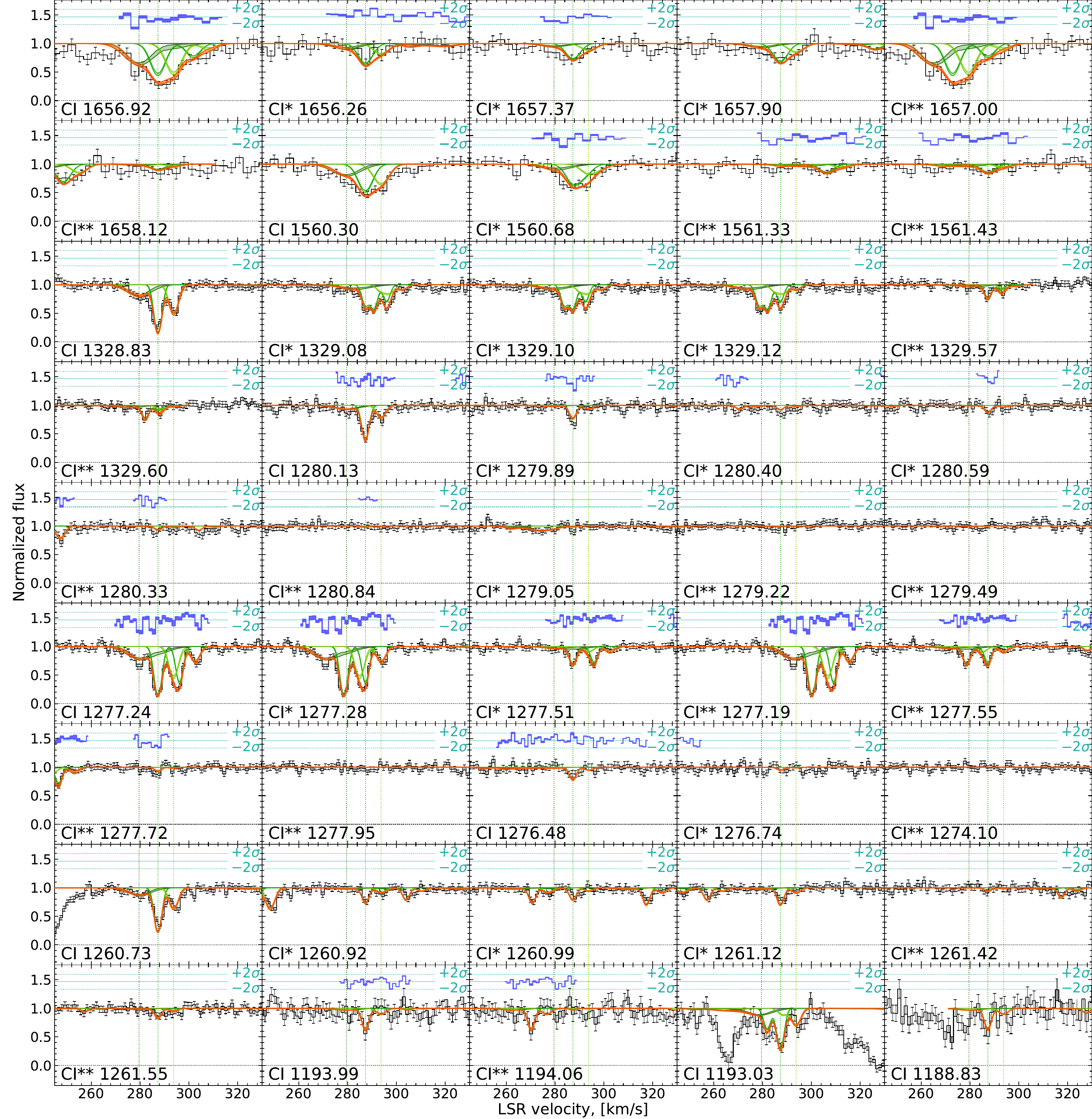}
    \caption{\CI\, absorption lines fit in the system towards Sk-67 5 in the LMC. Lines are the same as in Figure~\ref{fig:Sk67_2_CI}.}
    \label{fig:Sk67_5_CI}
\end{figure*}

\begin{figure*}
    \centering
    \includegraphics[width=\linewidth]{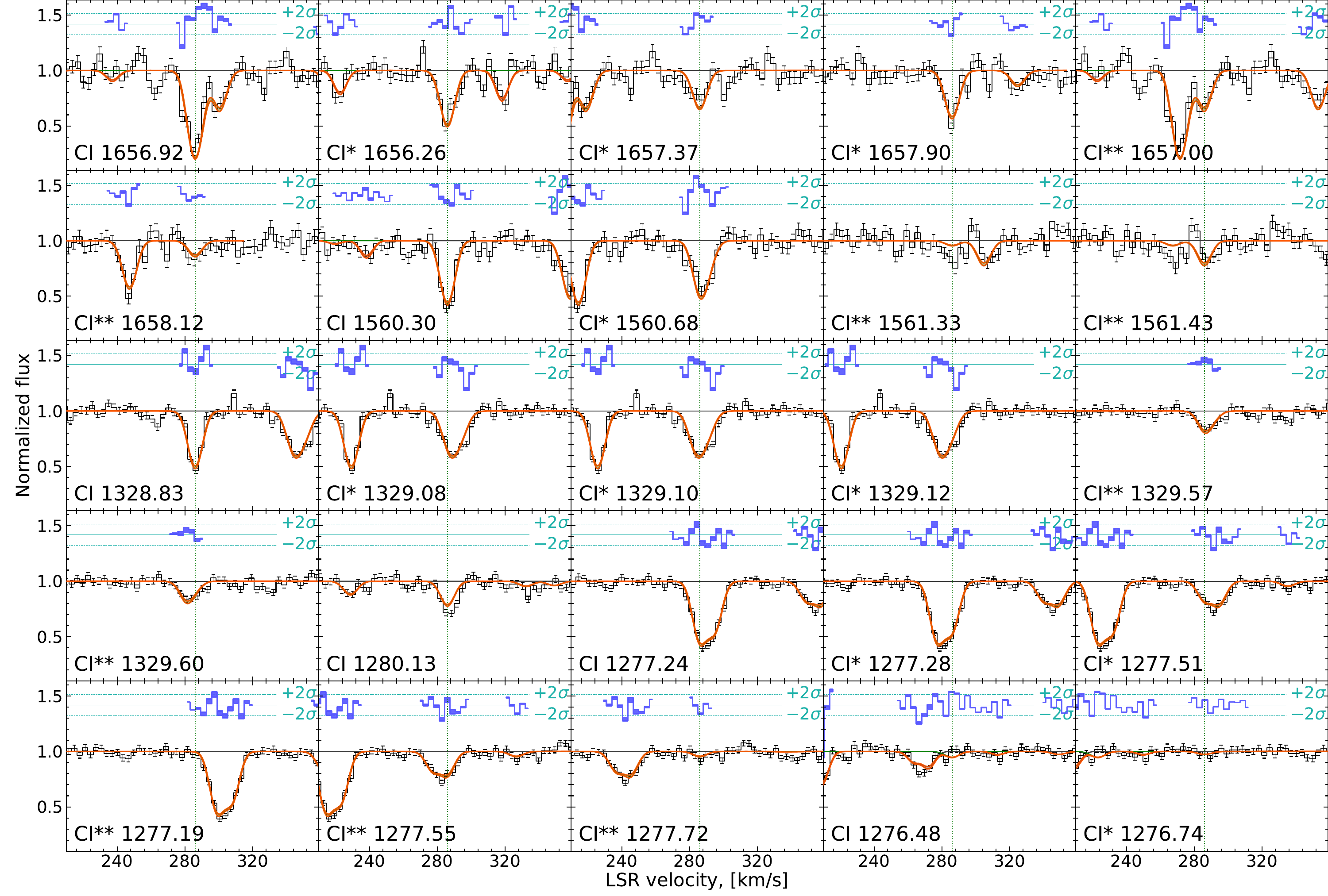}
    \caption{\CI\, absorption lines fit in the system towards Sk-67 20 in the LMC. Lines are the same as in Figure~\ref{fig:Sk67_2_CI}.}
    \label{fig:Sk67_20_CI}
\end{figure*}

\begin{figure*}
    \centering
    \includegraphics[width=\linewidth]{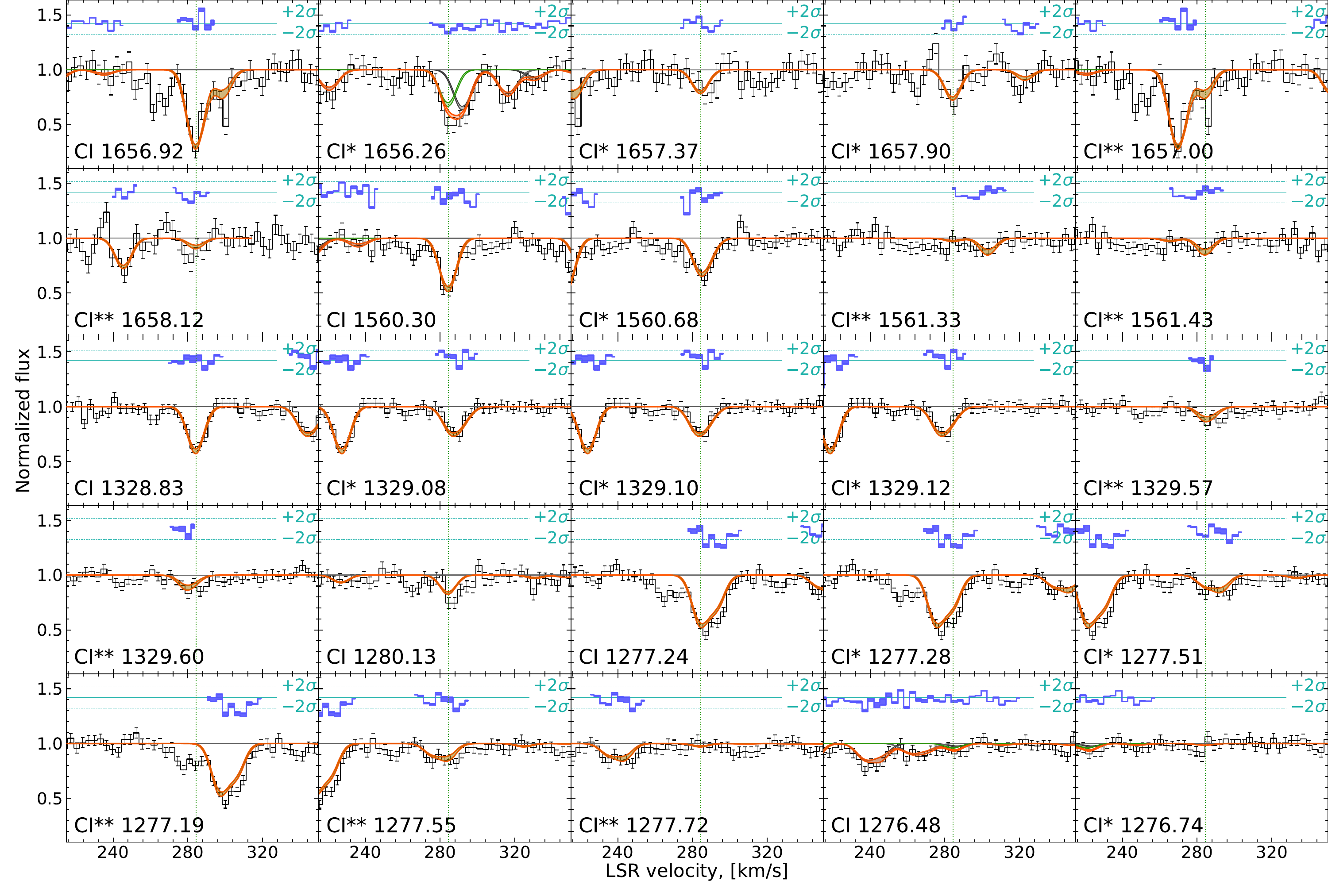}
    \caption{\CI\, absorption lines fit in the system towards PGMW 3070 in the LMC. Lines are the same as in Figure~\ref{fig:Sk67_2_CI}.}
    \label{fig:PGMW3070_CI}
\end{figure*}

\begin{figure*}
    \centering
    \includegraphics[width=\linewidth]{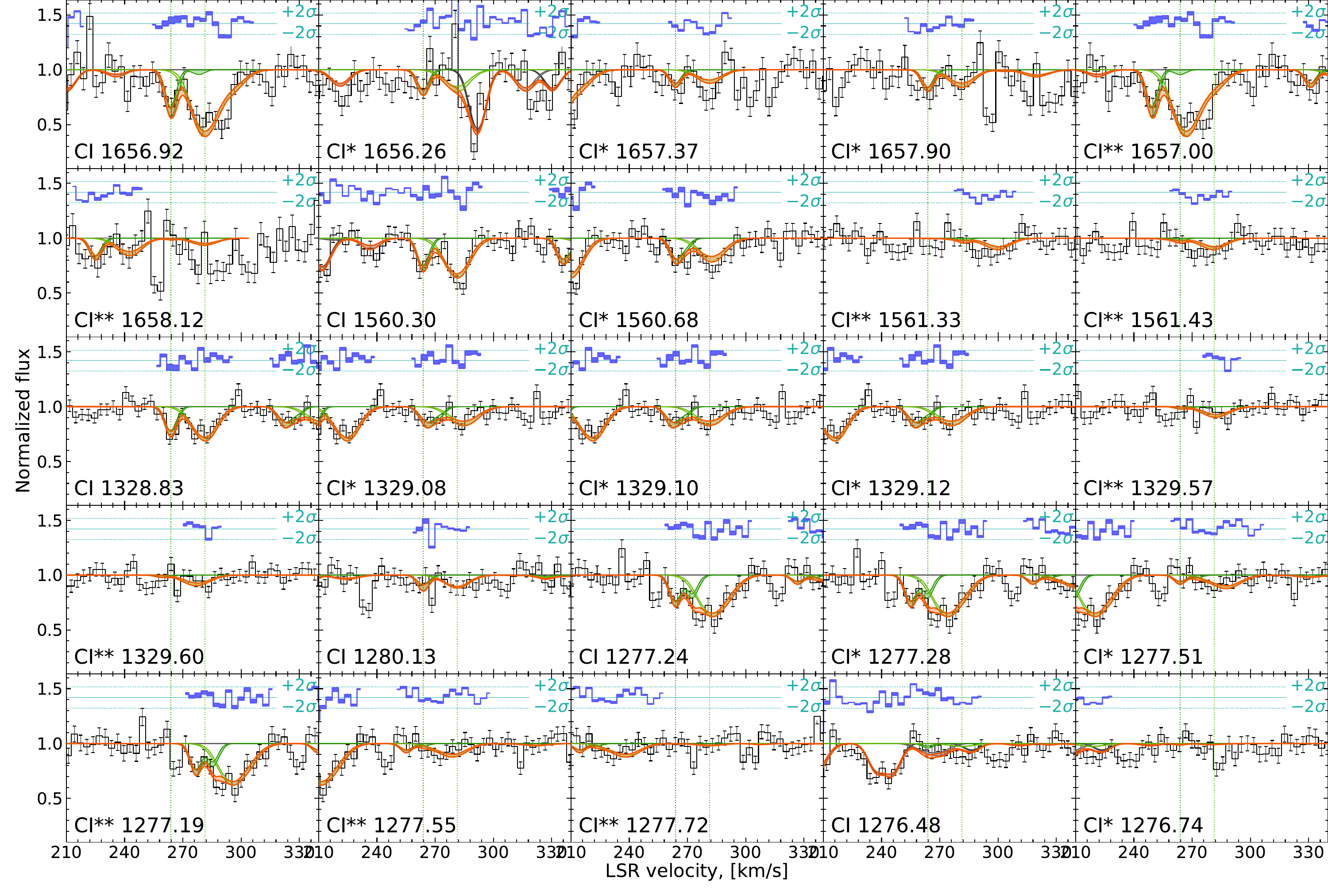}
    \caption{\CI\, absorption lines fit in the system towards LH10 3120 in the LMC. Lines are the same as in Figure~\ref{fig:Sk67_2_CI}.}
    \label{fig:LH10_3120_CI}
\end{figure*}

\begin{figure*}
    \centering
    \includegraphics[width=\linewidth]{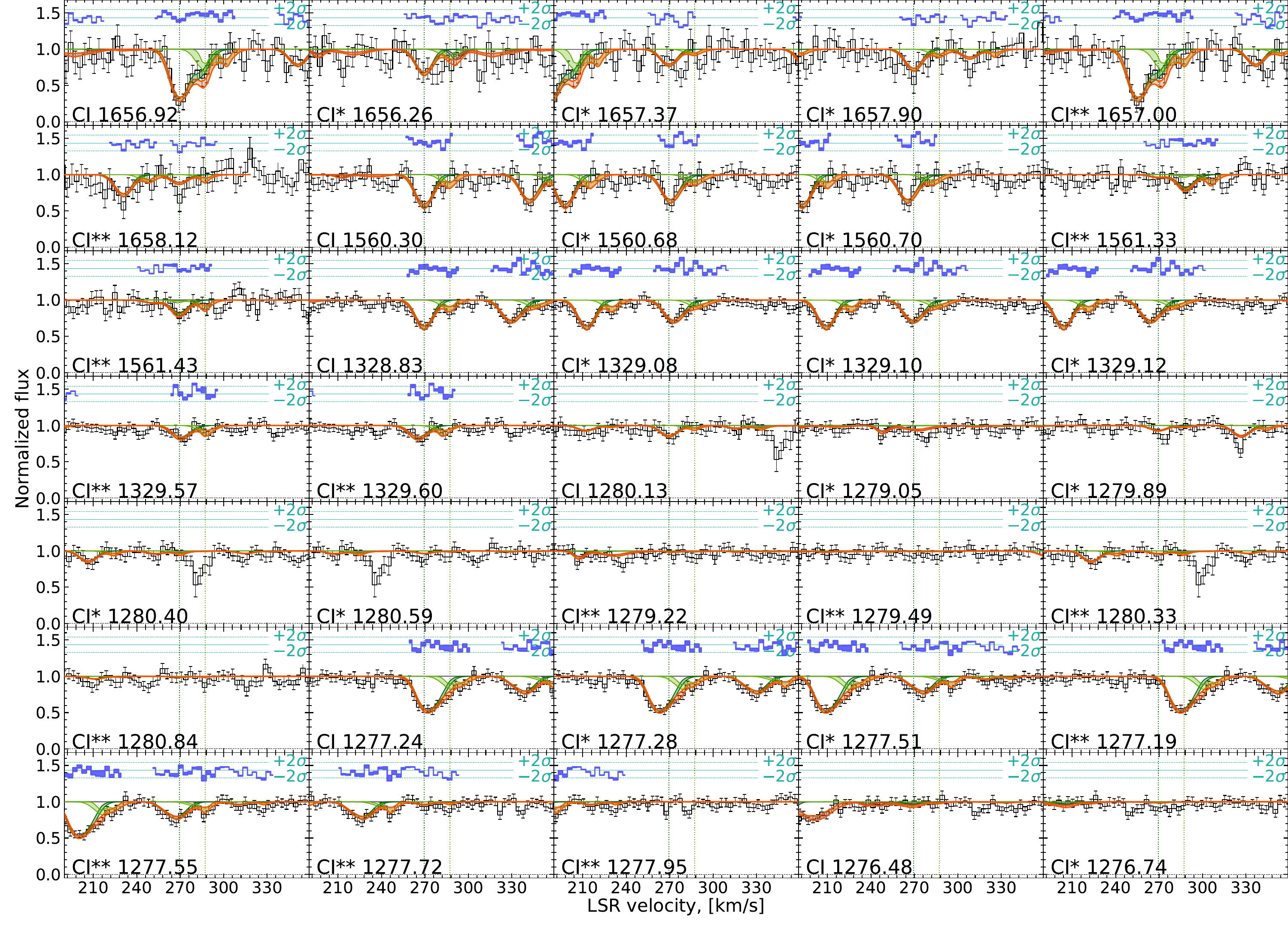}
    \caption{\CI\, absorption lines fit in the system towards PGMW 3223in the LMC. Lines are the same as in Figure~\ref{fig:Sk67_2_CI}.}
    \label{fig:PGMW3223_CI}
\end{figure*}

\begin{figure*}
    \centering
    \includegraphics[width=\linewidth]{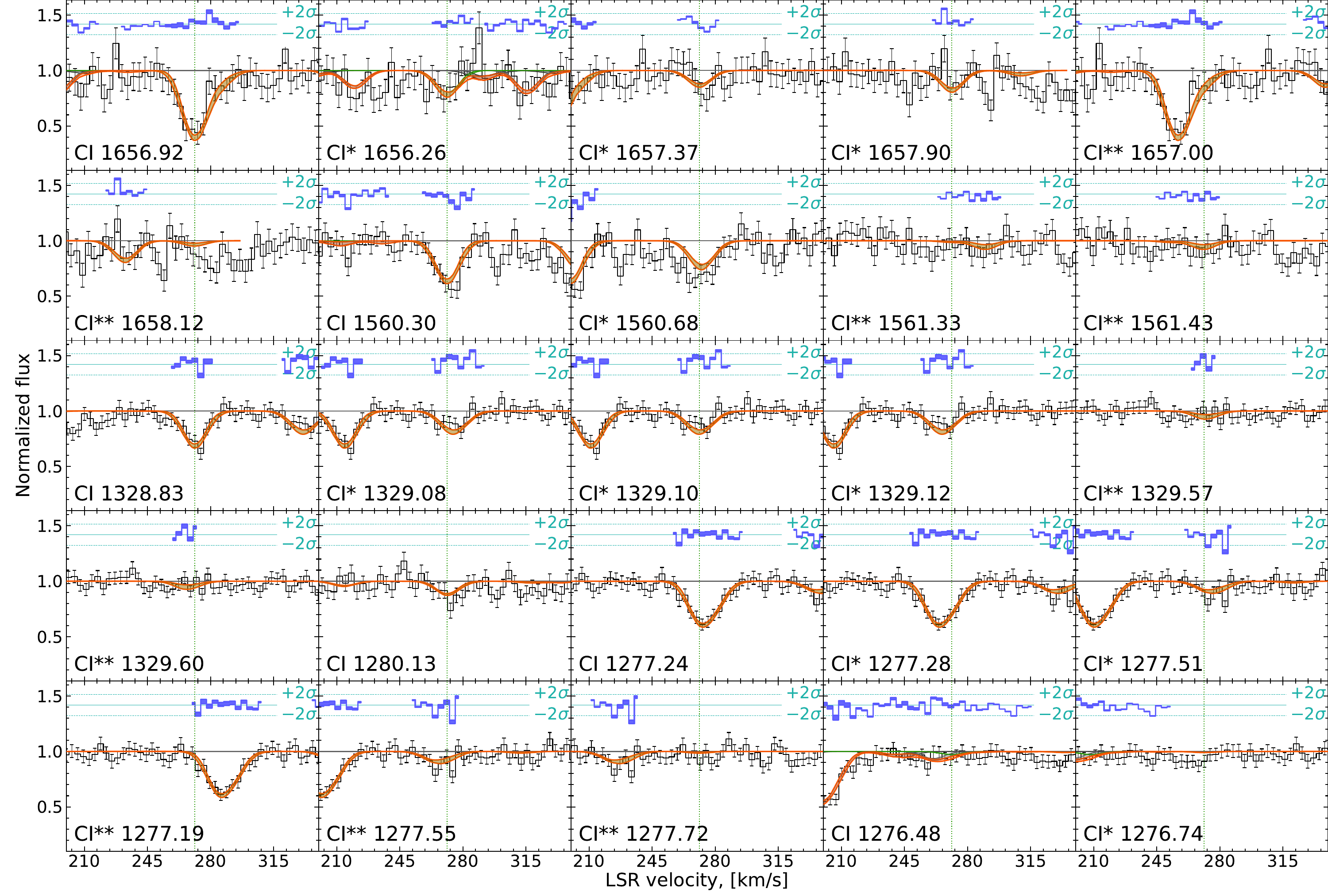}
    \caption{\CI\, absorption lines fit in the system towards Sk-66 35 in the LMC. Lines are the same as in Figure~\ref{fig:Sk67_2_CI}.}
    \label{fig:Sk66_35_CI}
\end{figure*}

\begin{figure*}
    \centering
    \includegraphics[width=\linewidth]{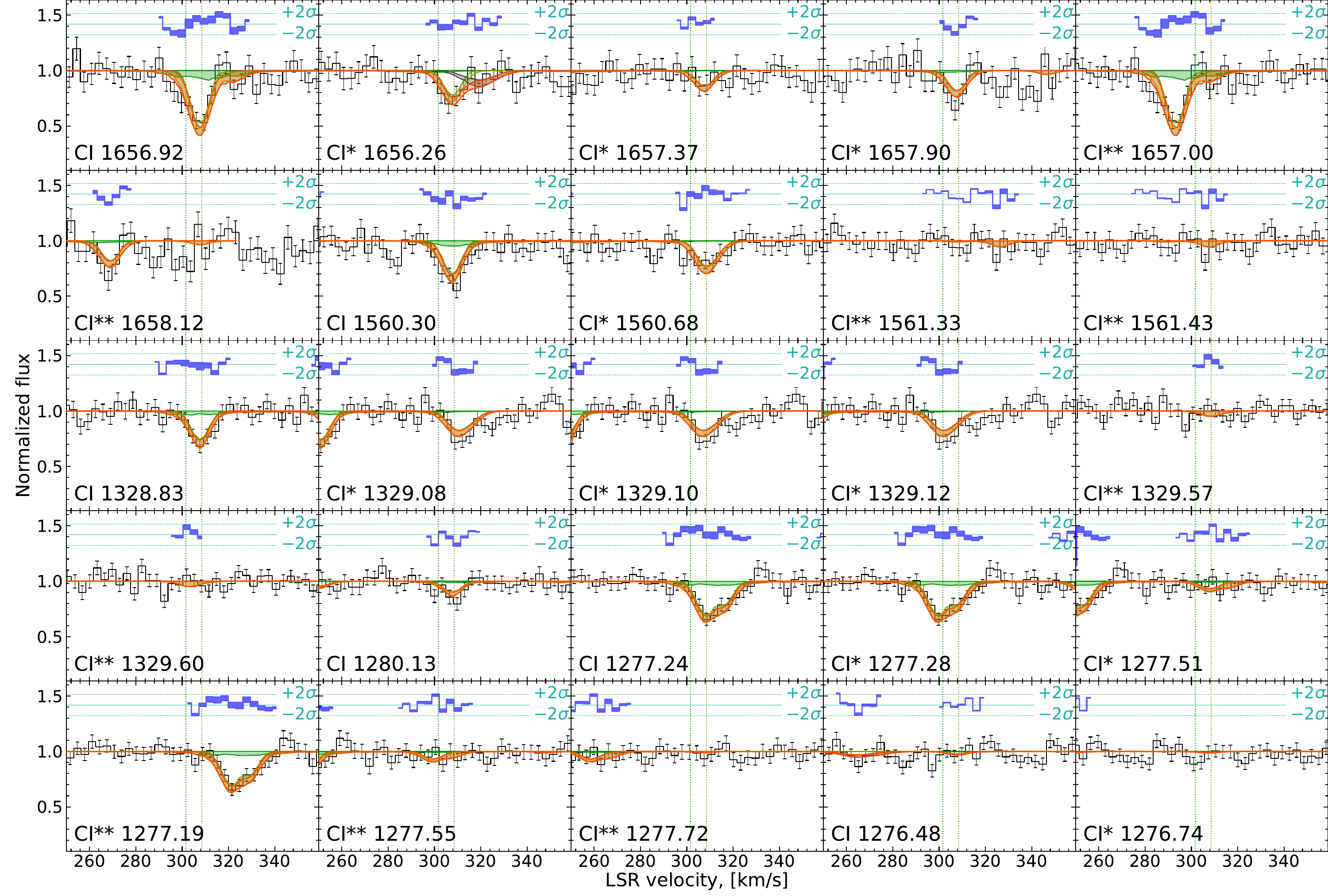}
    \caption{\CI\, absorption lines fit in the system towards Sk-66 51 in the LMC. Lines are the same as in Figure~\ref{fig:Sk67_2_CI}.}
    \label{fig:Sk66_51_CI}
\end{figure*}

\begin{figure*}
    \centering
    \includegraphics[width=\linewidth]{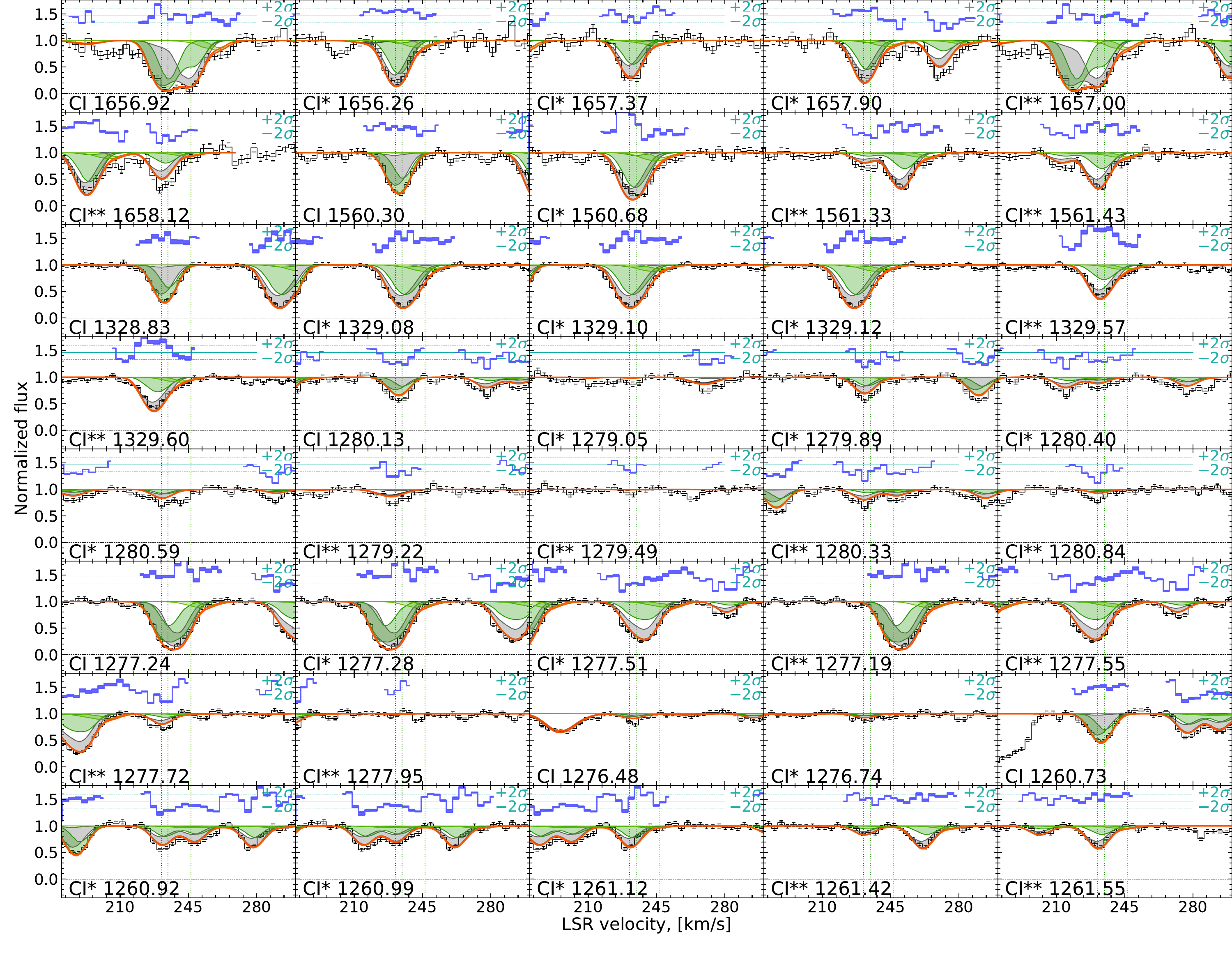}
    \caption{\CI\, absorption lines fit in the system towards Sk-70 79 in the LMC. Lines are the same as in Figure~\ref{fig:Sk67_2_CI}.}
    \label{fig:Sk70_79_CI}
\end{figure*}

\begin{figure*}
    \centering
    \includegraphics[width=\linewidth]{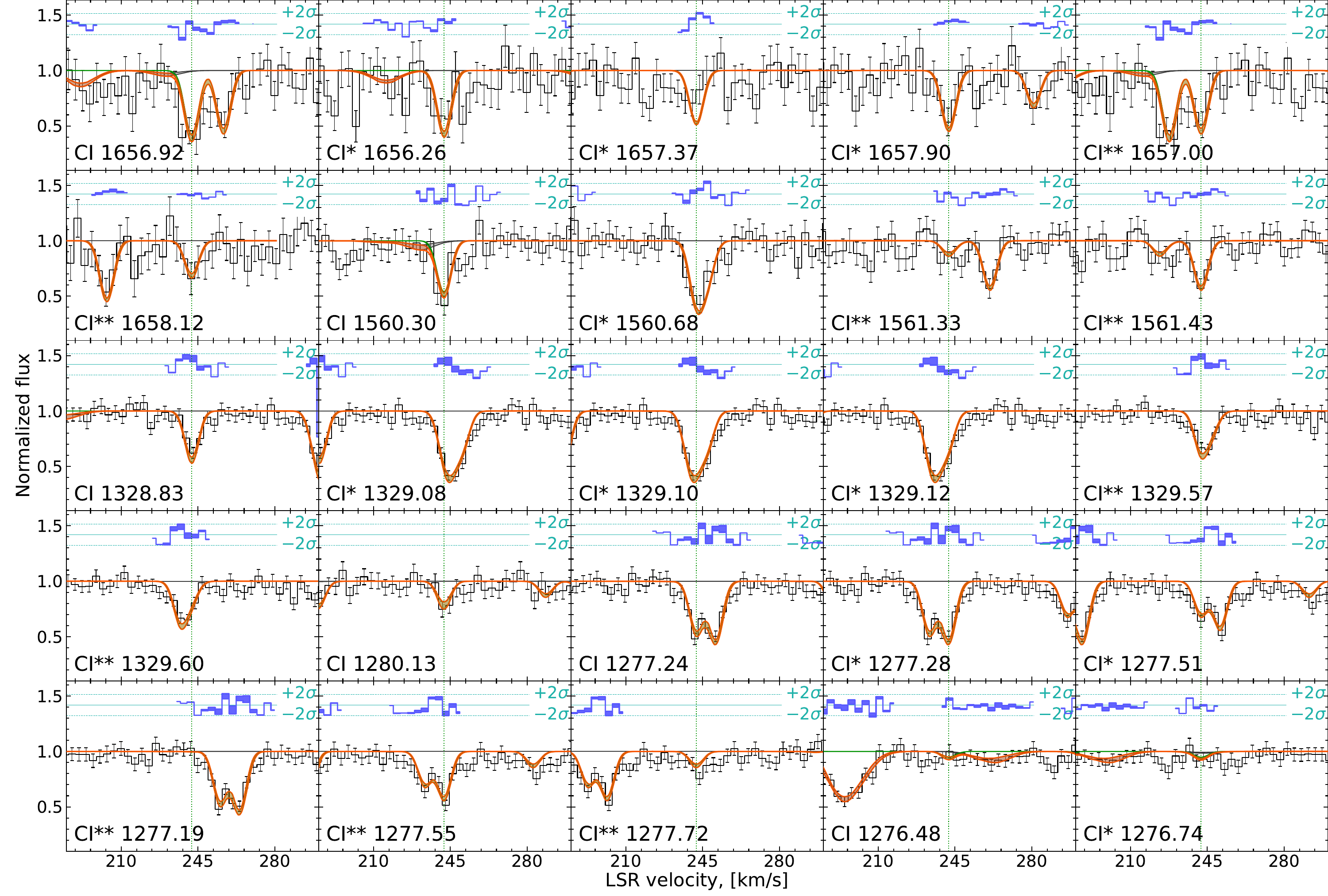}
    \caption{\CI\, absorption lines fit in the system towards Sk-68 52 in LMC. Lines are the same as in Figure~\ref{fig:Sk67_2_CI}.}
    \label{fig:Sk68_52_CI}
\end{figure*}

\begin{figure*}
    \centering
    \includegraphics[width=\linewidth]{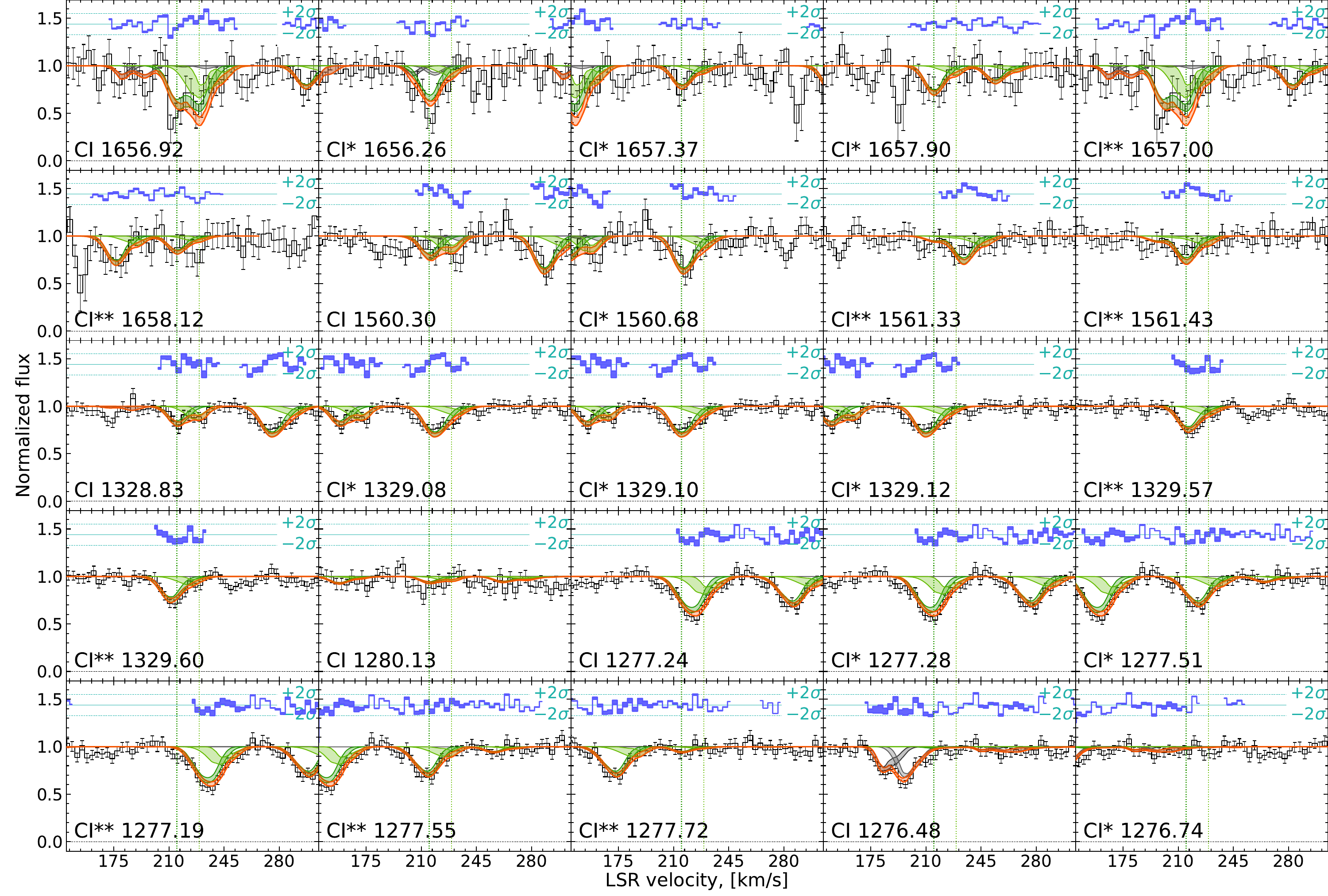}
    \caption{\CI\, absorption lines fit in the system towards Sk-71 8 in the LMC. Lines are the same as in Figure~\ref{fig:Sk67_2_CI}.}
    \label{fig:Sk71_8_CI}
\end{figure*}

\begin{figure*}
    \centering
    \includegraphics[width=\linewidth]{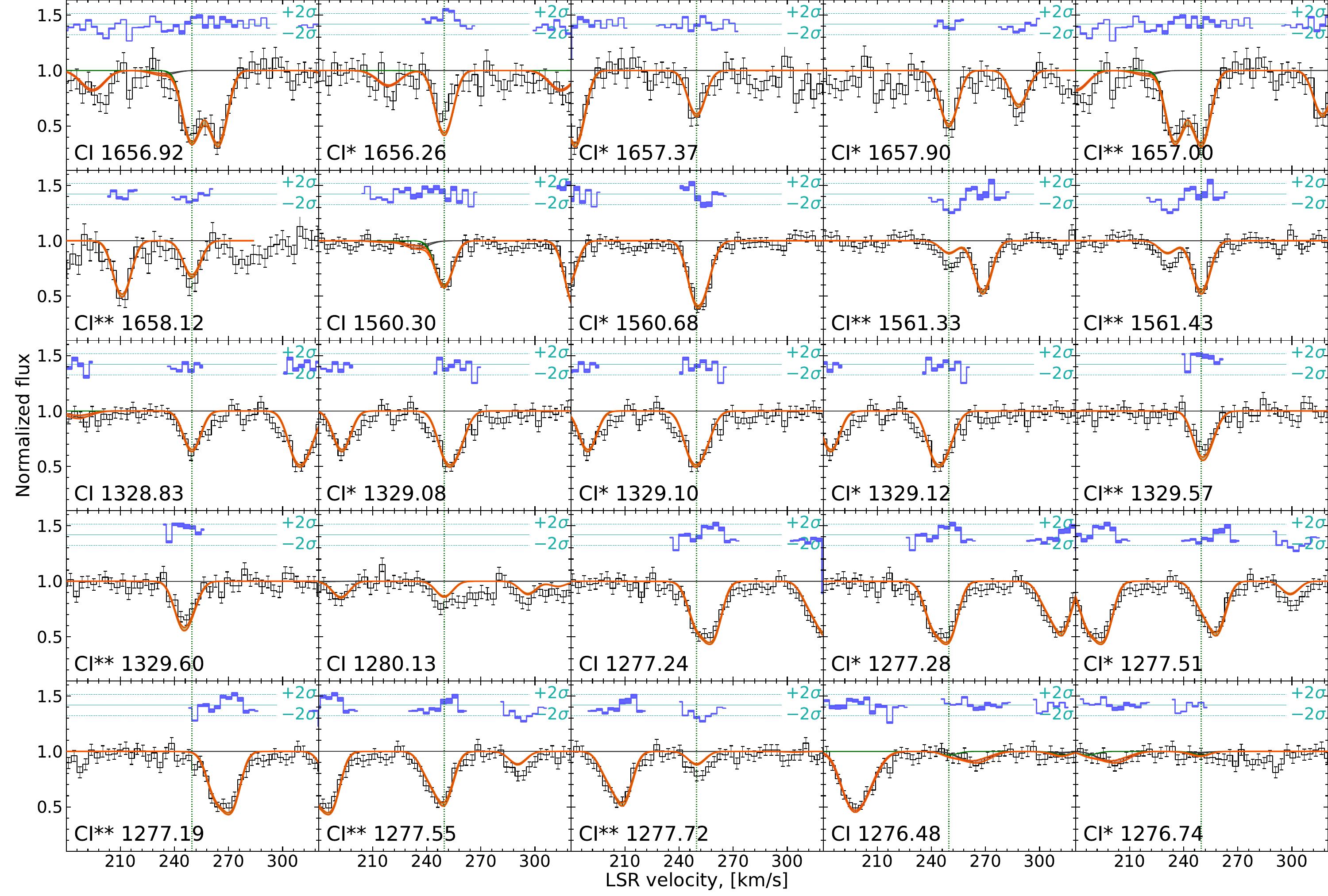}
    \caption{\CI\, absorption lines fit in the system towards Sk-69 106 in the LMC. Lines are the same as in Figure~\ref{fig:Sk67_2_CI}.}
    \label{fig:BI184_CI}
\end{figure*}

\begin{figure*}
    \centering
    \includegraphics[width=\linewidth]{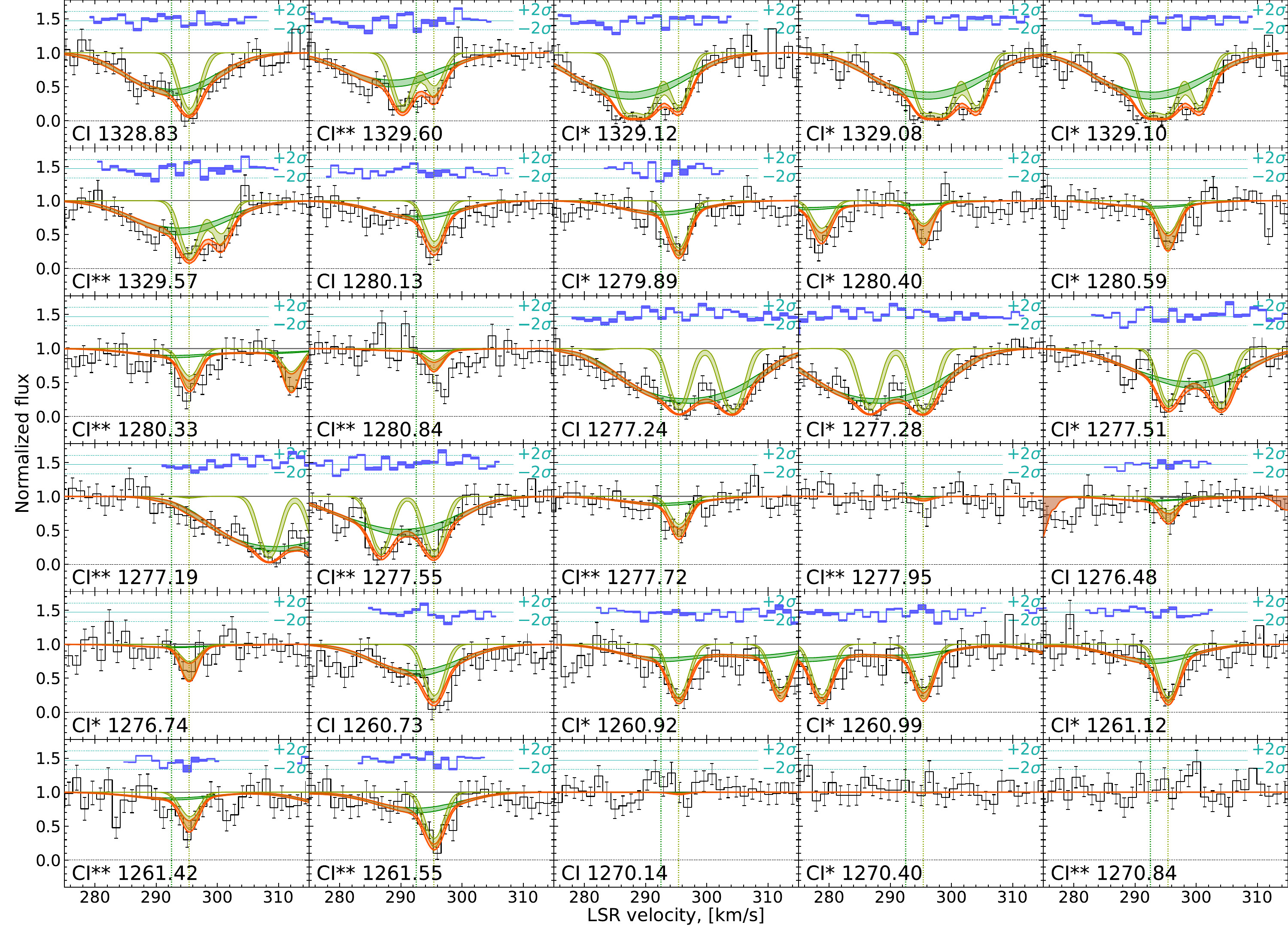}
    \caption{\CI\, absorption lines fit in the system towards Sk-68 73 in the LMC. Lines are the same as in Figure~\ref{fig:Sk67_2_CI}.}
    \label{fig:Sk68_73_CI}
\end{figure*}

\begin{figure*}
    \centering
    \includegraphics[width=\linewidth]{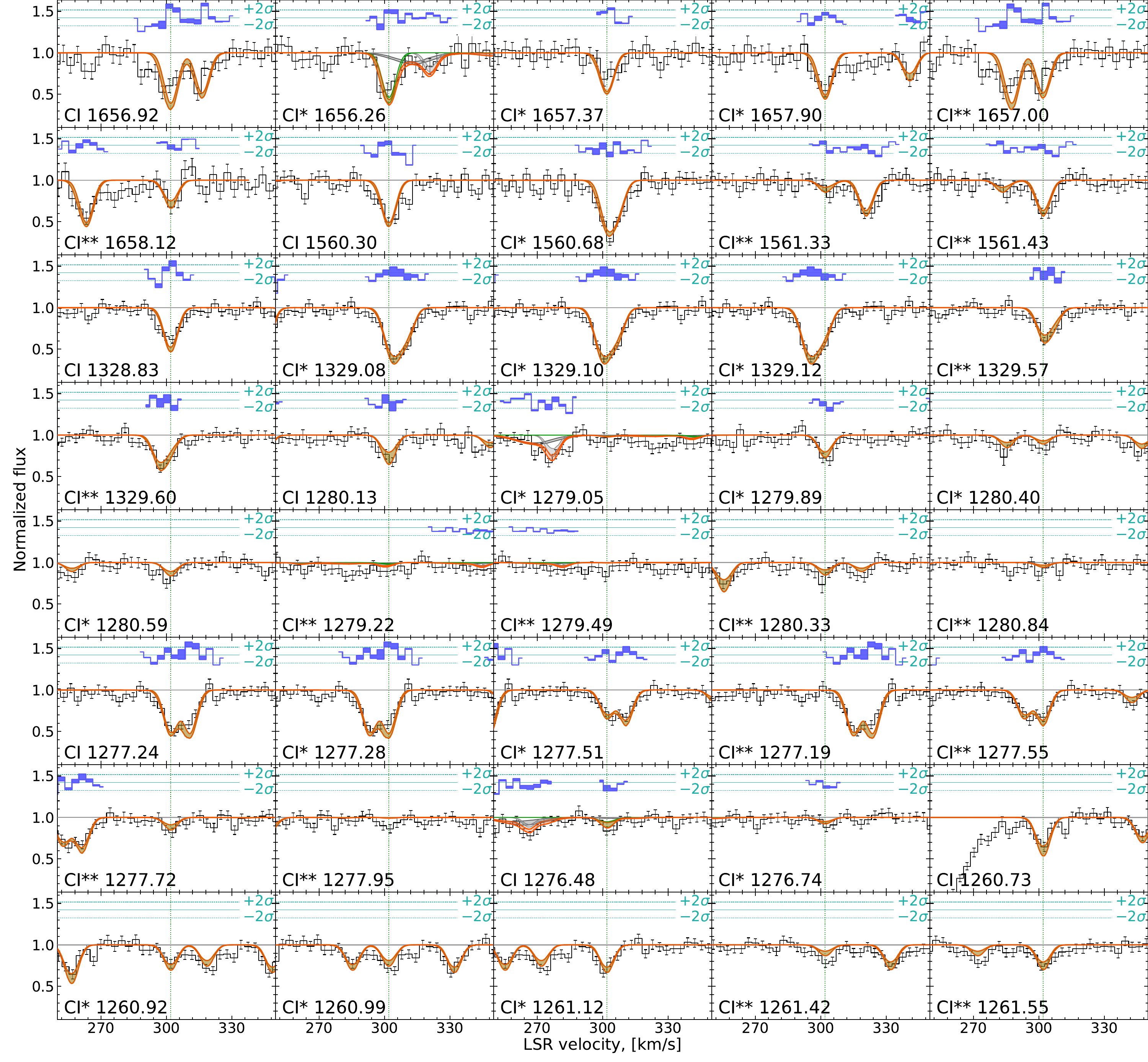}
    \caption{\CI\, absorption lines fit in the system towards Sk-67 105 in LMC. Lines are the same as in Figure~\ref{fig:Sk67_2_CI}.}
    \label{fig:Sk67_105_CI}
\end{figure*}

\begin{figure*}
    \centering
    \includegraphics[width=\linewidth]{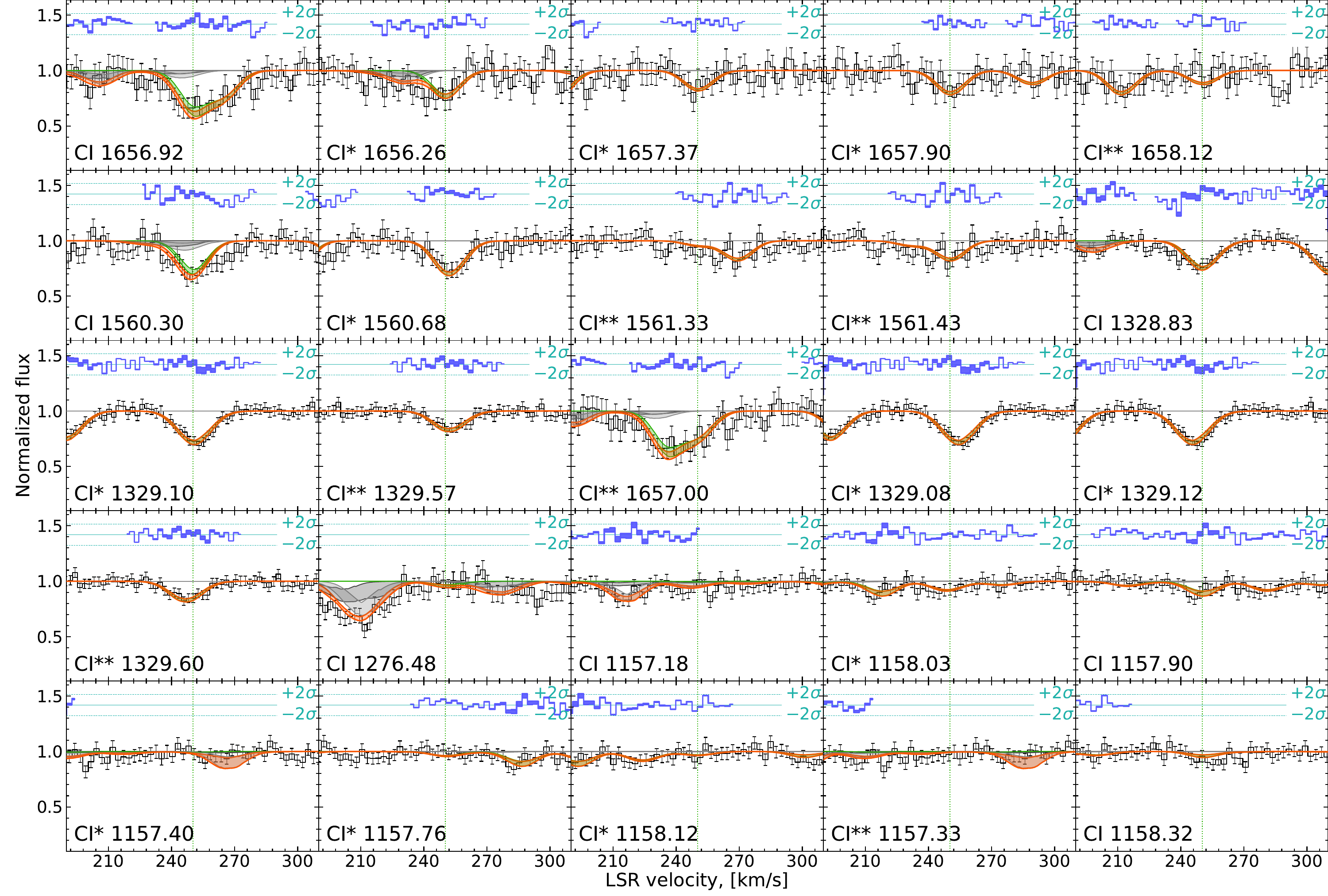}
    \caption{\CI\, absorption lines fit in the system towards BI 184 in the LMC. Lines are the same as in Figure~\ref{fig:Sk67_2_CI}.}
    \label{fig:BI184_CI}
\end{figure*}

\begin{figure*}
    \centering
    \includegraphics[width=\linewidth]{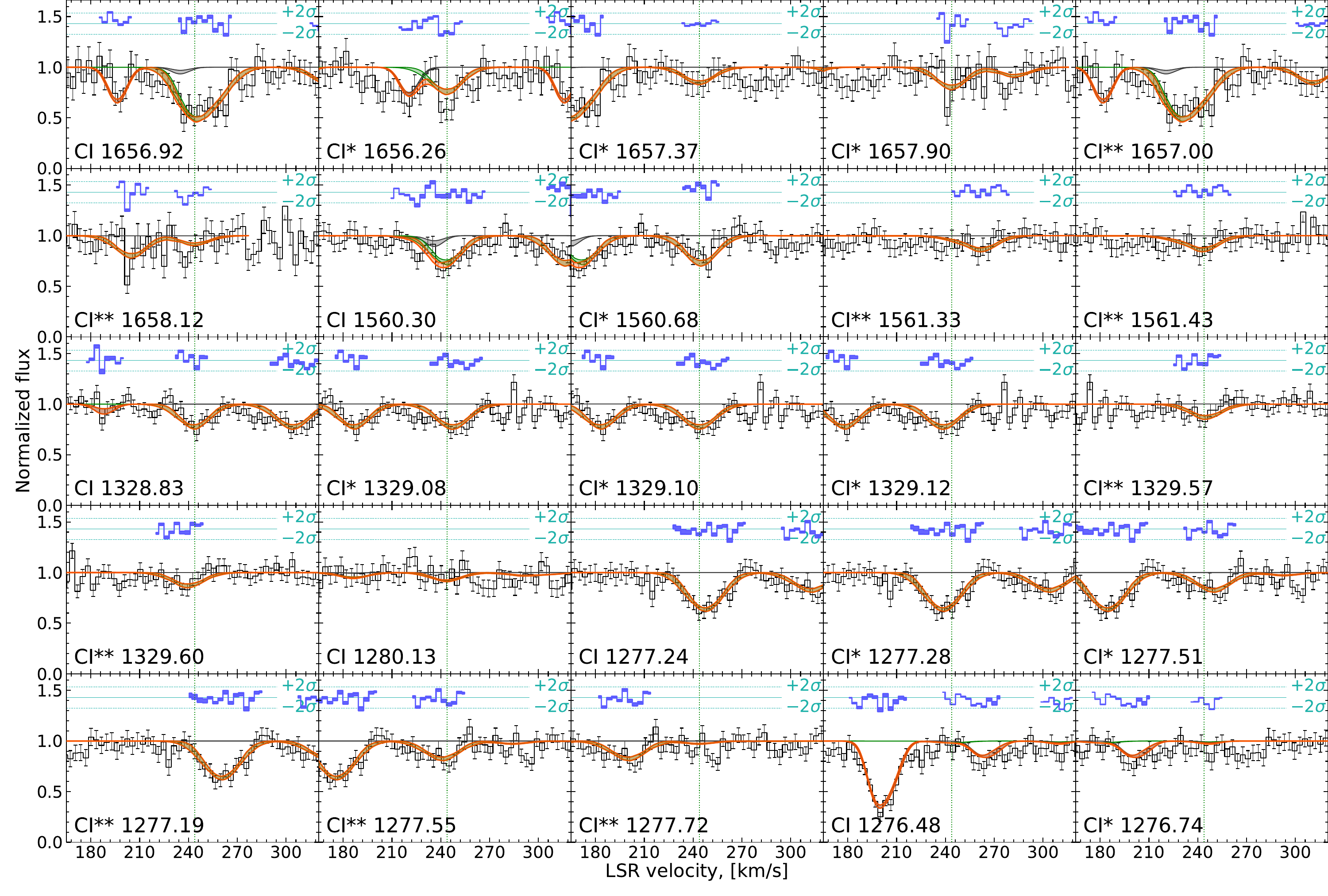}
    \caption{\CI\, absorption lines fit in the system towards Sk-71 45 in the LMC. Lines are the same as in Figure~\ref{fig:Sk67_2_CI}.}
    \label{fig:Sk71_45_CI}
\end{figure*}

\begin{figure*}
    \centering
    \includegraphics[width=\linewidth]{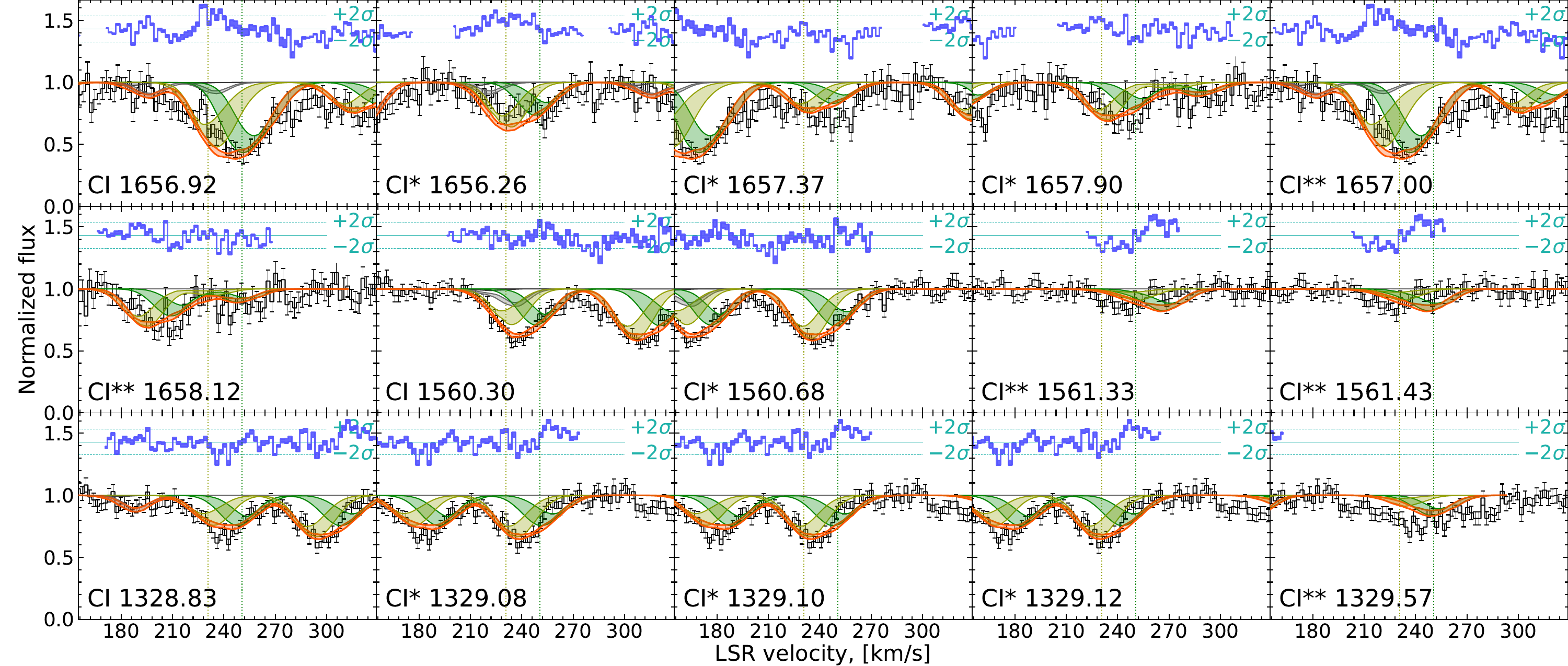}
    \caption{\CI\, absorption lines fit in the system towards Sk-71 46 in LMC. Lines are the same as in Figure~\ref{fig:Sk67_2_CI}.}
    \label{fig:Sk71_46_CI}
\end{figure*}

\begin{figure*}
    \centering
    \includegraphics[width=\linewidth]{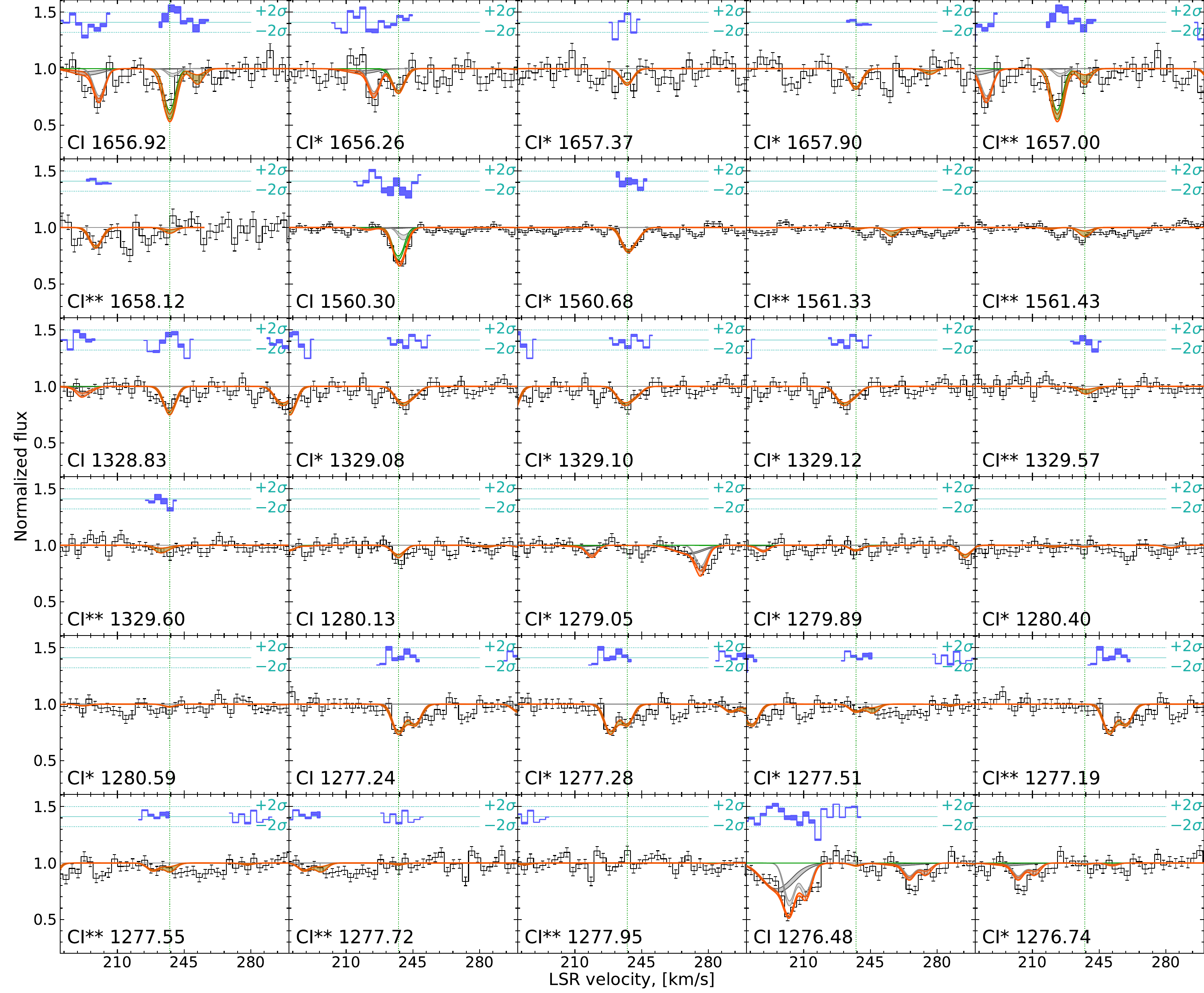}
    \caption{\CI\, absorption lines fit in the system towards Sk-69 191 in LMC. Lines are the same as in Figure~\ref{fig:Sk67_2_CI}.}
    \label{fig:Sk69_191_CI}
\end{figure*}

\begin{figure*}
    \centering
    \includegraphics[width=\linewidth]{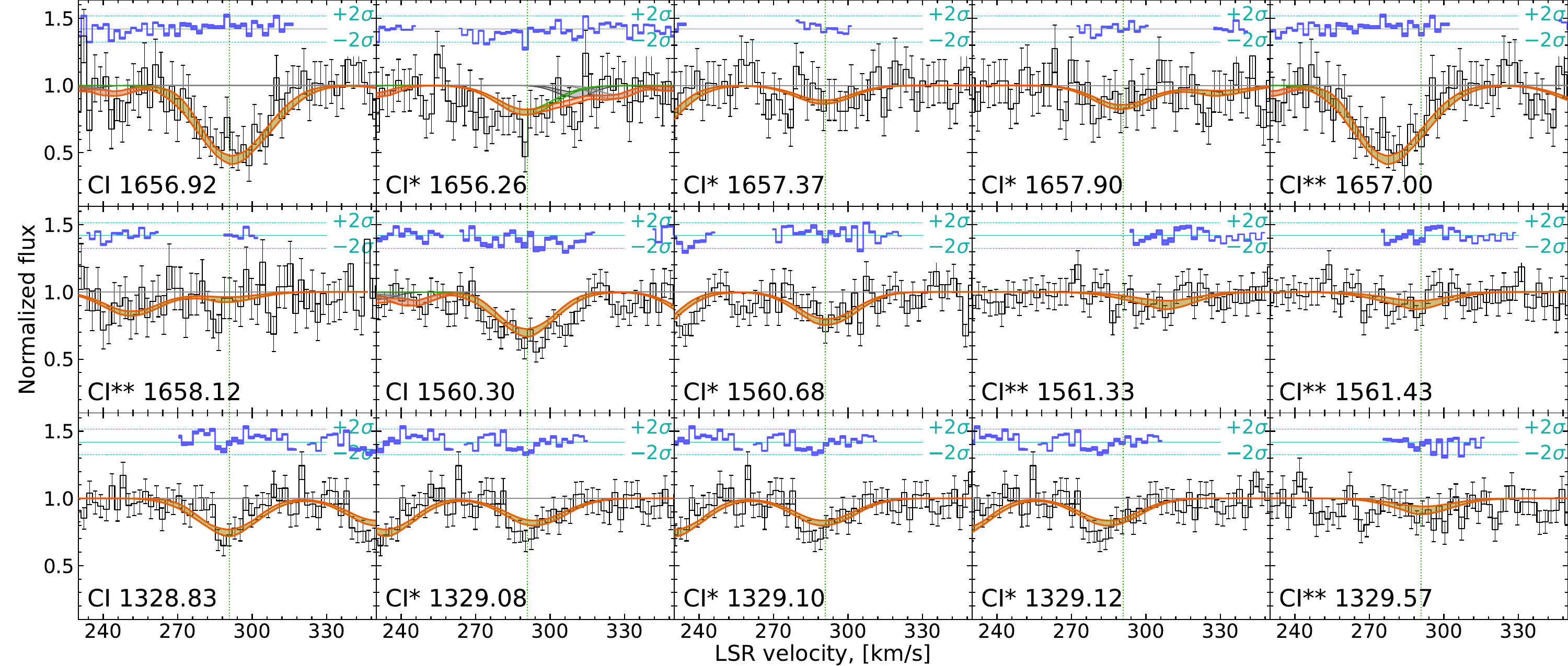}
    \caption{\CI\, absorption lines fit in the system towards BI 237 in the LMC. Lines are the same as in Figure~\ref{fig:Sk67_2_CI}.}
    \label{fig:BI237_CI}
\end{figure*}

\begin{figure*}
    \centering
    \includegraphics[width=\linewidth]{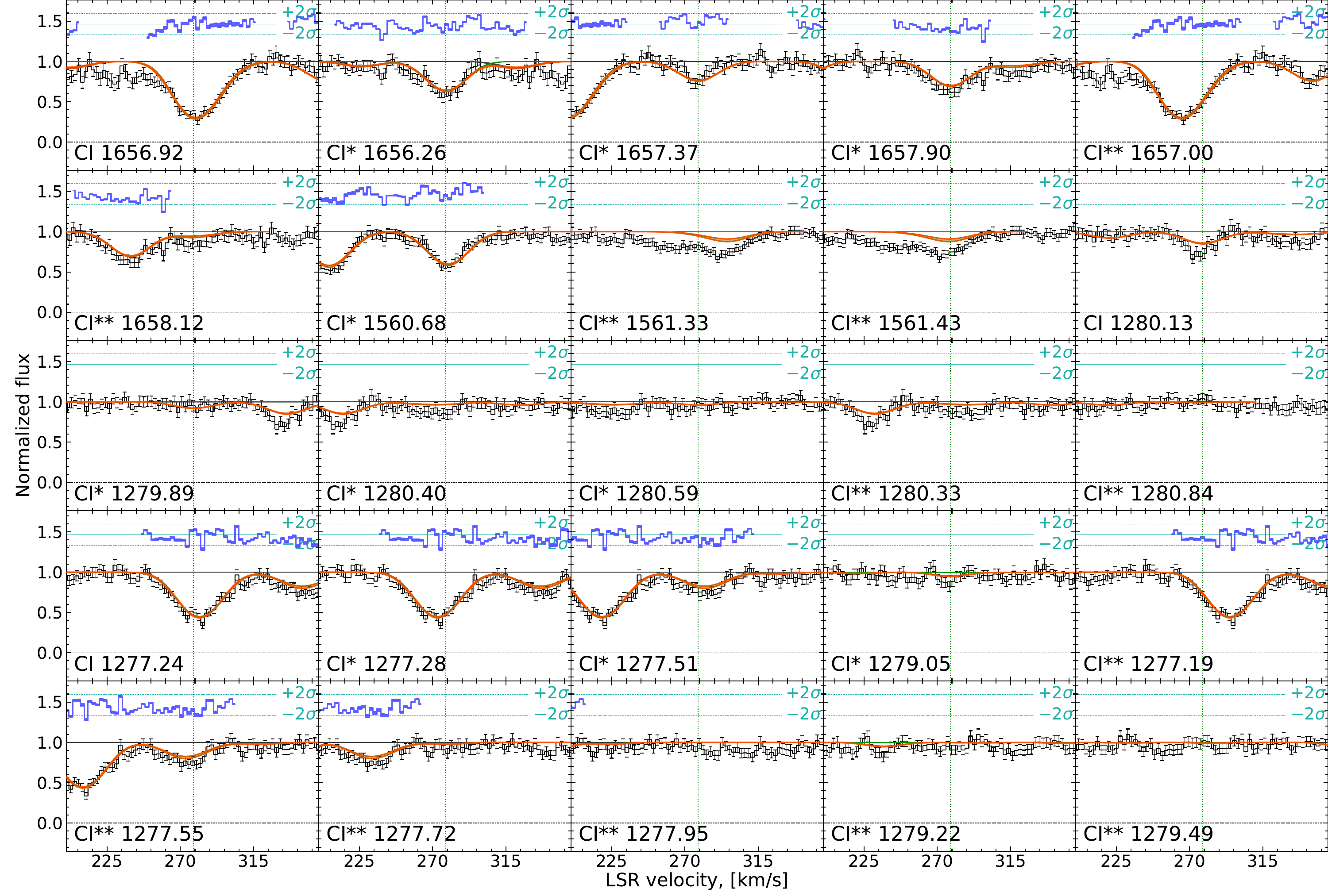}
    \caption{\CI\, absorption lines fit in the system towards Sk-68 129 in the LMC. Lines are the same as in Figure~\ref{fig:Sk67_2_CI}.}
    \label{fig:Sk68_129_CI}
\end{figure*}

\begin{figure*}
    \centering
    \includegraphics[width=\linewidth]{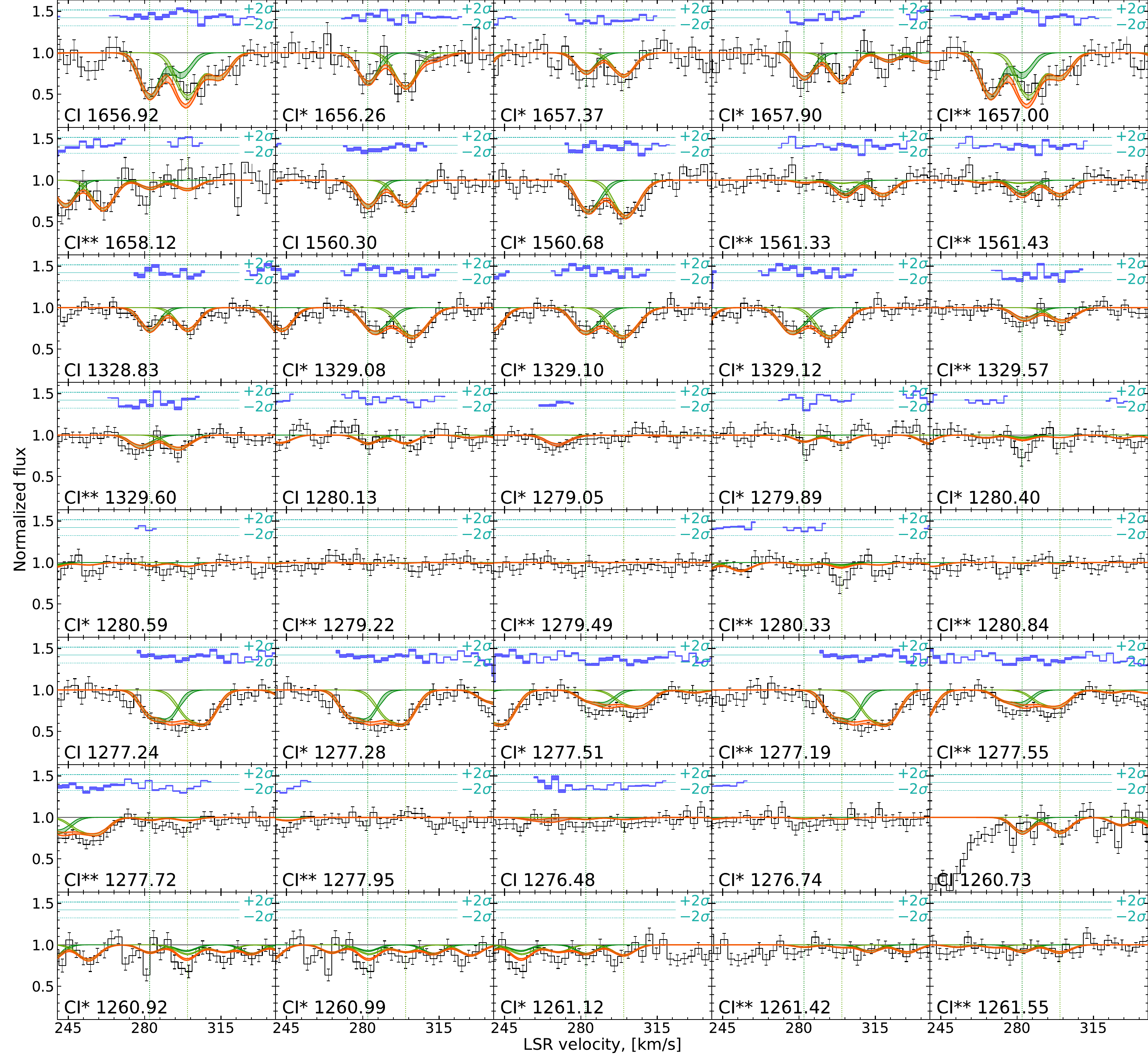}
    \caption{\CI\, absorption lines fit in the system towards Sk-66 172 in LMC. Lines are the same as in Figure~\ref{fig:Sk67_2_CI}.}
    \label{fig:Sk66_172_CI}
\end{figure*}

\begin{figure*}
    \centering
    \includegraphics[width=\linewidth]{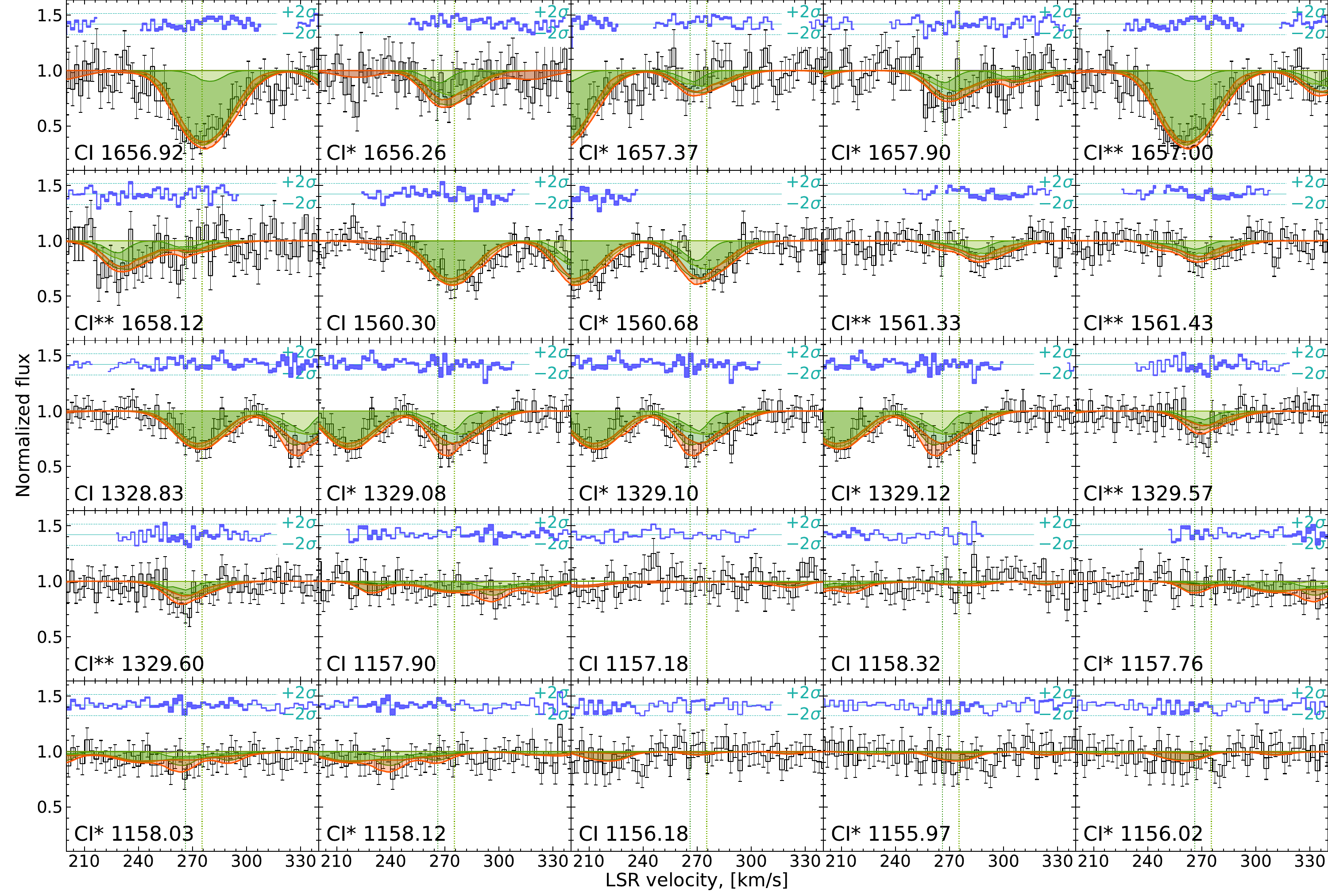}
    \caption{\CI\, absorption lines fit in the system towards BI 253 in the LMC. Lines are the same as in Figure~\ref{fig:Sk67_2_CI}.}
    \label{fig:BI253_CI}
\end{figure*}

\begin{figure*}
    \centering
    \includegraphics[width=\linewidth]{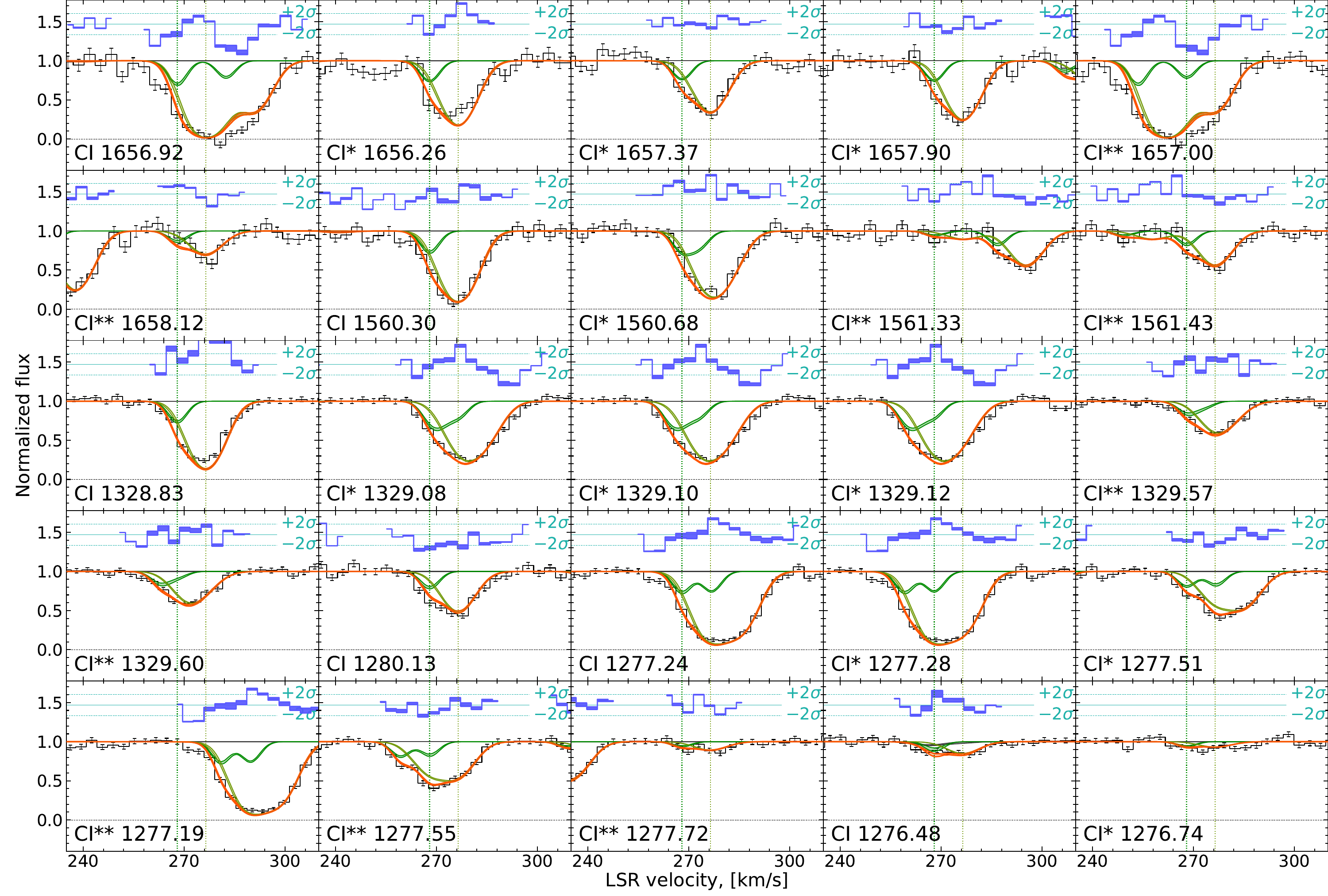}
    \caption{\CI\, absorption lines fit in the system towards Sk-68 135 in the LMC. Lines are the same as in Figure~\ref{fig:Sk67_2_CI}.}
    \label{fig:Sk68_135_CI}
\end{figure*}

\begin{figure*}
    \centering
    \includegraphics[width=\linewidth]{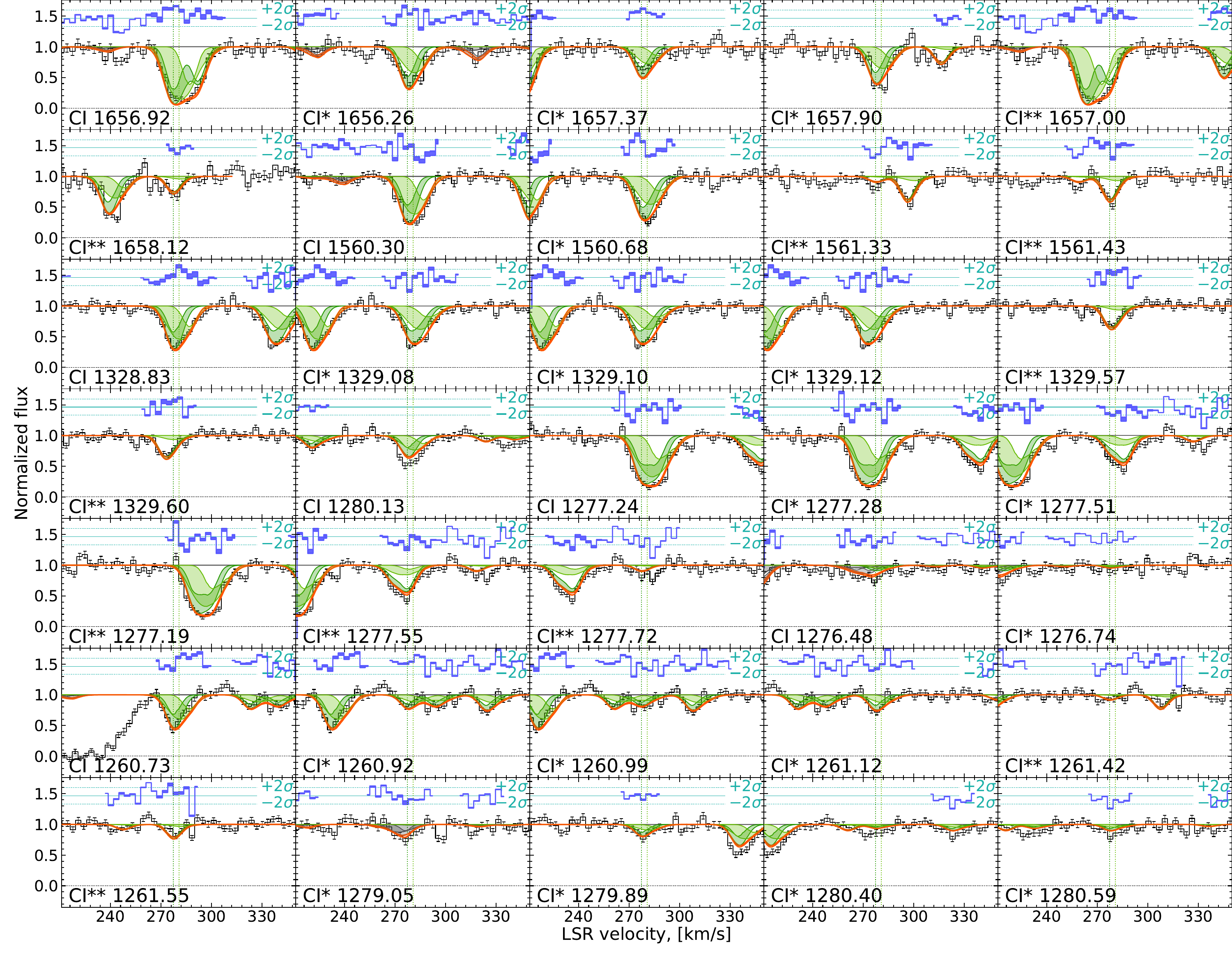}
    \caption{\CI\, absorption lines fit in the system towards Sk-69 246 in the LMC. Lines are the same as in Figure~\ref{fig:Sk67_2_CI}.}
    \label{fig:Sk69_246_CI}
\end{figure*}

\begin{figure*}
    \centering
    \includegraphics[width=\linewidth]{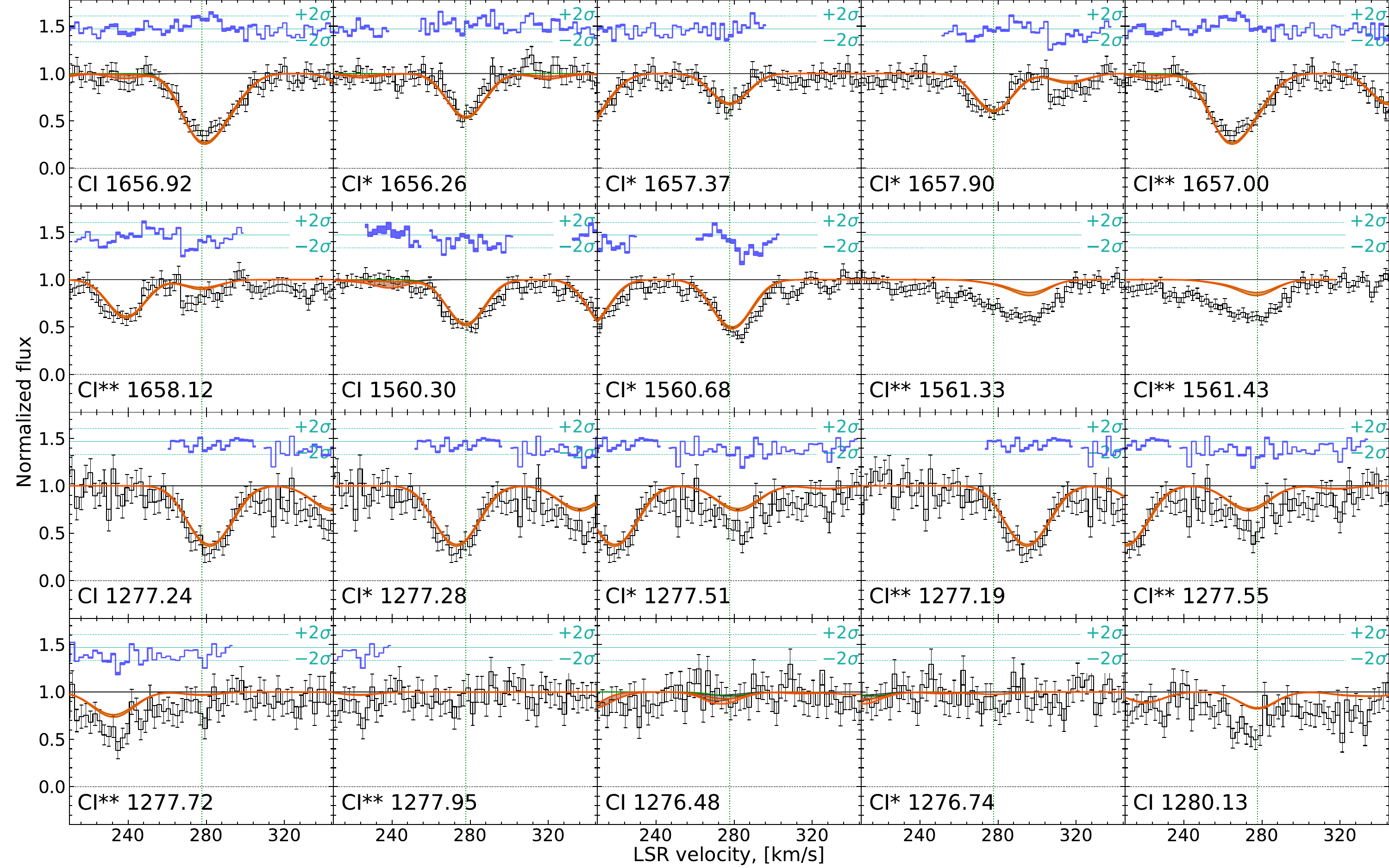}
    \caption{\CI\, absorption lines fit in the system towards Sk-68 140 in the LMC. Lines are the same as in Figure~\ref{fig:Sk67_2_CI}.}
    \label{fig:Sk68_140_CI}
\end{figure*}

\begin{figure*}
    \centering
    \includegraphics[width=\linewidth]{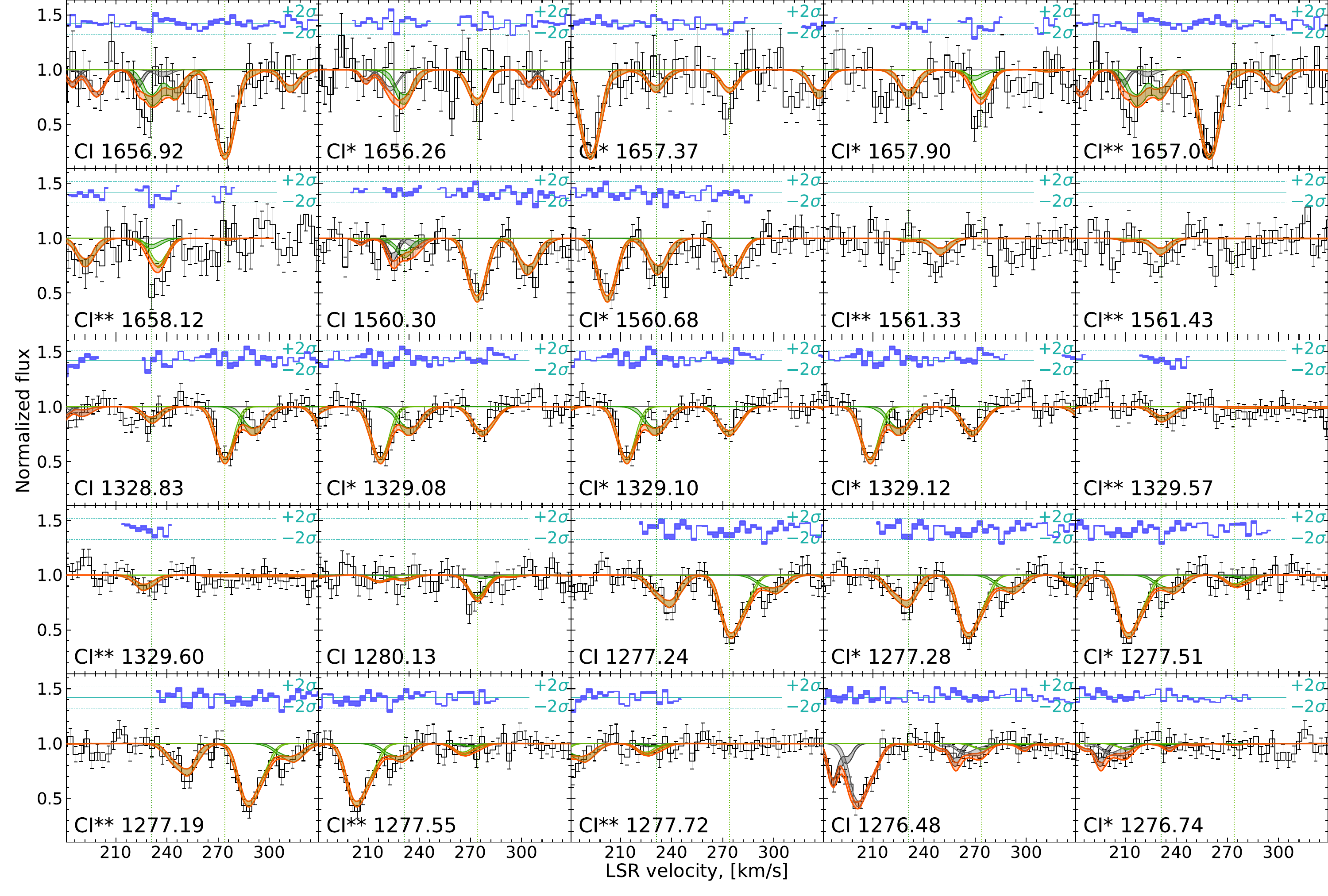}
    \caption{\CI\, absorption lines fit in the system towards Sk-71 50 in the LMC. Lines are the same as in Figure~\ref{fig:Sk67_2_CI}.}
    \label{fig:Sk71_50_CI}
\end{figure*}

\begin{figure*}
    \centering
    \includegraphics[width=\linewidth]{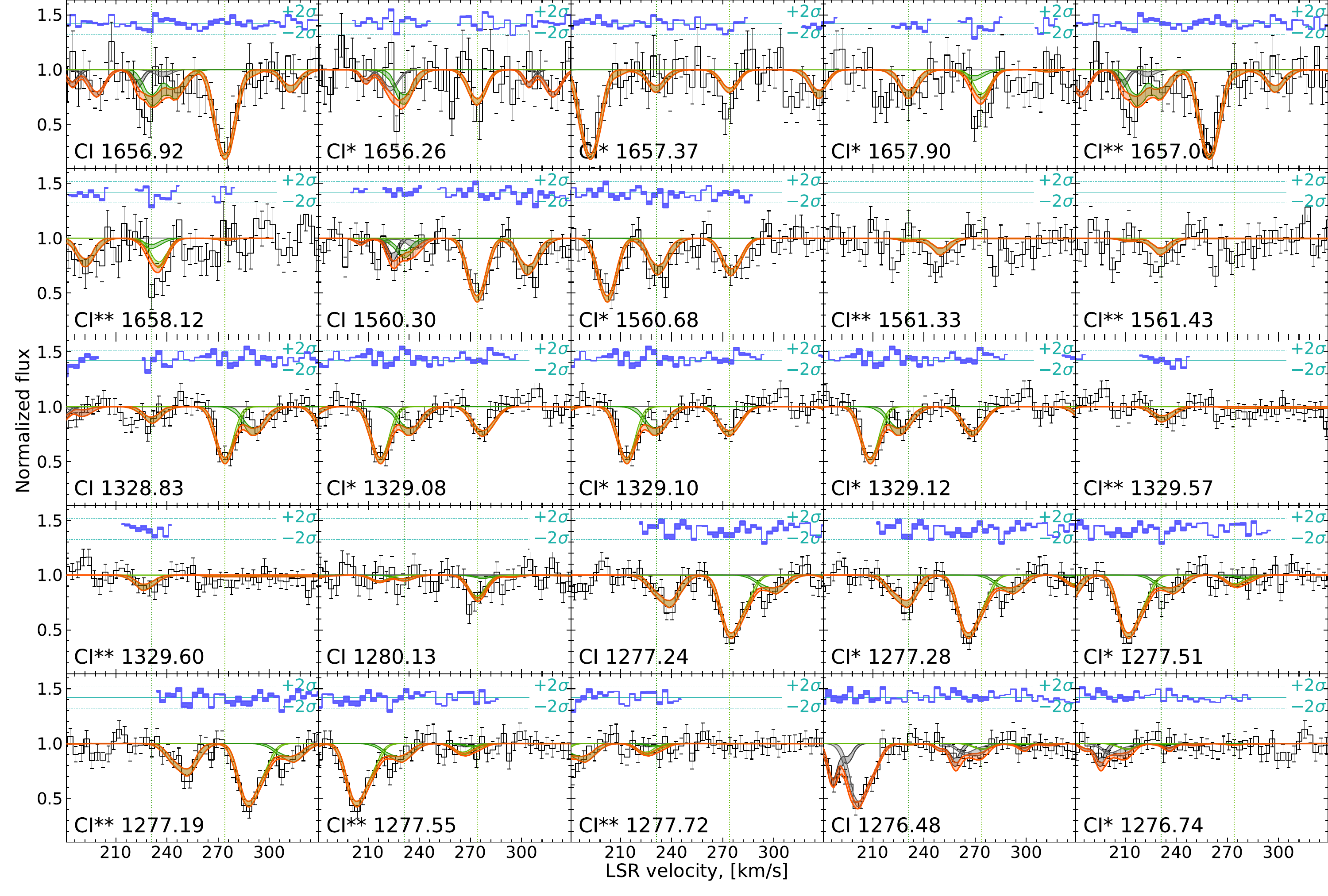}
    \caption{\CI\, absorption lines fit in the system towards Sk-71 50 in the LMC. Lines are the same as in Figure~\ref{fig:Sk67_2_CI}.}
    \label{fig:Sk71_50_CI}
\end{figure*}

\begin{figure*}
    \centering
    \includegraphics[width=\linewidth]{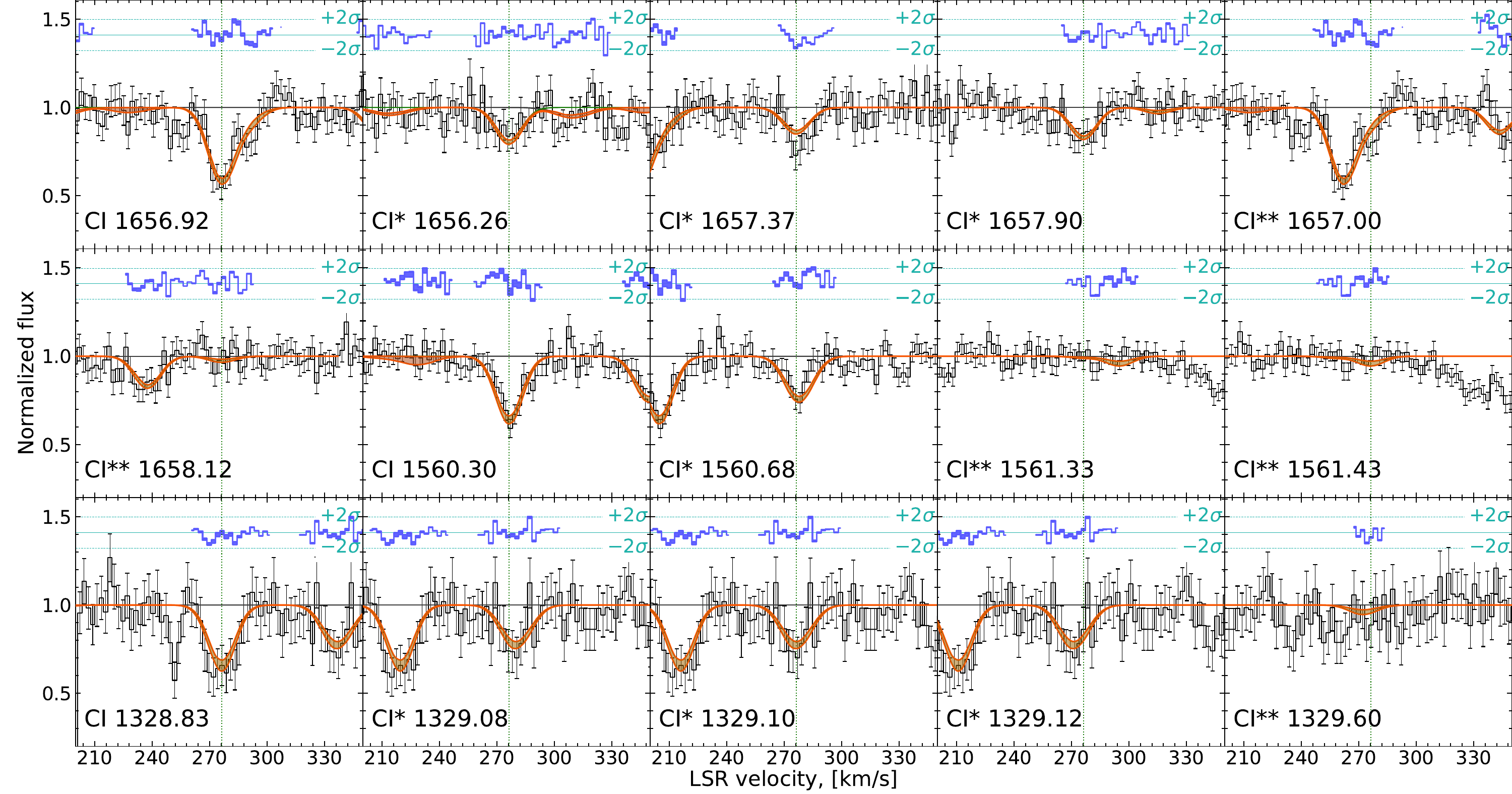}
    \caption{\CI\, absorption lines fit in the system towards Sk-69 279 in LMC. Lines are the same as in Figure~\ref{fig:Sk67_2_CI}.}
    \label{fig:Sk69_279_CI}
\end{figure*}

\begin{figure*}
    \centering
    \includegraphics[width=\linewidth]{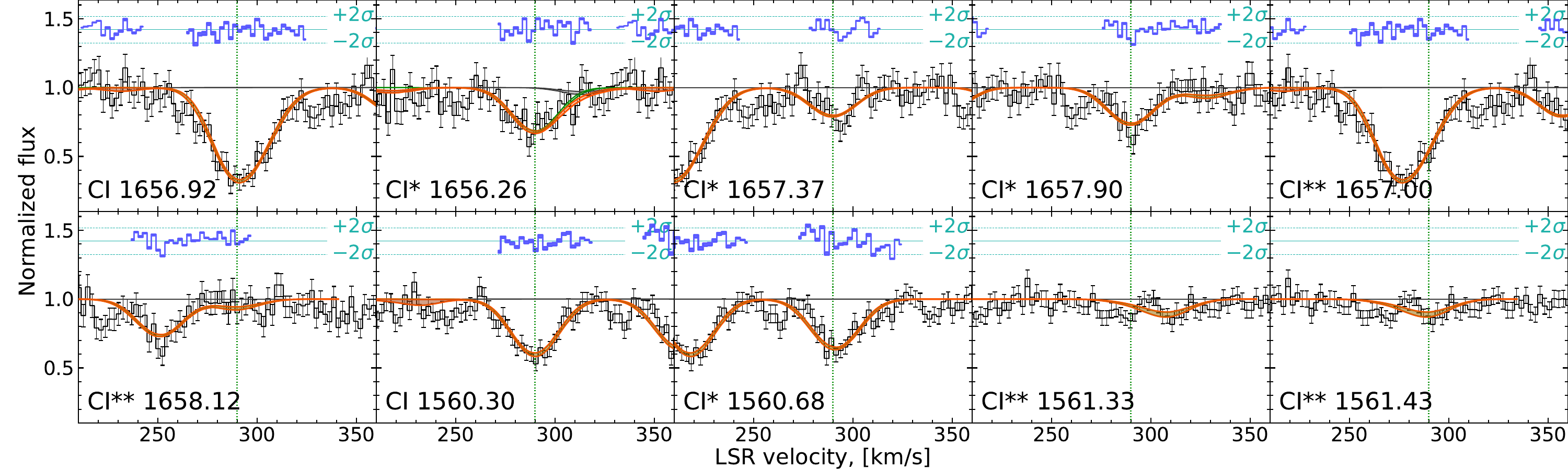}
    \caption{\CI\, absorption lines fit in the system towards Sk-68 155 in LMC. Lines are the same as in Figure~\ref{fig:Sk67_2_CI}.}
    \label{fig:Sk68_155_CI}
\end{figure*}

\begin{figure*}
    \centering
    \includegraphics[width=\linewidth]{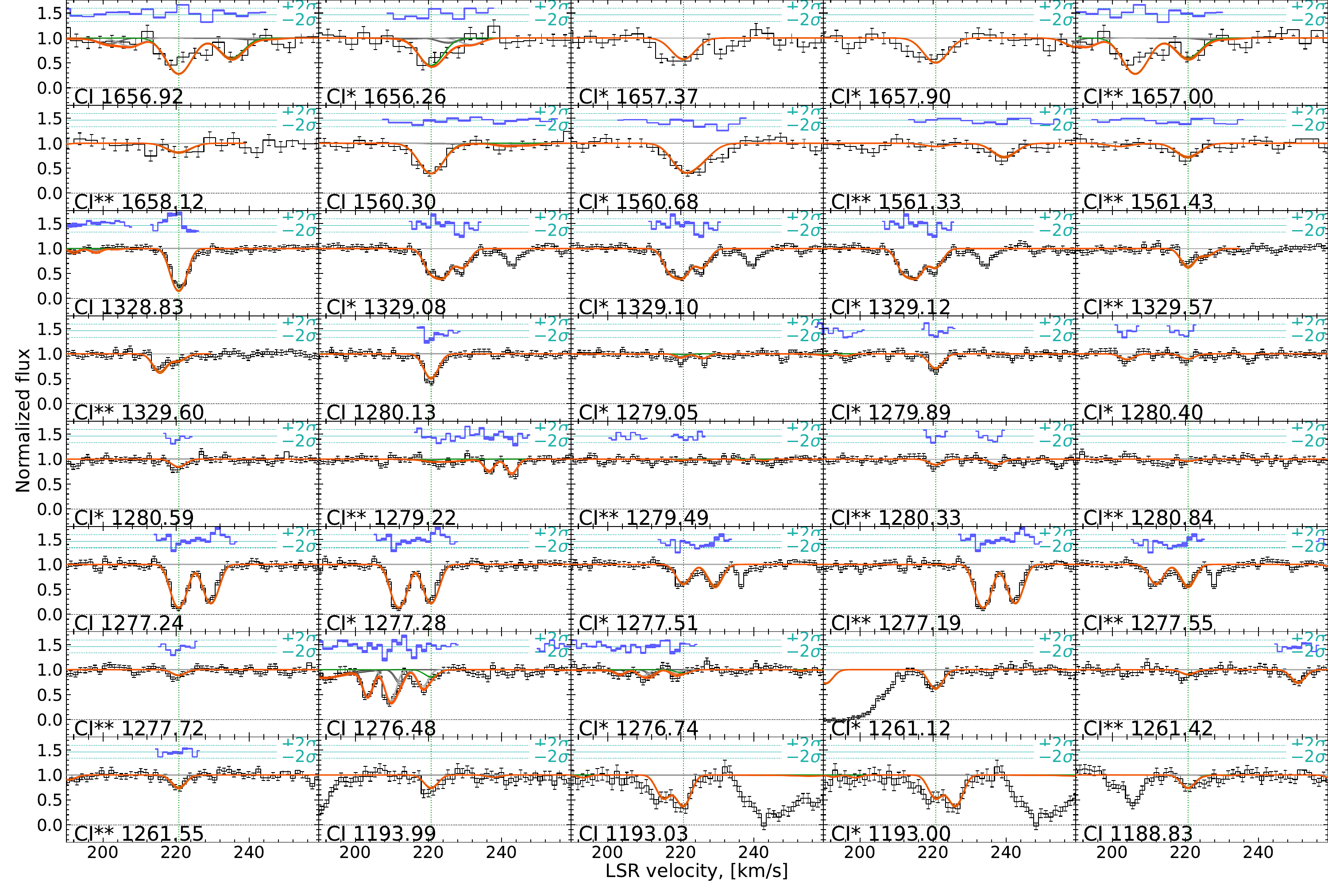}
    \caption{\CI\, absorption lines fit in the system towards Sk-70 115 in the LMC. Lines are the same as in Figure~\ref{fig:Sk67_2_CI}.}
    \label{fig:Sk70_115_CI}
\end{figure*}

\subsection{Small Magellanic Cloud}

\begin{figure*}
    \centering
    \includegraphics[width=\linewidth]{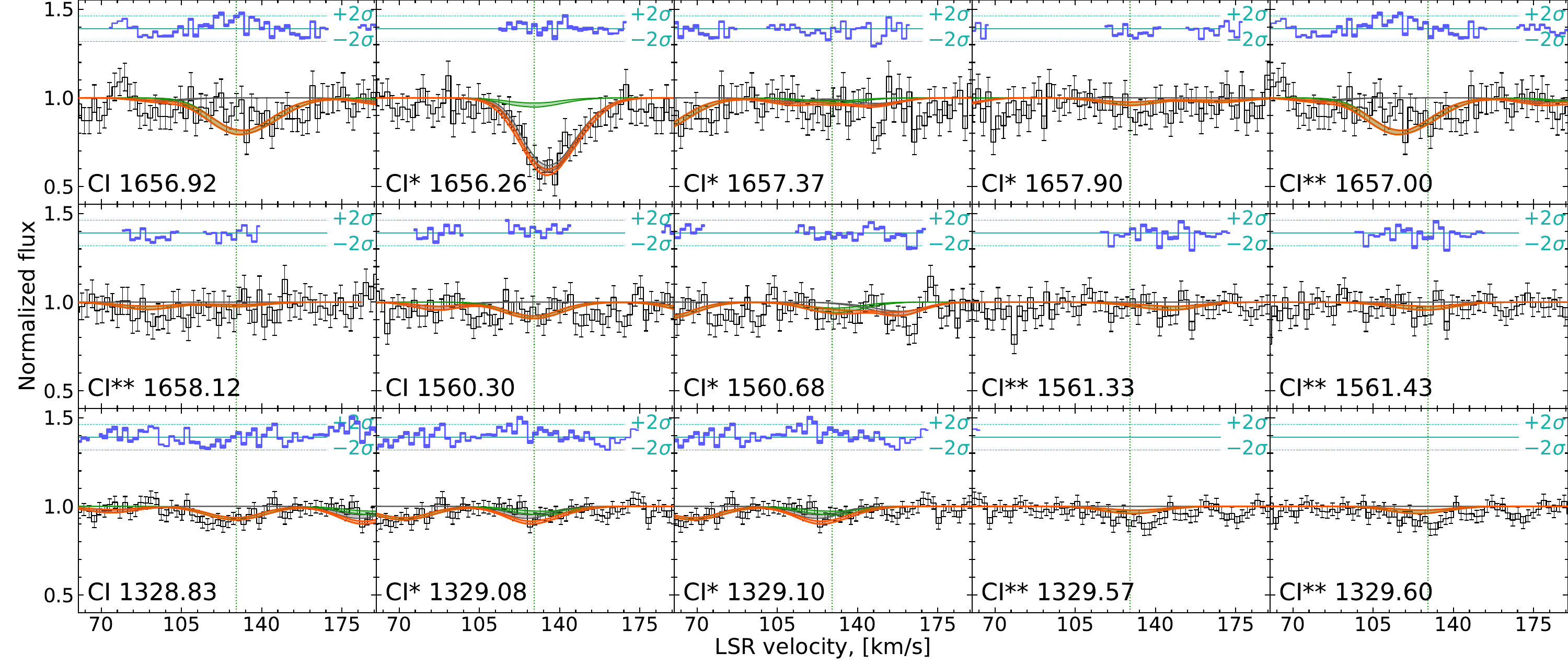}
    \caption{\CI\, absorption lines fit in the system towards AV 6 in the SMC. Lines are the same as in Figure~\ref{fig:Sk67_2_CI}}
    \label{fig:AV6_CI}
\end{figure*}

\begin{figure*}
    \centering
    \includegraphics[width=\linewidth]{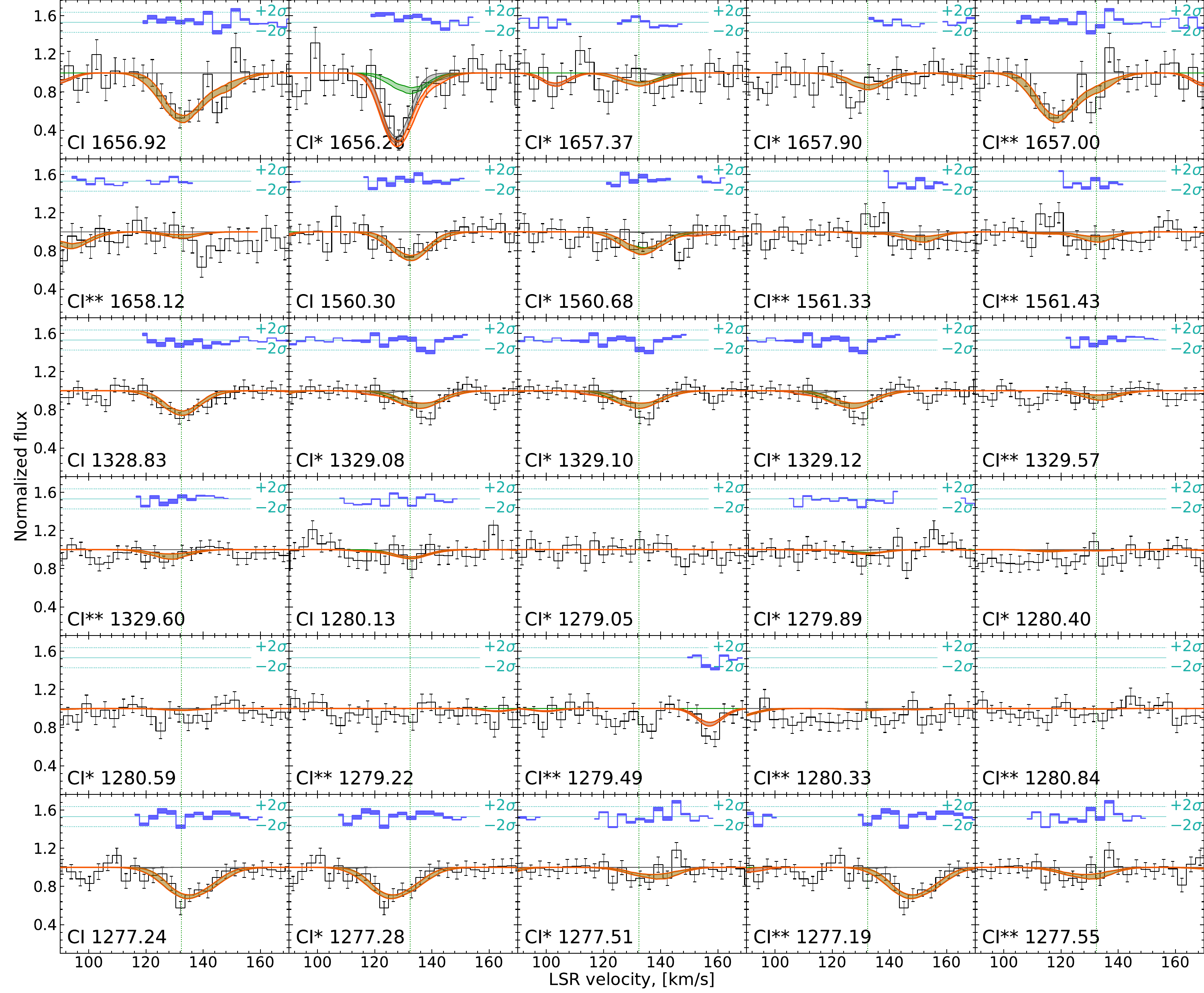}
    \caption{\CI\, absorption lines fit in the system towards AV 15 in the SMC. Lines are the same as in Figure~\ref{fig:Sk67_2_CI}}
    \label{fig:AV15_CI}
\end{figure*}

\begin{figure*}
    \centering
    \includegraphics[width=\linewidth]{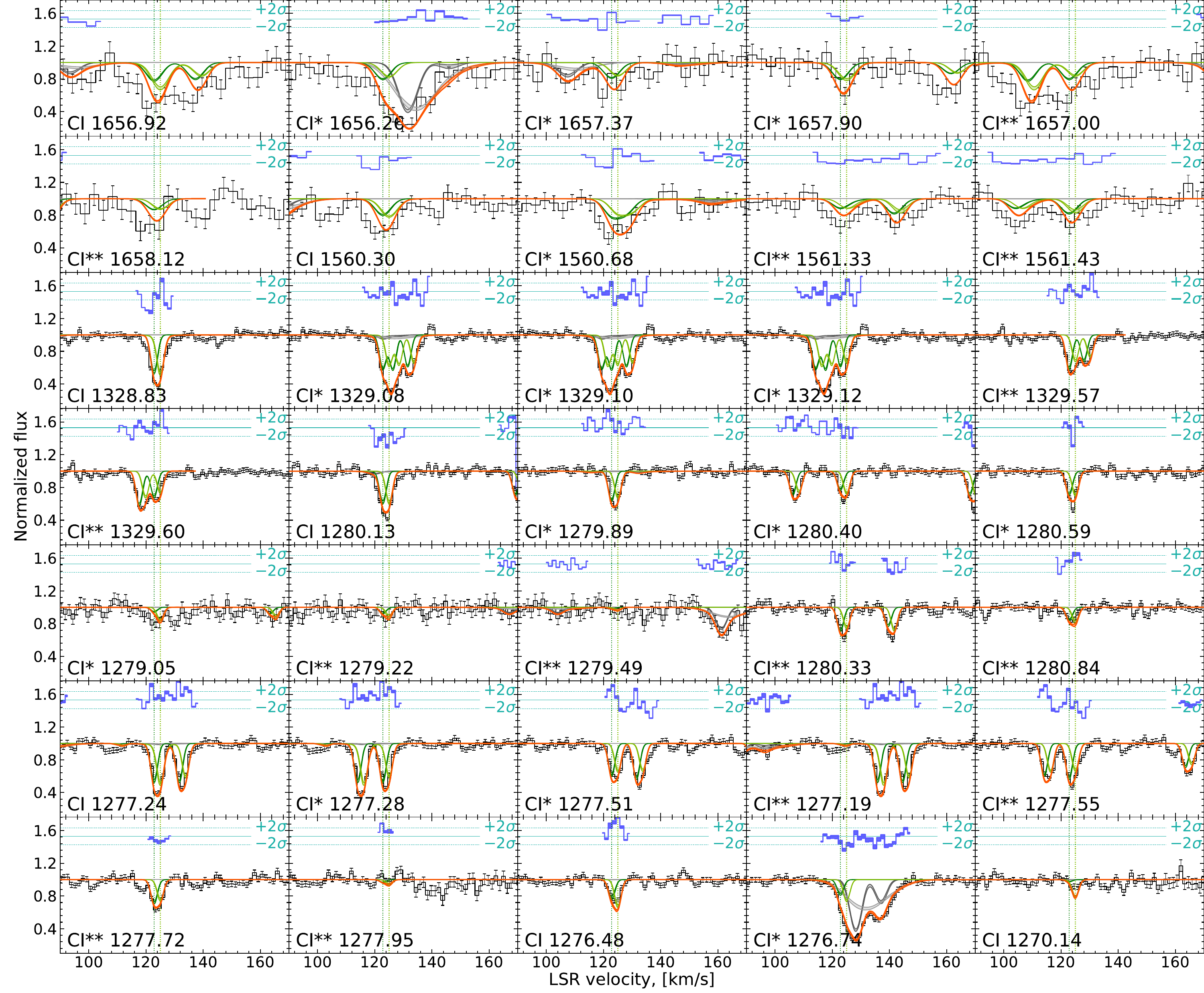}
    \caption{\CI\, absorption lines fit in the system towards AV 26 in the SMC. Lines are the same as in Figure~\ref{fig:Sk67_2_CI}}
    \label{fig:AV26_CI}
\end{figure*}

\begin{figure*}
    \centering
    \includegraphics[width=\linewidth]{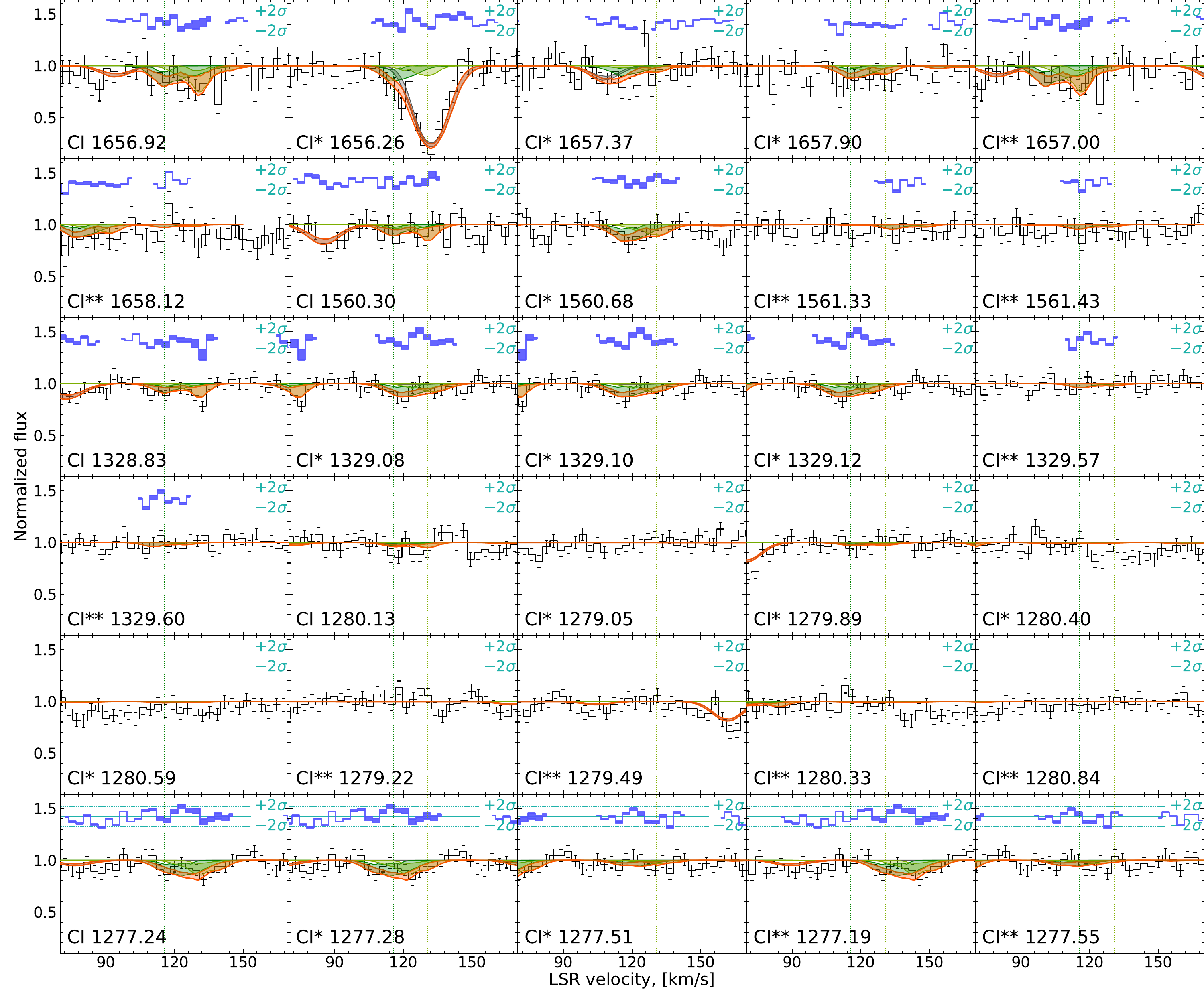}
    \caption{\CI\, absorption lines fit in the system towards AV 47 in the SMC. Lines are the same as in Figure~\ref{fig:Sk67_2_CI}}
    \label{fig:AV47_CI}
\end{figure*}

\begin{figure*}
    \centering
    \includegraphics[width=\linewidth]{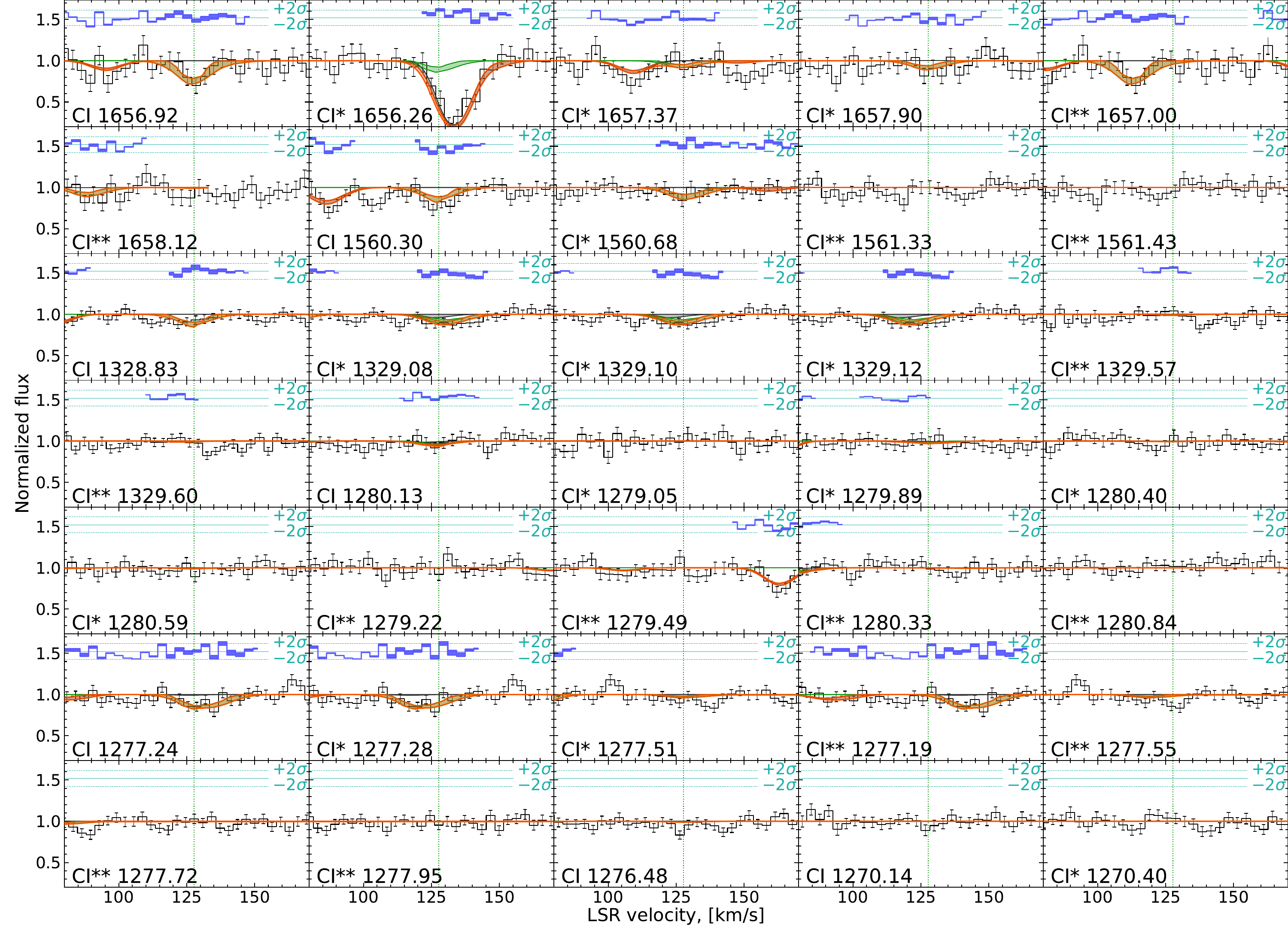}
    \caption{\CI\, absorption lines fit in the system towards AV 69 in SMC. Lines are the same as in Figure~\ref{fig:Sk67_2_CI}}
    \label{fig:AV69_CI}
\end{figure*}

\begin{figure*}
    \centering
    \includegraphics[width=\linewidth]{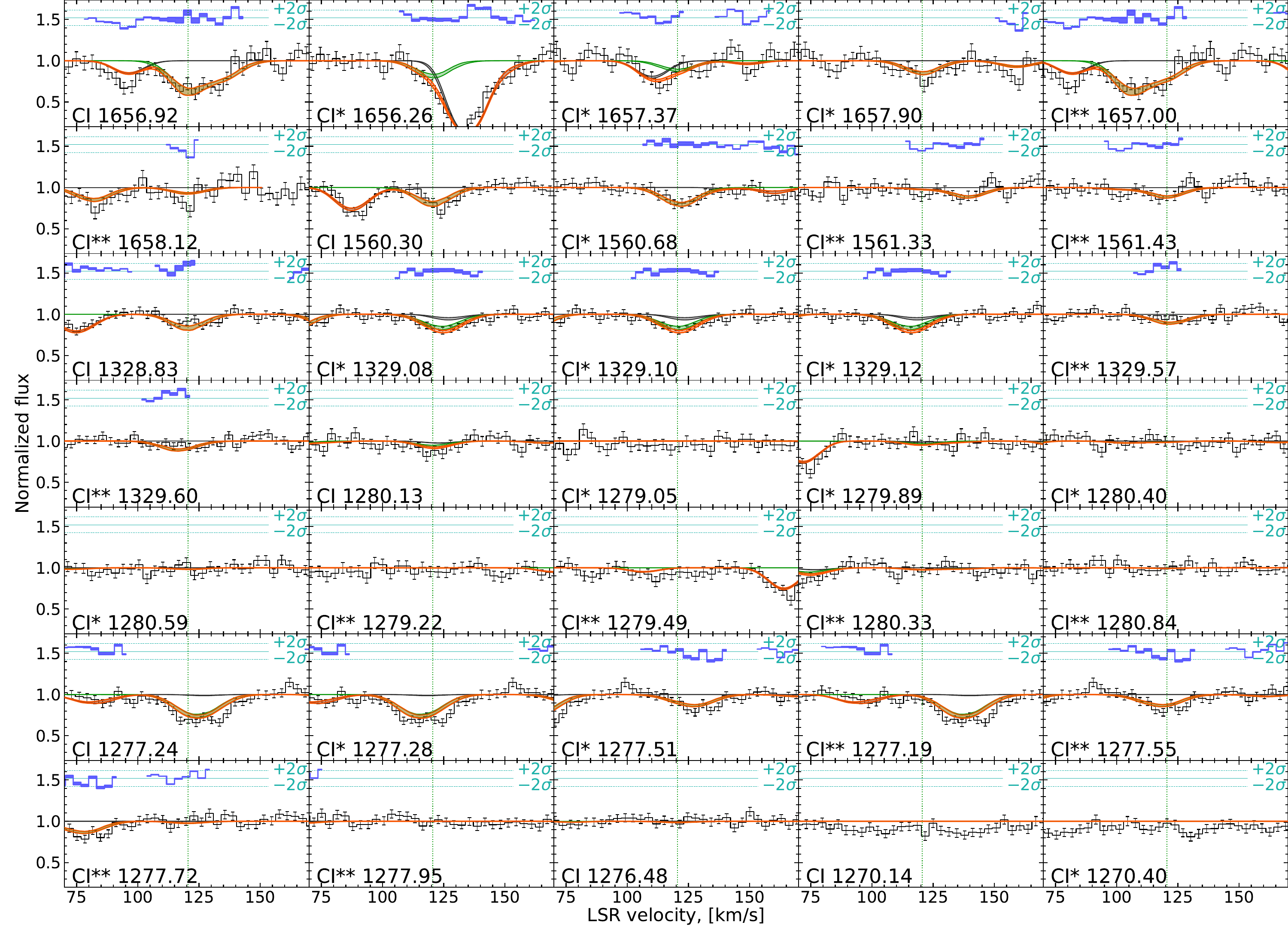}
    \caption{\CI\, absorption lines fit in the system towards AV 75 in the SMC. Lines are the same as in Figure~\ref{fig:Sk67_2_CI}}
    \label{fig:AV75_CI}
\end{figure*}

\begin{figure*}
    \centering
    \includegraphics[width=\linewidth]{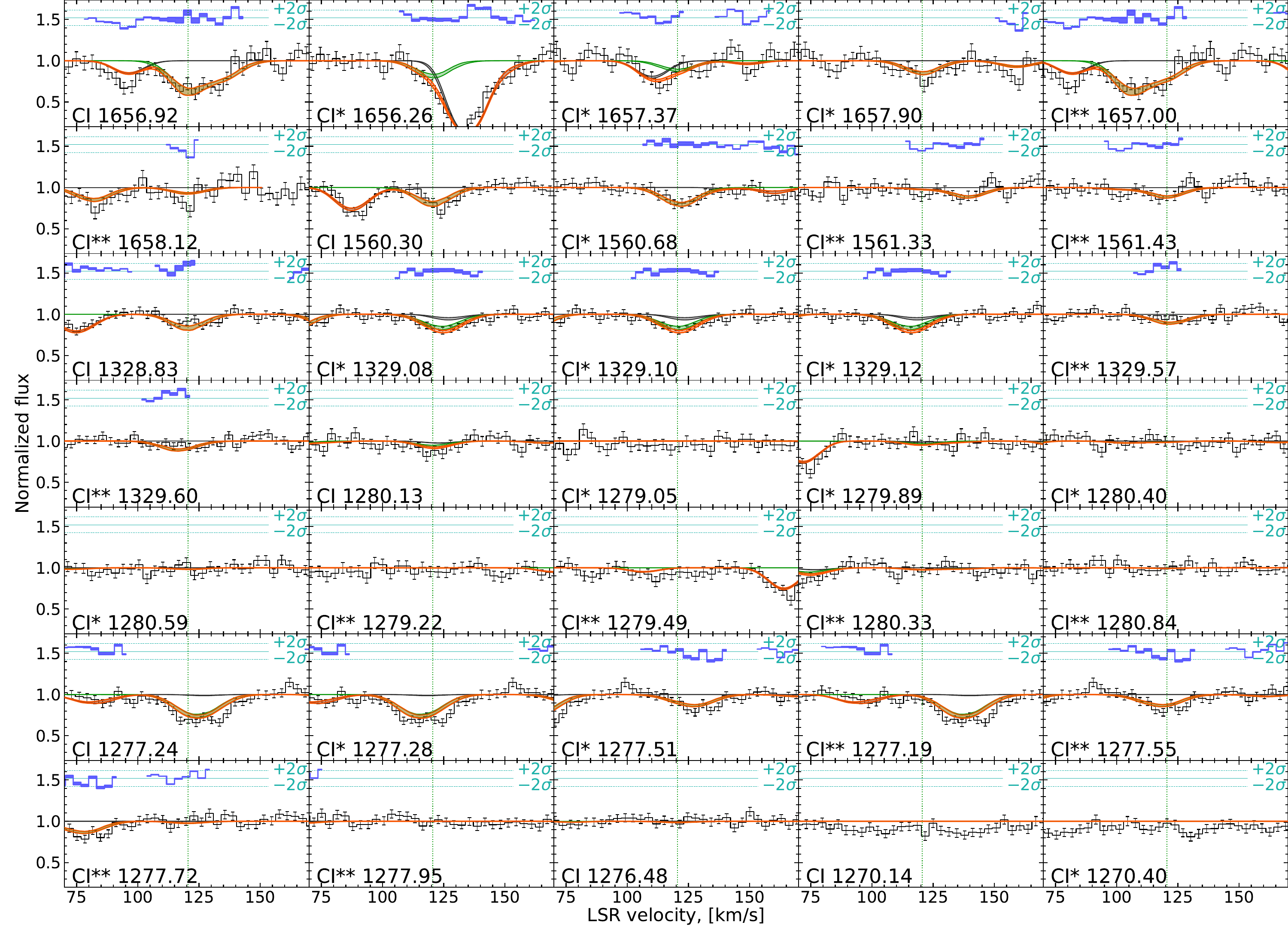}
    \caption{\CI\, absorption lines fit in the system towards AV 80 in the SMC. Lines are the same as in Figure~\ref{fig:Sk67_2_CI}}
    \label{fig:AV80_CI}
\end{figure*}

\begin{figure*}
    \centering
    \includegraphics[width=\linewidth]{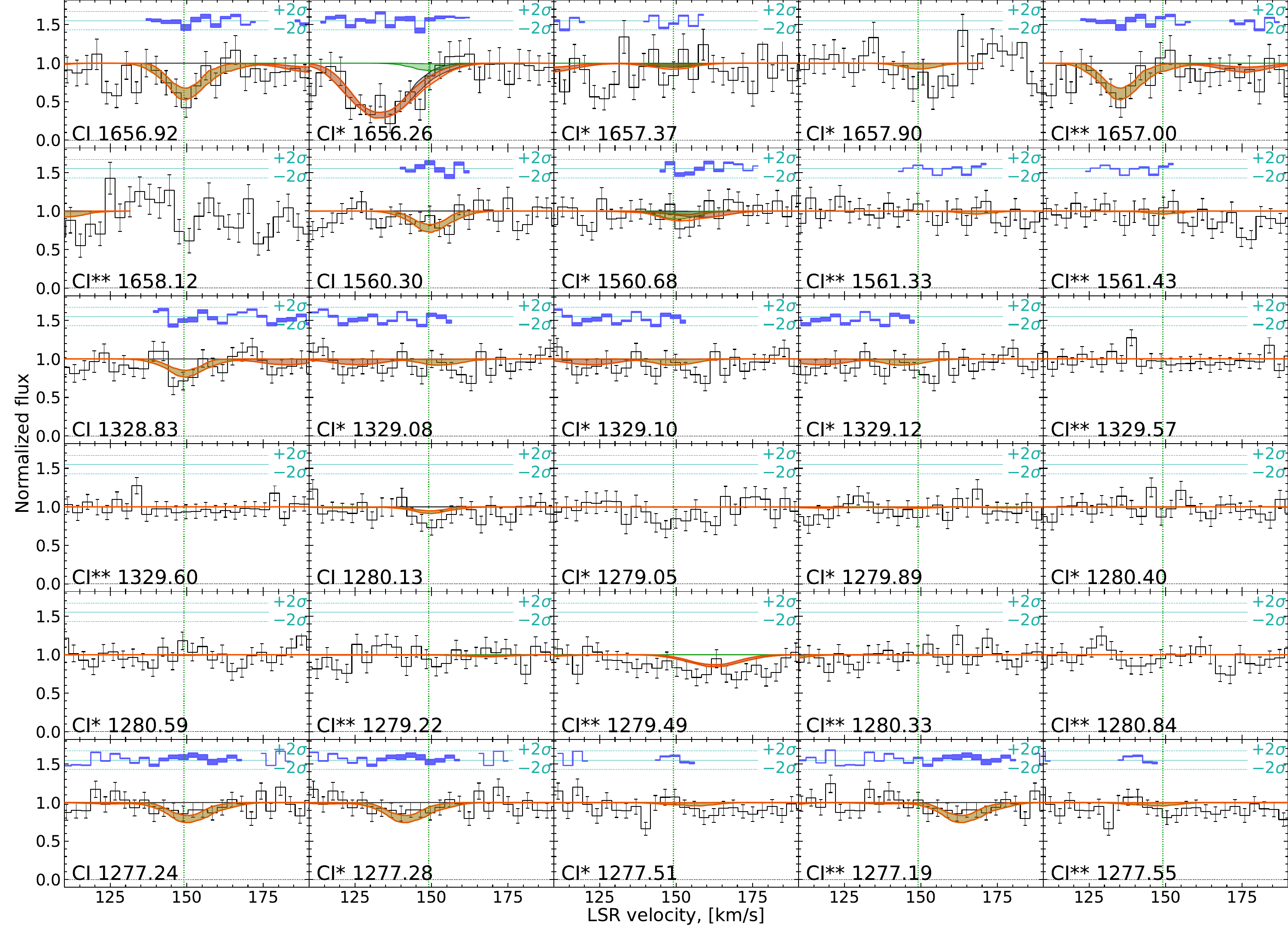}
    \caption{\CI\, absorption lines fit in the system towards AV 81 in the SMC. Lines are the same as in Figure~\ref{fig:Sk67_2_CI}}
    \label{fig:AV81_CI}
\end{figure*}

\begin{figure*}
    \centering
    \includegraphics[width=\linewidth]{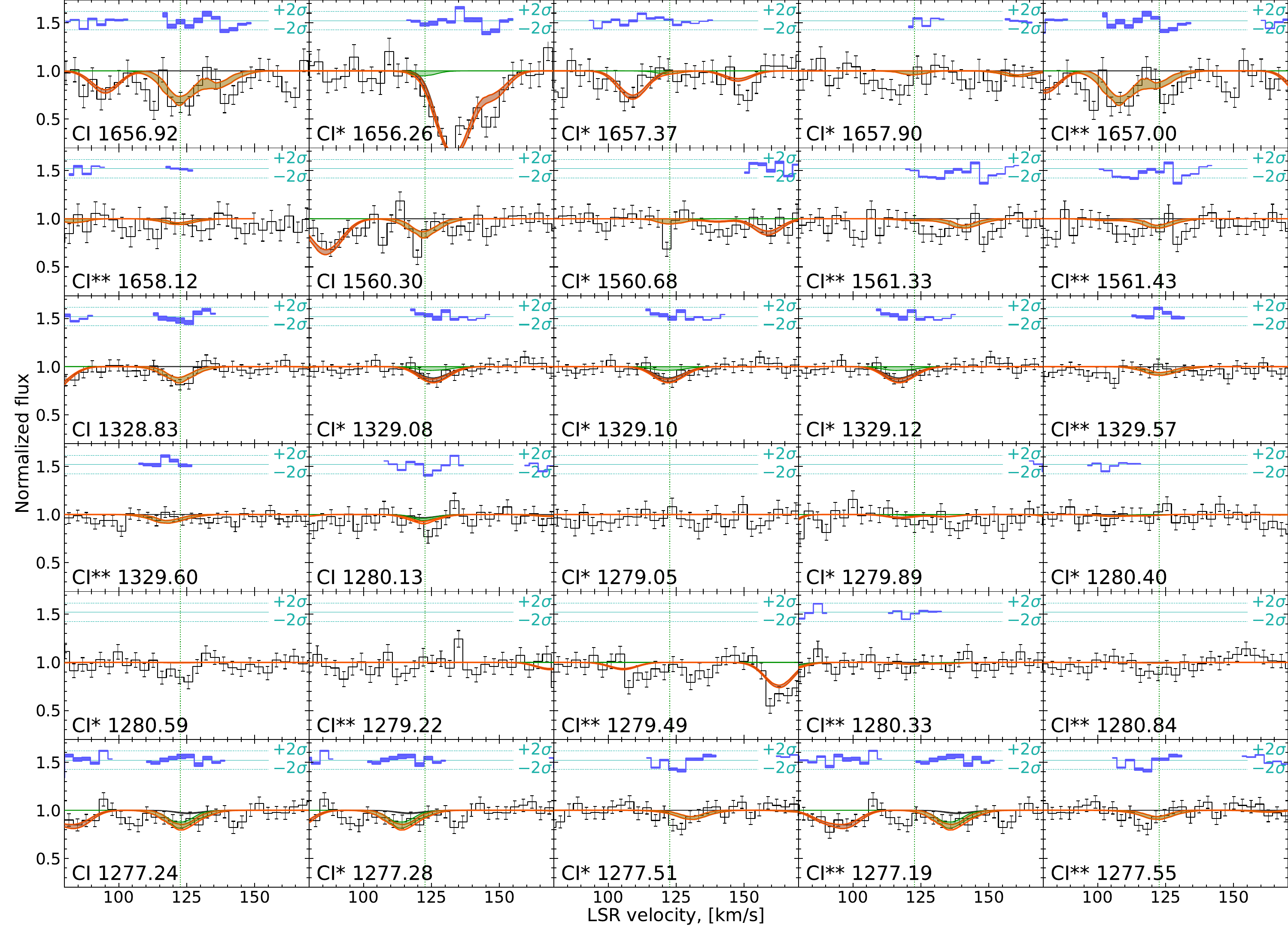}
    \caption{\CI\, absorption lines fit in the system towards AV 95 in SMC. Lines are the same as in Figure~\ref{fig:Sk67_2_CI}}
    \label{fig:AV95_CI}
\end{figure*}

\begin{figure*}
    \centering
    \includegraphics[width=\linewidth]{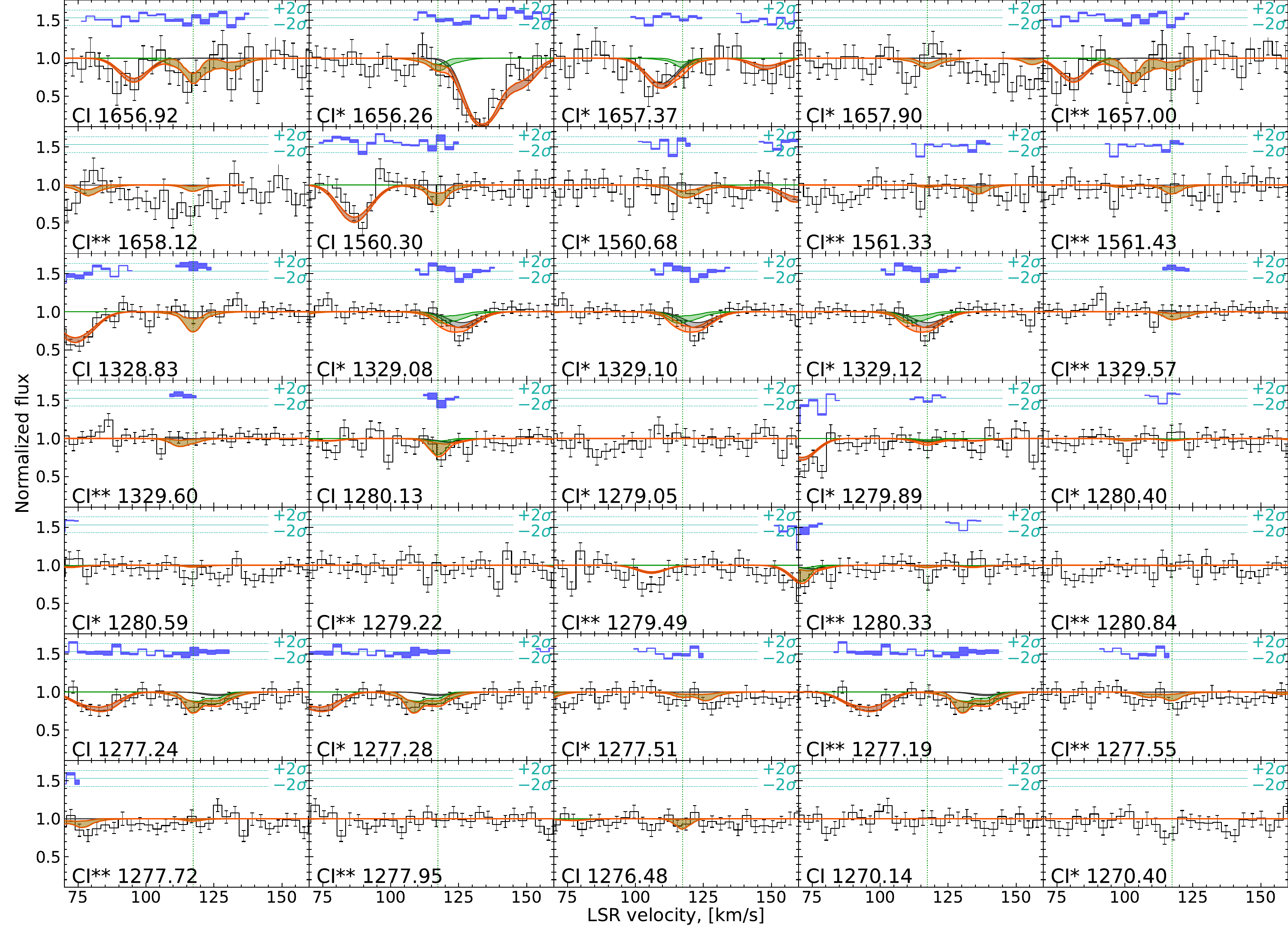}
    \caption{\CI\, absorption lines fit in the system towards AV 104 in the SMC. Lines are the same as in Figure~\ref{fig:Sk67_2_CI}}
    \label{fig:AV104_CI}
\end{figure*}

\begin{figure*}
    \centering
    \includegraphics[width=\linewidth]{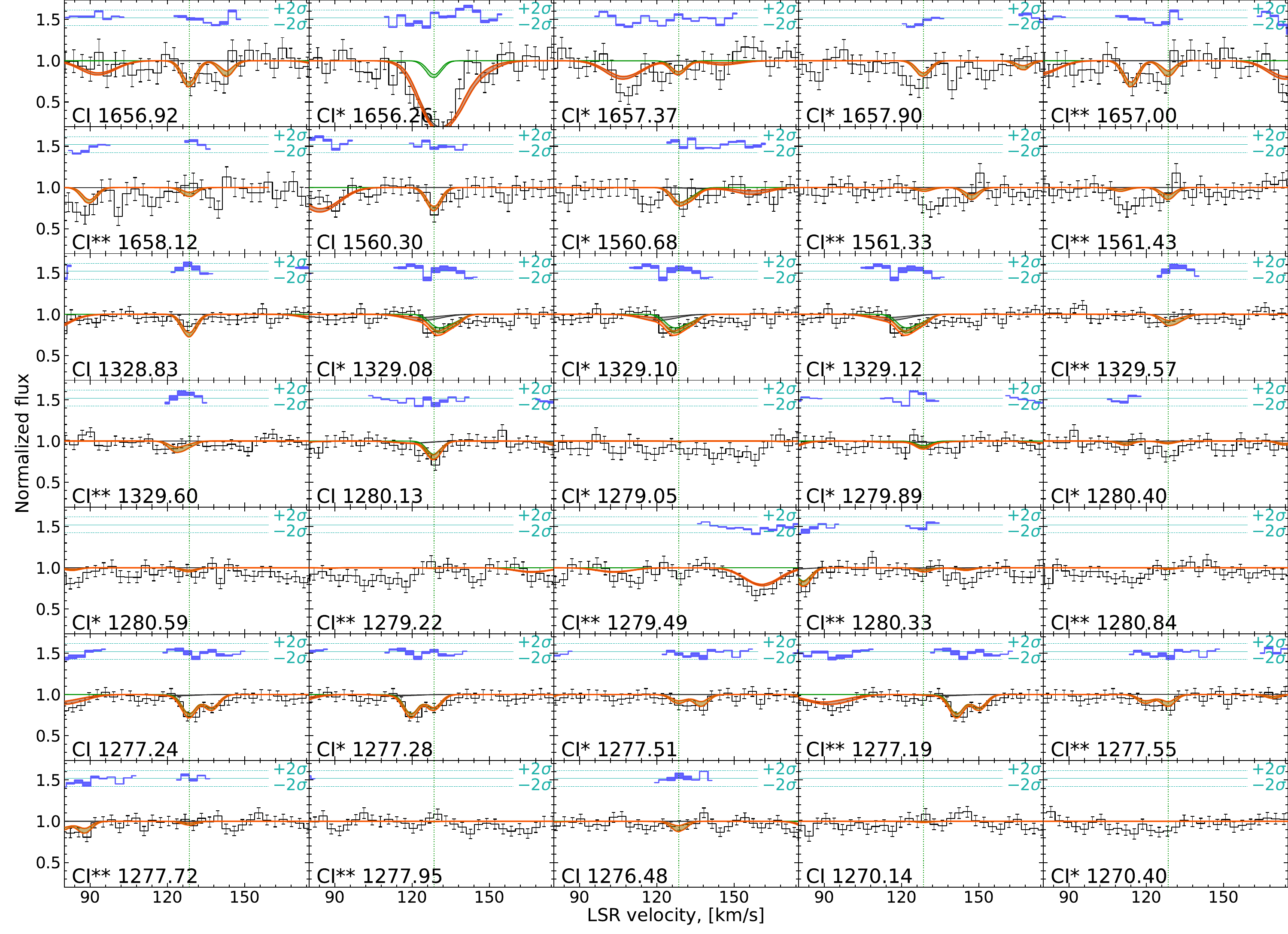}
    caption{\CI\, absorption lines fit in the system towards AV 170 in the SMC. Lines are the same as in Figure~\ref{fig:Sk67_2_CI}}
    \label{fig:AV170_CI}
\end{figure*}

\begin{figure*}
    \centering
    \includegraphics[width=\linewidth]{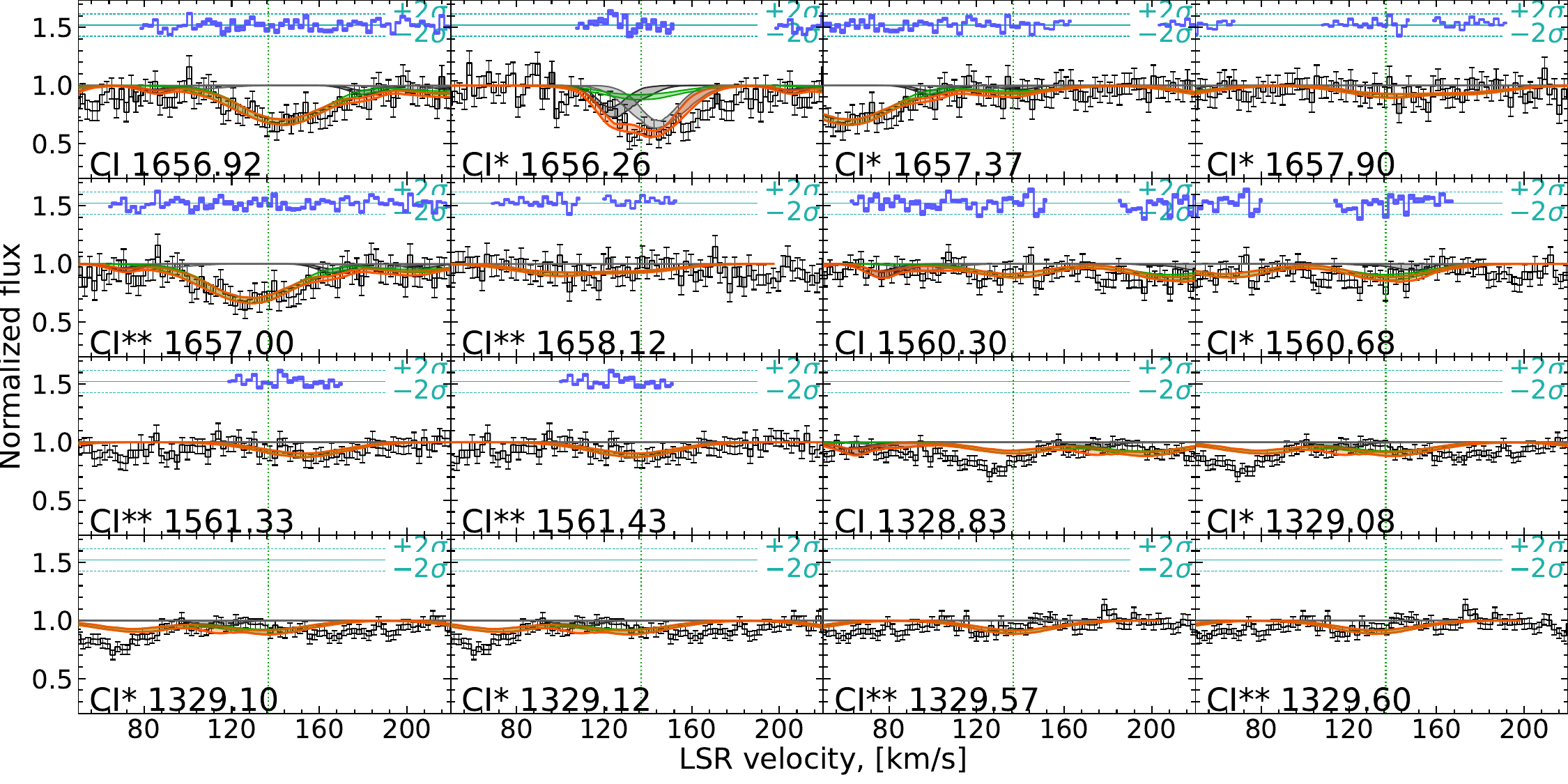}
    \caption{\CI\, absorption lines fit in the system towards AV 175 in the SMC. Lines are the same as in Figure~\ref{fig:Sk67_2_CI}}
    \label{fig:AV175_CI}
\end{figure*}

\begin{figure*}
    \centering
    \includegraphics[width=\linewidth]{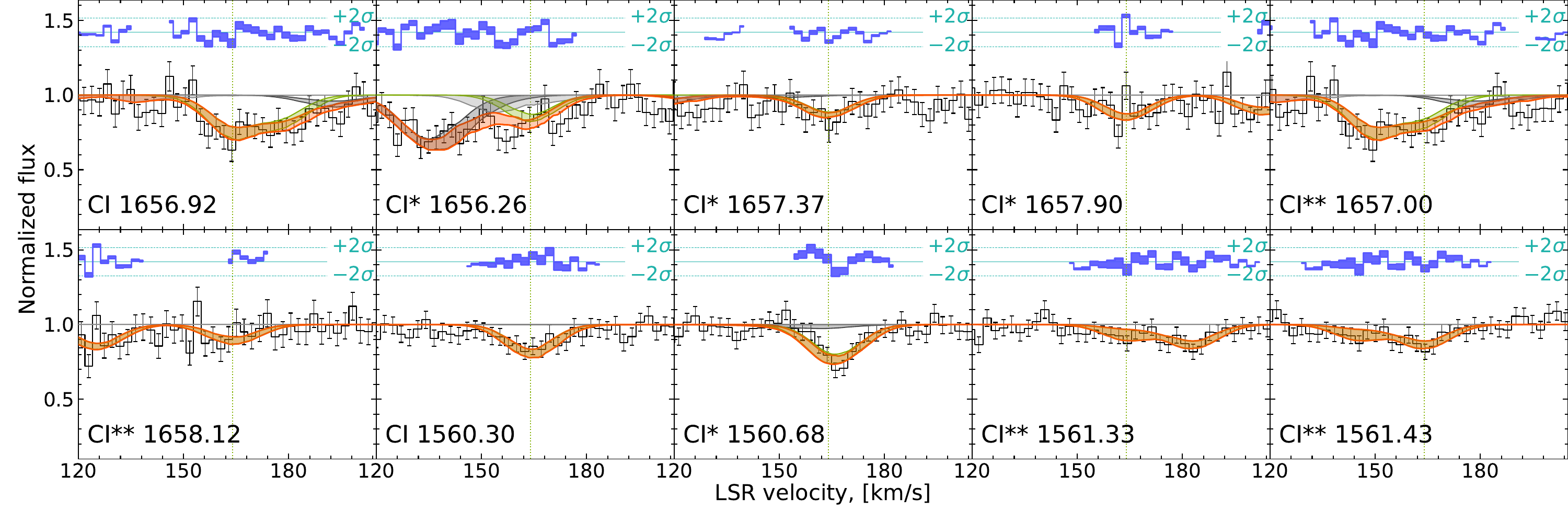}
    \caption{\CI\, absorption lines fit in the system towards AV 207 in SMC. Lines are the same as in Figure~\ref{fig:Sk67_2_CI}}
    \label{fig:AV207_CI}
\end{figure*}

\begin{figure*}
    \centering
    \includegraphics[width=\linewidth]{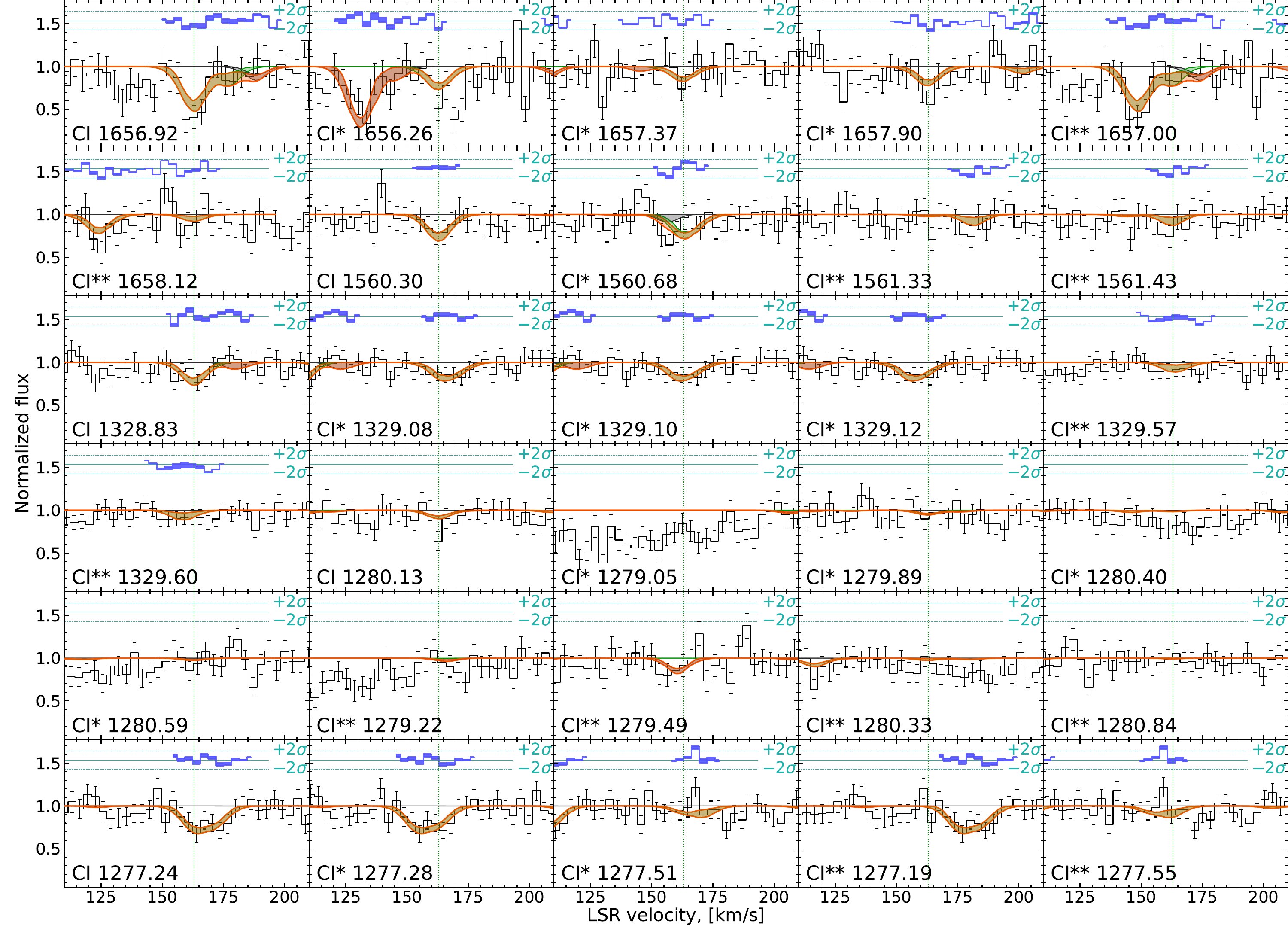}
    \caption{\CI\, absorption lines fit in the system towards AV 210 in the SMC. Lines are the same as in Figure~\ref{fig:Sk67_2_CI}}
    \label{fig:AV210_CI}
\end{figure*}

\begin{figure*}
    \centering
    \includegraphics[width=\linewidth]{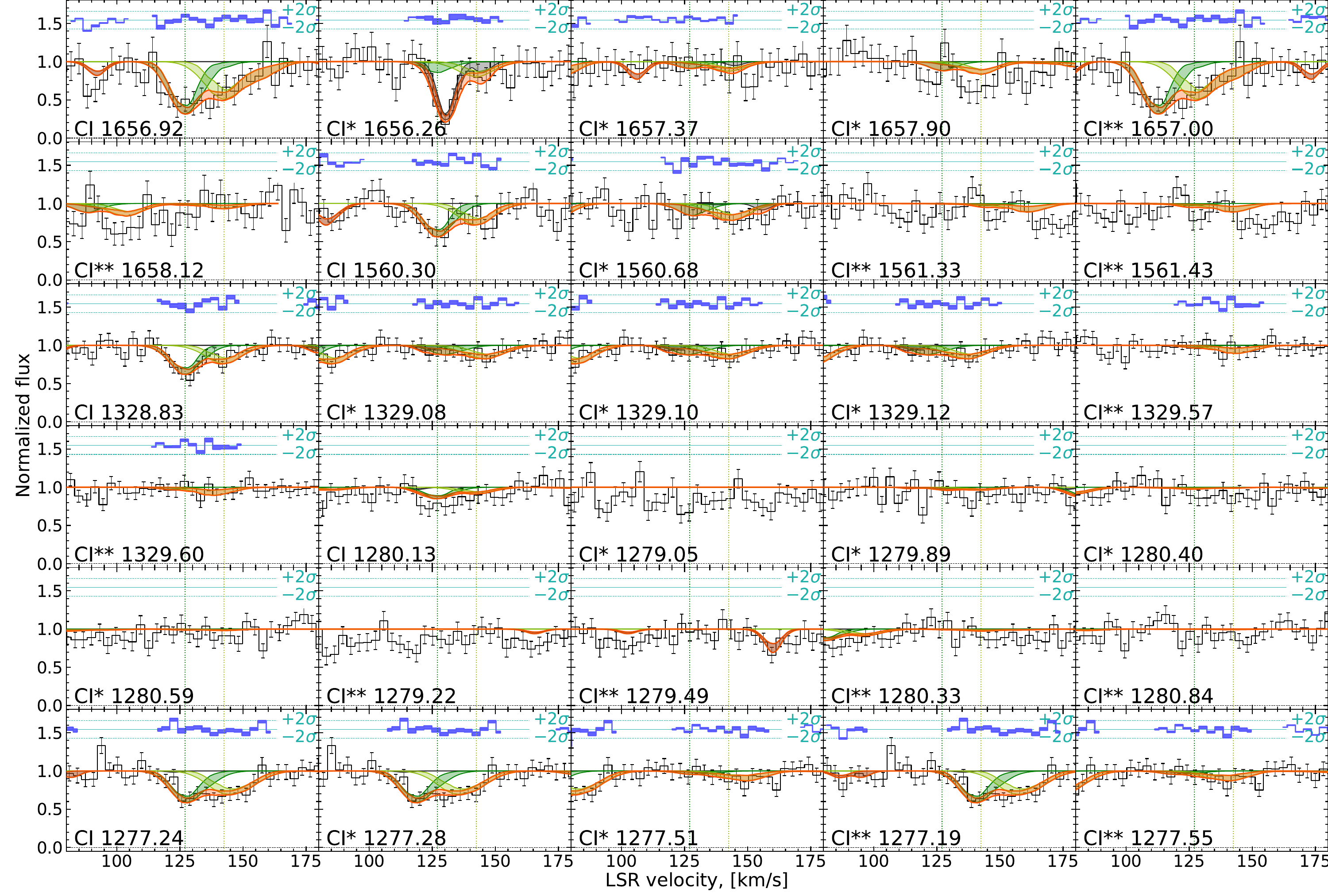}
    \caption{\CI\, absorption lines fit in the system towards AV 215 in the SMC. Lines are the same as in Figure~\ref{fig:Sk67_2_CI}}
    \label{fig:AV215_CI}
\end{figure*}

\begin{figure*}
    \centering
    \includegraphics[width=\linewidth]{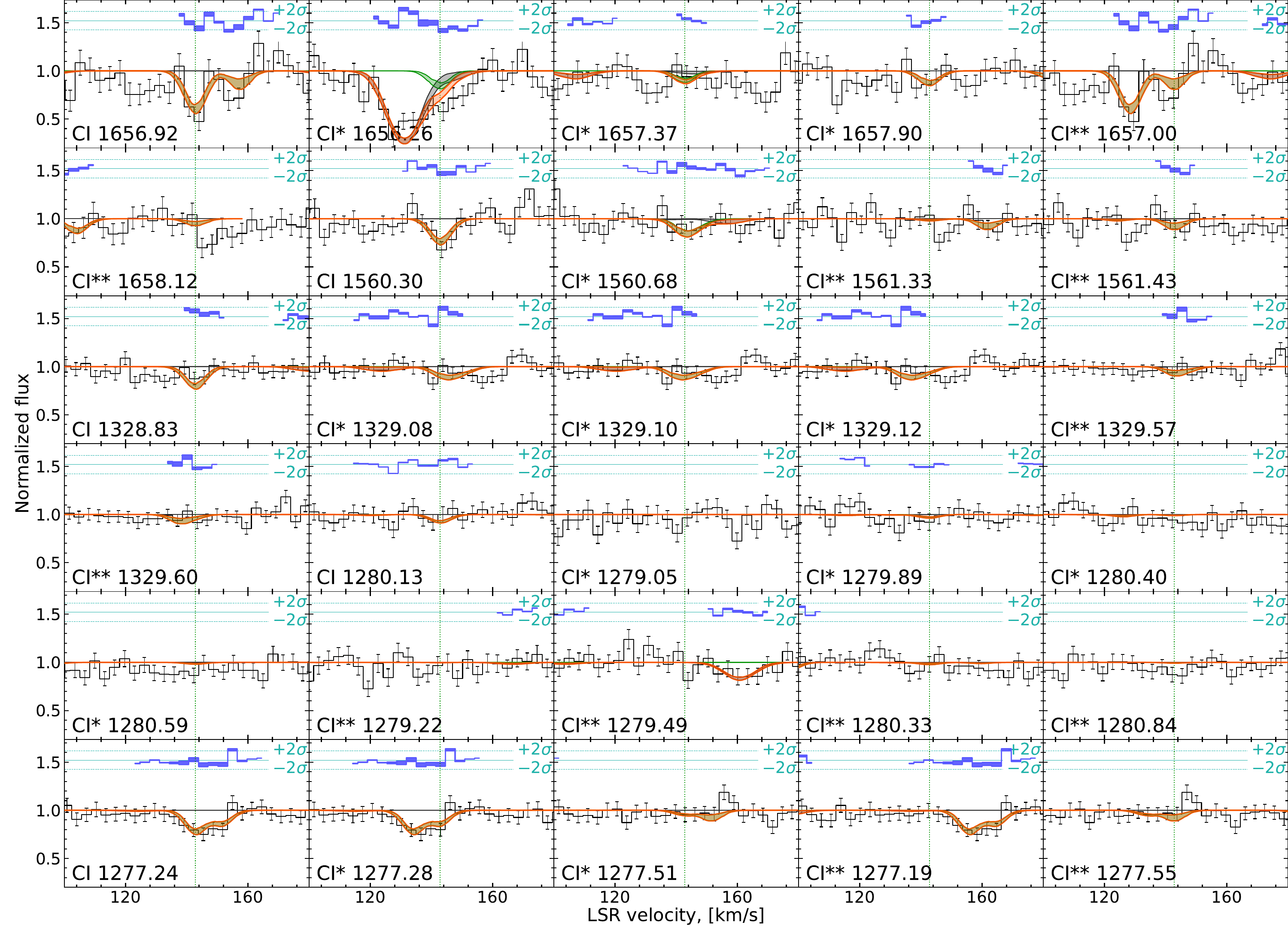}
    \caption{\CI\, absorption lines fit in the system towards AV 216 in the SMC. Lines are the same as in Figure~\ref{fig:Sk67_2_CI}}
    \label{fig:AV216_CI}
\end{figure*}

\begin{figure*}
    \centering
    \includegraphics[width=\linewidth]{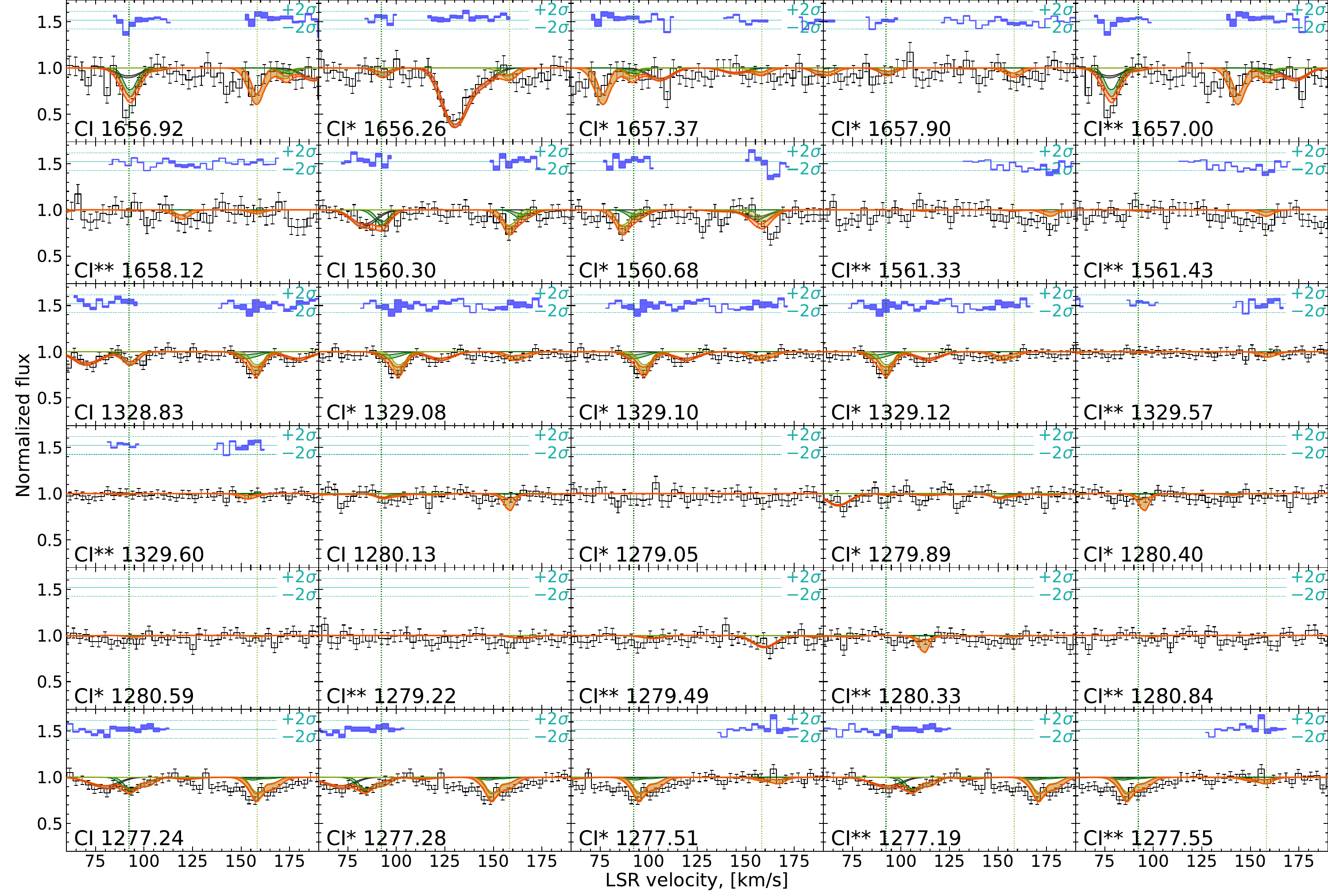}
    \caption{\CI\, absorption lines fit in the system towards AV 242 in SMC. Lines are the same as in Figure~\ref{fig:Sk67_2_CI}}
    \label{fig:AV242_CI}
\end{figure*}

\begin{figure*}
    \centering
    \includegraphics[width=\linewidth]{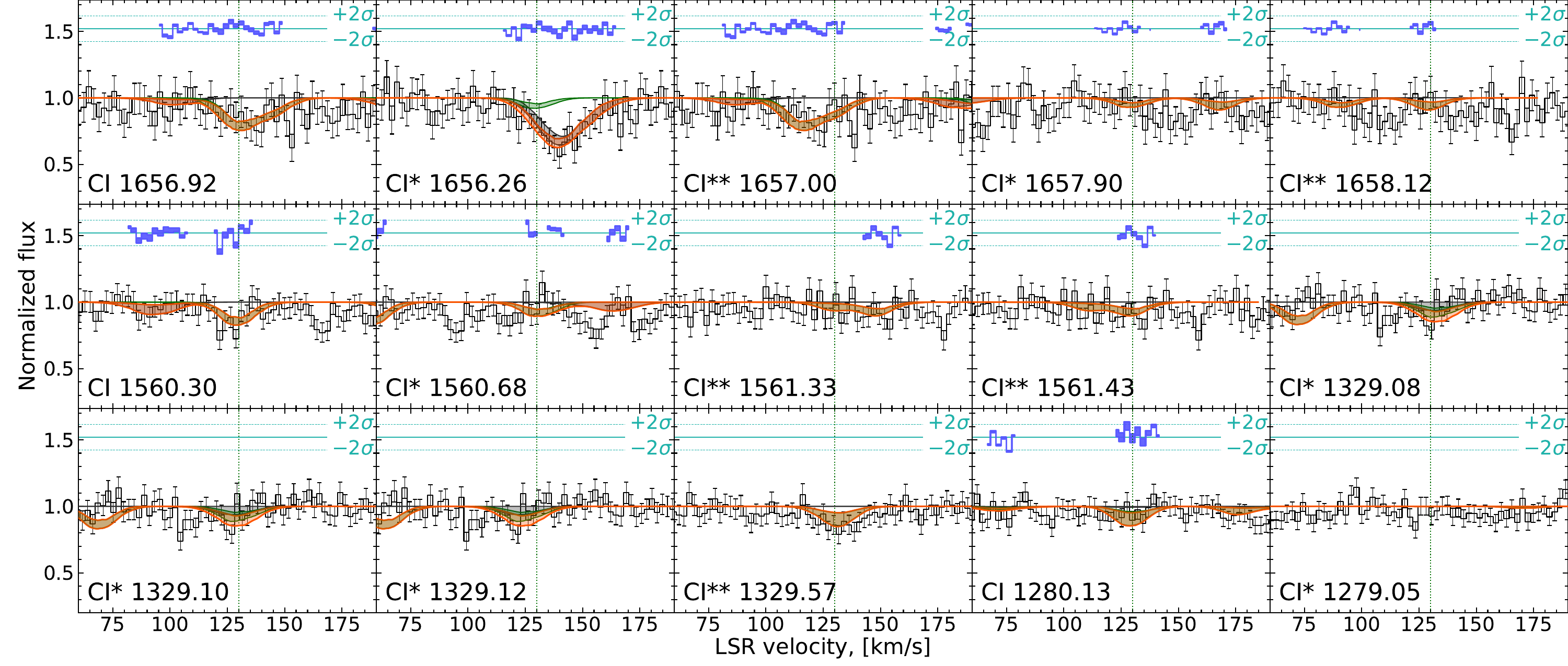}
    \caption{\CI\, absorption lines fit in the system towards AV 243 in SMC. Lines are the same as in Figure~\ref{fig:Sk67_2_CI}}
    \label{fig:AV243_CI}
\end{figure*}

\clearpage
\begin{figure*}
    \centering
    \includegraphics[width=\linewidth]{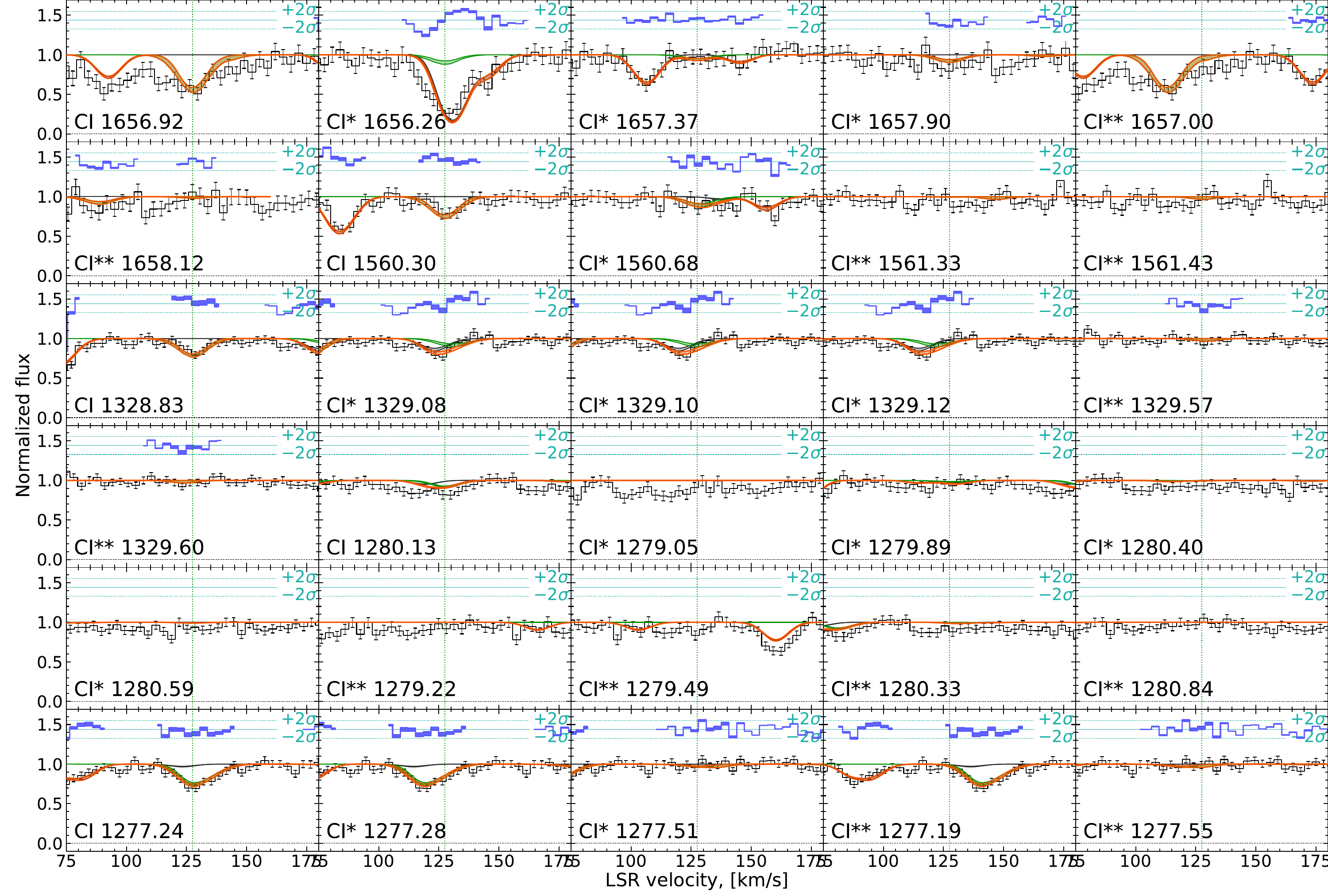}
    \caption{\CI\, absorption lines fit in the system towards AV 266 in the SMC. Lines are the same as in Figure~\ref{fig:Sk67_2_CI}}
    \label{fig:AV266_CI}
\end{figure*}

\begin{figure*}
    \centering
    \includegraphics[width=\linewidth]{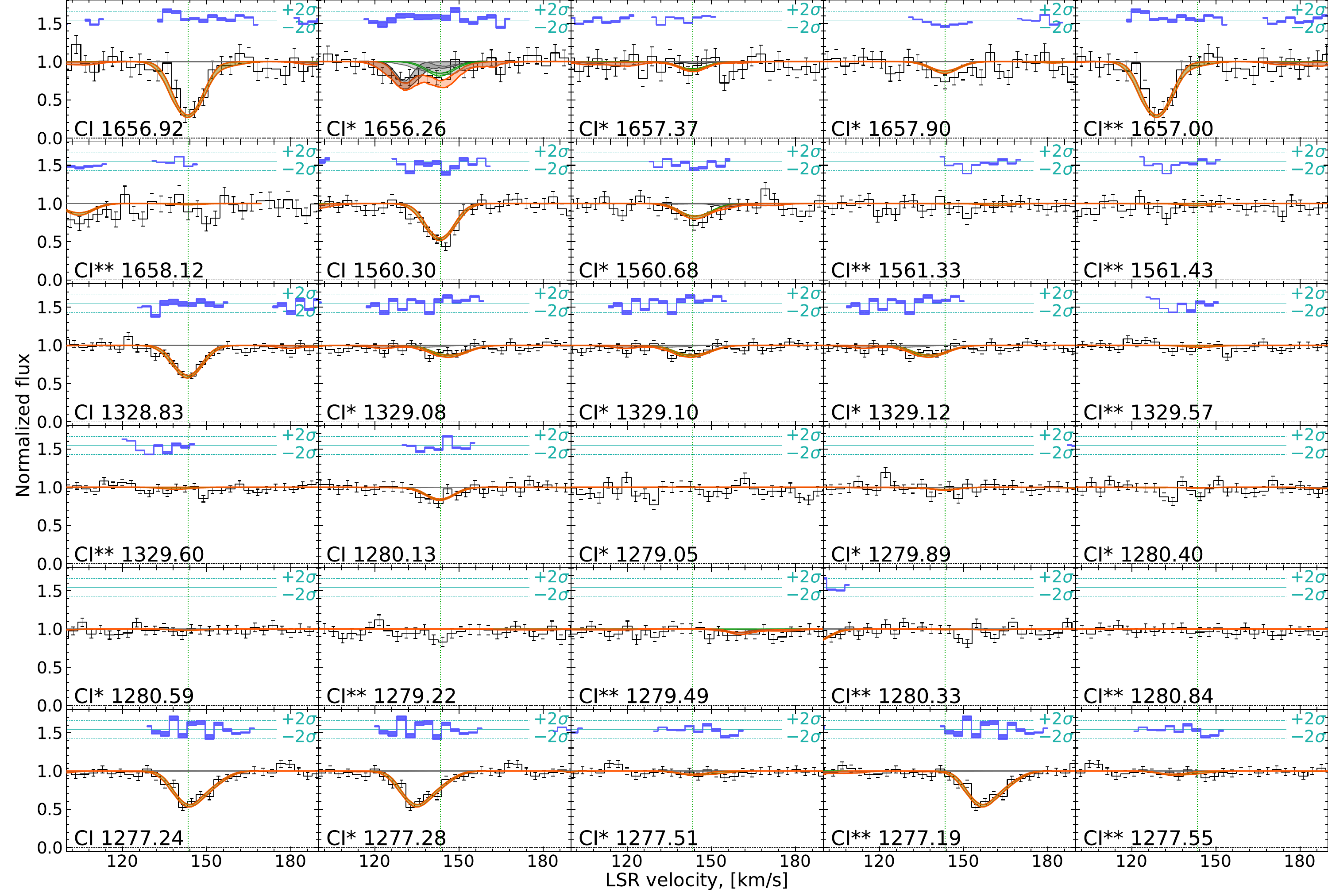}
    \caption{\CI\, absorption lines fit in the system towards AV 372 in the SMC. Lines are the same as in Figure~\ref{fig:Sk67_2_CI}}
    \label{fig:AV372_CI}
\end{figure*}

\clearpage
\begin{figure*}
    \centering
    \includegraphics[width=\linewidth]{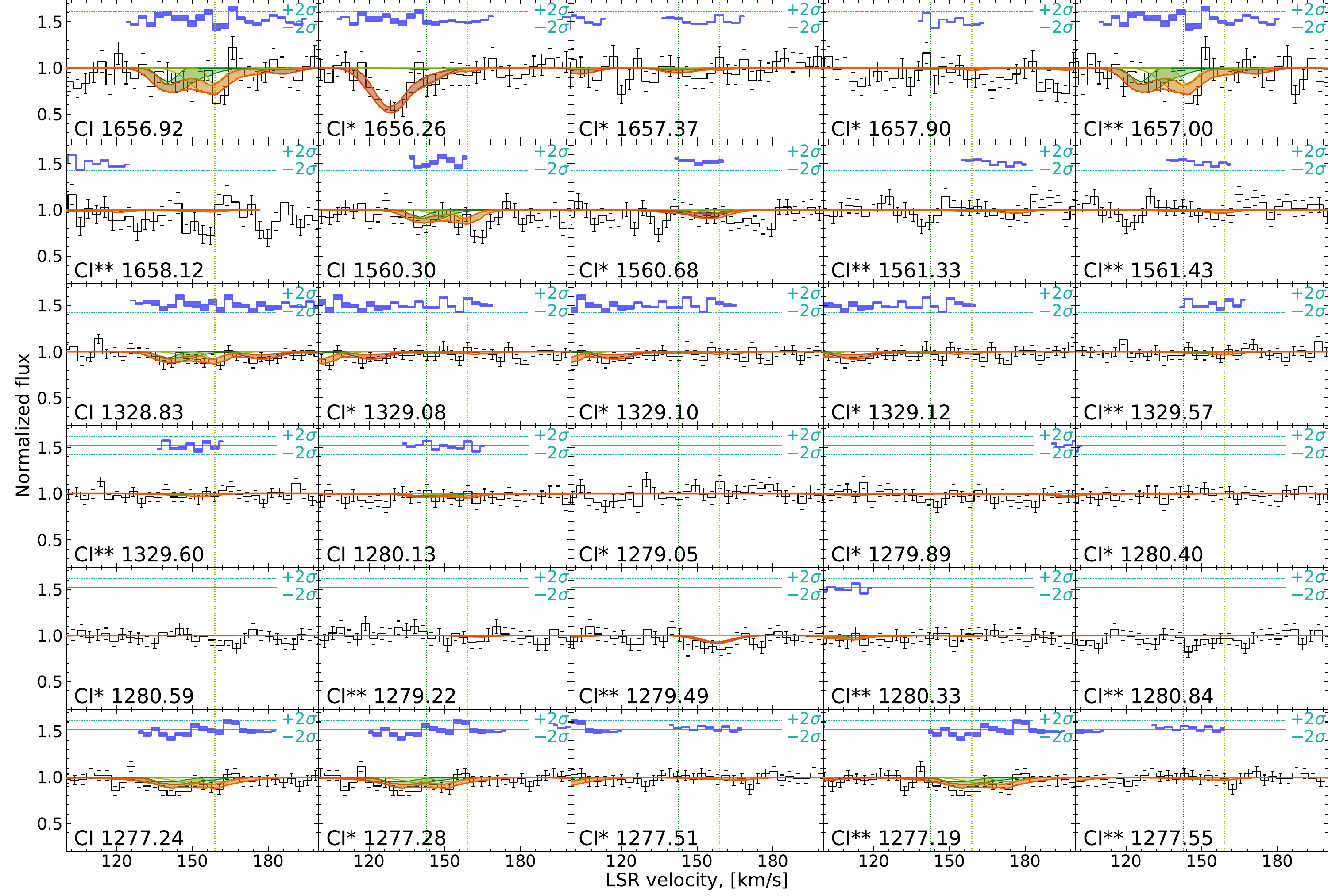}
    \caption{\CI\, absorption lines fit in the system towards AV 423 in SMC. Lines are the same as in Figure~\ref{fig:Sk67_2_CI}}
    \label{fig:AV423_CI}
\end{figure*}

\begin{figure*}
    \centering
    \includegraphics[width=\linewidth]{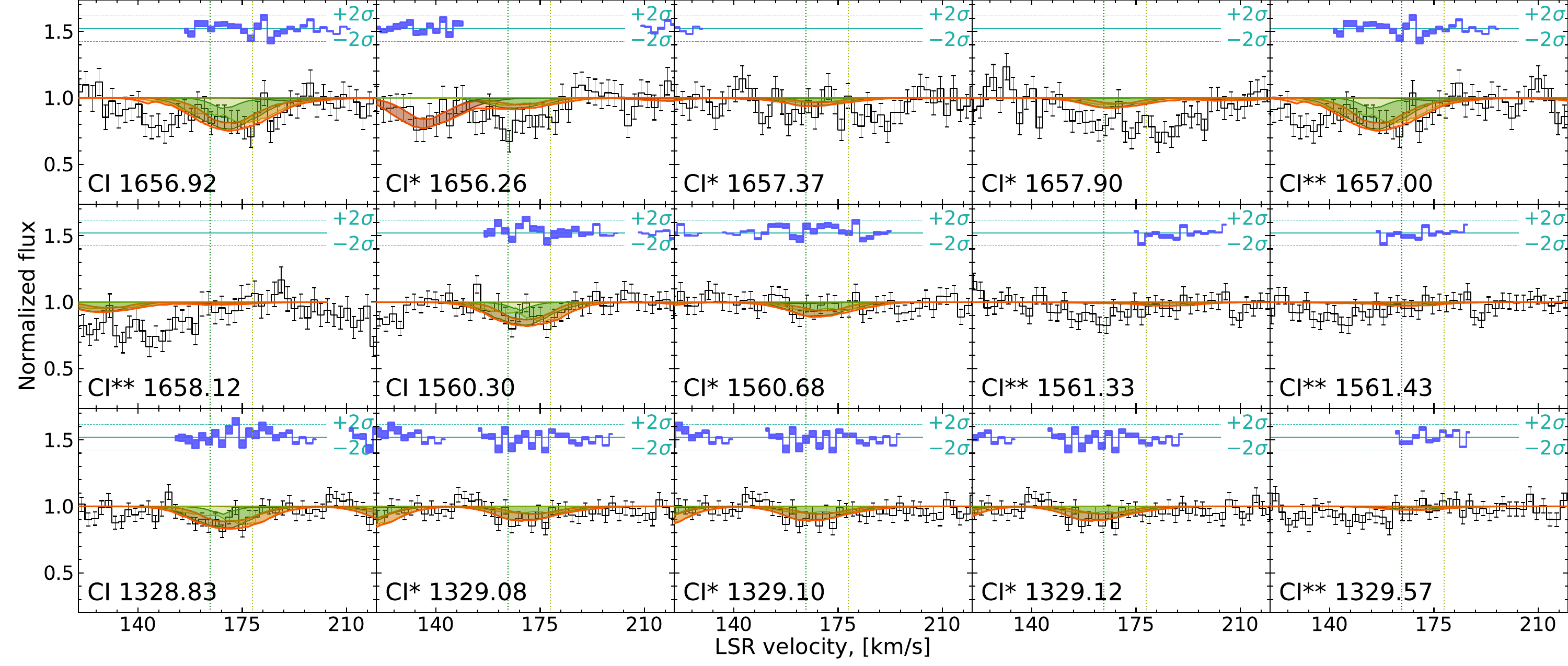}
    \caption{\CI\, absorption lines fit in the system towards AV 440 in SMC. Lines are the same as in Figure~\ref{fig:Sk67_2_CI}}
    \label{fig:AV440_CI}
\end{figure*}

\begin{figure*}
    \centering
    \includegraphics[width=\linewidth]{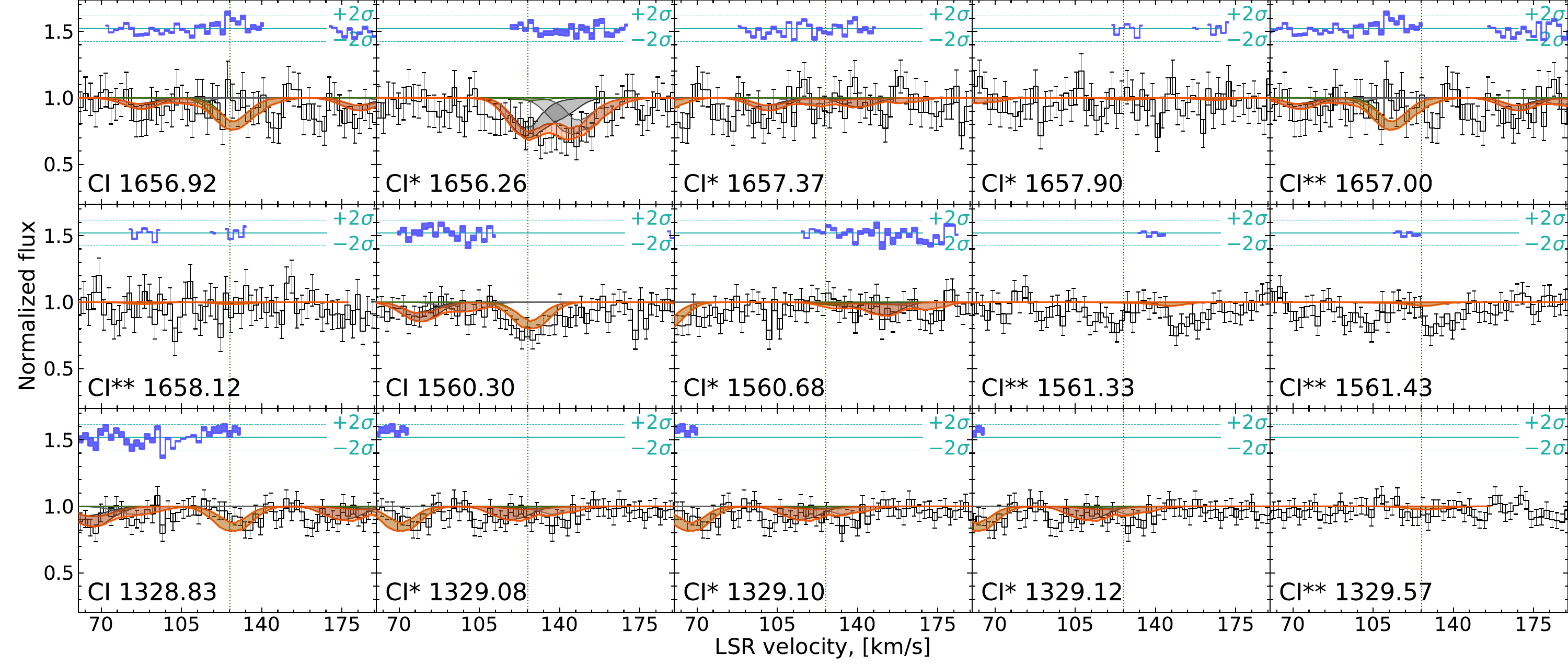}
    \caption{\CI\, absorption lines fit in the system towards AV 472 in the SMC. Lines are the same as in Figure~\ref{fig:Sk67_2_CI}}
    \label{fig:AV472_CI}
\end{figure*}

\begin{figure*}
    \centering
    \includegraphics[width=\linewidth]{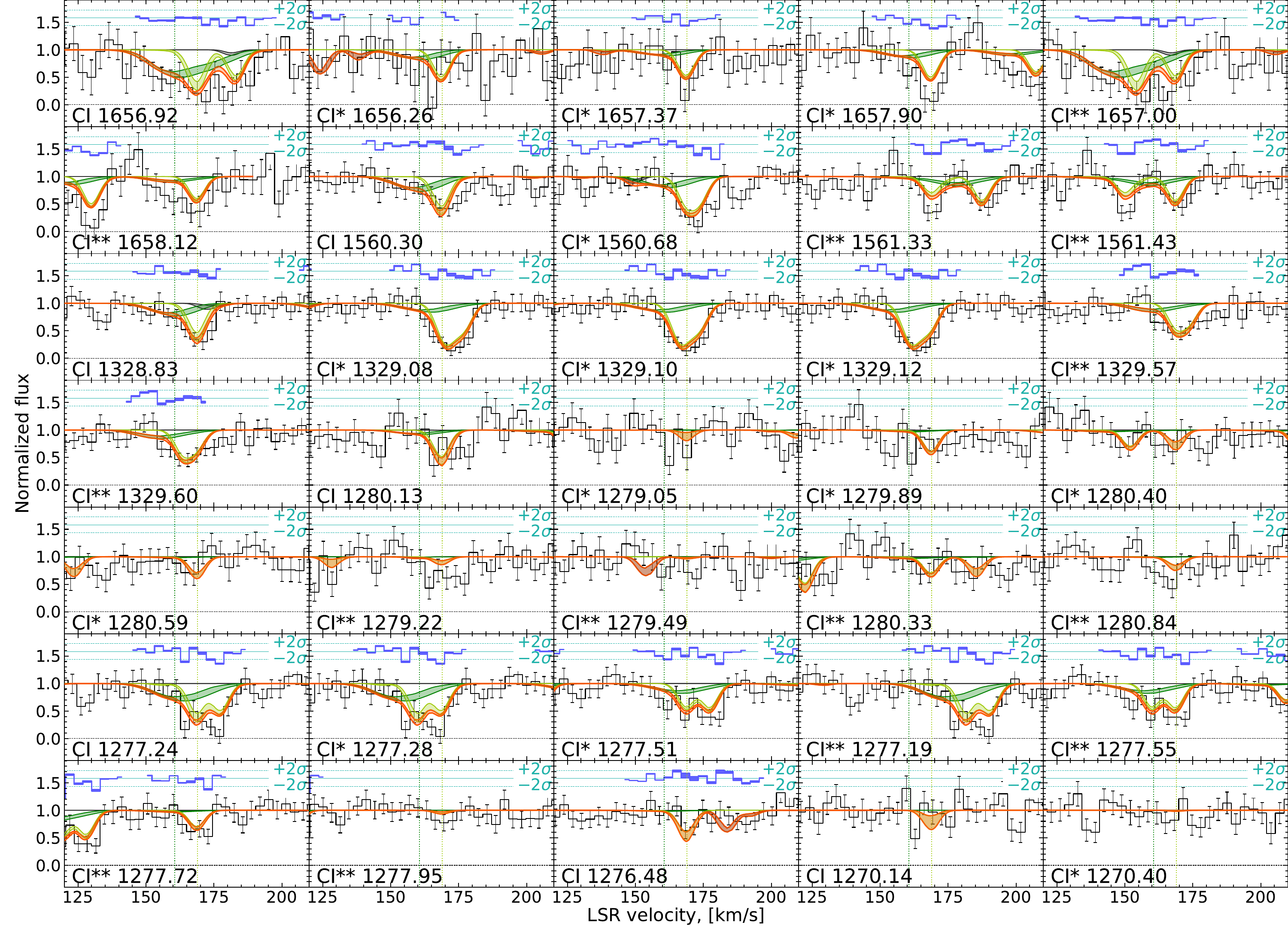}
    \caption{\CI\, absorption lines fit in the system towards AV 476 in the SMC. Lines are the same as in Figure~\ref{fig:Sk67_2_CI}}
    \label{fig:AV476_CI}
\end{figure*}

\begin{figure*}
    \centering
    \includegraphics[width=\linewidth]{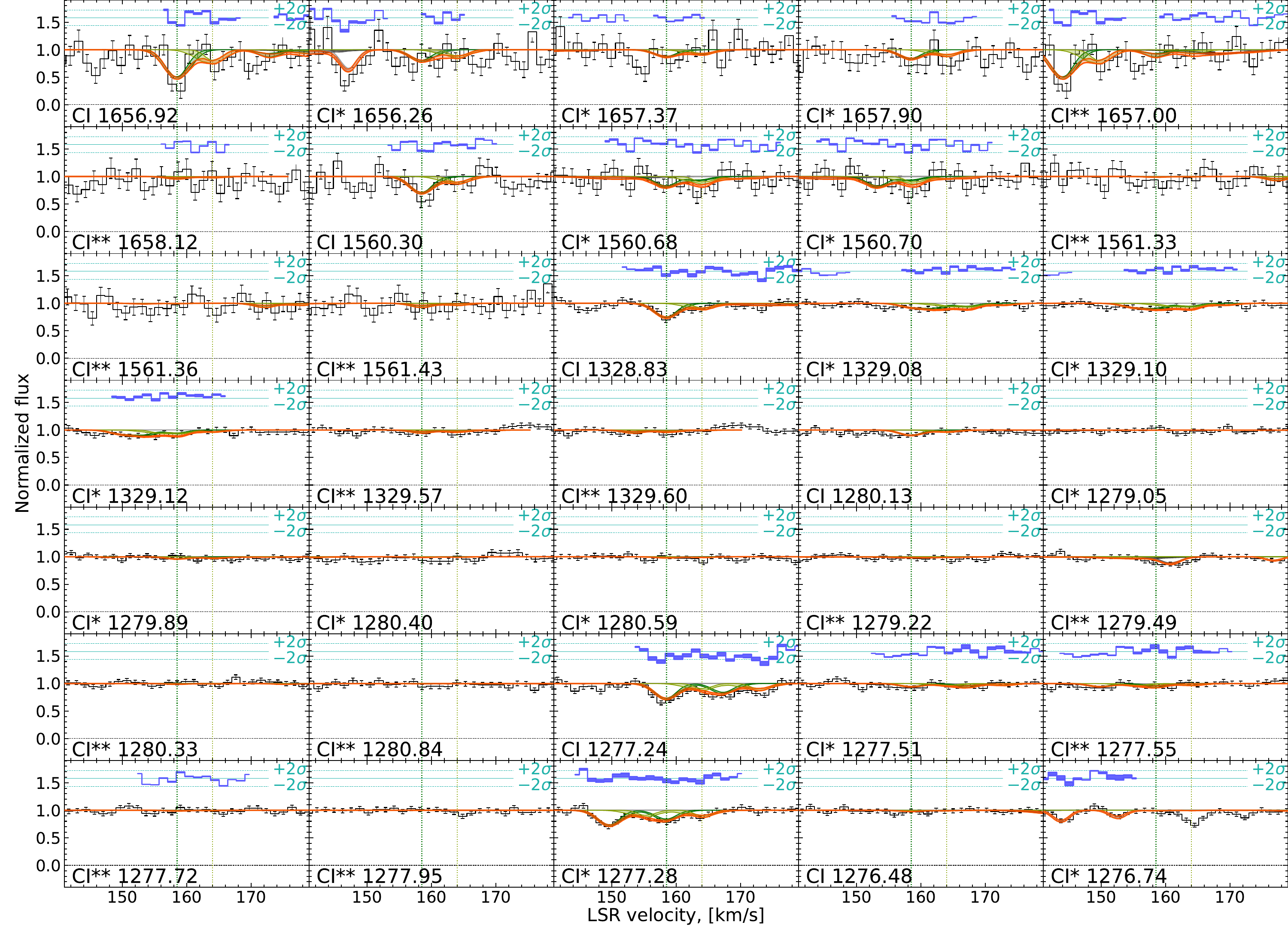}
    \caption{\CI\, absorption lines fit in the system towards AV 479 in the SMC. Lines are the same as in Figure~\ref{fig:Sk67_2_CI}}
    \label{fig:AV479_CI}
\end{figure*}

\begin{figure*}
    \centering
    \includegraphics[width=\linewidth]{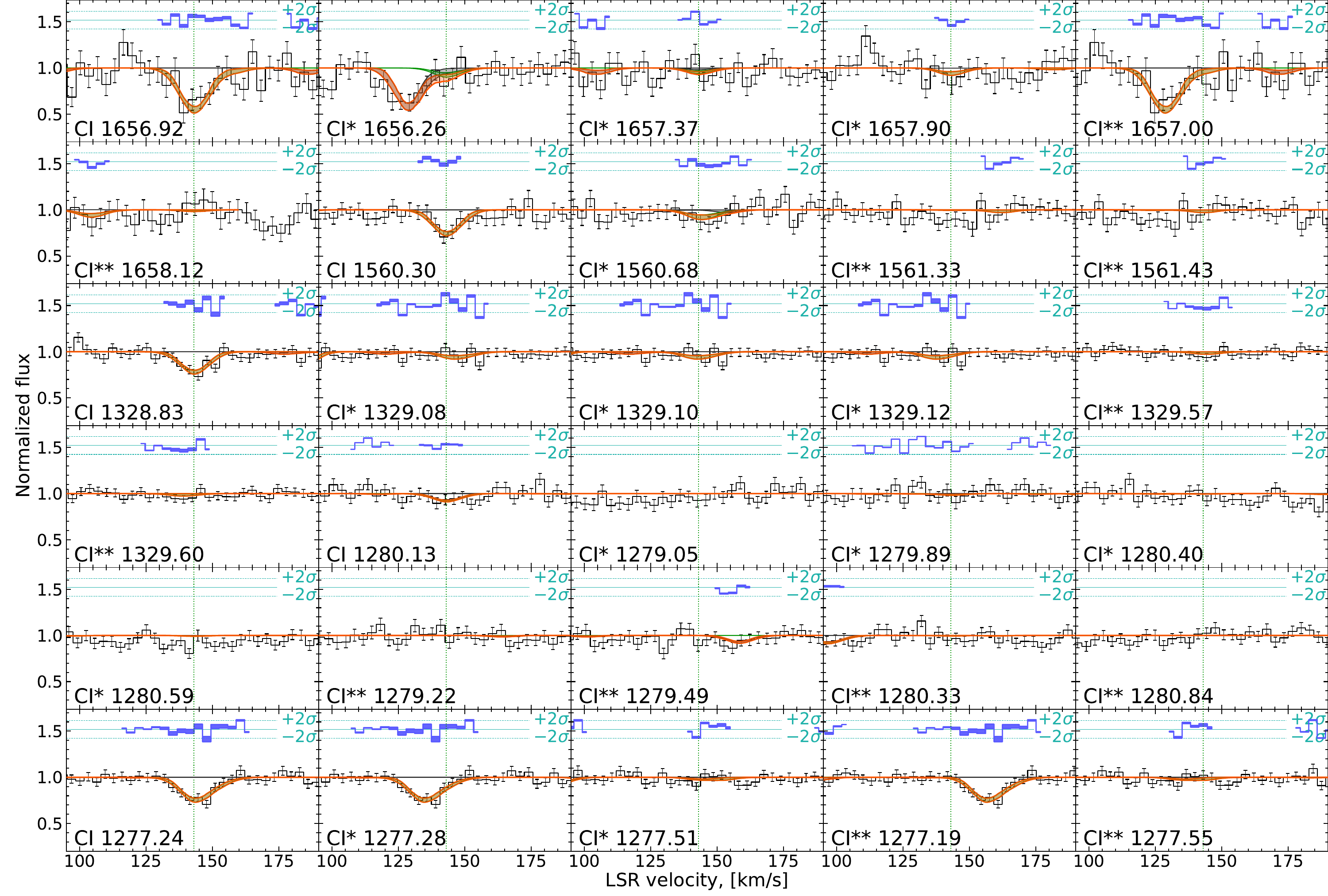}
    \caption{\CI\, absorption lines fit in the system towards AV 488 in SMC. Lines are the same as in Figure~\ref{fig:Sk67_2_CI}}
    \label{fig:AV488_CI}
\end{figure*}

\begin{figure*}
    \centering
    \includegraphics[width=\linewidth]{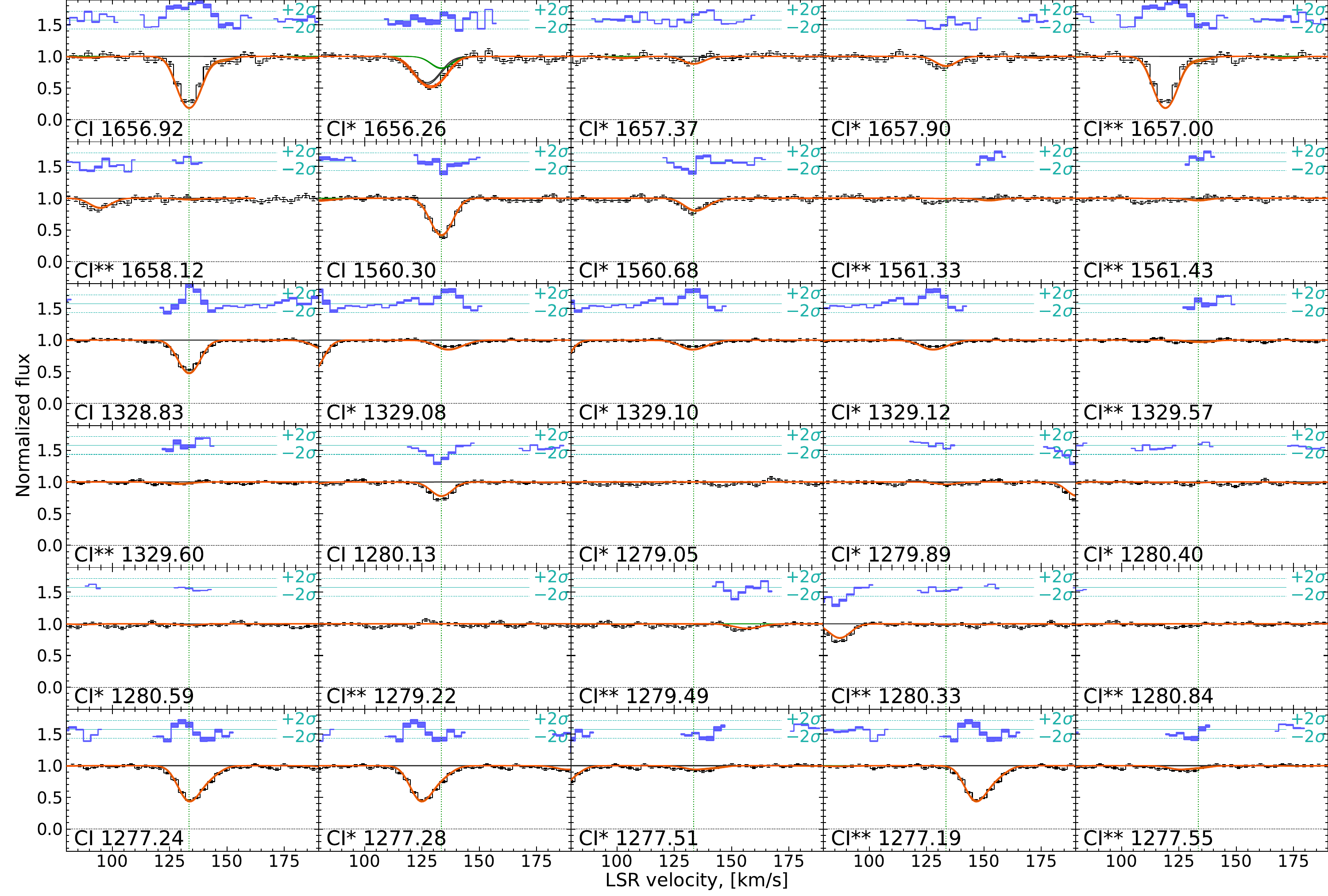}
    \caption{\CI\, absorption lines fit in the system towards AV 490 in the SMC. Lines are the same as in Figure~\ref{fig:Sk67_2_CI}}
    \label{fig:AV490_CI}
\end{figure*}

\begin{figure*}
    \centering
    \includegraphics[width=\linewidth]{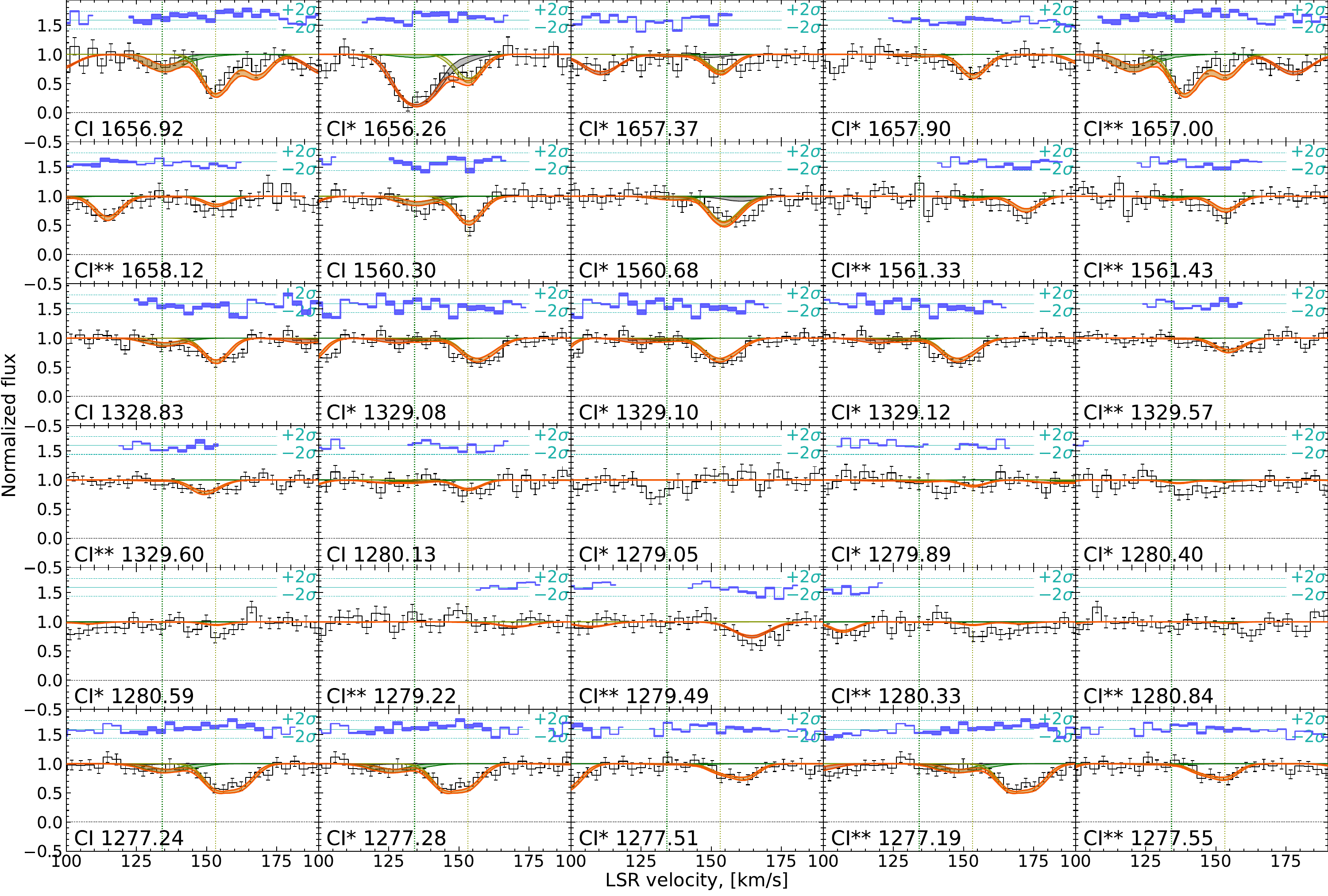}
    \caption{\CI\, absorption lines fit in the system towards Sk 191 in the SMC. Lines are the same as in Figure~\ref{fig:Sk67_2_CI}}
    \label{fig:Sk191_CI}
\end{figure*}

\clearpage

\section{Details on metal lines fit}
\label{sect:metal_fit}

In this section we show the fit to metal line profiles which are used to estimate metallicities in the systems discussed in the Section~\ref{sect:phys_properties}.

\subsection{Large Magellanic Cloud}

\begin{figure*}
    \begin{minipage}{0.49\linewidth}
        \centering
        \includegraphics[width=\linewidth]{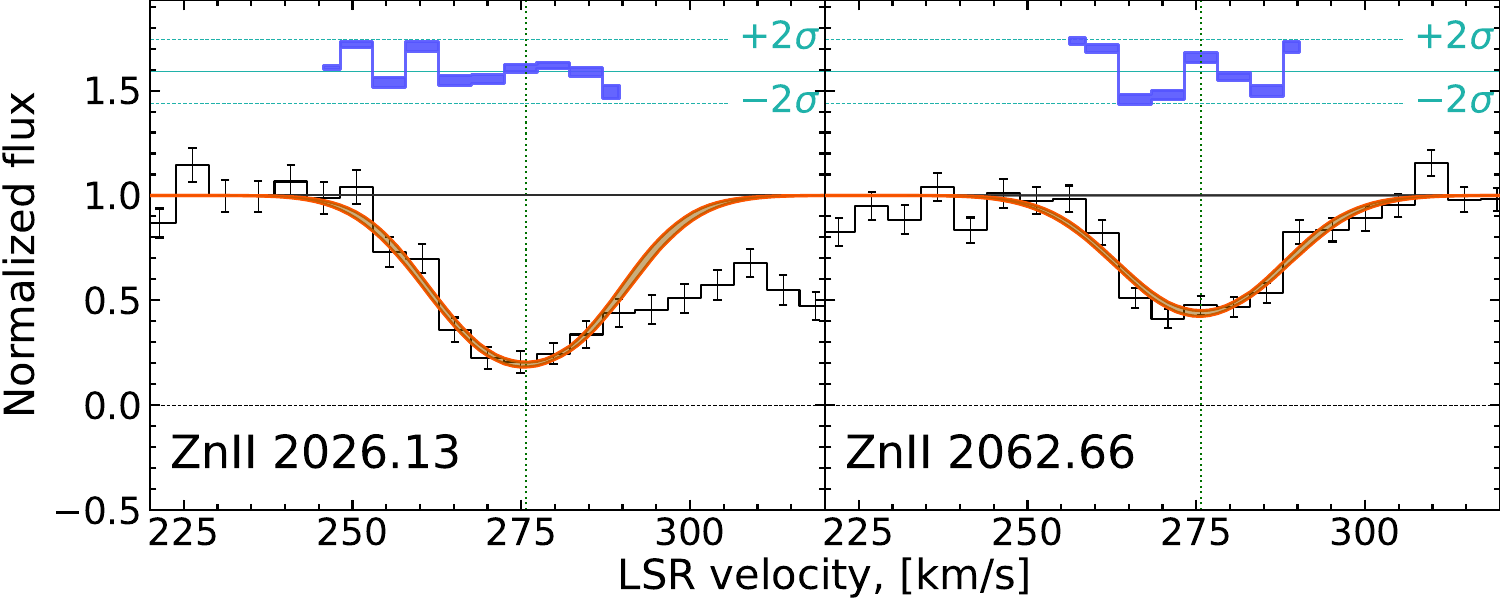}
        \caption{\ZnII\,
    absorption lines fit in the system towards Sk-67 2 in the LMC. Lines are the same as in Figure~\ref{fig:Sk67_2_CI}}
        \label{fig:Sk67_2_ZnII}
    \end{minipage}
    \hfill
    \begin{minipage}{0.49\linewidth}
        \centering
        \includegraphics[width=\linewidth]{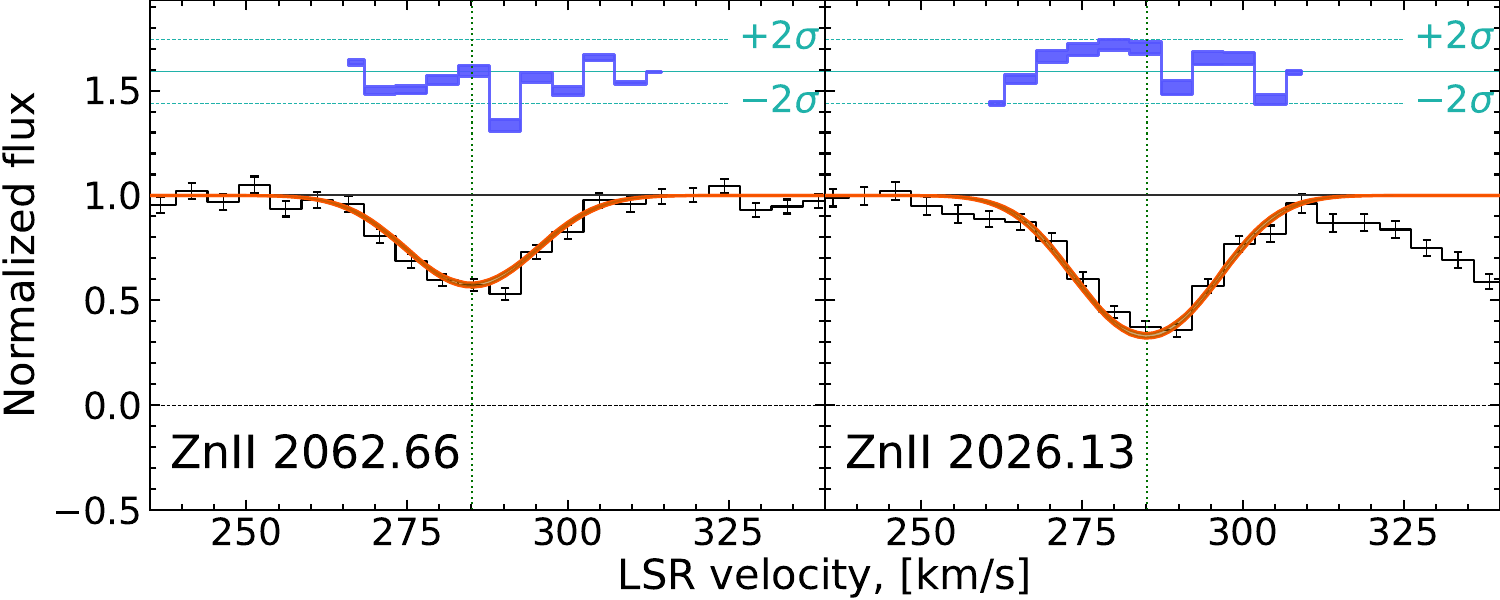}
        \caption{\ZnII\,absorption lines fit in the system towards Sk-67 5 in the LMC. Lines are the same as in Figure~\ref{fig:Sk67_2_CI}}
        \label{fig:Sk67_5_ZnII}
    \end{minipage}
\end{figure*}

\begin{figure*}
    \begin{minipage}{0.49\linewidth}
        \centering
        \includegraphics[width=\linewidth]{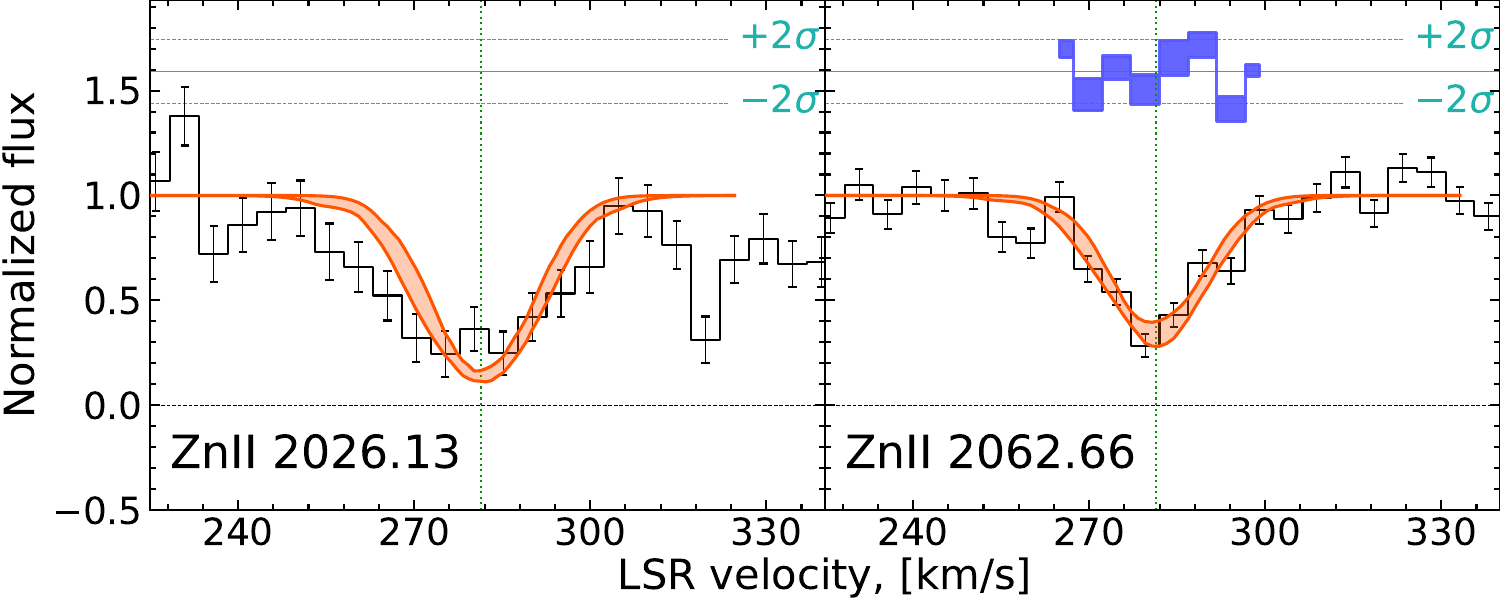}
        \caption{\ZnII\, absorption lines fit in the system towards LH10 3120 in the LMC. Lines are the same as in Figure~\ref{fig:Sk67_2_CI}}
        \label{fig:LH10_3120_ZnII}
    \end{minipage}
    \hfill
    \begin{minipage}{0.49\linewidth}
        \centering
        \includegraphics[width=\linewidth]{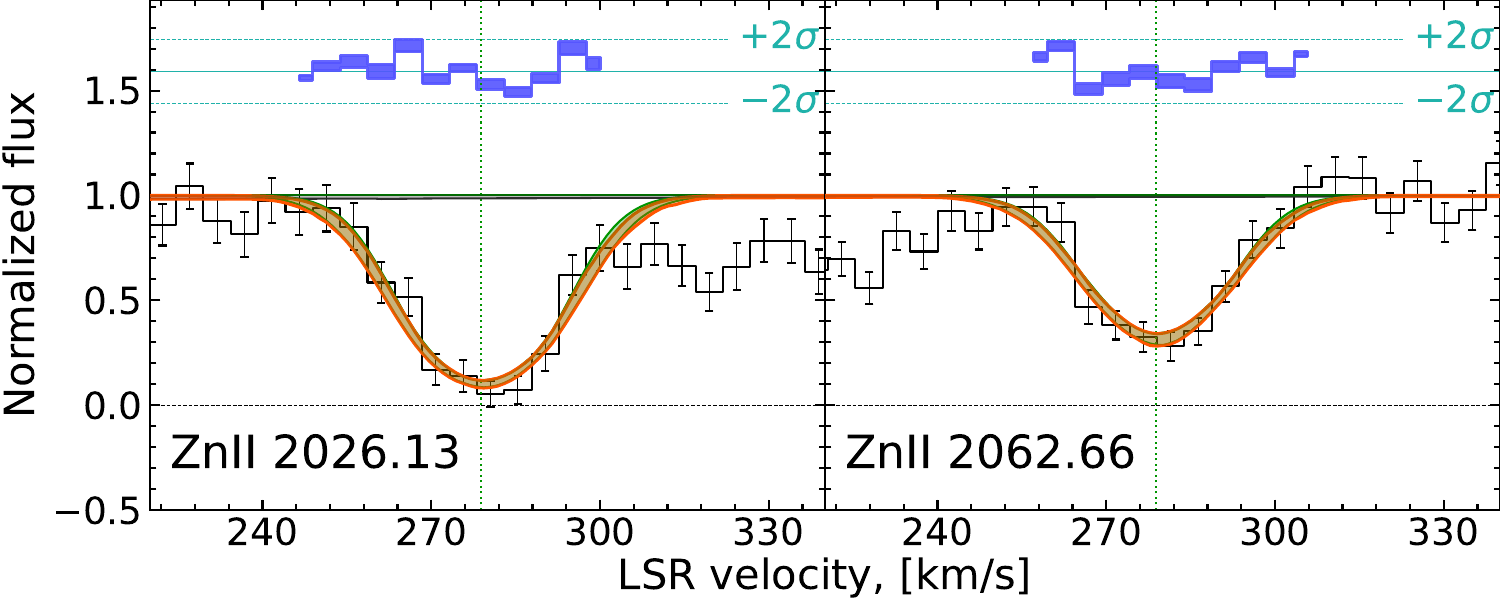}
        \caption{\ZnII\, absorption lines fit in the system towards PGMW 3223 in the LMC. Lines are the same as in Figure~\ref{fig:Sk67_2_CI}}
        \label{fig:PGMW3223_ZnII}
    \end{minipage}
\end{figure*}

\begin{figure*}
    \begin{minipage}{0.49\linewidth}
        \centering
        \includegraphics[width=\linewidth]{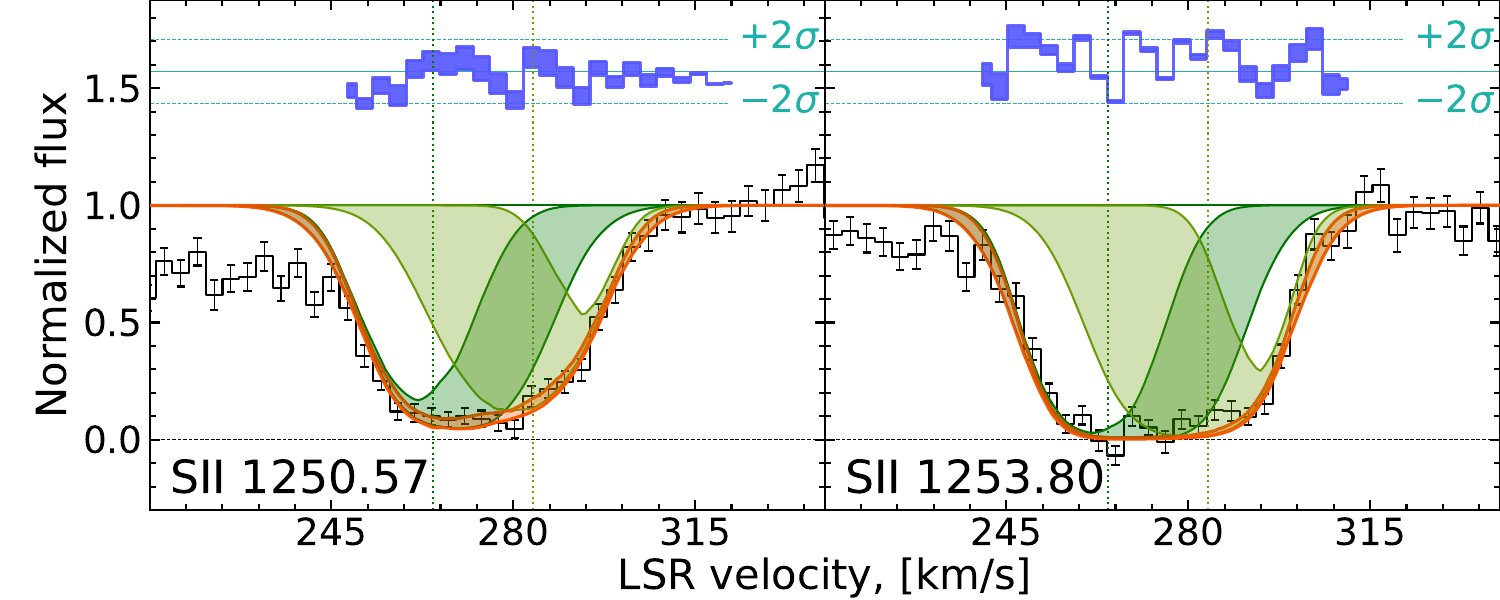}
        \caption{\SII\, absorption lines fit in the system towards Sk-66 35 in LMC. Lines are the same as in Figure~\ref{fig:Sk67_2_CI}}
        \label{fig:Sk66_35_SII}
    \end{minipage}
    \hfill
    \begin{minipage}{0.49\linewidth}
        \centering
        \includegraphics[width=\linewidth]{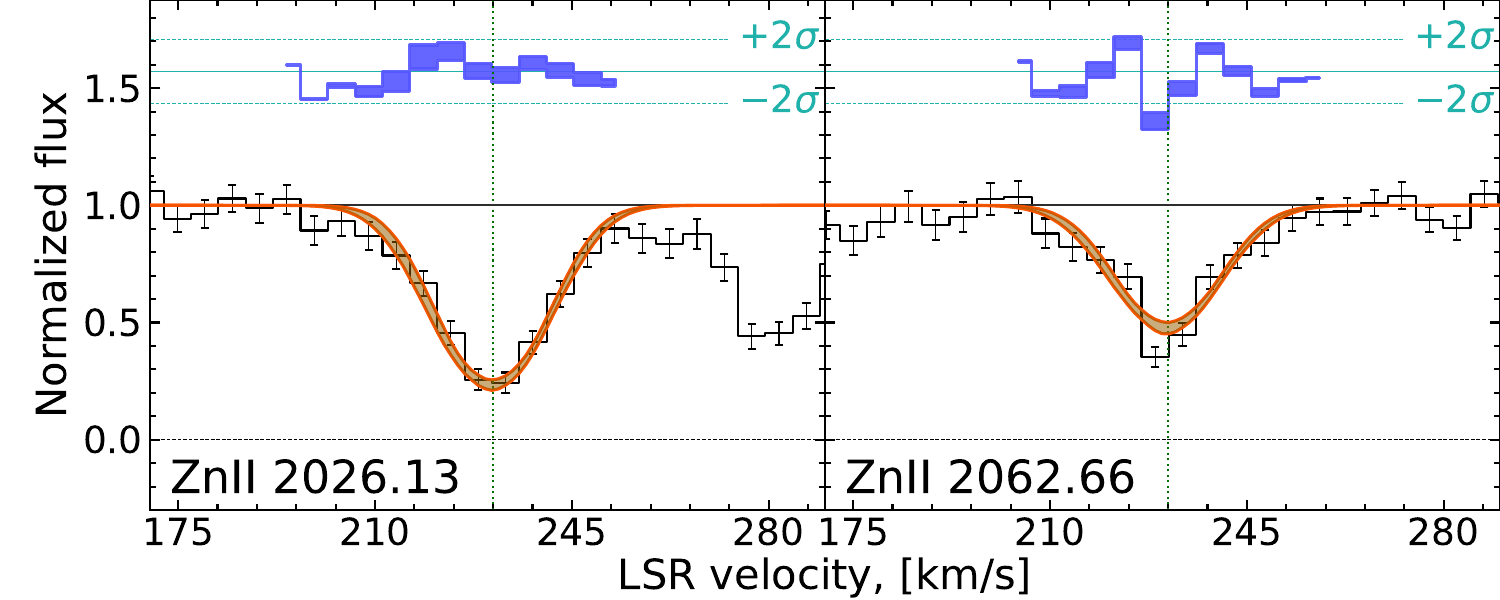}
        \caption{\ZnII\, absorption lines fit in the system towards Sk-70 79 in the LMC. Lines are the same as in Figure~\ref{fig:Sk67_2_CI}}
        \label{fig:Sk70_79_ZnII}
    \end{minipage}
\end{figure*}

\begin{figure*}
    \begin{minipage}{0.49\linewidth}
        \centering
        \includegraphics[width=\linewidth]{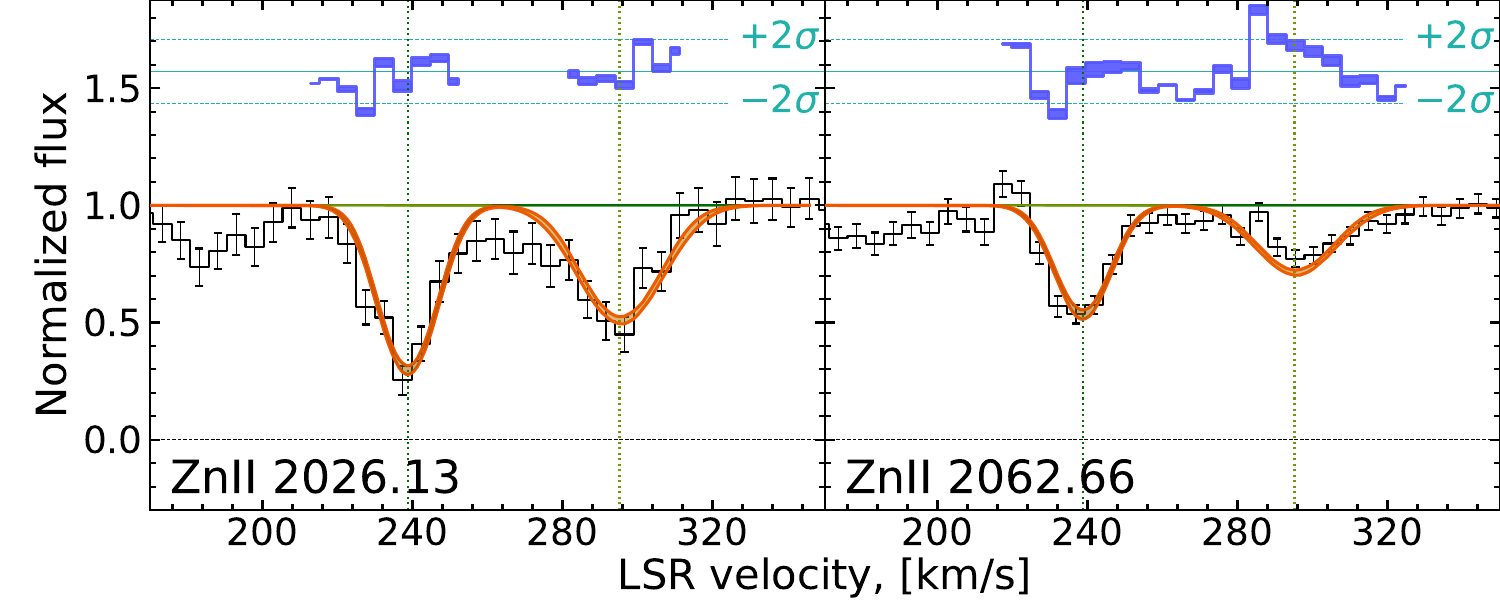}
        \caption{\ZnII\,
    absorption lines fit in the system towards Sk-68 52 in the LMC. Lines are the same as in Figure~\ref{fig:Sk67_2_CI}}
        \label{fig:Sk68_52_ZnII}
    \end{minipage}
    \hfill
    \begin{minipage}{0.49\linewidth}
        \centering
        \includegraphics[width=\linewidth]{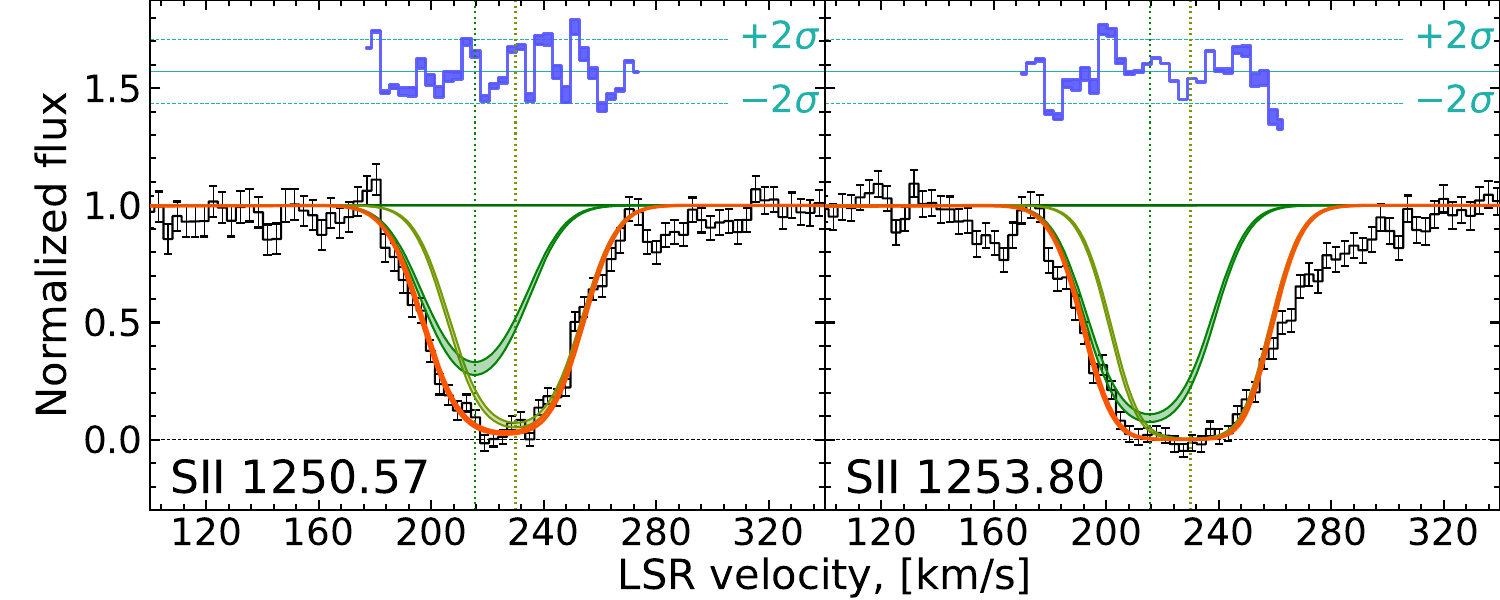}
        \caption{\SII\,absorption lines fit in the system towards Sk-71 8 in the LMC. Lines are the same as in Figure~\ref{fig:Sk67_2_CI}}
        \label{fig:Sk71_8_ZnII}
    \end{minipage}
\end{figure*}

\begin{figure*}
    \begin{minipage}{0.49\linewidth}
        \centering
        \includegraphics[width=\linewidth]{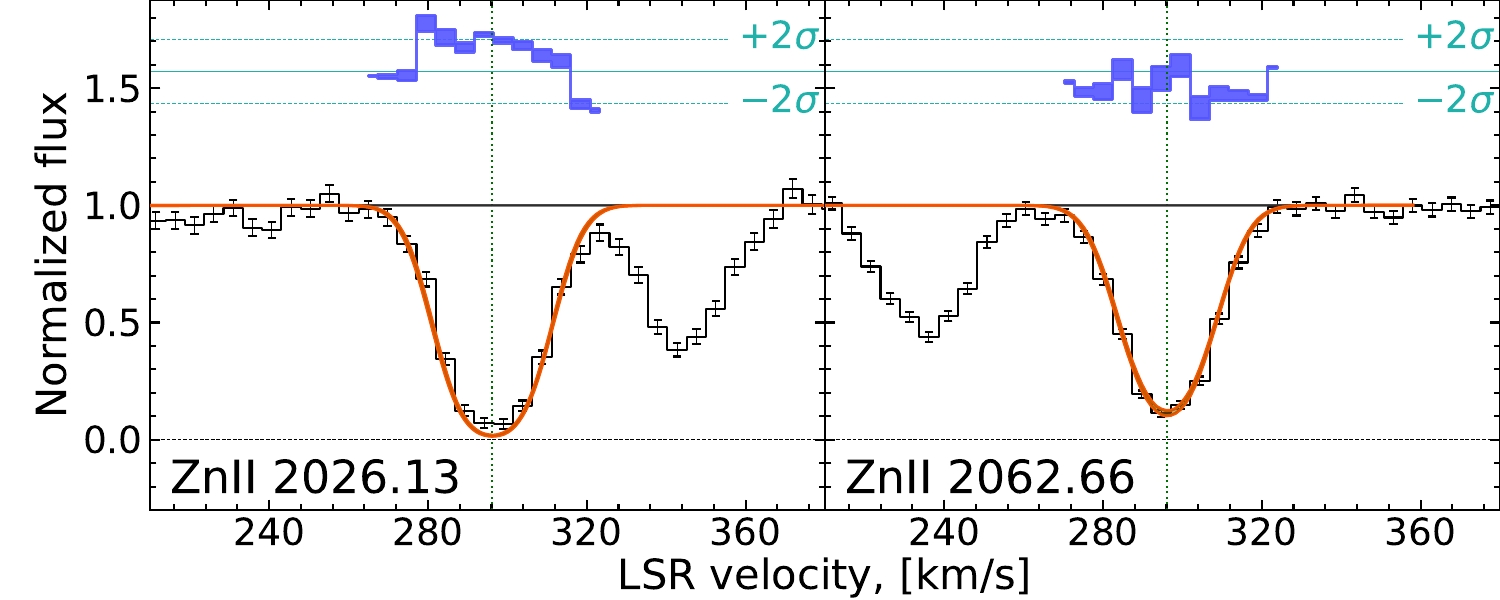}
        \caption{\ZnII\,
    absorption lines fit in the system towards Sk-68 73 in LMC. Lines are the same as in Figure~\ref{fig:Sk67_2_CI}}
        \label{fig:Sk68_73_ZnII}
    \end{minipage}
    \hfill
    \begin{minipage}{0.49\linewidth}
        \centering
        \includegraphics[width=\linewidth]{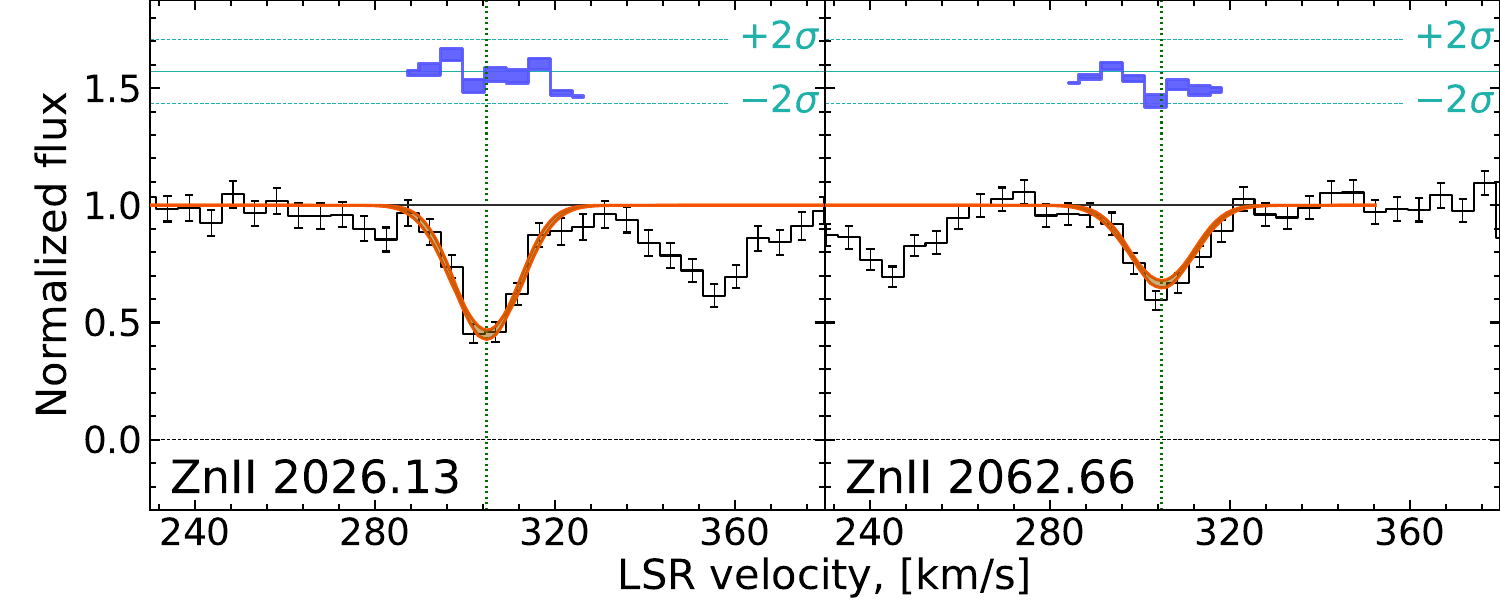}
        \caption{\ZnII\,absorption lines fit in the system towards Sk-67 105 in LMC. Lines are the same as in Figure~\ref{fig:Sk67_2_CI}}
        \label{fig:Sk67_105_ZnII}
    \end{minipage}
\end{figure*}

\begin{figure*}
    \begin{minipage}{0.49\linewidth}
        \centering
        \includegraphics[width=0.8\linewidth]{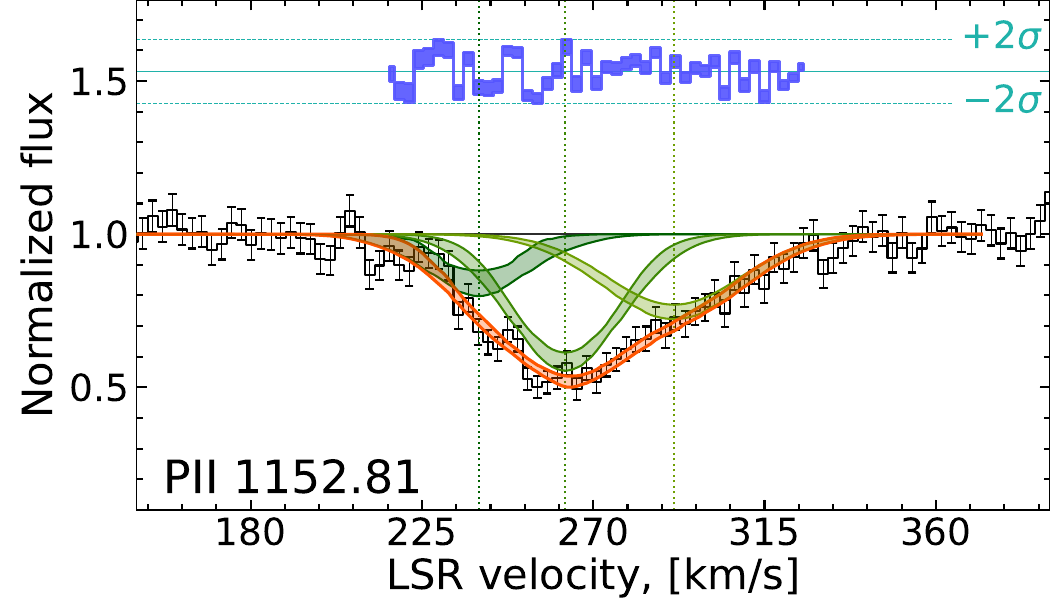}
        \caption{\PII\,
    absorption lines fit in the system towards BI 184 in the LMC. Lines are the same as in Figure~\ref{fig:Sk67_2_CI}}
        \label{fig:BI184_PII}
    \end{minipage}
    \hfill
    \begin{minipage}{0.49\linewidth}
        \centering
        \includegraphics[width=\linewidth]{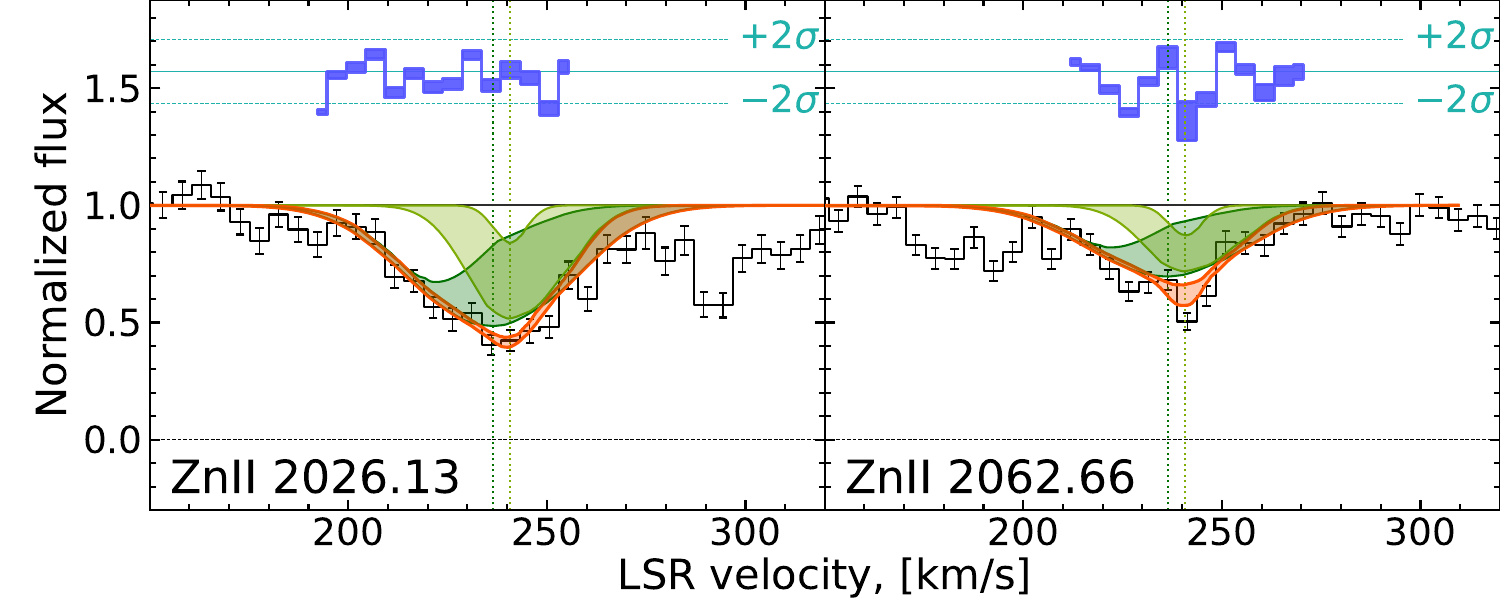}
        \caption{\ZnII\,absorption lines fit in the system towards Sk-71 45 in the LMC. Lines are the same as in Figure~\ref{fig:Sk67_2_CI}}
        \label{fig:Sk71_45_ZnII}
    \end{minipage}
\end{figure*}

\begin{figure*}
    \begin{minipage}{0.49\linewidth}
        \centering
        \includegraphics[width=\linewidth]{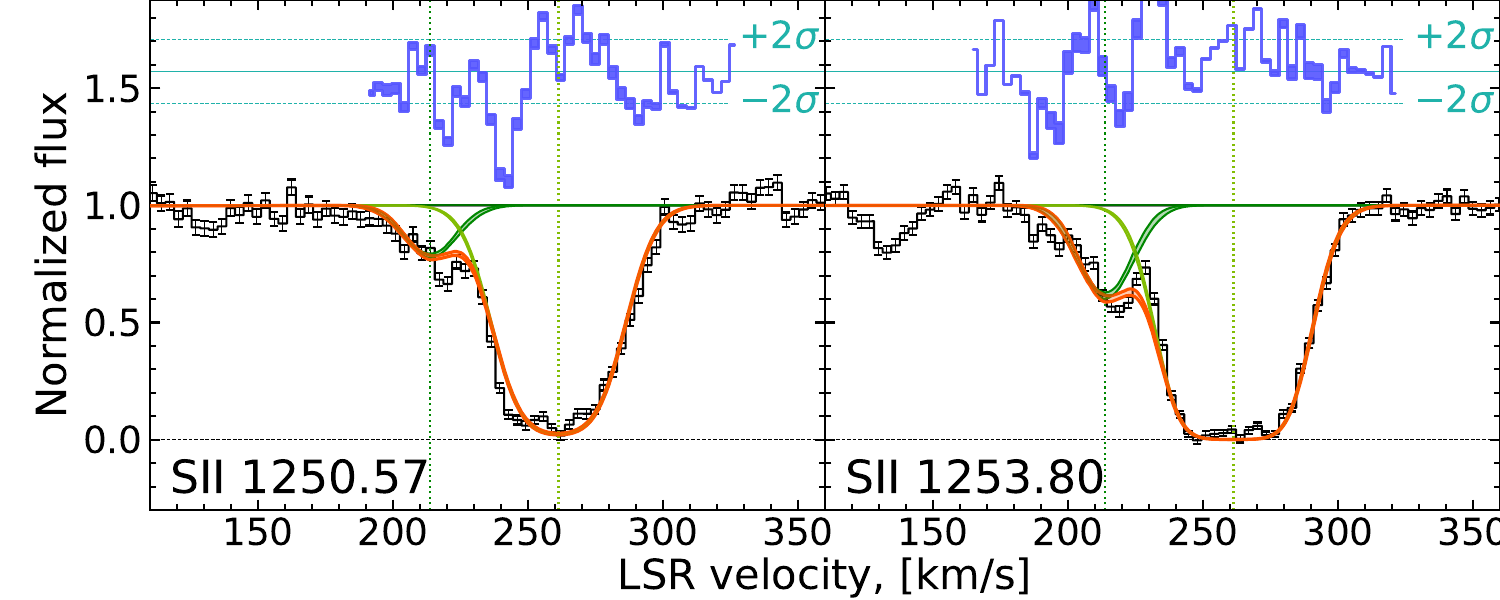}
        \caption{\SII\,
    absorption lines fit in the system towards Sk-69 191 in LMC. Lines are the same as in Figure~\ref{fig:Sk67_2_CI}}
        \label{fig:Sk69_191_SII}
    \end{minipage}
    \hfill
    \begin{minipage}{0.49\linewidth}
        \centering
        \includegraphics[width=\linewidth]{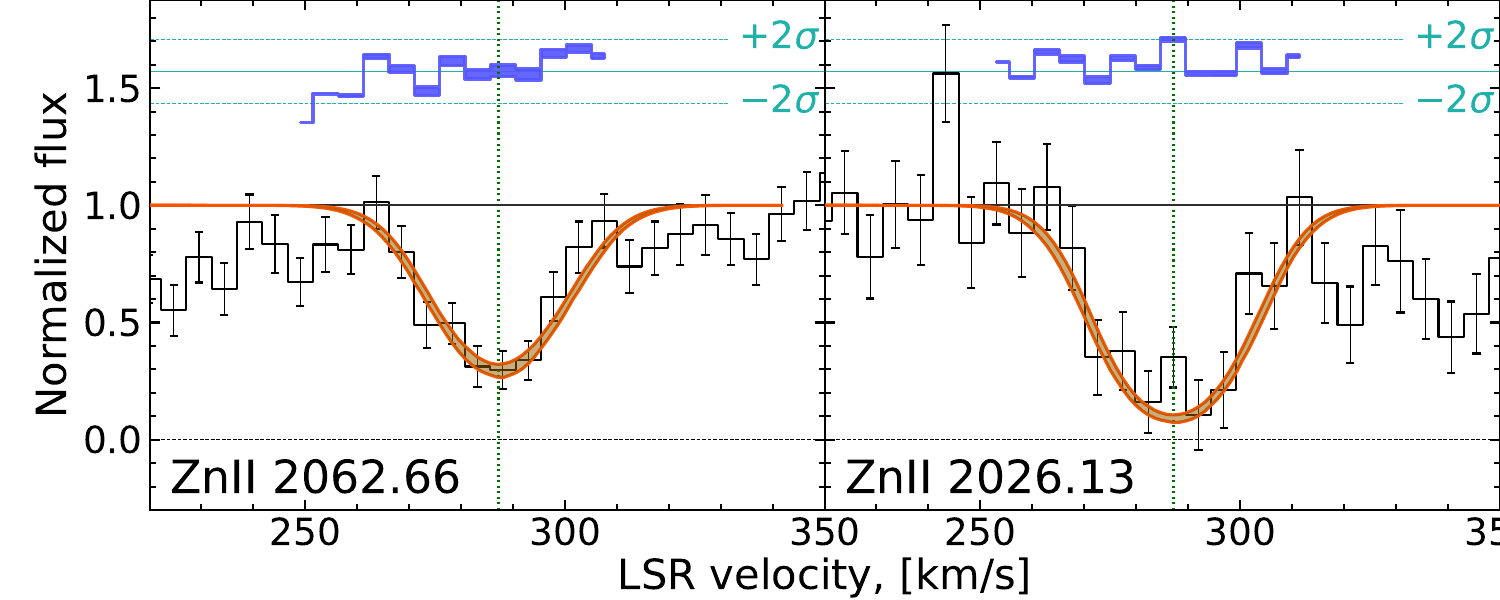}
        \caption{\ZnII\,absorption lines fit in the system towards BI237 in the LMC. Lines are the same as in Figure~\ref{fig:Sk67_2_CI}}
        \label{fig:BI237_ZnII}
    \end{minipage}
\end{figure*}

\begin{figure*}
    \begin{minipage}{0.49\linewidth}
        \centering
        \includegraphics[width=\linewidth]{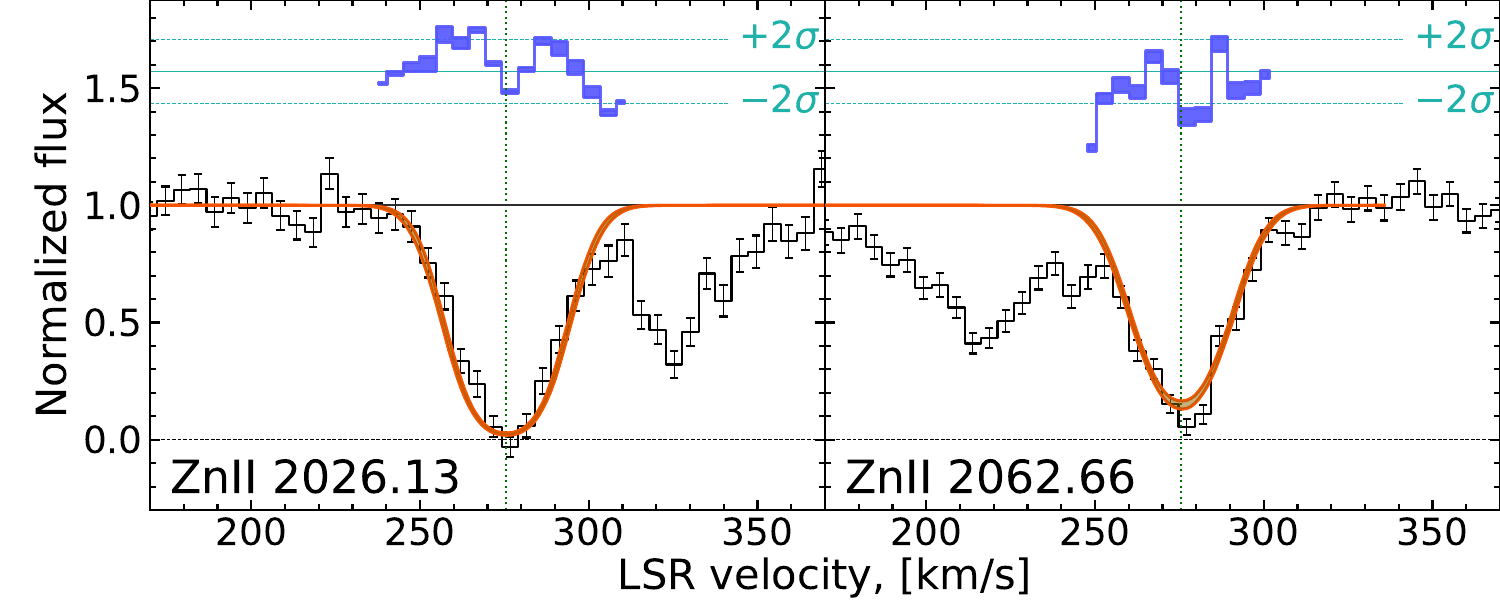}
        \caption{\ZnII\,
    absorption lines fit in the system towards Sk-68 129 in the LMC. Lines are the same as in Figure~\ref{fig:Sk67_2_CI}}
        \label{fig:Sk68_2129_ZnII}
    \end{minipage}
    \hfill
    \begin{minipage}{0.49\linewidth}
        \centering
        \includegraphics[width=\linewidth]{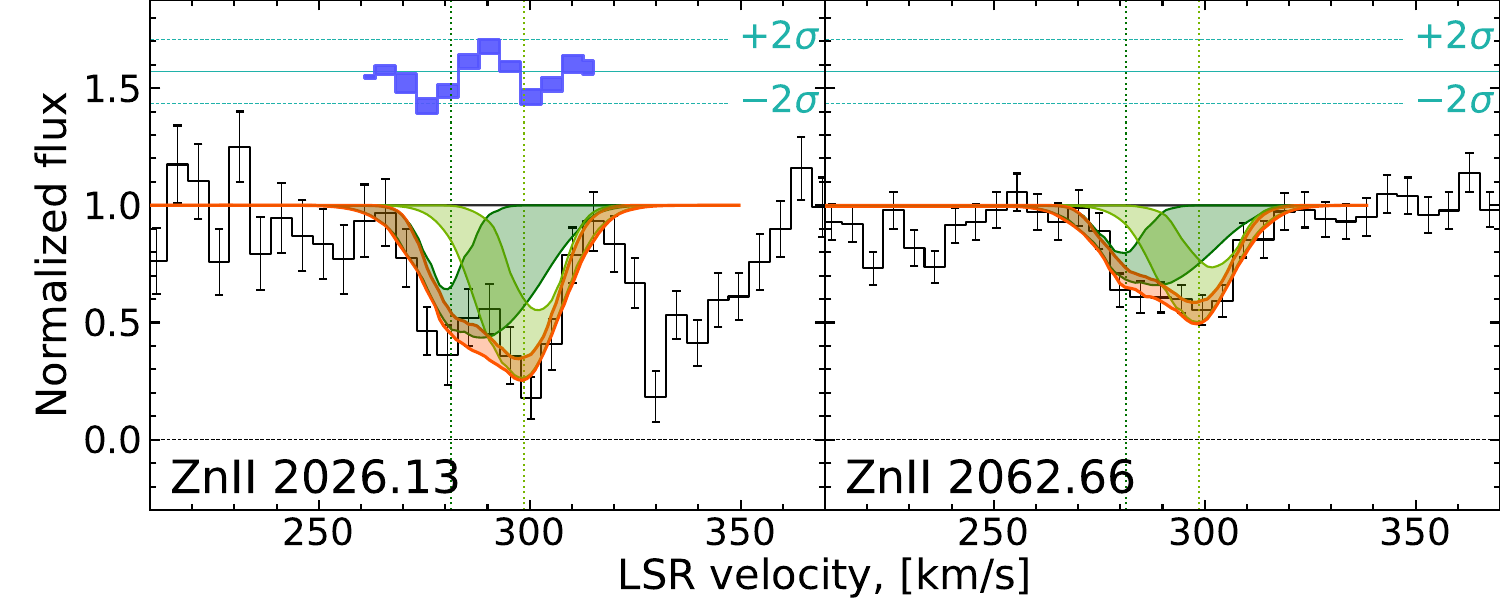}
        \caption{\ZnII\,absorption lines fit in the system towards Sk-66 172 in LMC. Lines are the same as in Figure~\ref{fig:Sk67_2_CI}}
        \label{fig:Sk66_172_ZnII}
    \end{minipage}
\end{figure*}

\begin{figure*}
    \begin{minipage}{0.49\linewidth}
        \centering
        \includegraphics[width=\linewidth]{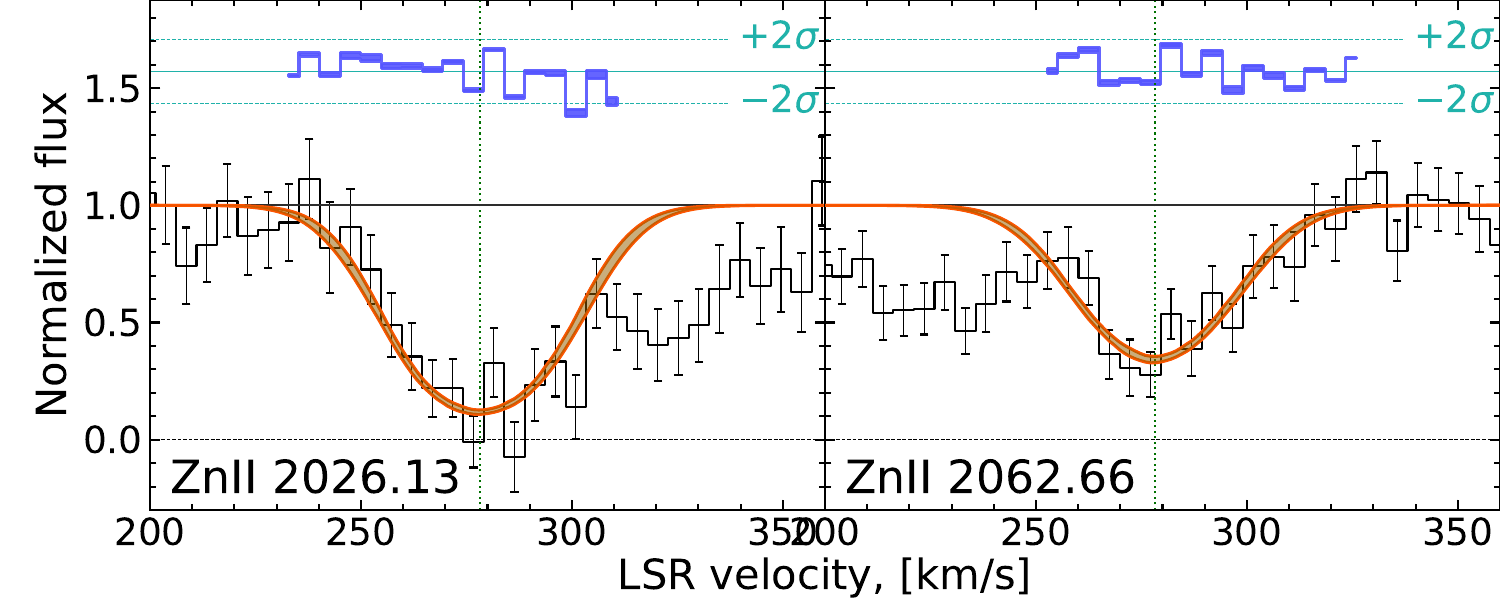}
        \caption{\ZnII\,
    absorption lines fit in the system towards BI 253 in the LMC. Lines are the same as in Figure~\ref{fig:Sk67_2_CI}}
        \label{fig:BI253_ZnII}
    \end{minipage}
    \hfill
    \begin{minipage}{0.49\linewidth}
        \centering
        \includegraphics[width=\linewidth]{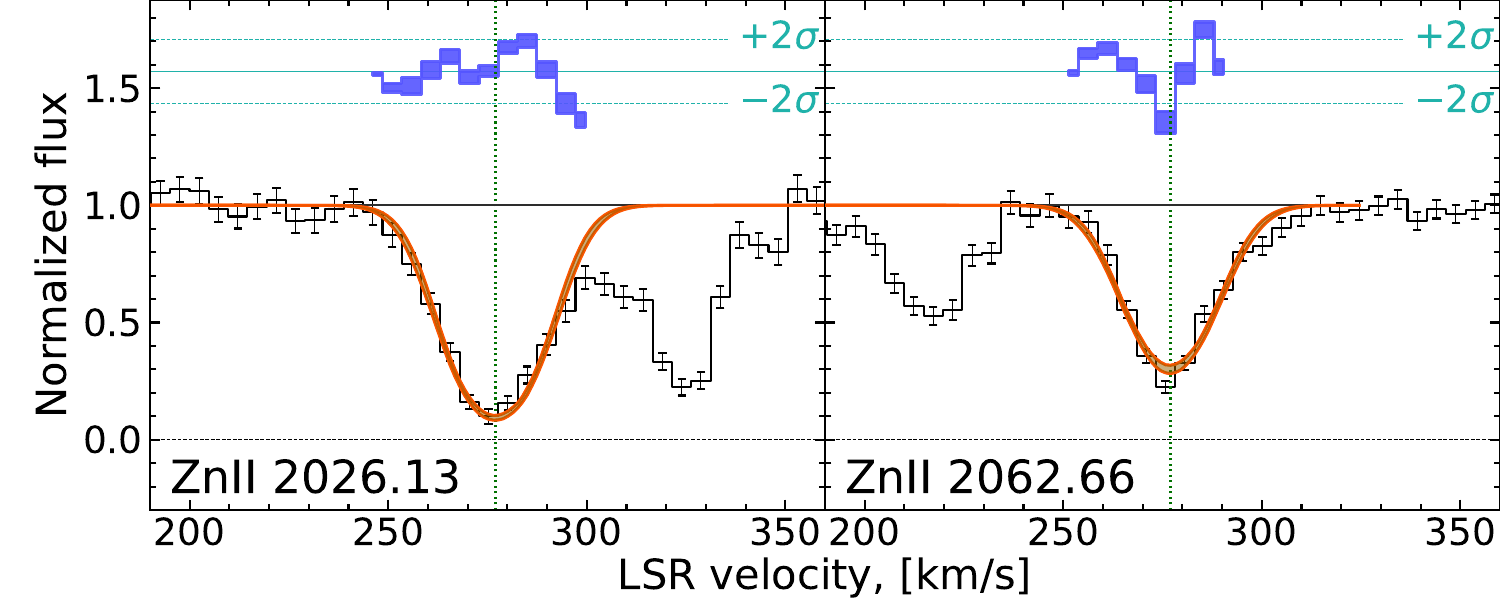}
        \caption{\ZnII\,absorption lines fit in the system towards Sk-68 135 in LMC. Lines are the same as in Figure~\ref{fig:Sk67_2_CI}}
        \label{fig:Sk68_135_ZnII}
    \end{minipage}
\end{figure*}

\begin{figure*}
    \begin{minipage}{0.49\linewidth}
        \centering
        \includegraphics[width=\linewidth]{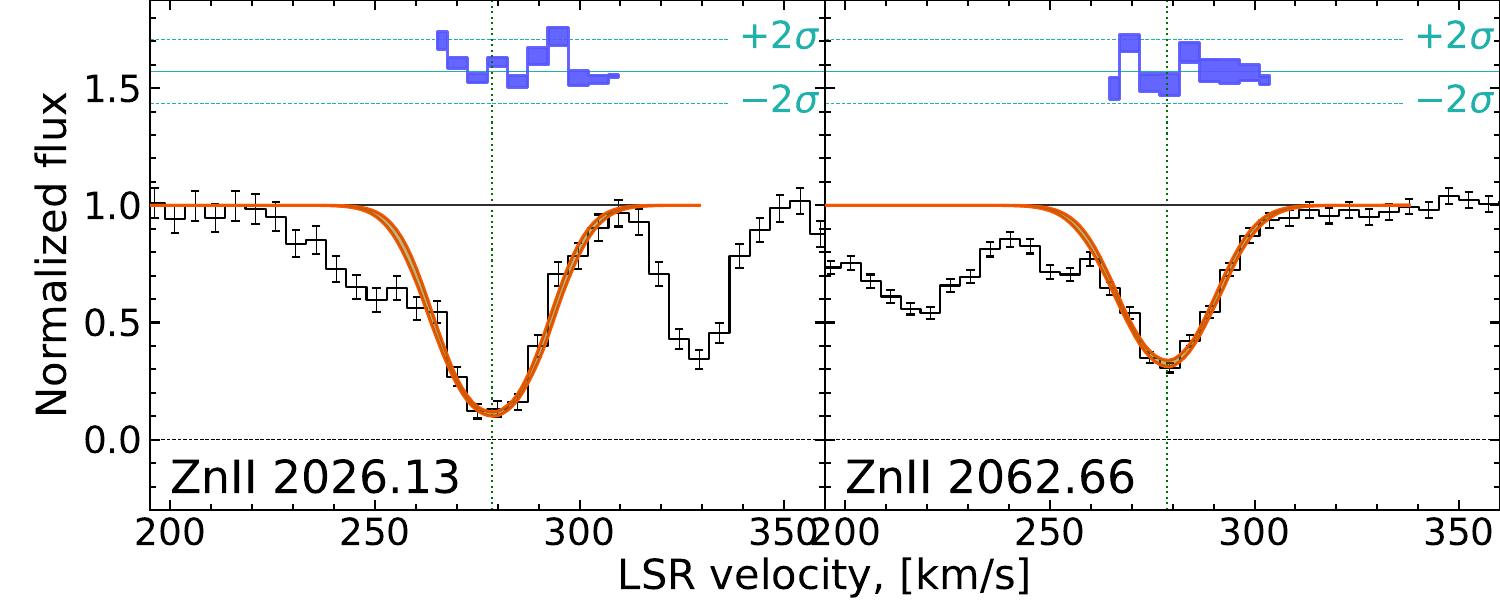}
        \caption{\ZnII\,
    absorption lines fit in the system towards Sk-69 246 in the LMC. Lines are the same as in Figure~\ref{fig:Sk67_2_CI}}
        \label{fig:Sk69_246_ZnII}
    \end{minipage}
    \hfill
    \begin{minipage}{0.49\linewidth}
        \centering
        \includegraphics[width=\linewidth]{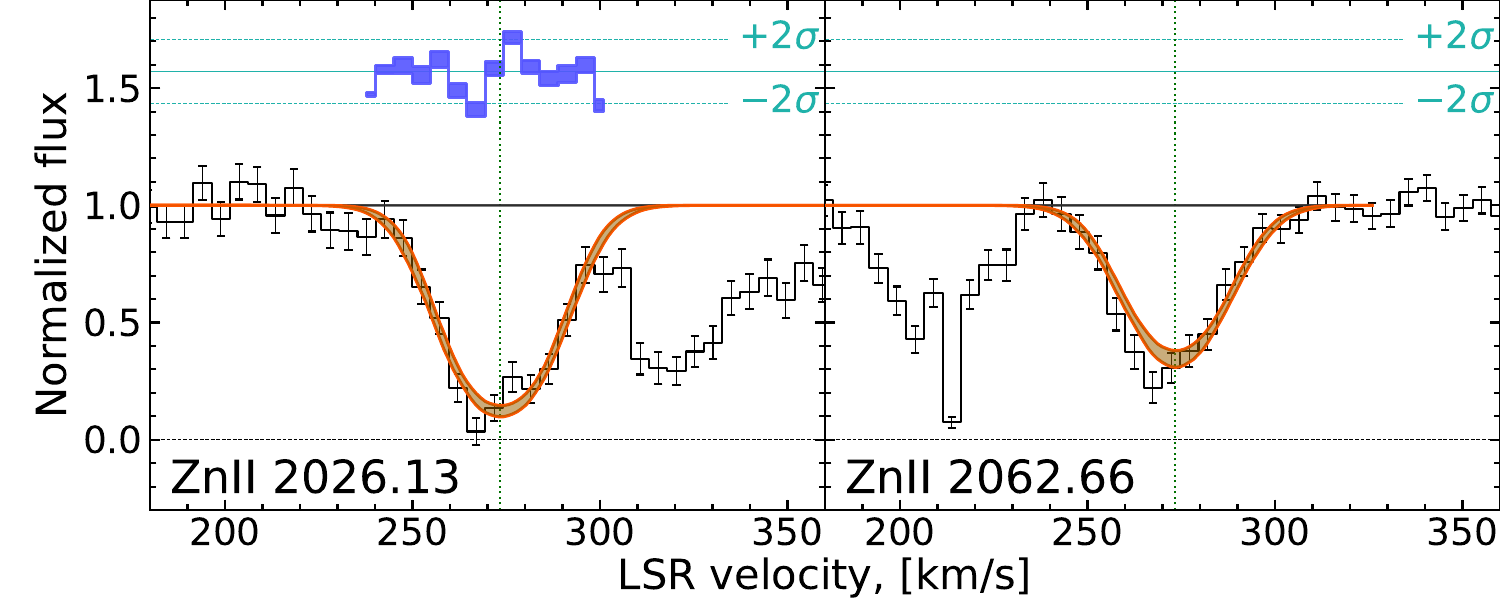}
        \caption{\ZnII\,absorption lines fit in the system towards Sk-68 140 in the LMC. Lines are the same as in Figure~\ref{fig:Sk67_2_CI}}
        \label{fig:Sk68_140_ZnII}
    \end{minipage}
\end{figure*}

\begin{figure*}
    \begin{minipage}{0.49\linewidth}
        \centering
        \includegraphics[width=\linewidth]{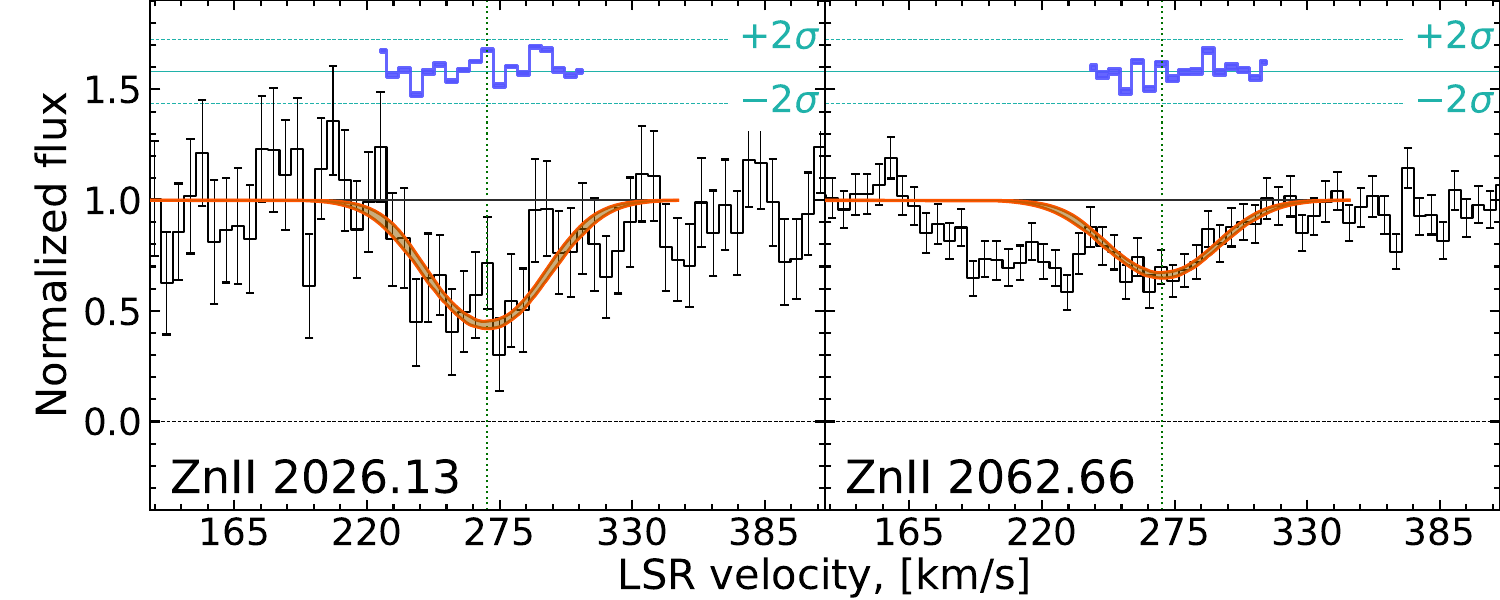}
        \caption{\ZnII\,
    absorption lines fit in the system towards Sk-71 50 in the LMC. Lines are the same as in Figure~\ref{fig:Sk67_2_CI}}
        \label{fig:Sk71_50_ZnII}
    \end{minipage}
    \hfill
    \begin{minipage}{0.49\linewidth}
        \centering
        \includegraphics[width=\linewidth]{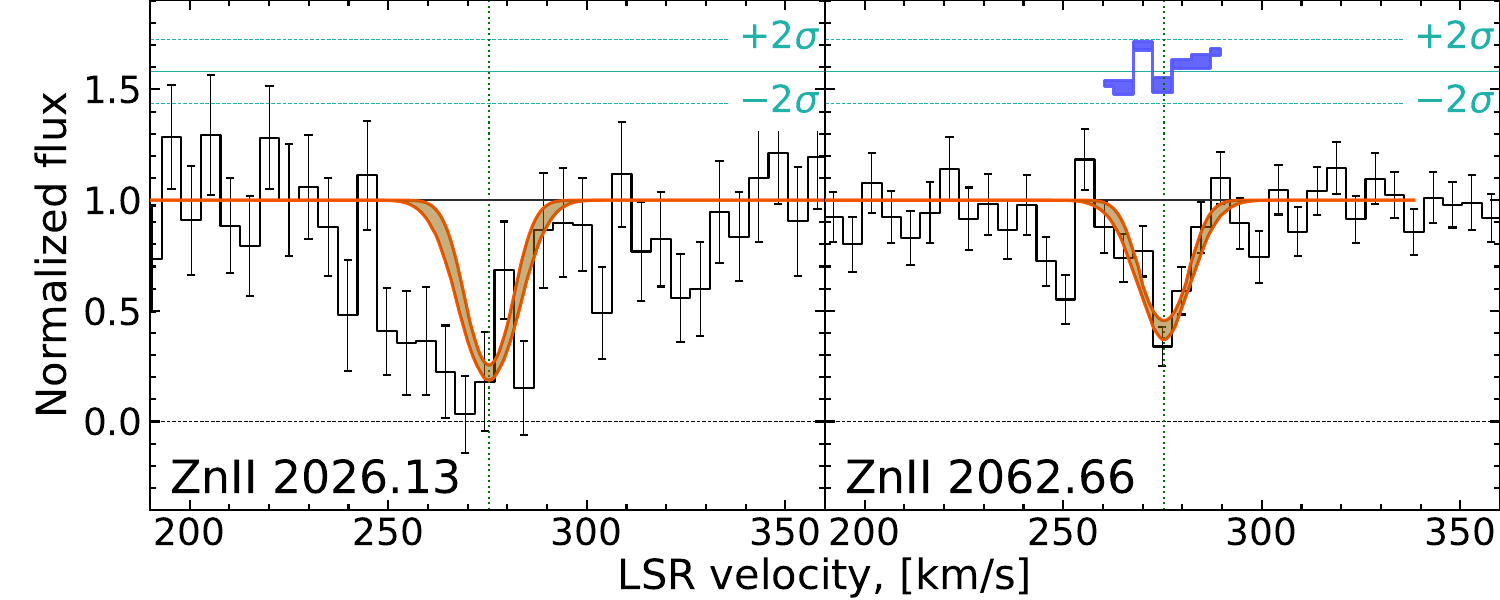}
        \caption{\ZnII\,absorption lines fit in the system towards Sk-69 279 in the LMC. Lines are the same as in Figure~\ref{fig:Sk67_2_CI}}
        \label{fig:Sk69_279_ZnII}
    \end{minipage}
\end{figure*}

\begin{figure*}
    \begin{minipage}{0.49\linewidth}
        \centering
        \includegraphics[width=\linewidth]{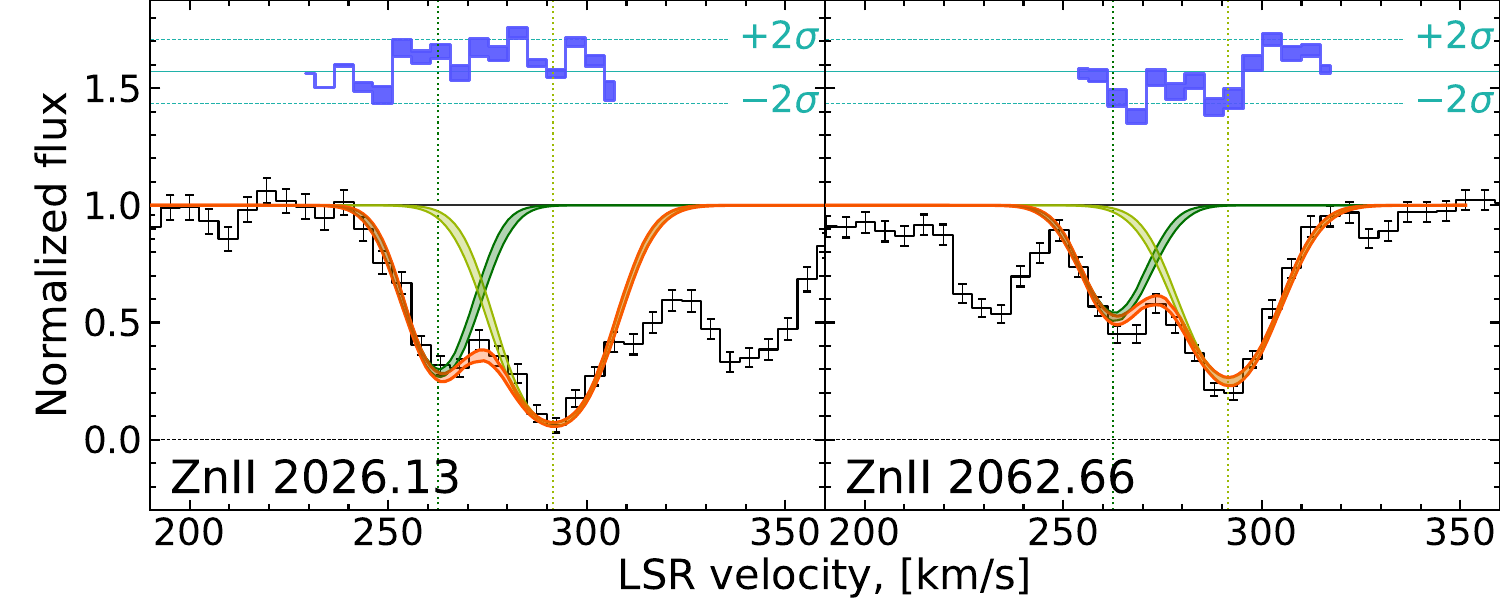}
        \caption{\ZnII\,
    absorption lines fit in the system towards Sk-68 155 in the LMC. Lines are the same as in Figure~\ref{fig:Sk67_2_CI}}
        \label{fig:Sk68_155_ZnII}
    \end{minipage}
    \hfill
    \begin{minipage}{0.49\linewidth}
        \centering
        \includegraphics[width=\linewidth]{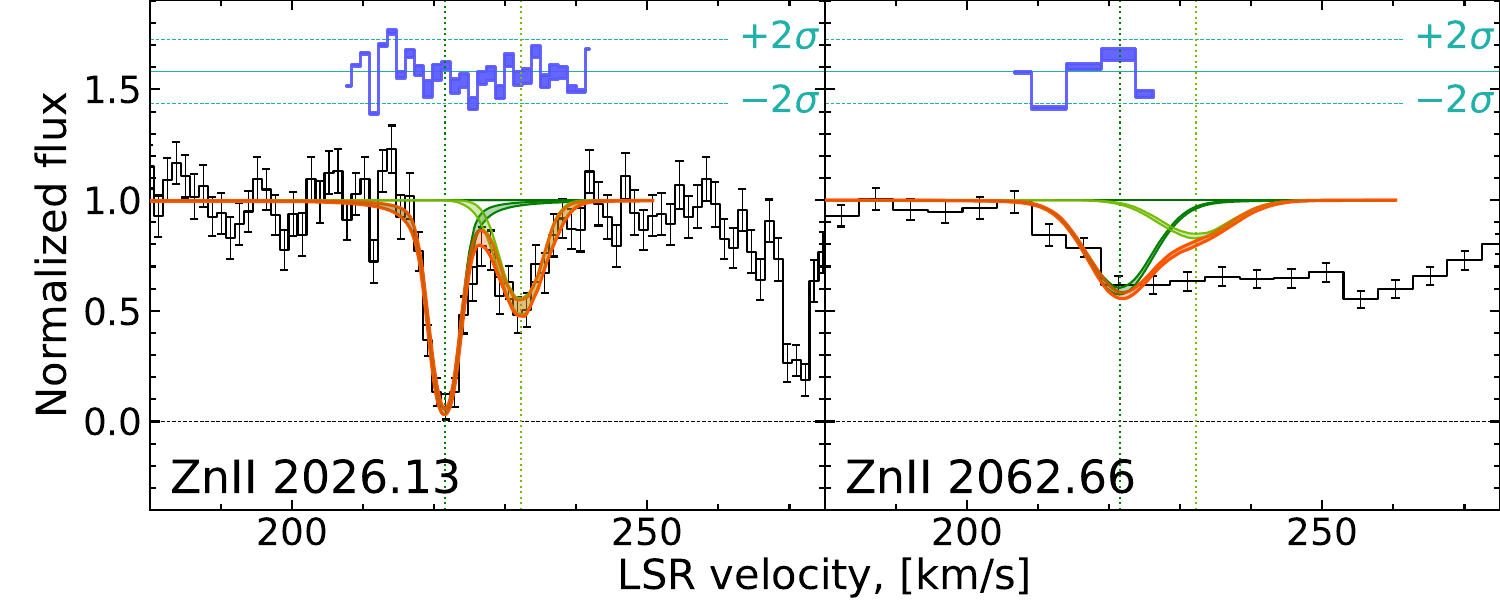}
        \caption{\ZnII\,absorption lines fit in the system towards Sk-70 115 in the LMC. Lines are the same as in Figure~\ref{fig:Sk67_2_CI}}
        \label{fig:Sk70_115_ZnII}
    \end{minipage}
\end{figure*}

\subsection{Small Magellanic Cloud}

\begin{figure*}
    \begin{minipage}{0.55\linewidth}
        \centering
        \includegraphics[width=\linewidth]{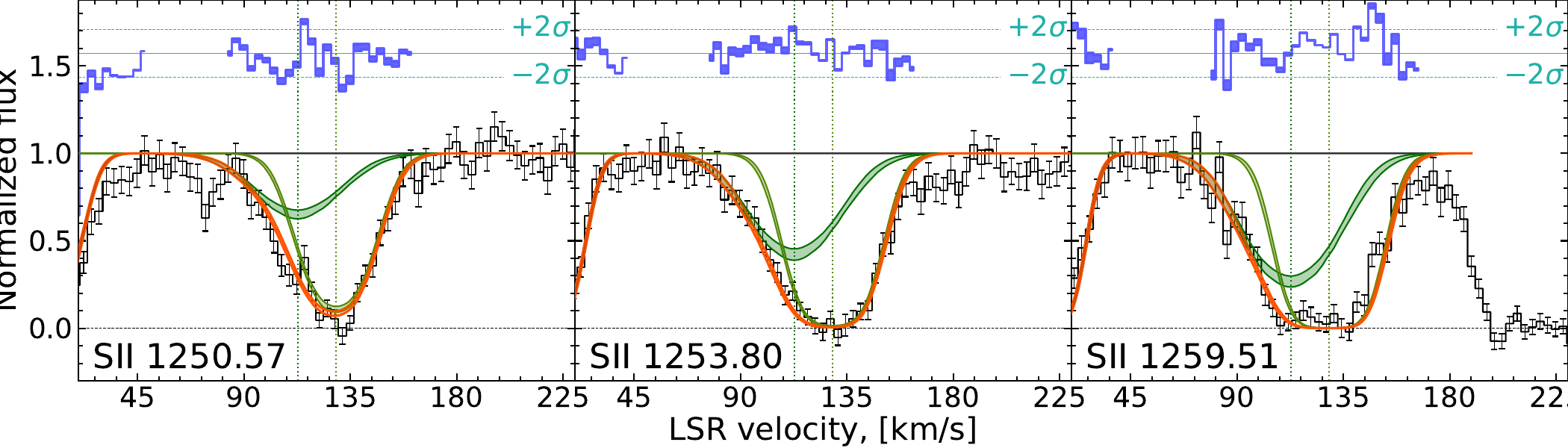}
        \caption{\SII\,
    absorption lines fit in the system towards AV 15 in the SMC. Lines are the same as in Figure~\ref{fig:Sk67_2_CI}}
        \label{fig:AV15_SII}
    \end{minipage}
    \hfill
    \begin{minipage}{0.4\linewidth}
        \centering
        \includegraphics[width=\linewidth]{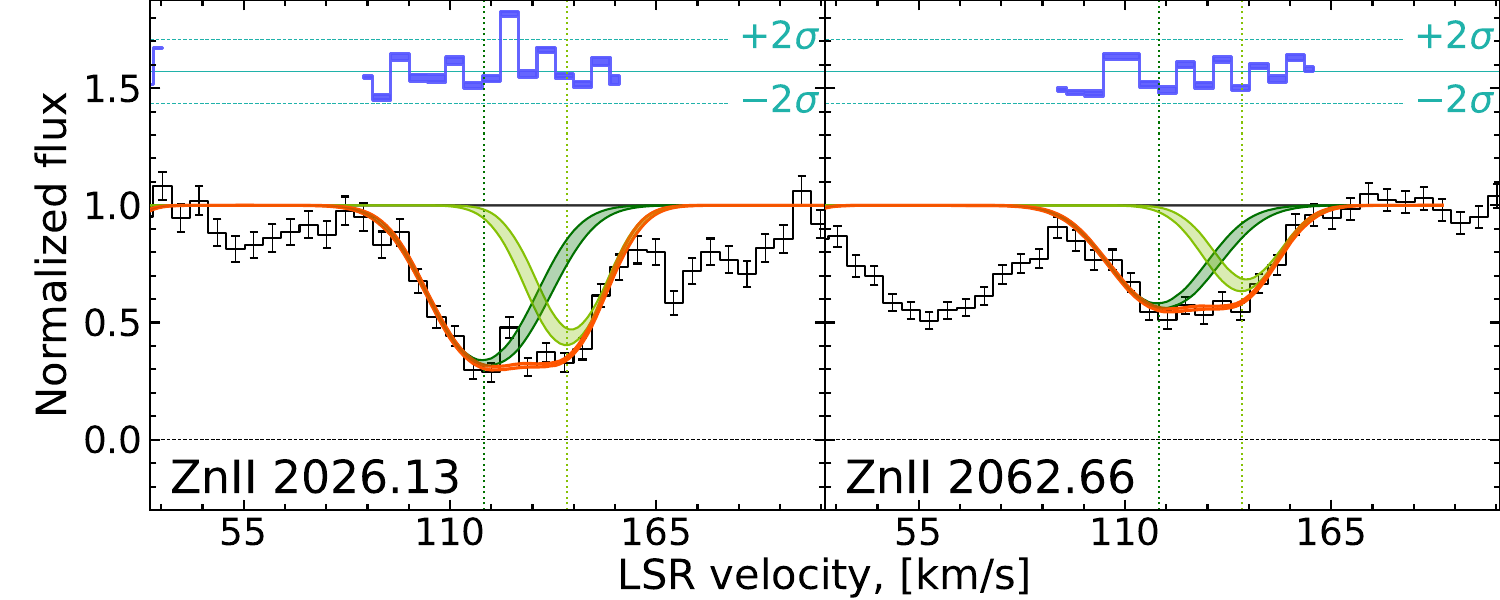}
        \caption{\ZnII\,absorption lines fit in the system towards AV 26 in the SMC. Lines are the same as in Figure~\ref{fig:Sk67_2_CI}}
        \label{fig:AV26_ZnII}
    \end{minipage}
\end{figure*}

\begin{figure*}
    \begin{minipage}{0.55\linewidth}
        \centering
        \includegraphics[width=\linewidth]{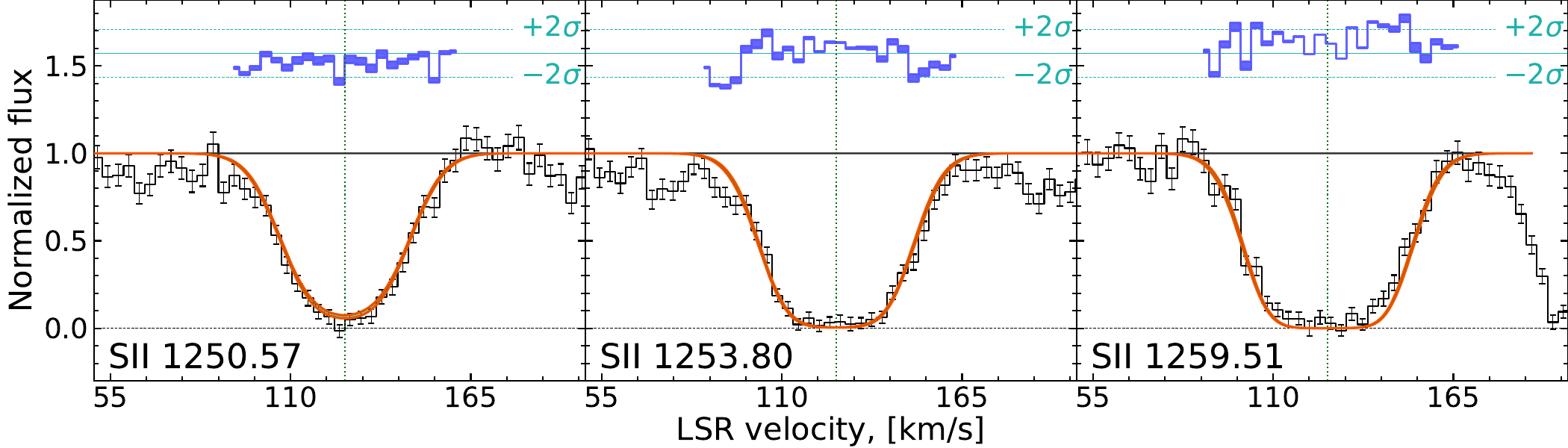}
        \caption{\SII\,absorption lines fit in the system towards AV 69 in the SMC. Lines are the same as in Figure~\ref{fig:Sk67_2_CI}}
        \label{fig:AV69_SII}
    \end{minipage}
    \hfill
    \begin{minipage}{0.4\linewidth}
        \centering
        \includegraphics[width=\linewidth]{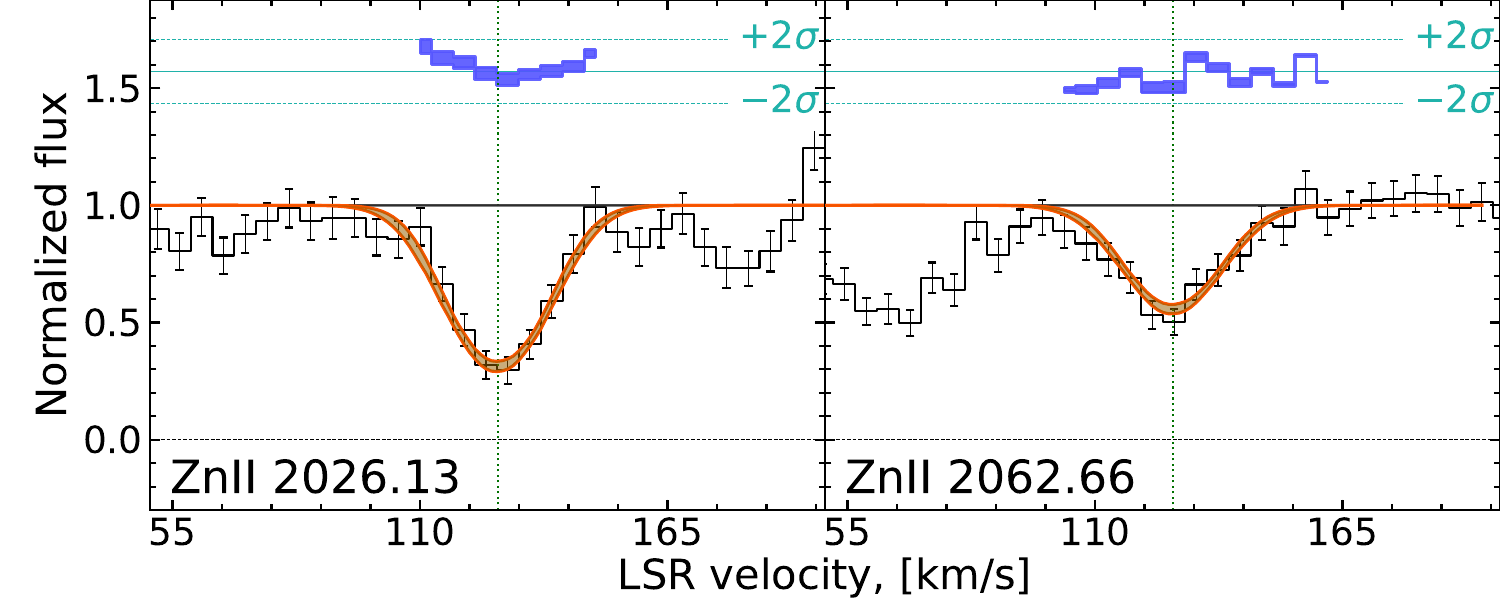}
        \caption{\ZnII\,
    absorption lines fit in the system towards AV 47 in the SMC. Lines are the same as in Figure~\ref{fig:Sk67_2_CI}}
        \label{fig:AV47_ZnII}
    \end{minipage} 
\end{figure*}

\begin{figure*}
    \begin{minipage}{0.55\linewidth}
        \centering
        \includegraphics[width=\linewidth]{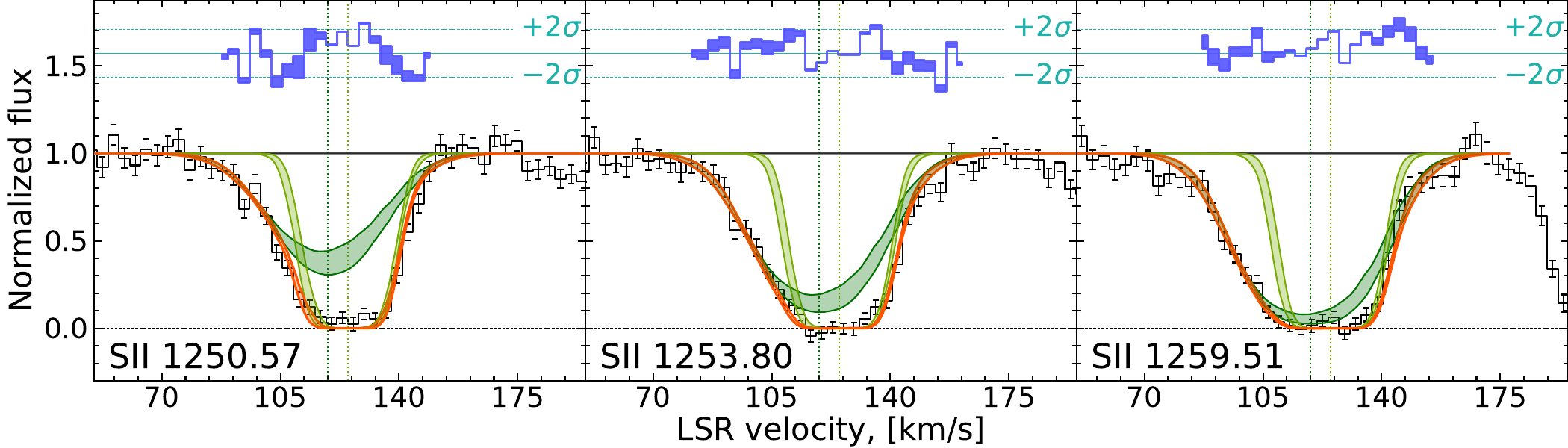}
        \caption{\SII\,absorption lines fit in the system towards AV 75 in the SMC. Lines are the same as in Figure~\ref{fig:Sk67_2_CI}}
        \label{fig:AV75_SII}
    \end{minipage}
    \hfill
    \begin{minipage}{0.4\linewidth}
        \centering
        \includegraphics[width=\linewidth]{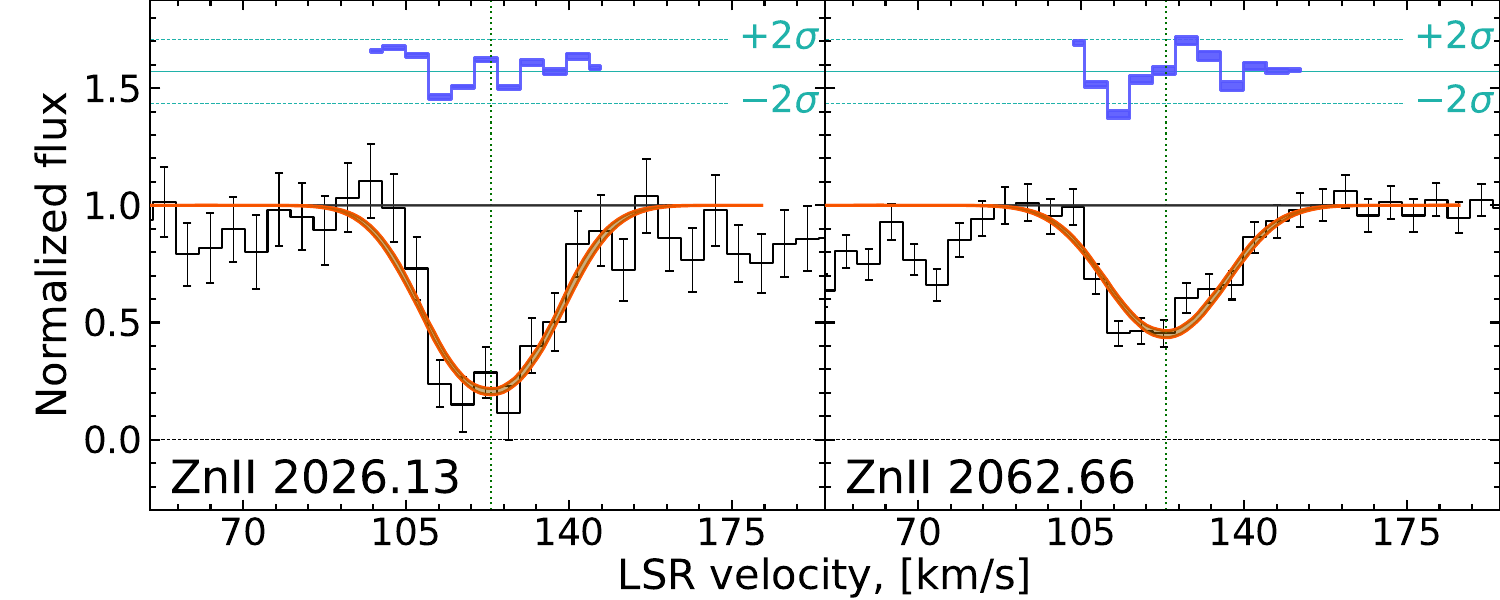}
        \caption{\ZnII\,
    absorption lines fit in the system towards AV 80 in the SMC. Lines are the same as in Figure~\ref{fig:Sk67_2_CI}}
        \label{fig:AV80_ZnII}
    \end{minipage} 
\end{figure*}

\begin{figure*}
    \begin{minipage}{0.49\linewidth}
        \centering
        \includegraphics[width=\linewidth]{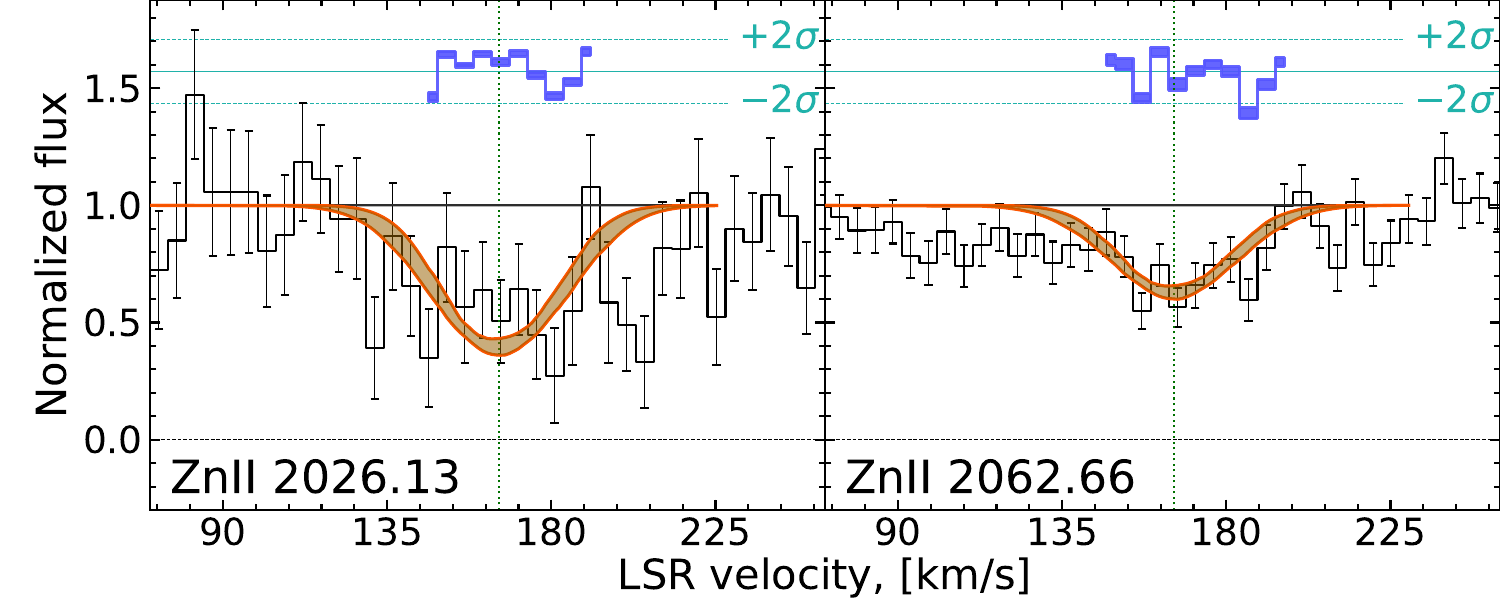}
        \caption{\ZnII\,absorption lines fit in the system towards AV 207 in the SMC. Lines are the same as in Figure~\ref{fig:Sk67_2_CI}}
        \label{fig:AV207_SII}
    \end{minipage}
    \hfill
    \begin{minipage}{0.49\linewidth}
        \centering
        \includegraphics[width=\linewidth]{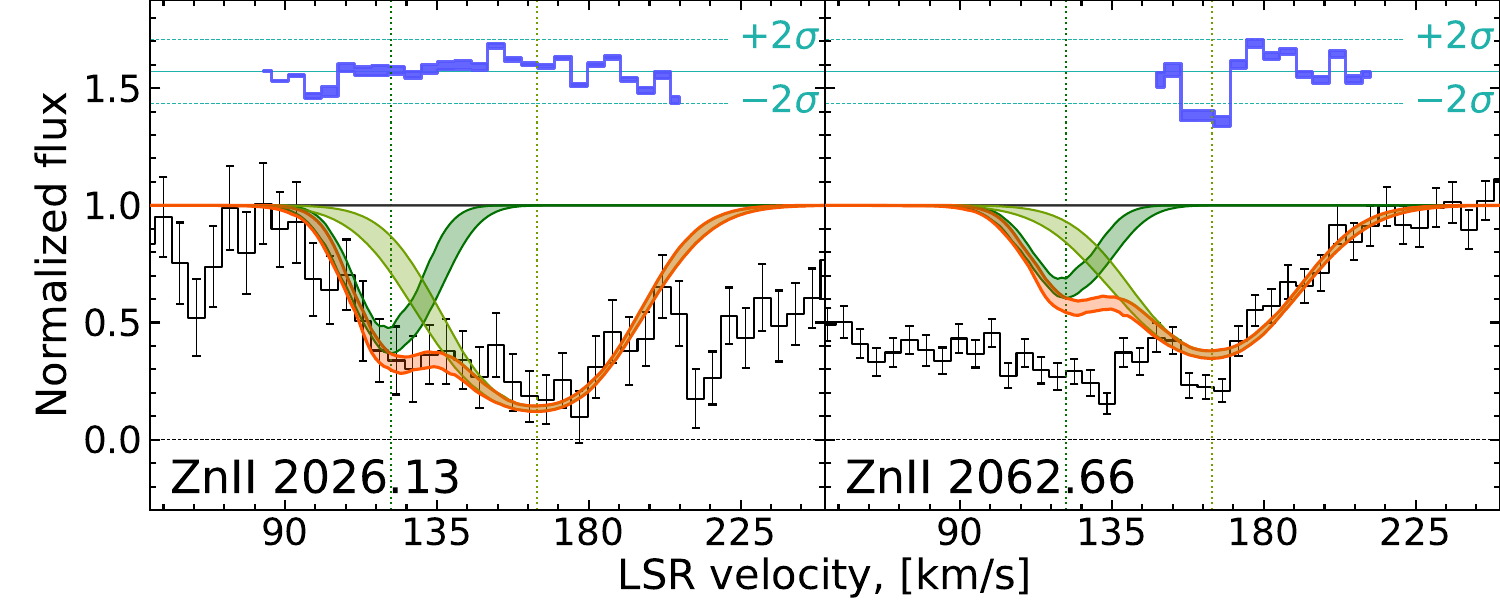}
        \caption{\ZnII\,
    absorption lines fit in the system towards AV 210 in the SMC. Lines are the same as in Figure~\ref{fig:Sk67_2_CI}}
        \label{fig:AV210_ZnII}
    \end{minipage} 
\end{figure*}

\begin{figure*}
    \begin{minipage}{0.49\linewidth}
        \centering
        \includegraphics[width=\linewidth]{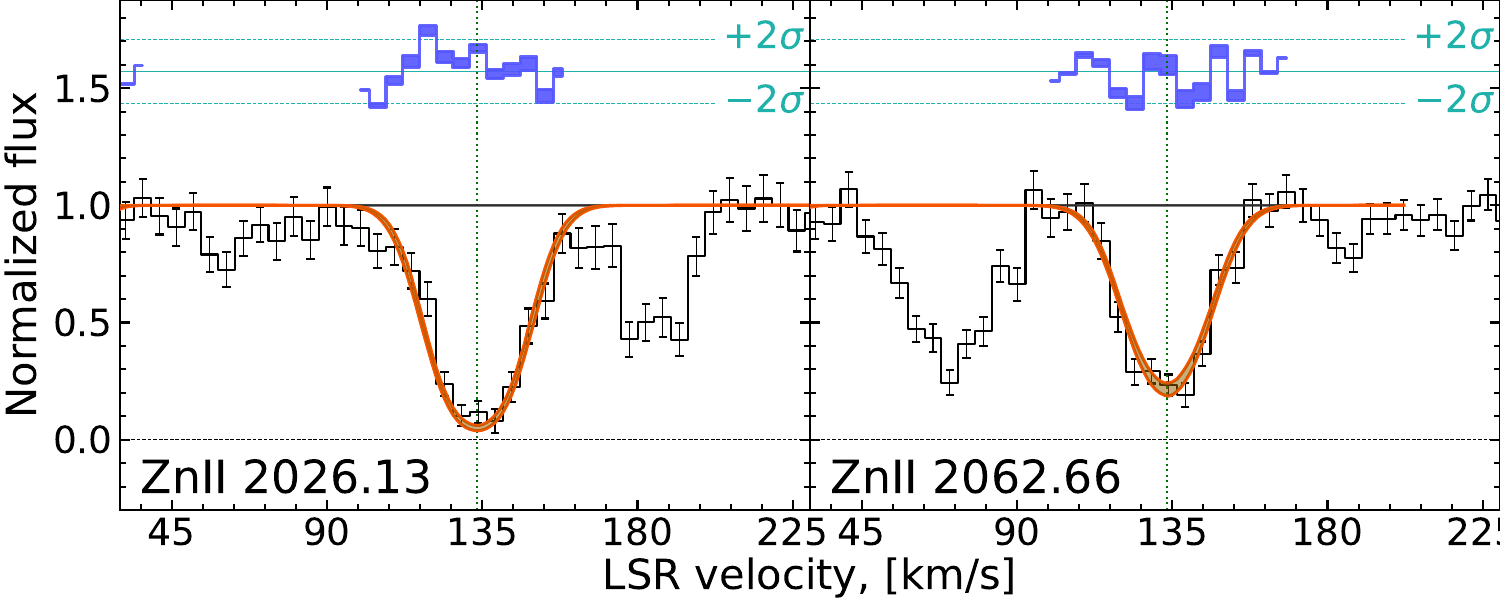}
        \caption{\ZnII\,absorption lines fit in the system towards AV 216 in the SMC. Lines are the same as in Figure~\ref{fig:Sk67_2_CI}}
        \label{fig:AV215_SII}
    \end{minipage}
    \hfill
    \begin{minipage}{0.49\linewidth}
        \centering
        \includegraphics[width=\linewidth]{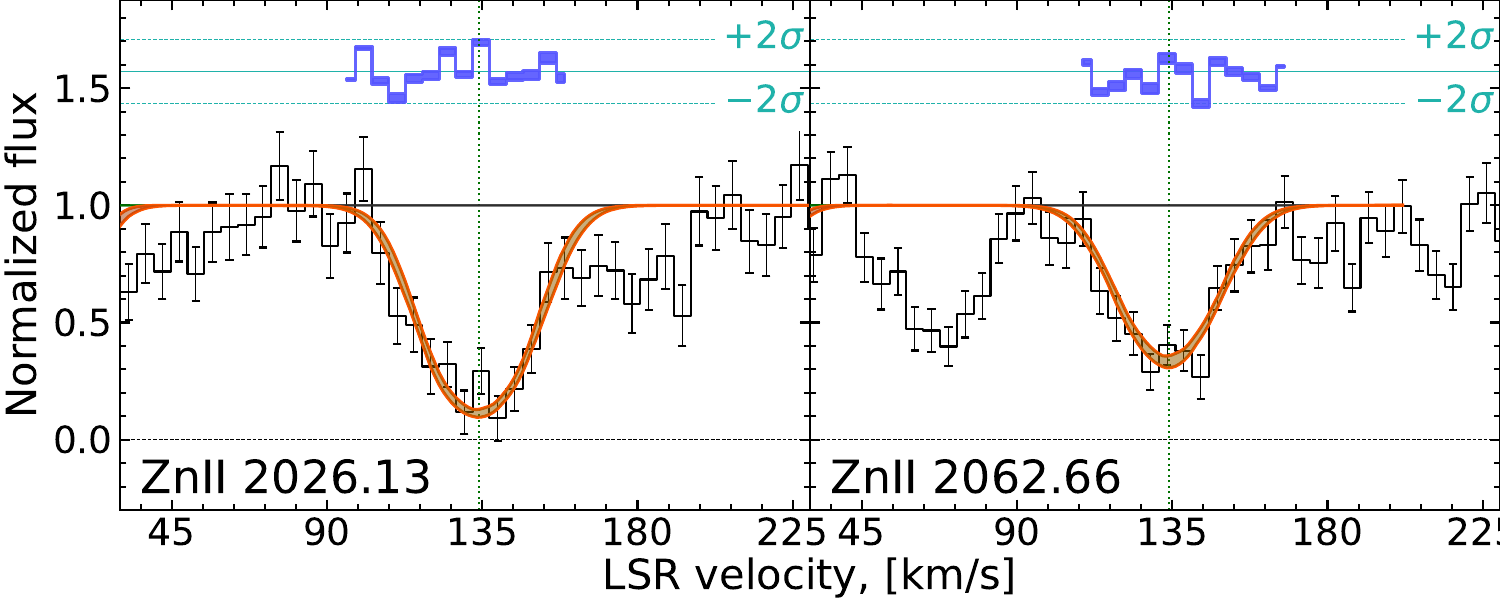}
        \caption{\ZnII\,
    absorption lines fit in the system towards AV 216 in the SMC. Lines are the same as in Figure~\ref{fig:Sk67_2_CI}}
        \label{fig:AV216_ZnII}
    \end{minipage} 
\end{figure*}

\begin{figure*}
    \begin{minipage}{0.49\linewidth}
        \centering
        \includegraphics[width=\linewidth]{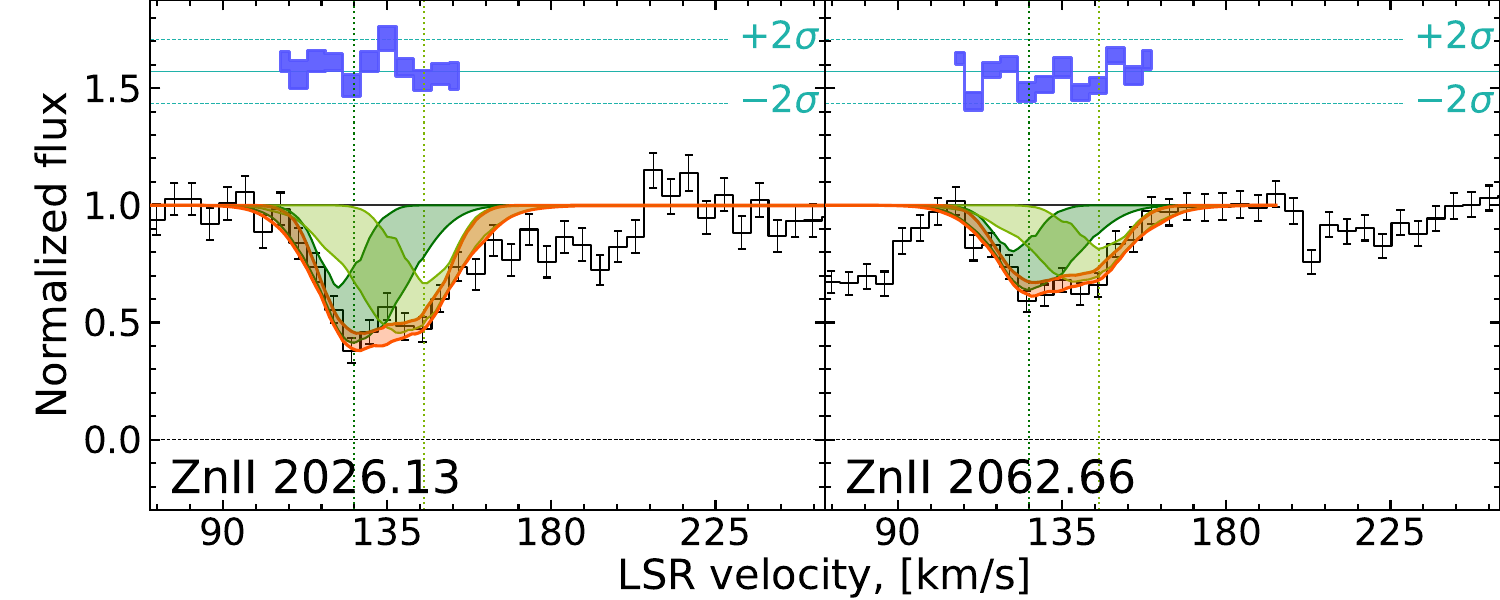}
        \caption{\ZnII\,absorption lines fit in the system towards AV 372 in the SMC. Lines are the same as in Figure~\ref{fig:Sk67_2_CI}}
        \label{fig:AV372_SII}
    \end{minipage}
    \hfill
    \begin{minipage}{0.49\linewidth}
        \centering
        \includegraphics[width=\linewidth]{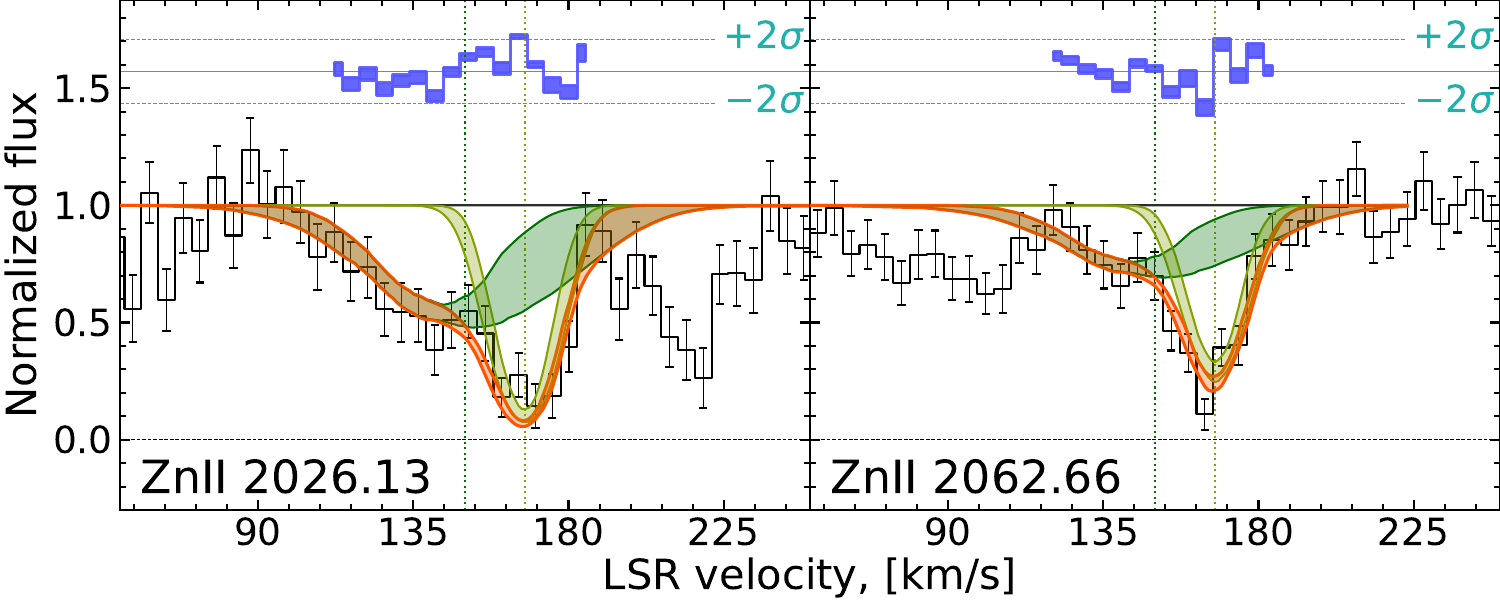}
        \caption{\ZnII\,
    absorption lines fit in the system towards AV 476 in the SMC. Lines are the same as in Figure~\ref{fig:Sk67_2_CI}}
        \label{fig:AV476_ZnII}
    \end{minipage} 
\end{figure*}

\begin{figure*}
    \begin{minipage}{0.49\linewidth}
        \centering
        \includegraphics[width=\linewidth]{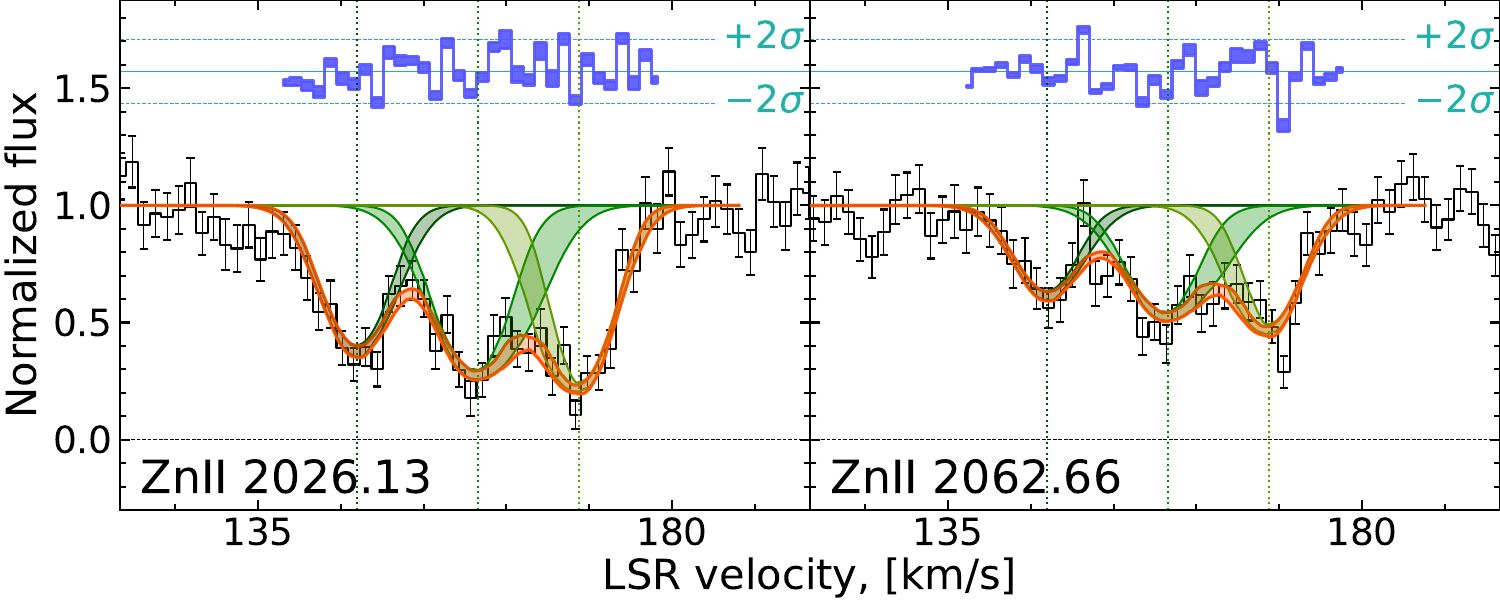}
        \caption{\ZnII\,absorption lines fit in the system towards AV 479 in the SMC. Lines are the same as in Figure~\ref{fig:Sk67_2_CI}}
        \label{fig:AV479_SII}
    \end{minipage}
    \hfill
    \begin{minipage}{0.49\linewidth}
        \centering
        \includegraphics[width=\linewidth]{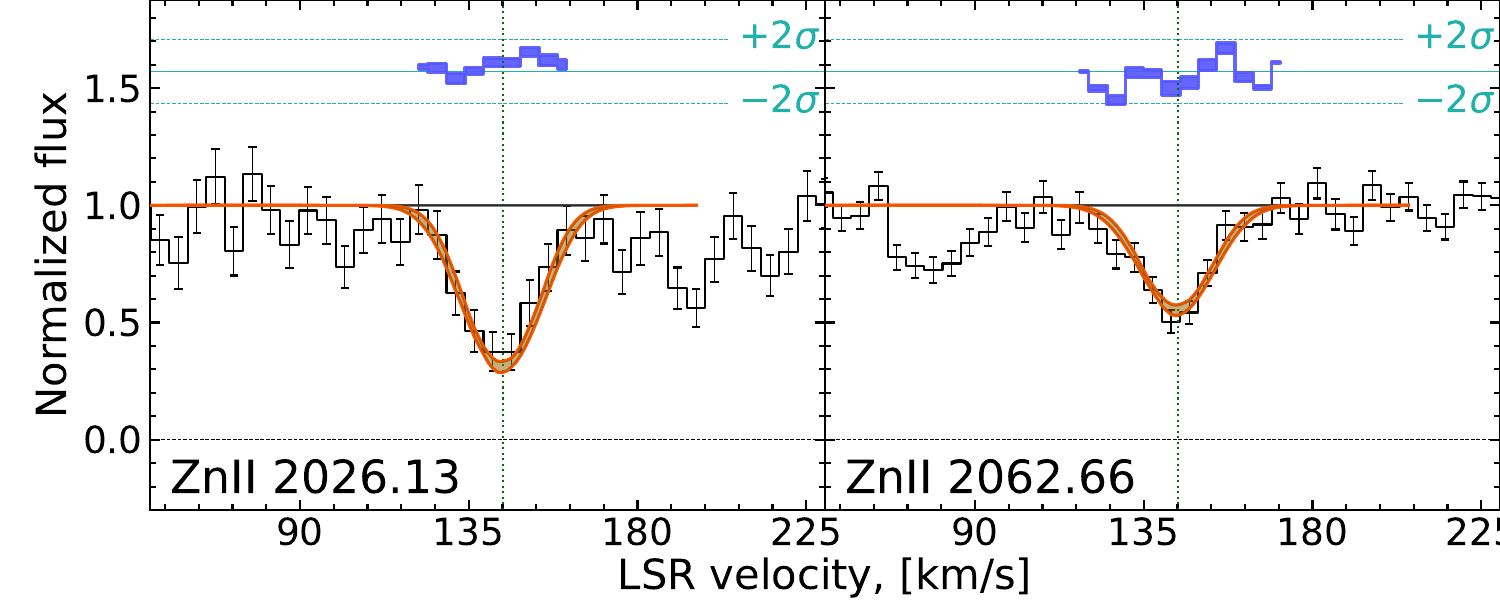}
        \caption{\ZnII\,
    absorption lines fit in the system towards AV 488 in the SMC. Lines are the same as in Figure~\ref{fig:Sk67_2_CI}}
        \label{fig:AV488_ZnII}
    \end{minipage} 
\end{figure*}

\begin{figure*}
    \begin{minipage}{0.49\linewidth}
        \centering
        \includegraphics[width=0.8\linewidth]{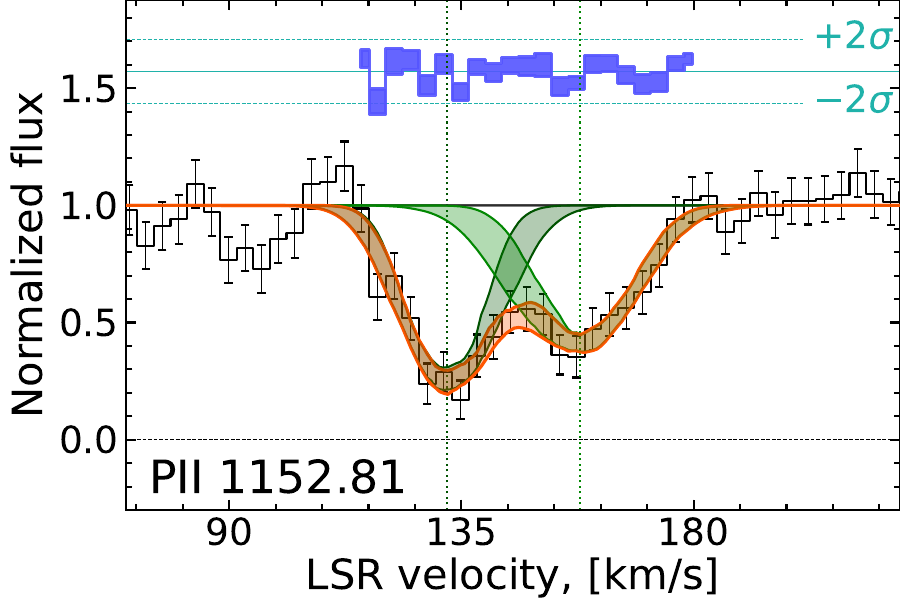}
        \caption{\PII\,absorption lines fit in the system towards AV 490 in the SMC. Lines are the same as in Figure~\ref{fig:Sk67_2_CI}}
        \label{fig:AV490_PII}
    \end{minipage}
    \hfill
    \begin{minipage}{0.49\linewidth}
        \centering
        \includegraphics[width=\linewidth]{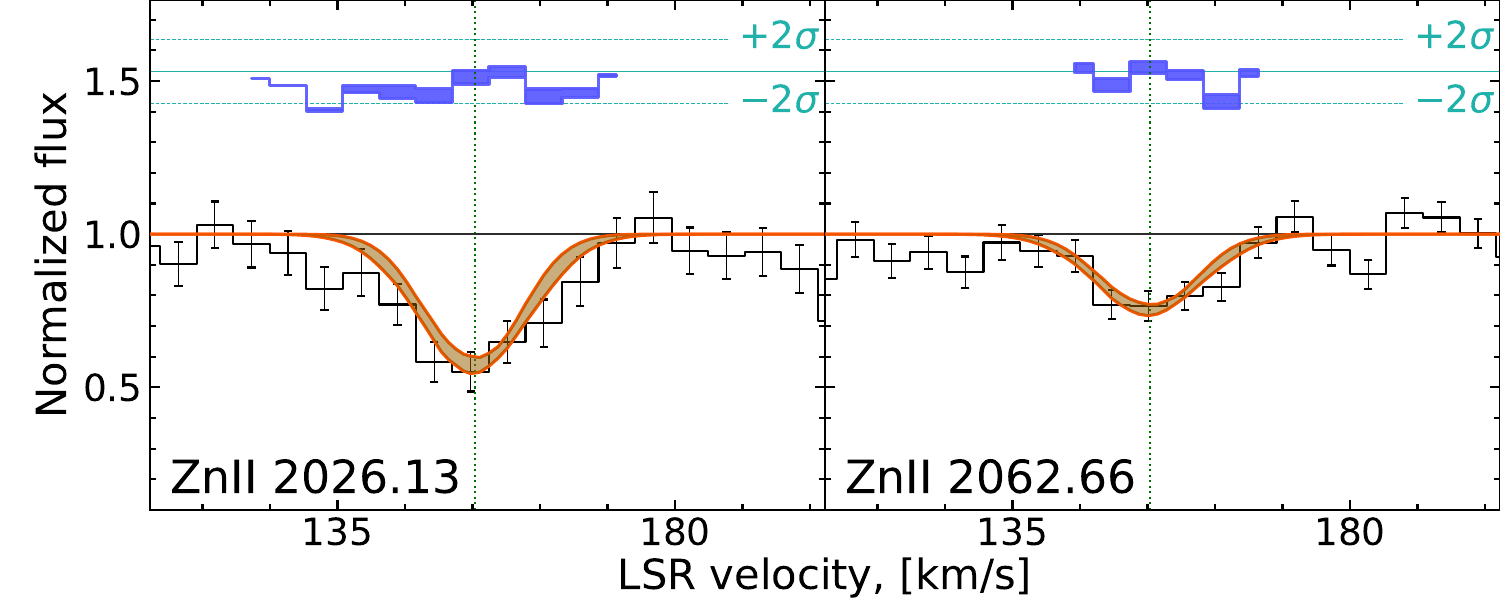}
        \caption{\ZnII\,
    absorption lines fit in the system towards Sk 191 in the SMC. Lines are the same as in Figure~\ref{fig:Sk67_2_CI}}
        \label{fig:Sk191_PII}
    \end{minipage} 
\end{figure*}

\clearpage

\section{Details on constraints of \lowercase{n}$_{\rm H}$ and $\chi$}
\label{sect:phys_constr}

In this section we present the constraints for each individual sightline on the number density and UV field (or CRIR) using the excitation of \CI\ fine-structure levels and two lowest rotational levels of H$_2$. The joint constraints are summarize in Tables~\ref{tab:LMC_phys_cond} and \ref{tab:SMC_phys_cond} for the LMC and SMC, respectively 

\begin{figure*}
    \begin{minipage}{0.49\linewidth}
        \centering
    \includegraphics[width=\linewidth]{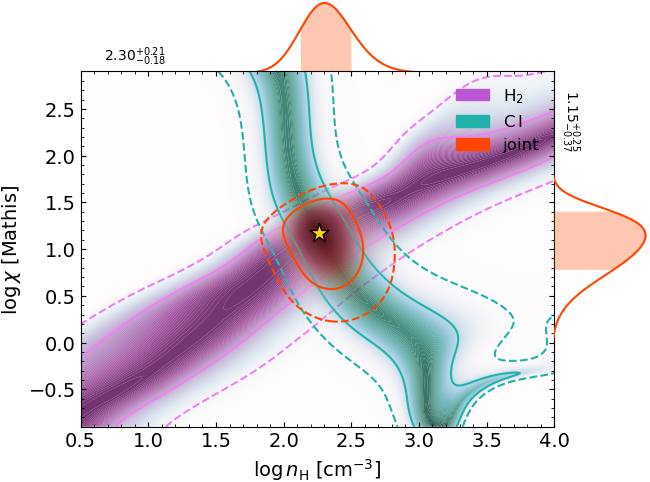}
    \caption{Estimate on the number density and UV field intensity for the system towards Sk-67 2 in the LMC. Lines are the same as for Figure~\ref{fig:n_chi_Sk675}.}
    \label{fig:n_chi_Sk672}
    \end{minipage}
    \hfill
    \begin{minipage}{0.49\linewidth}
        \centering
    \includegraphics[width=\linewidth]{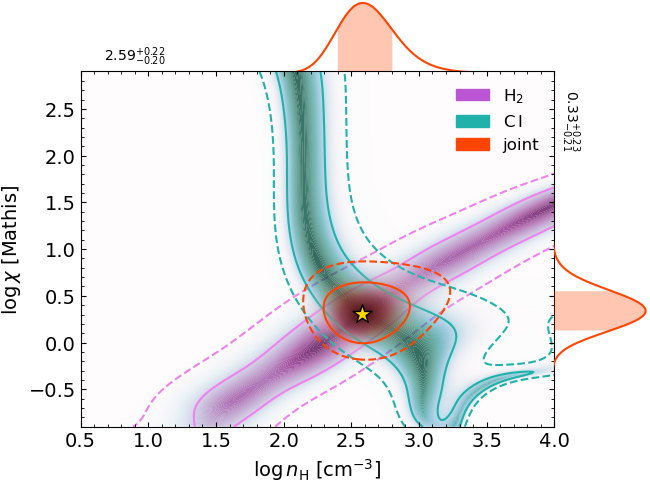}
    \caption{Estimate on the number density and UV field intensity for the system towards Sk-67 20 in the LMC. Lines are the same as for Figure~\ref{fig:n_chi_Sk675}.}
    \label{fig:n_chi_Sk6720}
    \end{minipage}
\end{figure*}

\begin{figure*}
\begin{minipage}{0.49\linewidth}
    \centering
    \includegraphics[width=\linewidth]{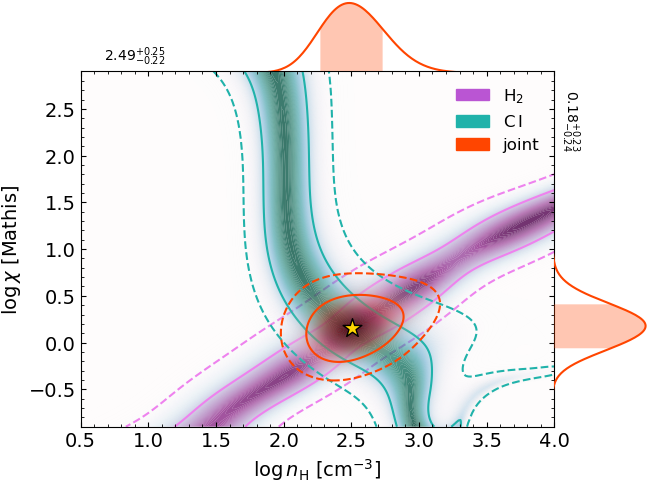}
    \caption{Estimate on the number density and UV field intensity for the system towards PGMW 3070 in the LMC. Lines are the same as for Figure~\ref{fig:n_chi_Sk675}.}
    \label{fig:n_chi_PGMW3070}
\end{minipage}
\hfill
\begin{minipage}{0.49\linewidth}
    \centering
    \includegraphics[width=\linewidth]{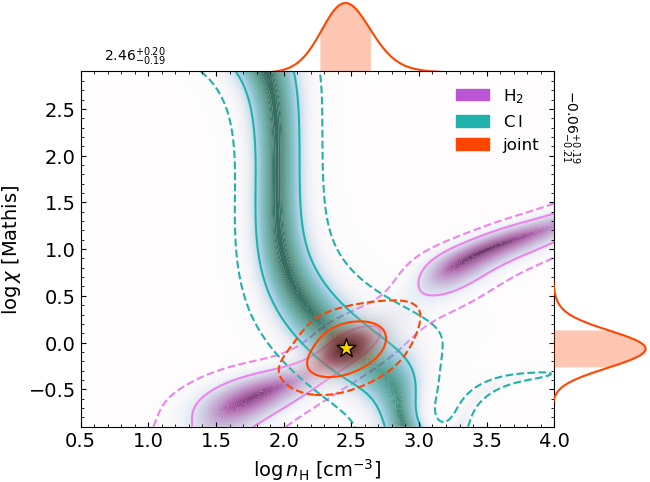}
    \caption{Estimate on the number density and UV field intensity for the system towards LH10 3120 in the LMC. Lines are the same as for Figure~\ref{fig:n_chi_Sk675}.}
    \label{fig:n_chi_LH103120}
\end{minipage}
\end{figure*}

\begin{figure*}
    \begin{minipage}{0.49\linewidth}
    \centering
    \includegraphics[width=\linewidth]{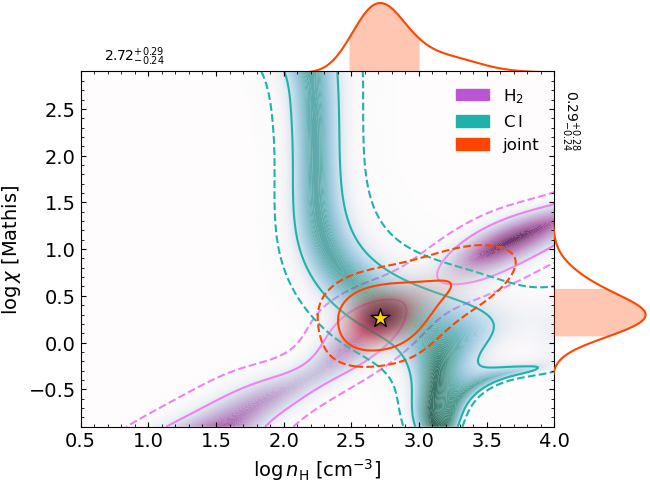}
    \caption{Estimate on the number density and UV field intensity for the system towards PGMW 3223 in the LMC. Lines are the same as for Figure~\ref{fig:n_chi_Sk675}.}
    \label{fig:n_chi_PGMW3223}
\end{minipage}
\hfill
\begin{minipage}{0.49\linewidth}
    \centering
    \includegraphics[width=\linewidth]{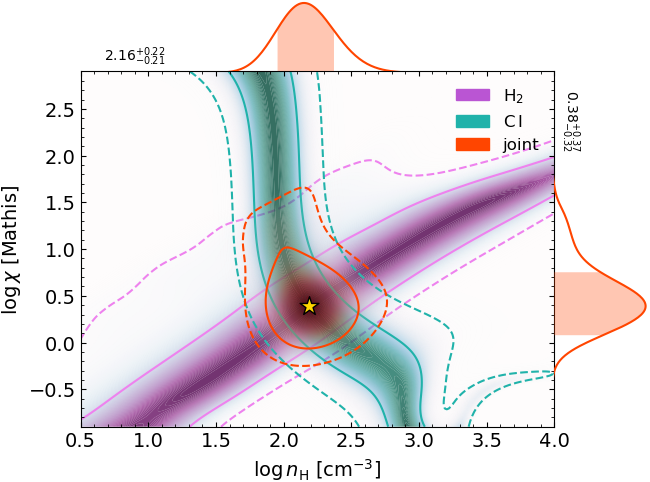}
    \caption{Estimate on the number density and UV field intensity for the system towards Sk-66 35 in the LMC. Lines are the same as for Figure~\ref{fig:n_chi_Sk675}.}
    \label{fig:n_chi_Sk6635}
\end{minipage}
\end{figure*}

\begin{figure*}
    \begin{minipage}{0.49\linewidth}
    \centering
    \includegraphics[width=\linewidth]{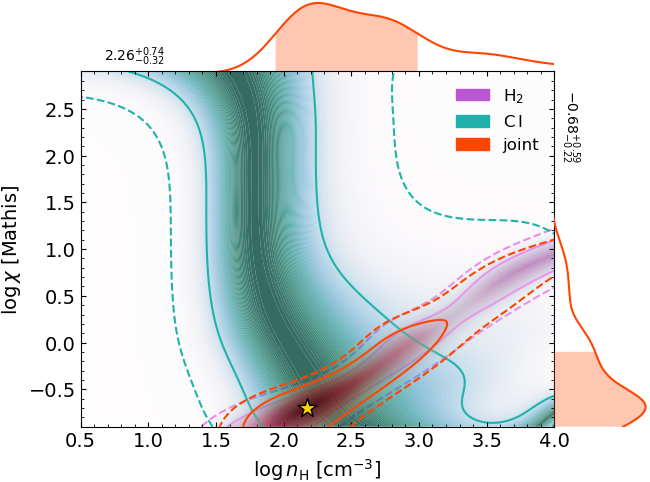}
    \caption{Estimate on the number density and UV field intensity for the system towards Sk-66 51 in the LMC. Lines are the same as for Figure~\ref{fig:n_chi_Sk675}.}
    \label{fig:n_chi_Sk6651}
\end{minipage}
\hfill
\begin{minipage}{0.49\linewidth}
    \centering
    \includegraphics[width=\linewidth]{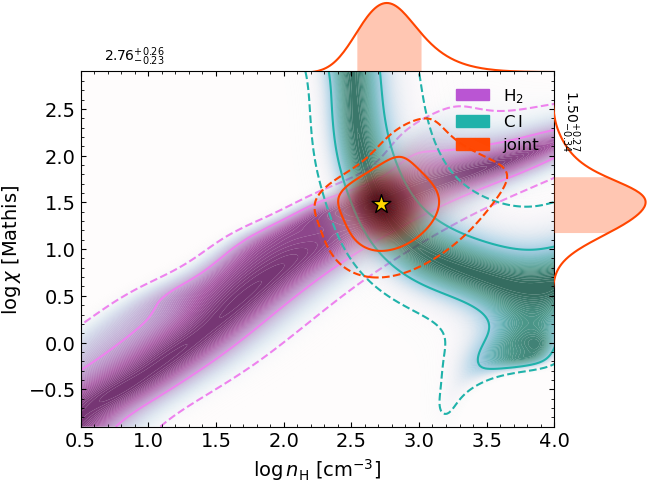}
    \caption{Estimate on the number density and UV field intensity for the system towards Sk-70 79 in the LMC. Lines are the same as for Figure~\ref{fig:n_chi_Sk675}.}
    \label{fig:n_chi_Sk7079}
\end{minipage}
\end{figure*}

\begin{figure*}
    \begin{minipage}{0.49\linewidth}
    \centering
    \includegraphics[width=\linewidth]{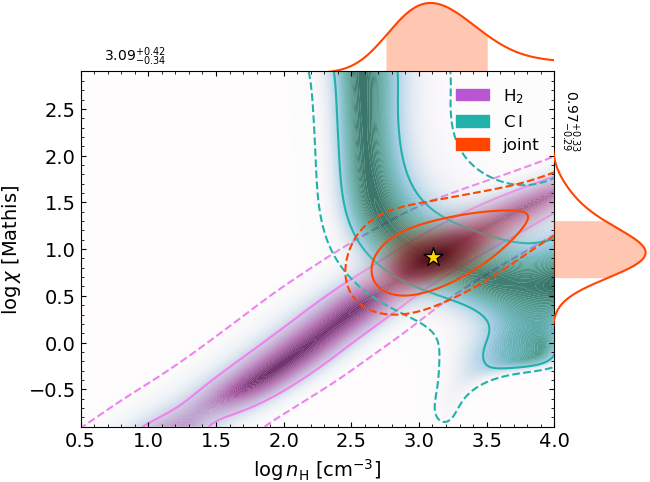}
    \caption{Estimate on the number density and UV field intensity for the system towards Sk-68 52 in the LMC. Lines are the same as for Figure~\ref{fig:n_chi_Sk675}.}
    \label{fig:n_chi_Sk6852}
\end{minipage}
\hfill
\begin{minipage}{0.49\linewidth}
    \centering
    \includegraphics[width=\linewidth]{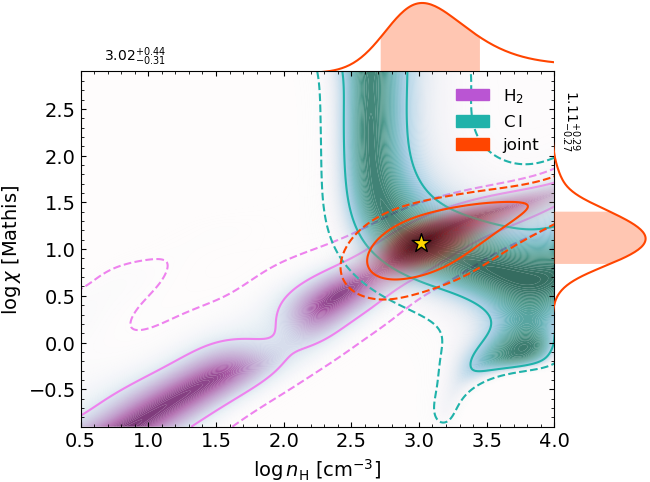}
    \caption{Estimate on the number density and UV field intensity for the system towards Sk-71 8 in the LMC. Lines are the same as for Figure~\ref{fig:n_chi_Sk675}.}
    \label{fig:n_chi_Sk718}
\end{minipage}
\end{figure*}

\begin{figure*}
    \begin{minipage}{0.49\linewidth}
    \centering
    \includegraphics[width=\linewidth]{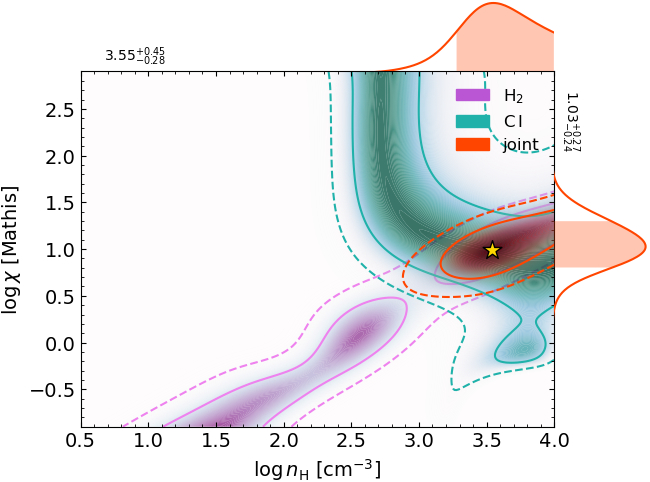}
    \caption{Estimate on the number density and UV field intensity for the system towards Sk-69 106 in the LMC. Lines are the same as for Figure~\ref{fig:n_chi_Sk675}.}
    \label{fig:n_chi_Sk69106}
\end{minipage}
\hfill
\begin{minipage}{0.49\linewidth}
    \centering
    \includegraphics[width=\linewidth]{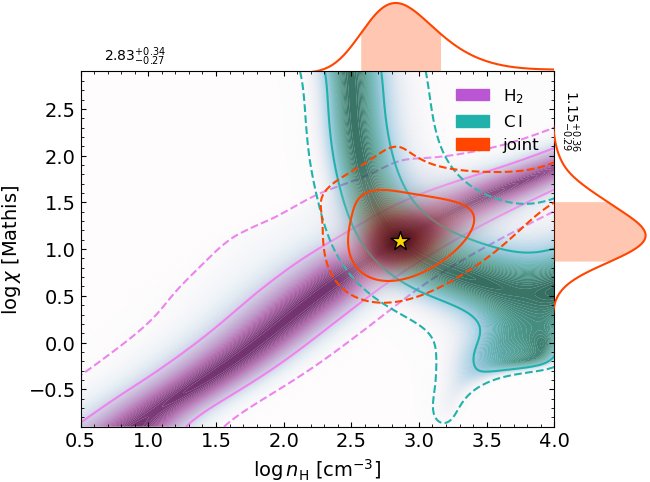}
    \caption{Estimate on the number density and UV field intensity for the system towards Sk-68 73 in the LMC. Lines are the same as for Figure~\ref{fig:n_chi_Sk675}.}
    \label{fig:n_chi_Sk6873}
\end{minipage}
\end{figure*}

\begin{figure*}
    \begin{minipage}{0.49\linewidth}
    \centering
    \includegraphics[width=\linewidth]{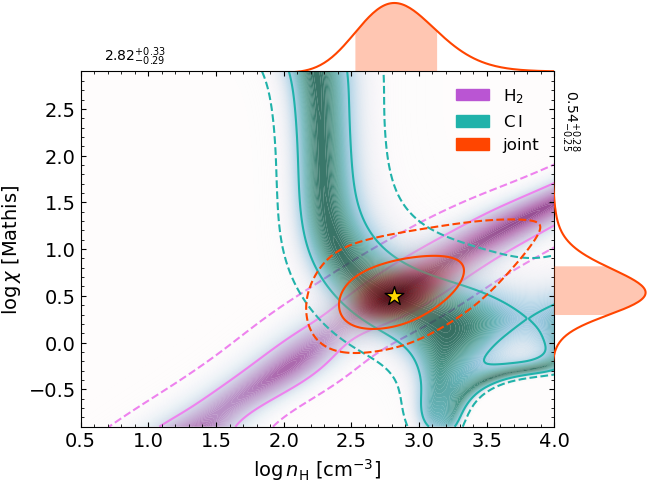}
    \caption{Estimate on the number density and UV field intensity for the system towards Sk-67 105 in the LMC. Lines are the same as for Figure~\ref{fig:n_chi_Sk675}.}
    \label{fig:n_chi_Sk67105}
\end{minipage}
\hfill
\begin{minipage}{0.49\linewidth}
    \centering
    \includegraphics[width=\linewidth]{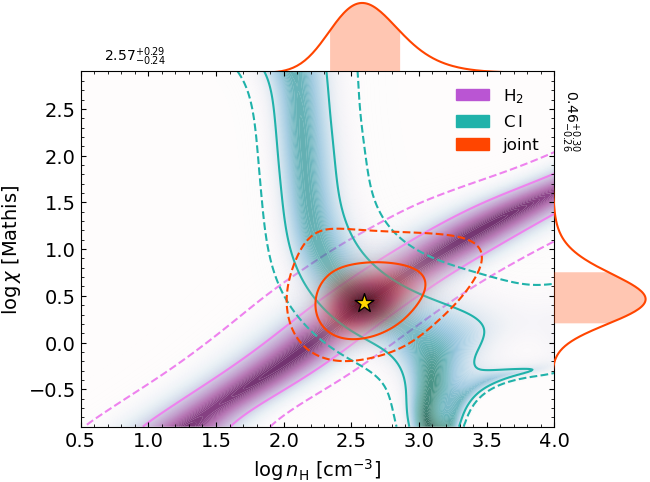}
    \caption{Estimate on the number density and UV field intensity for the system towards BI 184 in the LMC. Lines are the same as for Figure~\ref{fig:n_chi_Sk675}.}
    \label{fig:n_chi_BI184}
\end{minipage}
\end{figure*}

\begin{figure*}
    \begin{minipage}{0.49\linewidth}
    \centering
    \includegraphics[width=\linewidth]{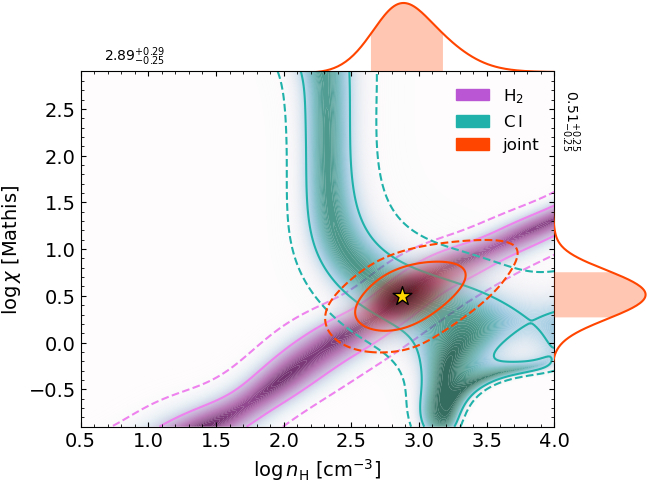}
    \caption{Estimate on the number density and UV field intensity for the system towards Sk-71 45 in the LMC. Lines are the same as for Figure~\ref{fig:n_chi_Sk675}.}
    \label{fig:n_chi_Sk7145}
\end{minipage}
\hfill
\begin{minipage}{0.49\linewidth}
    \centering
    \includegraphics[width=\linewidth]{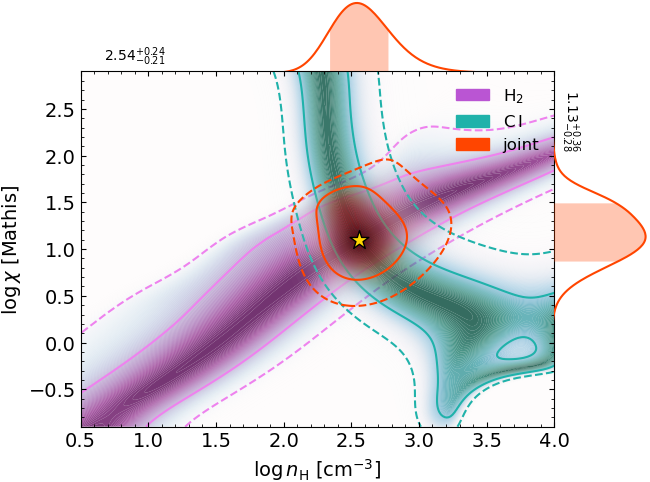}
    \caption{Estimate on the number density and UV field intensity for the system towards Sk-71 46 in the LMC. Lines are the same as for Figure~\ref{fig:n_chi_Sk675}.}
    \label{fig:n_chi_Sk7146}
\end{minipage}
\end{figure*}

\begin{figure*}
    \begin{minipage}{0.49\linewidth}
    \centering
    \includegraphics[width=\linewidth]{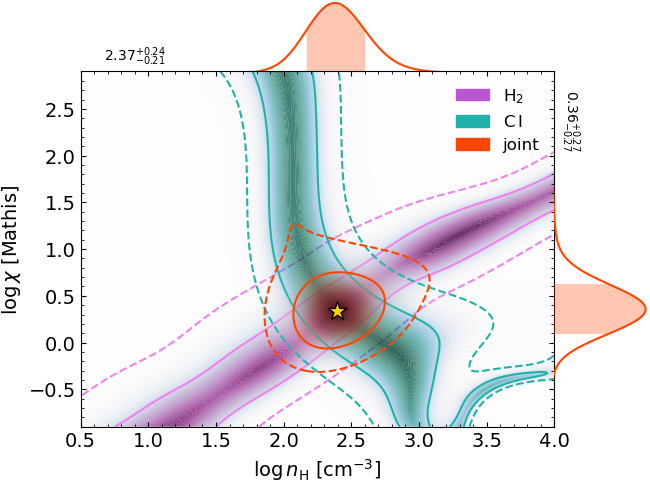}
    \caption{Estimate on the number density and UV field intensity for the system towards Sk-69 191 in the LMC. Lines are the same as for Figure~\ref{fig:n_chi_Sk675}.}
    \label{fig:n_chi_Sk69191}
\end{minipage}
\hfill
\begin{minipage}{0.49\linewidth}
    \centering
    \includegraphics[width=\linewidth]{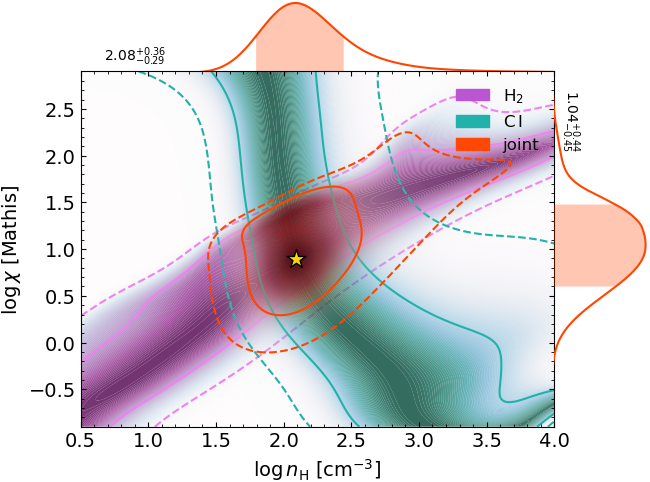}
    \caption{Estimate on the number density and UV field intensity for the system towards BI 237 in the LMC. Lines are the same as for Figure~\ref{fig:n_chi_Sk675}.}
    \label{fig:n_chi_BI237}
\end{minipage}
\end{figure*}

\begin{figure*}
    \begin{minipage}{0.49\linewidth}
    \centering
    \includegraphics[width=\linewidth]{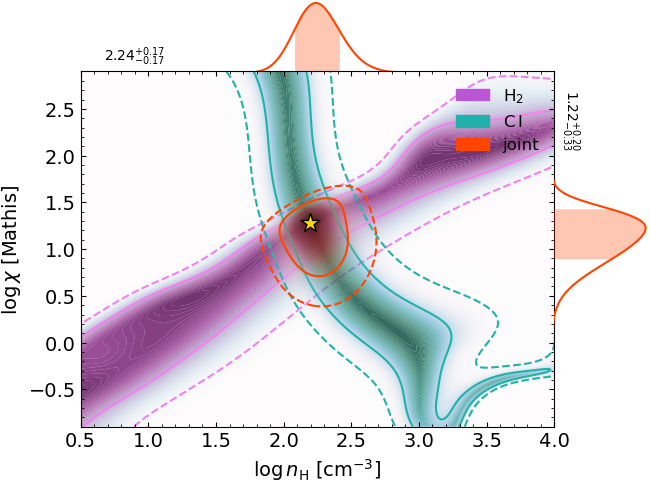}
    \caption{Estimate on the number density and UV field intensity for the system towards Sk-68 129 in the LMC. Lines are the same as for Figure~\ref{fig:n_chi_Sk675}.}
    \label{fig:n_chi_Sk68129}
\end{minipage}
\hfill
\begin{minipage}{0.49\linewidth}
    \centering
    \includegraphics[width=\linewidth]{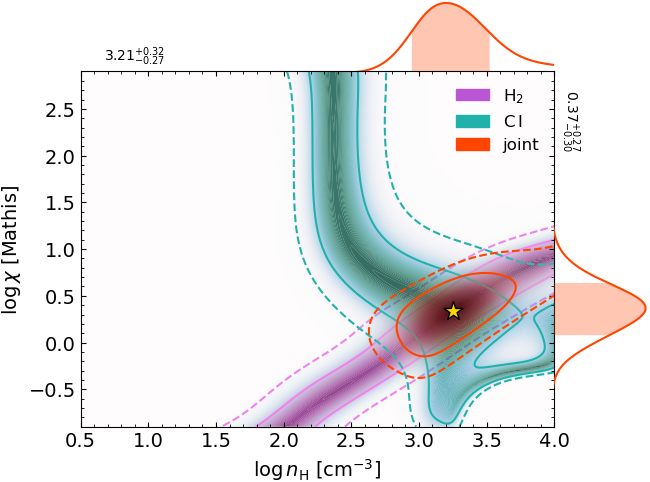}
    \caption{Estimate on the number density and UV field intensity for the system towards Sk-66 172 in the LMC. Lines are the same as for Figure~\ref{fig:n_chi_Sk675}.}
    \label{fig:n_chi_Sk66172}
\end{minipage}
\end{figure*}

\begin{figure*}
    \begin{minipage}{0.49\linewidth}
    \centering
    \includegraphics[width=\linewidth]{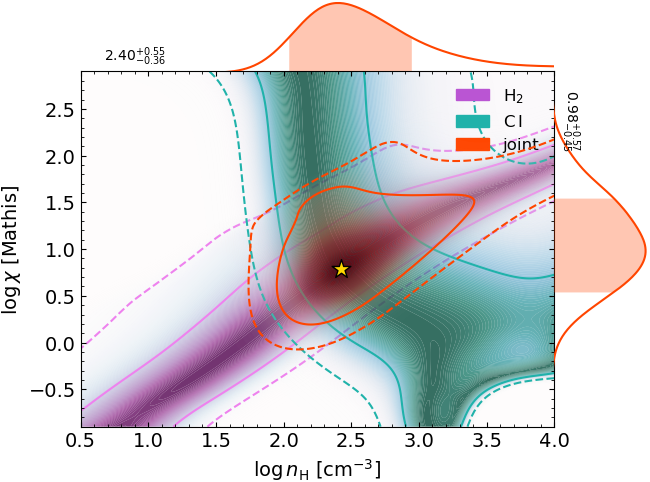}
    \caption{Estimate on the number density and UV field intensity for the system towards BI 253 in the LMC. Lines are the same as for Figure~\ref{fig:n_chi_Sk675}.}
    \label{fig:n_chi_BI253}
\end{minipage}
\hfill
\begin{minipage}{0.49\linewidth}
    \centering
    \includegraphics[width=\linewidth]{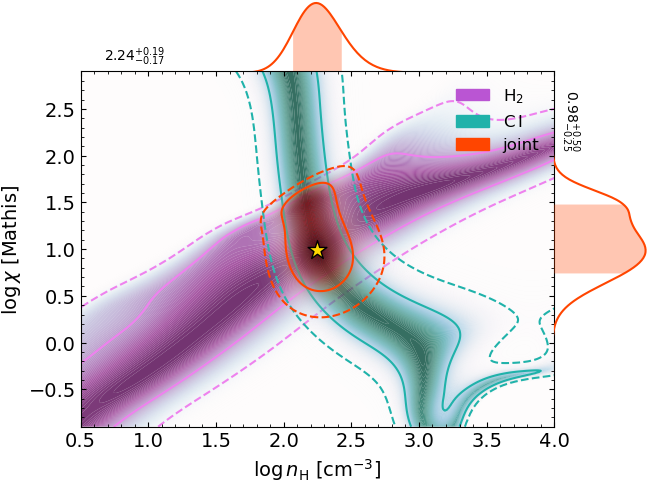}
    \caption{Estimate on the number density and UV field intensity for the system towards Sk-68 135 in the LMC. Lines are the same as for Figure~\ref{fig:n_chi_Sk675}.}
    \label{fig:n_chi_Sk68135}
\end{minipage}
\end{figure*}

\begin{figure*}
    \begin{minipage}{0.49\linewidth}
    \centering
    \includegraphics[width=\linewidth]{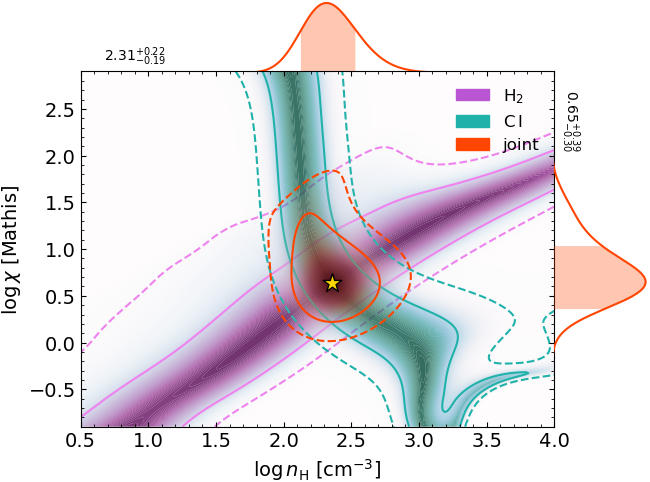}
    \caption{Estimate on the number density and UV field intensity for the system towards Sk-69 246 in the LMC. Lines are the same as for Figure~\ref{fig:n_chi_Sk675}.}
    \label{fig:n_chi_Sk69246}
\end{minipage}
\hfill
\begin{minipage}{0.49\linewidth}
    \centering
    \includegraphics[width=\linewidth]{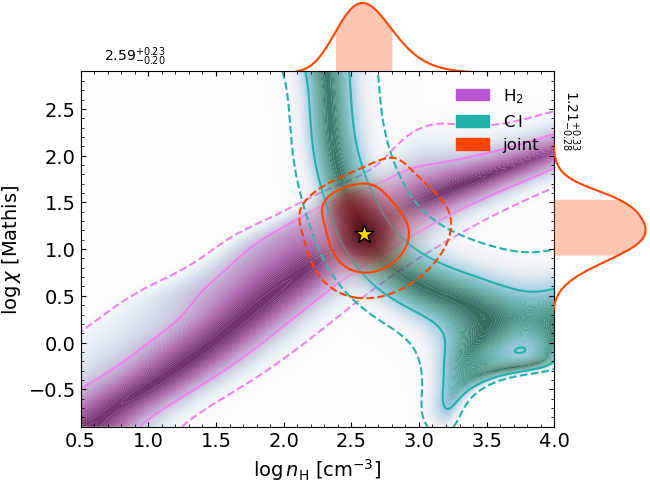}
    \caption{Estimate on the number density and UV field intensity for the system towards Sk-68 140 in the LMC. Lines are the same as for Figure~\ref{fig:n_chi_Sk675}.}
    \label{fig:n_chi_Sk68140}
\end{minipage}
\end{figure*}

\begin{figure*}
    \begin{minipage}{0.49\linewidth}
    \centering
    \includegraphics[width=\linewidth]{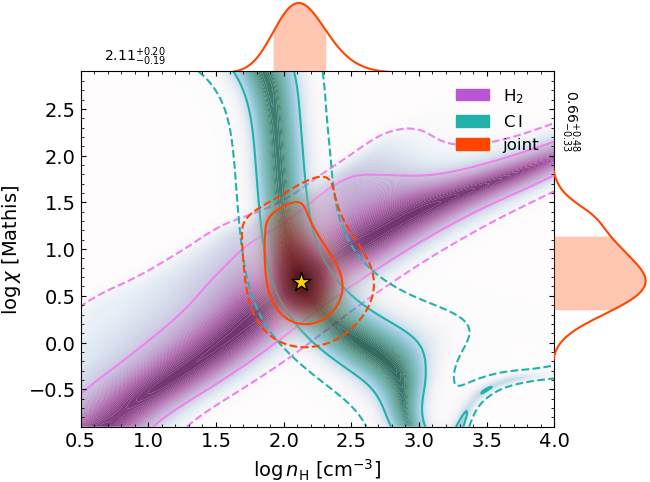}
    \caption{Estimate on the number density and UV field intensity for the system towards Sk-71 50 in the LMC. Lines are the same as for Figure~\ref{fig:n_chi_Sk675}.}
    \label{fig:n_chi_Sk7150}
\end{minipage}
\hfill
\begin{minipage}{0.49\linewidth}
    \centering
    \includegraphics[width=\linewidth]{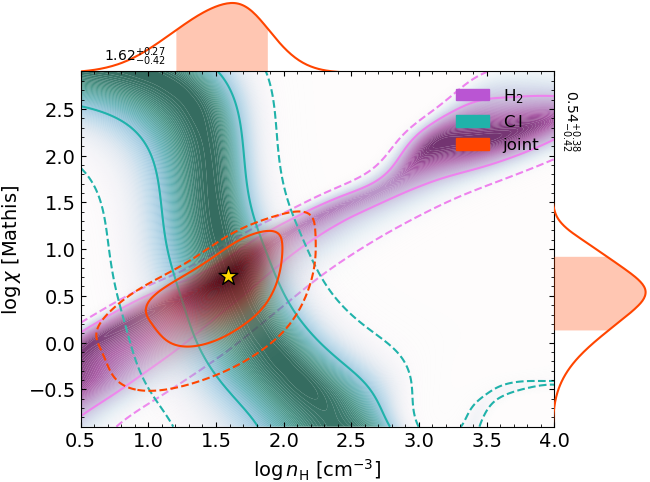}
    \caption{Estimate on the number density and UV field intensity for the system towards Sk-69 279 in the LMC. Lines are the same as for Figure~\ref{fig:n_chi_Sk675}.}
    \label{fig:n_chi_Sk69279}
\end{minipage}
\end{figure*}

\begin{figure*}
    \begin{minipage}{0.49\linewidth}
    \centering
    \includegraphics[width=\linewidth]{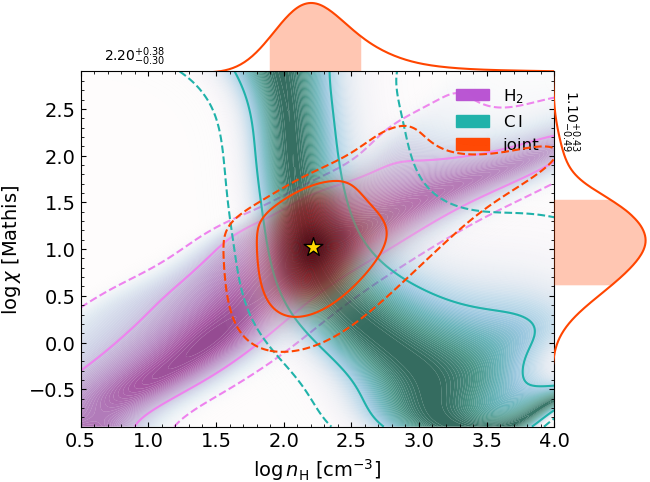}
    \caption{Estimate on the number density and UV field intensity for the system towards Sk-68 155 in the LMC. Lines are the same as for Figure~\ref{fig:n_chi_Sk675}.}
    \label{fig:n_chi_Sk68155}
\end{minipage}
\hfill
\begin{minipage}{0.49\linewidth}
    \centering
    \includegraphics[width=\linewidth]{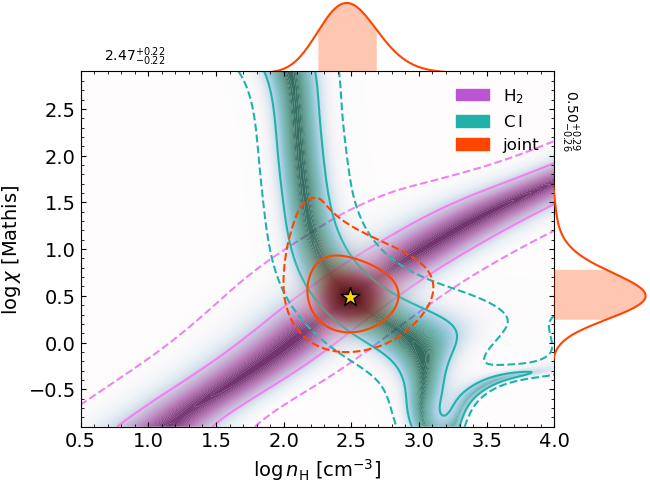}
    \caption{Estimate on the number density and UV field intensity for the system towards Sk-70 115 in the LMC. Lines are the same as for Figure~\ref{fig:n_chi_Sk675}.}
    \label{fig:n_chi_Sk70115}
\end{minipage}
\end{figure*}

\begin{figure*}
    \begin{minipage}{0.49\linewidth}
    \centering
    \includegraphics[width=\linewidth]{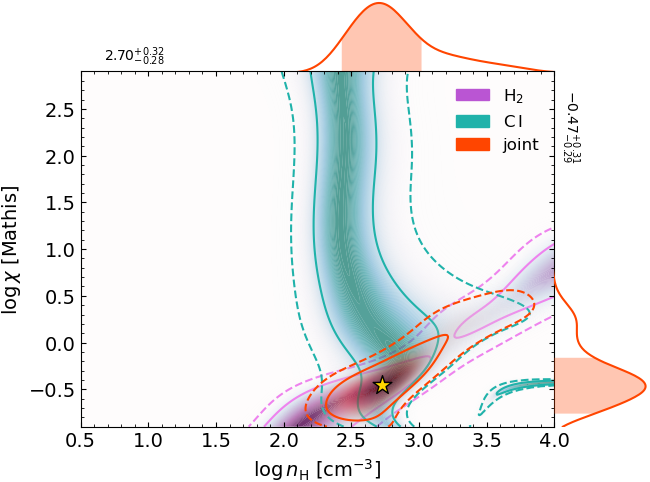}
    \caption{Estimate on the number density and UV field intensity for the system towards AV 15 in the SMC. Lines are the same as for Figure~\ref{fig:n_chi_Sk675}.}
    \label{fig:n_chi_AV15}
\end{minipage}
\hfill
\begin{minipage}{0.49\linewidth}
    \centering
    \includegraphics[width=\linewidth]{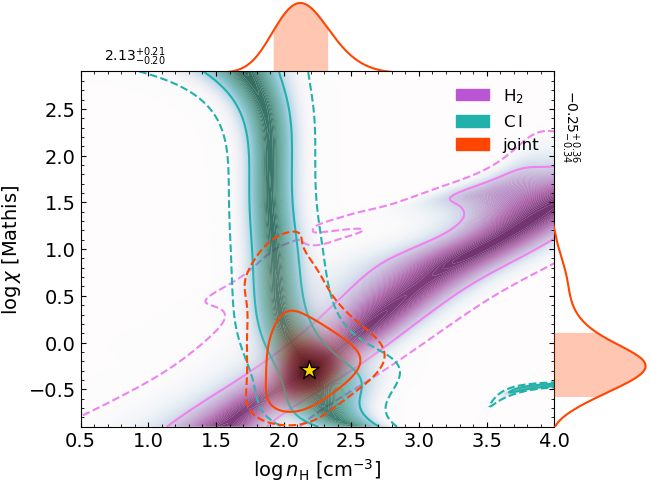}
    \caption{Estimate on the number density and UV field intensity for the system towards AV 26 in the SMC. Lines are the same as for Figure~\ref{fig:n_chi_Sk675}.}
    \label{fig:n_chi_AV26}
\end{minipage}
\end{figure*}

\begin{figure*}
    \begin{minipage}{0.49\linewidth}
    \centering
    \includegraphics[width=\linewidth]{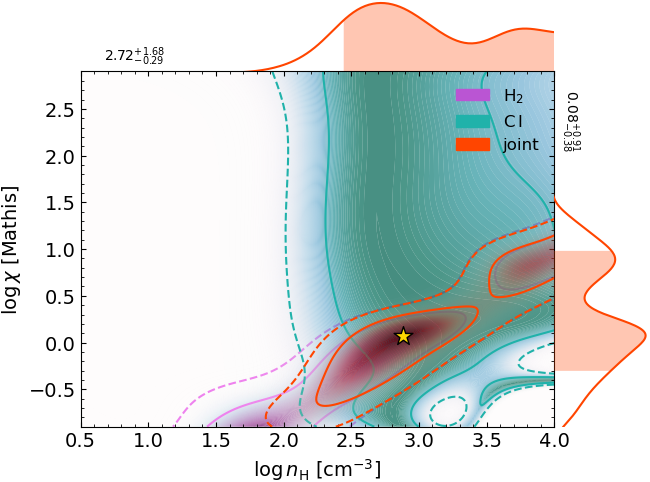}
    \caption{Estimate on the number density and UV field intensity for the system towards AV 47 in the SMC. Lines are the same as for Figure~\ref{fig:n_chi_Sk675}.}
    \label{fig:n_chi_AV47}
\end{minipage}
\hfill
\begin{minipage}{0.49\linewidth}
    \centering
    \includegraphics[width=\linewidth]{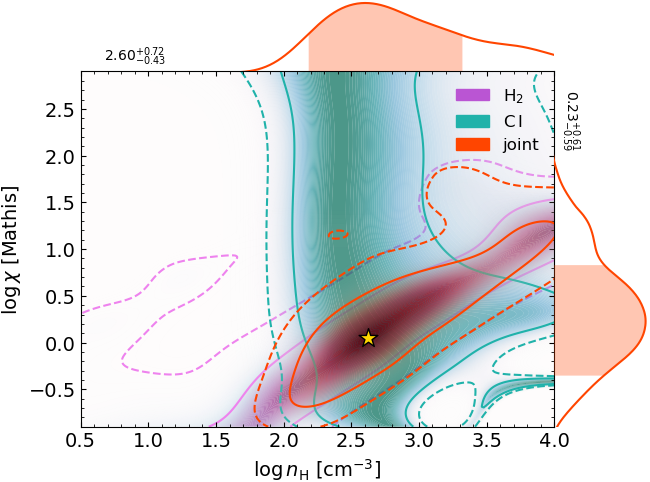}
    \caption{Estimate on the number density and UV field intensity for the system towards AV 69 in the SMC. Lines are the same as for Figure~\ref{fig:n_chi_Sk675}.}
    \label{fig:n_chi_AV69}
\end{minipage}
\end{figure*}

\begin{figure*}
    \begin{minipage}{0.49\linewidth}
    \centering
    \includegraphics[width=\linewidth]{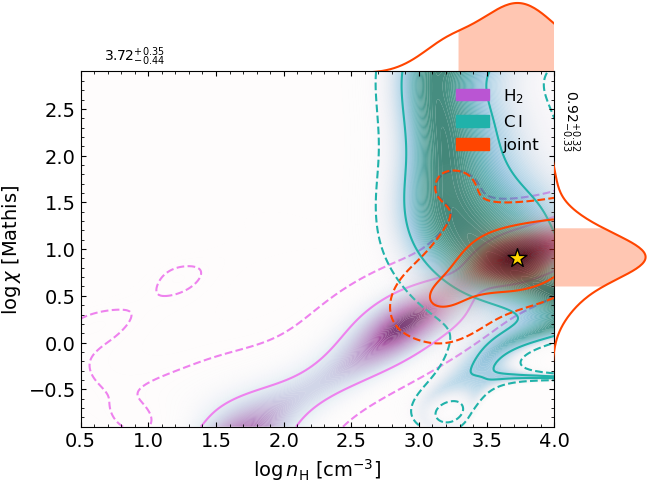}
    \caption{Estimate on the number density and UV field intensity for the system towards AV 75 in the SMC. Lines are the same as for Figure~\ref{fig:n_chi_Sk675}.}
    \label{fig:n_chi_AV75}
\end{minipage}
\hfill
\begin{minipage}{0.49\linewidth}
    \centering
    \includegraphics[width=\linewidth]{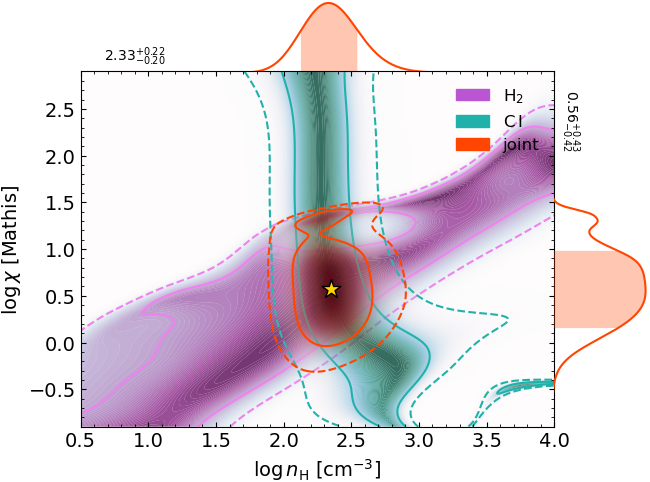}
    \caption{Estimate on the number density and UV field intensity for the system towards AV 80 in the SMC. Lines are the same as for Figure~\ref{fig:n_chi_Sk675}.}
    \label{fig:n_chi_AV80}
\end{minipage}
\end{figure*}

\begin{figure*}
    \begin{minipage}{0.49\linewidth}
    \centering
    \includegraphics[width=\linewidth]{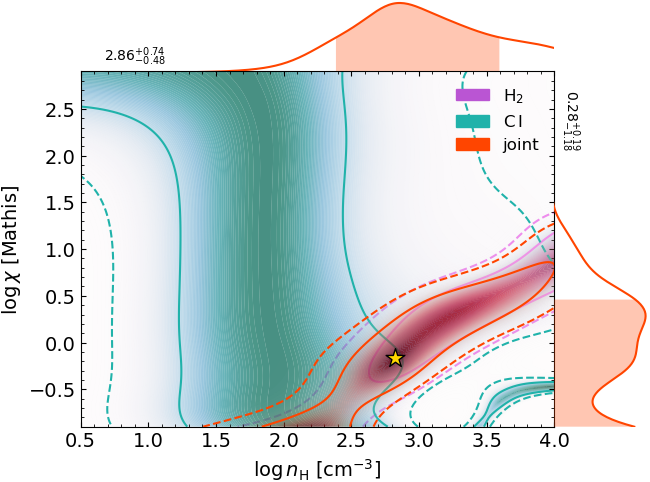}
    \caption{Estimate on the number density and UV field intensity for the system towards AV 81 in the SMC. Lines are the same as for Figure~\ref{fig:n_chi_Sk675}.}
    \label{fig:n_chi_AV81}
\end{minipage}
\hfill
\begin{minipage}{0.49\linewidth}
    \centering
    \includegraphics[width=\linewidth]{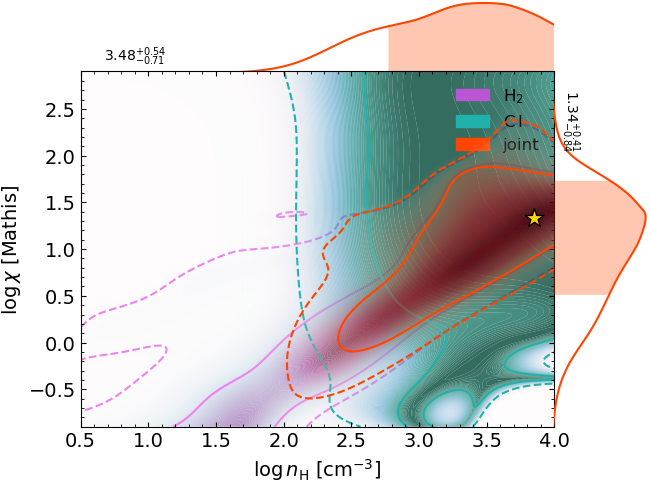}
    \caption{Estimate on the number density and UV field intensity for the system towards AV 207 in the SMC. Lines are the same as for Figure~\ref{fig:n_chi_Sk675}.}
    \label{fig:n_chi_AV207}
\end{minipage}
\end{figure*}

\begin{figure*}
    \begin{minipage}{0.49\linewidth}
    \centering
    \includegraphics[width=\linewidth]{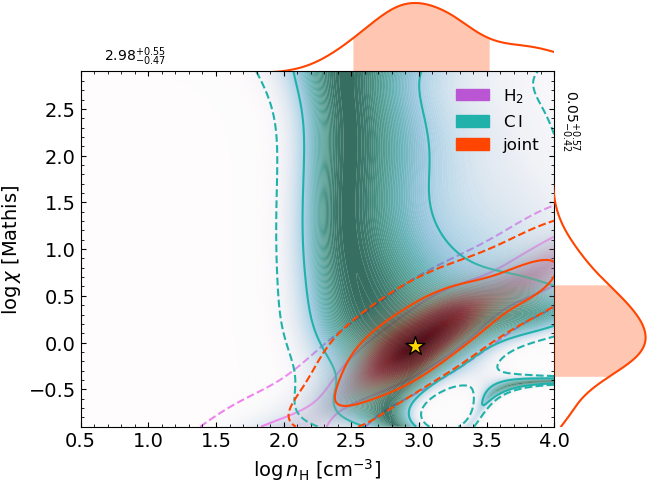}
    \caption{Estimate on the number density and UV field intensity for the system towards AV 210 in the SMC. Lines are the same as for Figure~\ref{fig:n_chi_Sk675}.}
    \label{fig:n_chi_AV210}
\end{minipage}
\hfill
\begin{minipage}{0.49\linewidth}
    \centering
    \includegraphics[width=\linewidth]{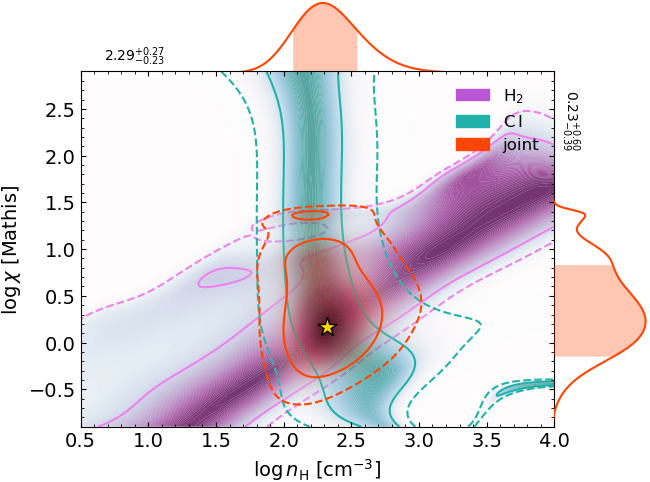}
    \caption{Estimate on the number density and UV field intensity for the system towards AV 215 in the SMC. Lines are the same as for Figure~\ref{fig:n_chi_Sk675}.}
    \label{fig:n_chi_AV215}
\end{minipage}
\end{figure*}

\begin{figure*}
    \begin{minipage}{0.49\linewidth}
    \centering
    \includegraphics[width=\linewidth]{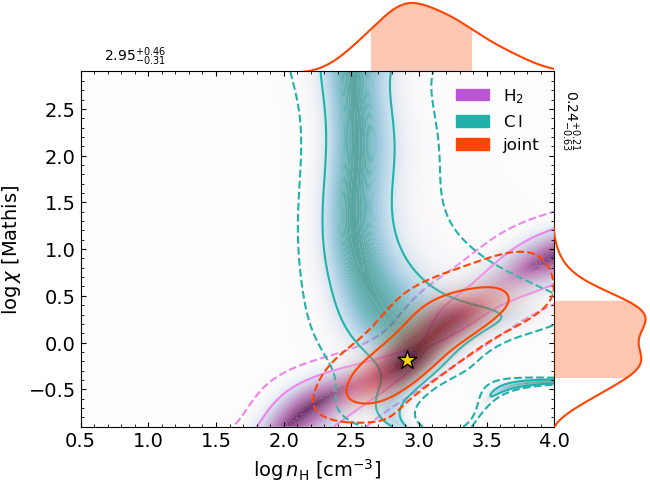}
    \caption{Estimate on the number density and UV field intensity for the system towards AV 216 in the SMC. Lines are the same as for Figure~\ref{fig:n_chi_Sk675}.}
    \label{fig:n_chi_AV216}
\end{minipage}
\hfill
\begin{minipage}{0.49\linewidth}
    \centering
    \includegraphics[width=\linewidth]{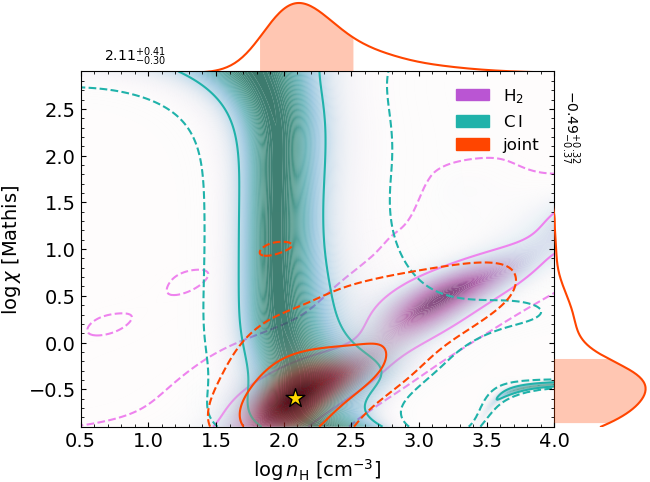}
    \caption{Estimate on the number density and UV field intensity for the system towards AV 266 in the SMC. Lines are the same as for Figure~\ref{fig:n_chi_Sk675}.}
    \label{fig:n_chi_AV266}
\end{minipage}
\end{figure*}

\begin{figure*}
    \begin{minipage}{0.49\linewidth}
    \centering
    \includegraphics[width=\linewidth]{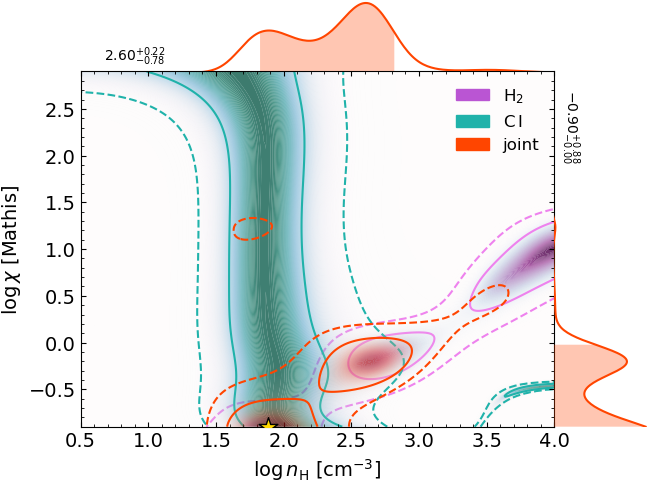}
    \caption{Estimate on the number density and UV field intensity for the system towards AV 372 in the SMC. Lines are the same as for Figure~\ref{fig:n_chi_Sk675}.}
    \label{fig:n_chi_AV372}
\end{minipage}
\hfill
\begin{minipage}{0.49\linewidth}
    \centering
    \includegraphics[width=\linewidth]{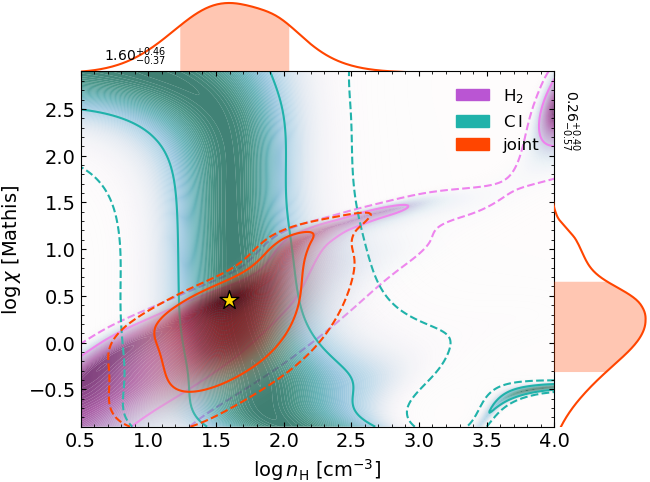}
    \caption{Estimate on the number density and UV field intensity for the system towards AV 476 in the SMC. Lines are the same as for Figure~\ref{fig:n_chi_Sk675}.}
    \label{fig:n_chi_AV476}
\end{minipage}
\end{figure*}

\begin{figure*}
    \begin{minipage}{0.49\linewidth}
    \centering
    \includegraphics[width=\linewidth]{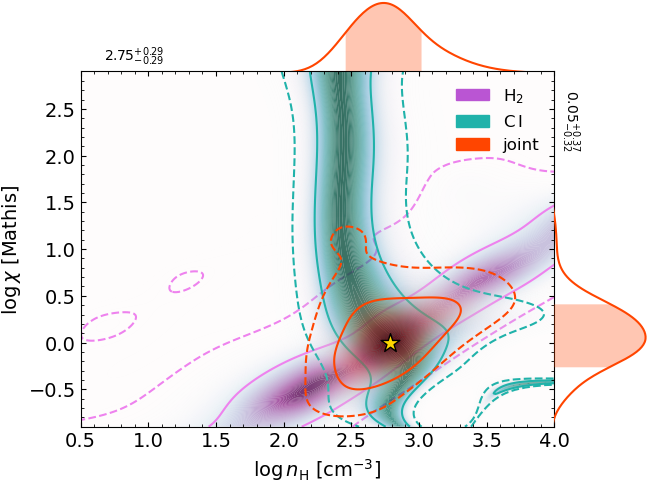}
    \caption{Estimate on the number density and UV field intensity for the system towards AV 479 in the SMC. Lines are the same as for Figure~\ref{fig:n_chi_Sk675}.}
    \label{fig:n_chi_AV479}
\end{minipage}
\hfill
\begin{minipage}{0.49\linewidth}
    \centering
    \includegraphics[width=\linewidth]{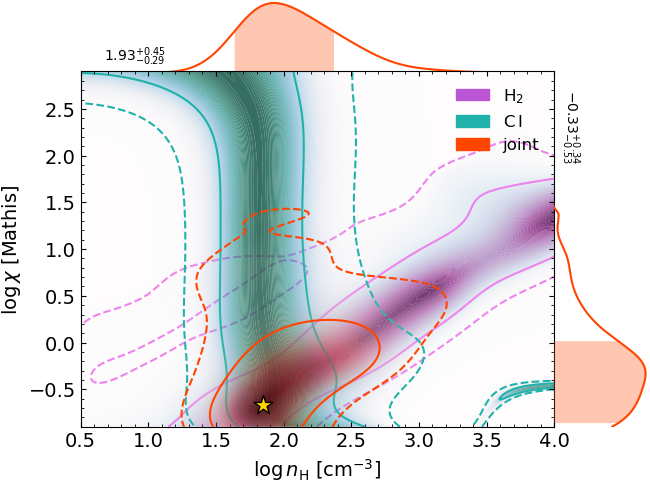}
    \caption{Estimate on the number density and UV field intensity for the system towards AV 488 in the SMC. Lines are the same as for Figure~\ref{fig:n_chi_Sk675}.}
    \label{fig:n_chi_AV488}
\end{minipage}
\end{figure*}

\begin{figure*}
    \begin{minipage}{0.49\linewidth}
    \centering
    \includegraphics[width=\linewidth]{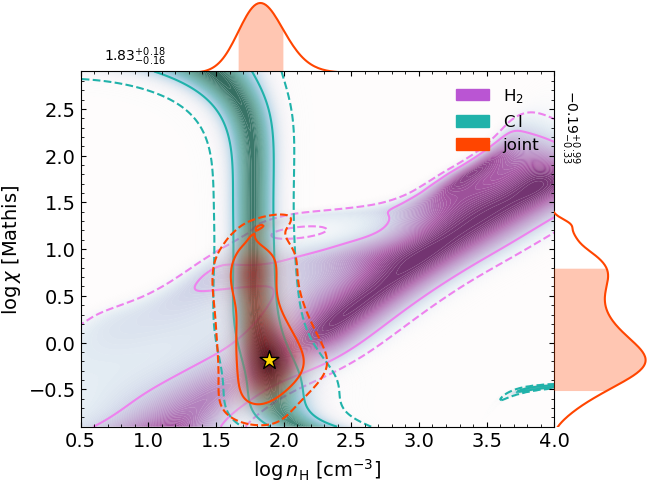}
    \caption{Estimate on the number density and UV field intensity for the system towards AV 15 in the SMC. Lines are the same as for Figure~\ref{fig:n_chi_Sk675}.}
    \label{fig:n_chi_AV490}
\end{minipage}
\hfill
\begin{minipage}{0.49\linewidth}
    \centering
    \includegraphics[width=\linewidth]{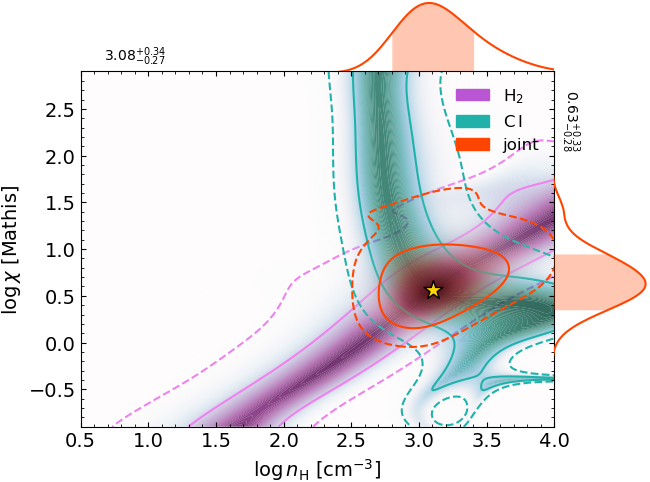}
    \caption{Estimate on the number density and UV field intensity for the system towards Sk 191 in the SMC. Lines are the same as for Figure~\ref{fig:n_chi_Sk675}.}
    \label{fig:n_chi_Sk191}
\end{minipage}
\end{figure*}

%**********************************************

\end{document}